\documentclass[11pt,a4paper]{iopart}

\usepackage[utf8]{inputenc}
\usepackage{url}
\usepackage{graphicx}
\usepackage{iopams}
\usepackage{setstack}
\usepackage{color}
\usepackage{bm}
\usepackage{accents}
\usepackage{hyperref}
\usepackage{ctable}
\usepackage{cite}

\hypersetup{
    colorlinks=true,
    linkcolor=blue,
    filecolor=blue,      
    citecolor=blue,
    urlcolor=blue
}

\usepackage{multirow}

\makeatletter
\long\def\dddddot#1{%
  {\mathop {#1}\limits ^{\vbox to-1.4\ex@ {\kern -\tw@ \ex@ \hbox {\normalfont .....}\vss }}}%
}
\long\def\multidots#1#2{%
  \count@=0
  {{\mathop {#2}\limits ^{\vbox to-1.4\ex@ {\kern -\tw@ \ex@ \hbox {\normalfont %
  \loop%
  \ifnum#1>\count@%
  .%
  \advance\count@ by1%
  \repeat%
  }\vss }}}}%
}
\makeatother

\newcommand{\udt}[3]{#1^{#2}_{\phantom{#2}#3}}
\newcommand{\udut}[4]{#1^{#2\phantom{#3}#4}_{\phantom{#2}#3\phantom{#4}}}

\newcommand{\dut}[3]{#1_{#2}^{\phantom{#2}#3}}
\newcommand{\dudt}[4]{#1_{#2\phantom{#3}#4}^{\phantom{#2}#3}}

\newcommand{\qm}[1]{``#1''}
\newcommand{\lc}[1]{\accentset{\circ}{#1}}

\begin{document}

\title[Constraining Teleparallel Gravity through Gaussian Processes]{Constraining Teleparallel Gravity through Gaussian Processes}

\author{Rebecca Briffa$^{1,2}$, Salvatore Capozziello$^{3,4}$, Jackson Levi Said$^{1,2}$\footnote{Author to whom any correspondence should be addressed.}, Jurgen Mifsud$^{1,2}$ and Emmanuel N. Saridakis$^{5,6,7}$}
\address{$^1$ Institute of Space Sciences and Astronomy, University of Malta, Msida, MSD 2080, Malta.}
\address{$^2$ Department of Physics, University of Malta, Msida, MSD 2080, Malta.}
\address{$^3$ Universit\`{a} degli studi di Napoli \qm{Federico II}, Dipartimento di Fisica \qm{Ettore Pancini}, Complesso Universitario di Monte S. Angelo, Via Cintia Edificio 6, 80126 Napoli, Italy.}
\address{$^4$ Istituto Nazionale di Fisica Nucleare, Sezione di Napoli, Complesso Universitario di Monte S. Angelo, Via Cintia Edificio 6, 80126 Napoli, Italy.}
\address{$^5$ National Observatory of Athens, Lofos Nymfon, 11852 Athens, 
Greece.}
\address{$^6$ CAS Key Laboratory for Researches in Galaxies and Cosmology, Department of Astronomy, University of Science and Technology of China, Hefei, Anhui 230026, P.R. China.}
\address{$^7$ School of Astronomy, School of Physical Sciences, 
University of Science and Technology of China, Hefei 230026, P.R. China.}
\eads{\mailto{rebecca.briffa.16@um.edu.mt}, \mailto{capozzie@na.infn.it}, \mailto{jackson.said@um.edu.mt}, \mailto{jurgen.mifsud@um.edu.mt} and \mailto{msaridak@phys.uoa.gr}}

\date{\today}

\begin{abstract}
We apply Gaussian processes (GP) in order to impose constraints on teleparallel gravity and its $f(T)$ extensions. We use available $H(z)$ observations from (i) cosmic chronometers data (CC);
(ii) Supernova Type Ia (SN) data from the compressed Pantheon release together with the CANDELS and CLASH Multi-Cycle Treasury programs; and (iii) baryonic acoustic oscillation (BAO) datasets from the Sloan Digital Sky Survey. For the involved covariance functions, we consider four widely used choices, namely the square exponential, Cauchy, Mat\'{e}rn and rational quadratic kernels, which are consistent with one another within 1$\sigma$ confidence levels. Specifically, we use the GP approach to reconstruct a model-independent determination of the Hubble constant $H_0$, for each of these kernels and dataset combinations. These analyses are complemented with three recently announced literature values of $H_0$, namely (i) Riess $H_0^{\rm R} = 74.22 \pm 1.82 \,{\rm km\, s}^{-1} {\rm Mpc}^{-1}$; (ii) H0LiCOW Collaboration $H_0^{\rm HW} = 73.3^{+1.7}_{-1.8} \,{\rm km\, s}^{-1} {\rm Mpc}^{-1}$; and (iii) Carnegie-Chicago Hubble Program $H_0^{\rm TRGB} = 69.8 \pm 1.9 \,{\rm km\, s}^{-1} {\rm Mpc}^{-1}$. Additionally, we investigate the transition redshift between the decelerating and accelerating cosmological phases through the GP reconstructed deceleration parameter. Furthermore, we reconstruct the model-independent evolution of the dark energy equation of state, and finally reconstruct the allowed $f(T)$ functions. As a result, the $\Lambda$CDM model lies inside the allowed region at 1$\sigma$ in all the examined kernels and datasets, however a negative slope for $f(T)$ versus $T$ is slightly favored.
\end{abstract}

\maketitle
%
%


\section{Introduction}

The Standard Cosmological Model, namely the $\Lambda$CDM model, is supported by unprecedented overwhelming evidences at all cosmological scales \cite{misner1973gravitation,Clifton:2011jh}, incorporating sectors beyond the standard model of particle physics. For galactic structure, dark matter takes the form of a stabilizing agent \cite{Baudis:2016qwx,Bertone:2004pz} and is realised through some form of cold dark matter. On larger scales, dark energy materialises through the cosmological constant \cite{Peebles:2002gy,Copeland:2006wr} and it is responsible for the observational fact of a late-time accelerating Universe \cite{Riess:1998cb,Perlmutter:1998np}. Despite great efforts, internal consistency issues persist with the cosmological constant \cite{RevModPhys.61.1}, while direct measurements of dark matter is becoming all the more elusive \cite{Gaitskell:2004gd}.

Recently, the predictive power of the $\Lambda$CDM scenario has been called into question, since its effectiveness here has become an open issue within the so-called $H_0$ tension \cite{DiValentino:2020zio}. This encompasses the discrepancy between model-independent measurements of the Hubble parameter at current times \cite{Riess:2019cxk,Wong:2019kwg} and its predicted value given by early Universe observations \cite{Aghanim:2018eyx,Ade:2015xua}, where such tension is appearing to be a growing feature in many measurements \cite{Riess:2020sih,Pesce:2020xfe,deJaeger:2020zpb}. While measurements from the tip of the red giant branch (TRGB, Carnegie-Chicago Hubble Program) point to a lower $H_0$ tension, the precise value of the tension may only be solved by future observations which may be produced by novel techniques such as the LISA mission \cite{Baker:2019nia,2017arXiv170200786A,Barack:2018yly} where gravitational astronomy could be systematically adopted \cite{Graef:2018fzu,Abbott:2017xzu}.

The above theoretical and observational issues might be resolved by theories beyond general relativity (GR) where gravity is modified. Theories beyond GR \cite{DeFelice:2010aj,Capozziello:2011et} are usually described within the curvature framework, using the Levi-Civita connection \cite{misner1973gravitation}. In a wide sense, they can be considered ``extensions" of standard GR because the Einstein theory has to be recovered as a particular case of a wide class of models \cite{Capozziello:2011et}. On the other hand, the body of work where torsion rather than curvature is considered has been drastically increasing in recent years and has produced a number of interesting models \cite{Aldrovandi:2013wha,Cai:2015emx,Krssak:2018ywd}.

Teleparallel gravity (TG) embodies the class of theories where the teleparallel connection is adopted \cite{Weitzenbock1923}, i.e. those theories where torsion is used to describe gravitation. Analogous to the Levi-Civita connection, the teleparallel connection is torsion-full and still respecting the metricity condition. All curvature quantities identically vanish, e.g. the Ricci scalar $\lc{R}$ (over-circles represent quantities calculated with the Levi-Civita connection) will vanish i.e. $\lc{R}=0$. In TG the Einstein-Hilbert Lagrangian can be exchanged by the torsion scalar $T$ to produce the same dynamical equations. This substitution is called the \textit{Teleparallel equivalent of General Relativity} (TEGR), and differs from GR by a boundary term $B$ in its Lagrangian.

In TEGR, the total divergence quantities appearing in GR are encapsulated in the boundary term which have a meaningful impact on the possible extended theories of gravity that can be produced in this framework. A pivotal case where this fact is especially prescient is in Lovelock's theorem \cite{Lovelock:1971yv} which is much broader in TG \cite{Gonzalez:2015sha,Bahamonde:2019shr} and may contain an infinite number of possible contributions. TG also has a number of other attractive features such as its likeness to Yang-Mills theory \cite{Aldrovandi:2013wha} which gives it a resemblance to particle physics theories, as well as the possibility of a definition of the gravitational energy-momentum tensor \cite{Blixt:2018znp,Blixt:2019mkt,Capozziello:2017xla,Capozziello:2018qcp} and that it does not require a Gibbons--Hawking--York boundary term to produce a well-defined Hamiltonian structure, among others.

Following the same rationale as $f(\lc{R})$ gravity \cite{DeFelice:2010aj,Capozziello:2011et}, TEGR can be directly generalised to produce $f(T)$ gravity \cite{Ferraro:2006jd,Ferraro:2008ey,Bengochea:2008gz,Linder:2010py,Chen:2010va,Bahamonde:2019zea,Ualikhanova:2019ygl} (see Ref.\cite{Cai:2015emx} for a review). This is a second-order theory that has shown promise in several key observational tests \cite{Cai:2015emx,Nesseris:2013jea,Farrugia:2016qqe,Finch:2018gkh,Farrugia:2016xcw,Iorio:2012cm,Ruggiero:2015oka,Deng:2018ncg,Yan:2019gbw,LeviSaid:2020mbb,Paliathanasis:2017htk}. $f(T)$ gravity is intrinsically distinct from $f(\lc{R})$ gravity in that to recover the latter one must consider an arbitrary inclusion of not only the torsion scalar but also the boundary term in $f(T,B)$ gravity \cite{Bahamonde:2015zma,Capozziello:2018qcp,Bahamonde:2016grb,Paliathanasis:2017flf,Farrugia:2018gyz,Bahamonde:2016cul,Bahamonde:2016cul,Wright:2016ayu}. In fact, this recovery only occurs for the subset for which the limit $f(T,B) = f(-T+B) = f(\lc{R})$ applies, which instigated a number of interesting observational studies \cite{Farrugia:2020fcu,Capozziello:2019msc,Farrugia:2018gyz,Bahamonde:2015zma,Paliathanasis:2017flf,Bahamonde:2016grb,Bahamonde:2016cul,Bahamonde:2015zma,Escamilla-Rivera:2019ulu,Franco:2020lxx}. Another interesting extension to TEGR is $f(T,T_G)$ gravity where $T_G$ is the teleparallel equivalent of the Gauss-Bonnet term \cite{Kofinas:2014daa,Capozziello:2016eaz,delaCruz-Dombriz:2018nvt,delaCruz-Dombriz:2017lvj}.

Having the above in mind, the central question that arises is on the specific viable modified teleparallel theory that should be considered. Gaussian processes (GP) have the great advantage that they can offer one avenue by which gravitational theories can be reconstructed using observational data, without imposing stringent physical model assumptions \cite{10.5555/1162254}. This has a particularly appealing interest for the development of theories beyond GR since it offers a more structured approach to construct data-driven models. In modified teleparallel gravity this was applied in Ref.\cite{Cai:2019bdh}, where cosmic chronometers and baryonic acoustic oscillation Hubble data \cite{Zhang:2016tto} was used to produce a background reconstruction for the arbitrary $f(T)$ function.

In this paper, we probe $f(T)$ gravity models using Hubble data in conjunction with GP. The cosmological dynamics is introduced in section \ref{sec:TG} for both TEGR and its $f(T)$ gravity generalisation. GP are considered in section \ref{sec:GPH0}, where GP are used to determine the value of $H_0$. In section \ref{sec:GPfT}, we perform a GP analysis for $f(T)$ gravity, from which any preferred deviations from the $\Lambda$CDM model will become evident. Finally, we summarise our core conclusions in section \ref{sec:conclusions}.

\section{\texorpdfstring{$f(T)$}{f(T)} cosmology}
\label{sec:TG}

\subsection{Teleparallel gravity and its \texorpdfstring{$f(T)$}{f(T)} extension}

In TG, the Levi-Civita connection $\lc{\Gamma}_{\mu\nu}^{\sigma}$ (we recall that over-circles denote quantities determined by the Levi-Civita connection) is replaced with the teleparallel connection $\Gamma^{\sigma}_{\mu\nu}$ \cite{Hayashi:1979qx,nakahara2003geometry,ortin2004gravity}. With this change of connection, the Riemann tensor vanishes identically. In such theories, one uses the tetrads $\udt{e}{a}{\mu}$ rather than the metric tensor $g_{\mu\nu}$. These act as a soldering agent between the general manifold (Greek indices) and the local Minkowski space (Latin indices) \cite{Aldrovandi:2013wha}. Thus, tetrads can be used to raise the Minkowski metric to the general metric through the relations
\begin{equation}\label{metric_tetrad_rel}
    g_{\mu\nu} = \udt{e}{a}{\mu}\udt{e}{b}{\nu}\eta_{ab}\,,\hspace{2cm}\eta_{ab} = \dut{e}{a}{\mu}\dut{e}{b}{\nu}g_{\mu\nu}\,,
\end{equation}
where the inverse tetrads $\dut{e}{a}{\mu}$ must also satisfy the orthogonality conditions
\begin{equation}
    \udt{e}{a}{\mu}\dut{e}{b}{\mu} = \delta^a_b\,,\hspace{2cm}\udt{e}{a}{\mu}\dut{e}{a}{\nu} = \delta^{\nu}_{\mu}\,,
\end{equation}
for consistency. The teleparallel connection can then be defined as \cite{Weitzenbock1923}
\begin{equation}
    \Gamma^{\sigma}_{\mu\nu}:= \dut{e}{a}{\mu}\partial_{\mu}\udt{e}{a}{\nu} + \dut{e}{a}{\sigma}\udt{\omega}{a}{b\mu}\udt{e}{b}{\nu}\,,
\end{equation}
where $\udt{\omega}{a}{b\mu}$ denotes the spin connection. In terms of a linear affine connection that is both curvatureless and satisfies metricity, this is the most general realisation \cite{Aldrovandi:2013wha}. The role of the spin connection is to preserve the invariance of the theory under local Lorentz transformations (LLTs) \cite{Krssak:2015oua}. This represents the possible local Lorentz transformations (three boosts and three rotations). Spin connections also appear in GR but are hidden in its internal structure except for rare circumstances such as in its spinor formalism \cite{chandrasekhar1993classical,misner1973gravitation}. The core difference is that, in TG, the spin connection components are entirely inertial and thus there will always exist a frame where they vanish \cite{Krssak:2018ywd}.

For any of the six LLTs $\udt{\Lambda}{a}{b}$, the spin connection components can be written as $\udt{\omega}{a}{b\mu}=\udt{\Lambda}{a}{c}\partial_{\mu}\dut{\Lambda}{b}{c}$ 
\cite{Aldrovandi:2013wha}. For a particular metric ansatz, there will exist an infinite number of possible tetrads that satisfy Eq.(\ref{metric_tetrad_rel}) due to the active role played by LLTs through the spin connection. Thus, it is the combination of a tetrad and its associated spin connection that represent gravitational and inertial fundamental dynamical object in TG. Any action in TG will naturally lead to ten possible independent field equations as in regular GR. However, in TG, we have an additional six potential field equations which must also be satisfied. These represent the six LLTs and are determined by considering the anti-symmetric field equations \cite{Li:2010cg}. The anti-symmetric field equations vanish due to the symmetries of the energy-momentum tensor and offer an avenue to relate the tetrad and spin connection components directly \cite{Krssak:2018ywd}. The gauge in which this occurs, that is where the spin connection vanished, is called the Weitzenb\"{o}ck connection or gauge \cite{Weitzenbock1923}.

The Riemann tensor gives a fundamental measure of curvature in GR. In the framework of TG, this is replaced by the torsion tensor \cite{Cai:2015emx}
\begin{equation}
    \udt{T}{\sigma}{\mu\nu} := 2\Gamma^{\sigma}_{[\mu\nu]}\,,
\end{equation}
where square brackets represent anti-symmetry, and where the torsion tensor is related to the gravitational field strength within the theory. As in GR, there are also two other helpful tensorial quantities in TG. First, the contorsion tensor is defined as the difference between the teleparallel connection and its Levi-Civita counterpart
\begin{equation}
    \udt{K}{\sigma}{\mu\nu} := \Gamma^{\sigma}_{\mu\nu} - \lc{\Gamma}^{\sigma}_{\mu\nu} =\frac{1}{2}\left(\dudt{T}{\mu}{\sigma}{\nu} 
+ \dudt{T}{\nu}{\sigma}{\mu} - \udt{T}{\sigma}{\mu\nu}\right)\,,
\end{equation}
and plays a crucial role in relating TG with its standard gravity analogs. Secondly, in TG, one can define a so-called superpotential
\begin{equation}
    \dut{S}{a}{\mu\nu}:=\frac{1}{2}\left(\udt{K}{\mu\nu}{a} -  
\dut{e}{a}{\nu}\udt{T}{\alpha\mu}{\alpha} + 
\dut{e}{a}{\mu}\udt{T}{\alpha\nu}{\alpha}\right)\,,
\end{equation}
which has been connected to the gauge current representation of the theory \cite{Blixt:2018znp,Blixt:2019mkt} but this questions remains largely open \cite{Aldrovandi:2003pa,Koivisto:2019jra}.

Contracting the torsion tensor with its superpotential directly leads to the torsion scalar
\begin{equation}\label{torsion_scalar}
    T:=\dut{S}{a}{\mu\nu}\udt{T}{a}{\mu\nu}\,,
\end{equation}
which is dependent on the teleparallel connection in an analogous way to the dependence of the Ricci scalar on the Levi-Civita connection. Given the intrinsic use of the contorsion tensor means that the Ricci and torsion scalars can be related to each other, and turn out to differ by a total divergence term \cite{Bahamonde:2015zma,Farrugia:2016qqe}
\begin{equation}
    R=\lc{R} + T -\frac{2}{e}\partial_{\mu}\left(e\udut{T}{\sigma}{\sigma}{\mu}\right) = 0\,,
\end{equation}
where $e=\det\left(\udt{e}{a}{\mu}\right)=\sqrt{-g}$ represents the tetrad determinant, and $R$ is the Ricci scalar as calculated with the teleparallel connection, which vanishes, while $\lc{R}$ is the regular Ricci scalar in standard gravity. Straightforwardly, this means that the Ricci and torsion scalars are equal up to a boundary term
\begin{equation}
    \lc{R} = -T + \frac{2}{e}\partial_{\mu} 
\left(e\udut{T}{\sigma}{\sigma}{\mu}\right):=-T+B\,.
\end{equation}
This point alone guarantees that the Ricci and torsion scalars will produce the same dynamical equations. Thus, TEGR can be defined as the theory in which the Lagrangian is simply the torsion scalar $T$, or where the action is represented by
\begin{equation}
    \mathcal{S}_{\rm TEGR} = -\frac{1}{2\kappa^2}\int {\rm d}^4 x\; eT + 
\int {\rm d}^4 x\; e L_{\rm m}\,,
\end{equation}
where $\kappa^2=8\pi G$ is the gravitational coupling and $L_{\rm m}$ is the matter Lagrangian.

Following the same reasoning as $f(\lc{R})$ gravity \cite{DeFelice:2010aj,Capozziello:2011et}, the TEGR Lagrangian can straightforwardly be elevated to a generalised $f(T)$ gravity framework
\cite{Ferraro:2006jd,Ferraro:2008ey,Bengochea:2008gz,Linder:2010py,Chen:2010va}. In this context, the action will then be given as 
\begin{equation}\label{f_T_Lagrangian}
    \mathcal{S}_{\tilde{f}(T)} =  \frac{1}{2\kappa^2}\int {\rm d}^4 x\; e \tilde{f}(T) + \int {\rm d}^4 x\; e L_{\rm m}\,,
\end{equation}
which produces second order equations of motion. This is only possible to a weakened Lovelock theorem in TG \cite{Lovelock:1971yv,Gonzalez:2015sha,Bahamonde:2019shr}. This can recover $\Lambda$CDM in the limit where $\tilde{f}(T)=-T+\Lambda$. On another note, $f(T)$ gravity also shares a number of interesting properties with GR such as having the same number of associated polarisation modes of its gravitational wave signature \cite{Bamba:2013ooa,Farrugia:2018gyz,Cai:2018rzd,Abedi:2017jqx,Chen:2019ftv}, and being  Gauss-Ostrogradsky ghost free (since it remains second-order) \cite{Krssak:2018ywd,ortin2004gravity}. Performing a variation with respect to the tetrad finally gives the field equations as 
\begin{eqnarray}\label{ft_FEs}
    e^{-1} &\partial_{\nu}\left(e\dut{e}{a}{\rho}\dut{S}{\rho}{\mu\nu}\right)\tilde{f}_T 
-  \dut{e}{a}{\lambda} \udt{T}{\rho}{\nu\lambda}\dut{S}{\rho}{\nu\mu} \tilde{f}_T + 
\frac{1}{4}\dut{e}{a}{\mu}\tilde{f}(T) \nonumber\\
    & + \dut{e}{a}{\rho}\dut{S}{\rho}{\mu\nu}\partial_{\nu}\left(T\right)\tilde{f}_{TT} + \dut{e}{b}{\lambda}\udt{\omega}{b}{a\nu}\dut{S}{\lambda}{\nu\mu}\tilde{f}_T = \kappa^2 \dut{e}{a}{\rho} \dut{\Theta}{\rho}{\mu}\,,
\end{eqnarray}
where subscripts denote derivatives, and $\dut{\Theta}{\rho}{\nu}$ is the regular energy-momentum tensor. In the ensuing work, we consider the flat Friedmann--Lema\^{i}tre--Robertson--Walker (FLRW) cosmology which has been shown to be compatible with the Weitzenb\"{o}ck gauge \cite{Bahamonde:2016grb,Bahamonde:2016cul,Escamilla-Rivera:2019ulu} which we shall be adopting. Moreover, we also choose to consider $f(T)$ as an extension through the transformation
\begin{equation}
    \tilde{f}(T) \rightarrow -T + f(T)\,,
\end{equation}
where $f(T)$ will appear as an extension to TEGR.

\subsection{\texorpdfstring{$f(T)$}{f(T)} cosmology}

The flat homogeneous and isotropic FLRW metric represented by
\begin{equation}\label{FLRW_metric}
     \mathrm{d}s^2=-\mathrm{d}t^2+a^2(t) \left(\mathrm{d}x^2+\mathrm{d}y^2+\mathrm{d}z^2\right)\,,
\end{equation}
can be produced by the tetrad choice
\begin{equation}
    \udt{e}{a}{\mu}={\rm diag}\left(1,\,a(t),\,a(t),\,a(t)\right)\,,
\end{equation}
where $a(t)$ is the scale factor which is compatible with the Weitzenb\"{o}ck gauge, i.e. $\udt{\omega}{a}{b\mu}=0$ in the $f(T)$ context \cite{Krssak:2015oua,Tamanini:2012hg}. Using the torsion scalar definition in Eq.(\ref{torsion_scalar}) immediately gives
\begin{equation}\label{Tor_sca_flrw}
    T=6H^2\,,
\end{equation}
where the boundary term will be $B=6\left(3H^2+\dot{H}\right)$, which straightforwardly gives the expected standard Ricci scalar for the flat FLRW setting, i.e. $\lc{R}=-T+B=6\left(\dot{H}+2H^2\right)$ (note the use of the standard metric convention \cite{Kofinas:2014owa}, rather than the one used in Refs. \cite{Bengochea:2008gz,Linder:2010py,Cai:2015emx} which differs by a sign difference in $T$). Evaluating the field equations in Eq.(\ref{ft_FEs}) results in the Friedmann equations
\begin{eqnarray}
3H^2 &= \kappa^2 \left(\rho_{\rm m}+\rho_{\rm eff}\right)\,,\label{Friedmann_eq}\\
3H^2 &+ 2\dot{H} = -\kappa^2\left(p_{\rm m}+p_{\rm eff}\right)\,,\label{Friedmann_eq2}
\end{eqnarray}
where $\rho_{\rm m}$ and $p_{\rm m}$ represent the energy density and pressure of the matter content respectively, and where $f(T)$ gravity can be interpreted as an effective fluid with components
\begin{eqnarray}
    \rho_{\rm{eff}} &:= \frac{1}{2\kappa^2}\left(2Tf_T - f\right)\,,\\
    p_{\rm{eff}} &:= -\frac{1}{\kappa^2}\left[2\dot{H}\left(f_T+2Tf_{TT}\right)\right] - \rho_{\rm{eff}}\,.
\end{eqnarray}
The effective fluid coincidentally also satisfies the standard conservation equation
\begin{equation}
    \dot{\rho}_{{\rm eff}} + 3H\left(\rho_{{\rm eff}}+p_{{\rm eff}}\right) = 0\,,
\end{equation}
and can be used to define an effective equation of state (EoS) giving \cite{Bahamonde:2016cul,Escamilla-Rivera:2019ulu}
\begin{eqnarray}\label{EoS_func}
    \omega_{\rm{eff}}  := \frac{p_{\rm{eff}}}{\rho_{\rm{eff}}} = -1 +\left(1+\omega_m\right)\frac{\left(T+f-2Tf_T\right)\left(f_T+2Tf_{TT}\right)}{\left(-1+f_T+2Tf_{TT}\right)\left(-f+2Tf_T\right)}\,.
\end{eqnarray}
An interesting point to highlight is that the $\Lambda$CDM scenario is recovered when $f(T)=\Lambda$.

\begin{table}[t]
\setlength{\tabcolsep}{2.5pt}
\caption{\label{tab:se_kernel} Different GP reconstructions of $H_0$ with the square exponential kernel function of Eq.(\ref{eq:square_exp}). The reconstructed values of $H_0$ are complemented by their distance (in units of $\sigma$) from literature priors.}

\centering
\begin{tabular}{@{}c c c c c c}
    \br
    {\small Data set(s)} & {\small $H_0$} & {\small $d(H_0,H_0^{\rm R})$} & {\small $d(H_0,H_0^{\rm TRGB})$} & {\small $d(H_0,H_0^{\rm HW})$} & {\small $d(H_0,H_0^{\rm P18})$} \\
    \mr
    {\footnotesize CC} & $67.539 \pm 4.7720$ & -1.3037 & -0.4408 & -1.1334 & 0.0290 \\
    {\footnotesize CC+SN} & $67.001 \pm 1.6531$ & -3.2253 & -1.1183 & -2.6165 & -0.2309 \\
    {\footnotesize CC+SN+BAO} & $66.197 \pm 1.4639$ & -3.8407 & -1.5127 & -3.1132 & -0.7776 \\
    \mr
    {\footnotesize CC+$H_0^{\rm R}$} & $73.782 \pm 1.3743$ & -0.12556 & 1.7106 & 0.2166 & 4.3640 \\
    {\footnotesize CC+SN+$H_0^{\rm R}$} & $72.022 \pm 1.0756$ & -1.1271 & 1.0265 & -0.6220 & 3.8969 \\
    {\footnotesize CC+SN+BAO+$H_0^{\rm R}$} & $71.180 \pm 1.0245$ & -1.6279 & 0.6447 & -1.0457 & 3.3155 \\
    \mr
    {\footnotesize CC+$H_0^{\rm TRGB}$} & $69.604 \pm 1.7557$ & -1.9599 & -0.0760 & -1.4908 & 1.2076 \\
    {\footnotesize CC+SN+$H_0^{\rm TRGB}$} & $68.468 \pm 1.2212$ & -2.9695 & -0.5942 & -2.2641 & 0.8096 \\
    {\footnotesize CC+SN+BAO+$H_0^{\rm TRGB}$} & $67.811 \pm 1.1470$ & -3.4070 & -0.9036 & -2.6233 & 0.3284 \\
    \mr
    {\footnotesize CC+$H_0^{\rm HW}$} & $72.966 \pm 1.6636$ & -0.4863 & 1.2617 & -0.1382 & 3.2043 \\
    {\footnotesize CC+SN+$H_0^{\rm HW}$} & $70.850 \pm 1.1991$ & -1.7111 & 0.4710 & -1.1550 & 2.6555 \\
    {\footnotesize CC+SN+BAO+$H_0^{\rm HW}$} & $69.911 \pm 1.1276$ & -2.2717 & 0.0506 & -1.6280 & 2.0355 \\
    \br
\end{tabular}
\end{table}

\section{Calculating \texorpdfstring{$H_0$}{H0} with Gaussian processes}
\label{sec:GPH0}

GP generalise the concept of a Gaussian distribution for a finite set of data points to a continuous distribution over a specified range for a function \cite{10.5555/971143,10.5555/1162254}. Thus, given a set of Gaussian distributed data points, GP provide an iterative process by which to produce the most likely underlying continuous function that describes the data together with its associated confidence bounds without assuming a prescribed form of the function \cite{Seikel:2013fda}. In this section, we briefly review the GP approach of reconstructing these underlying functions and apply this to Hubble parameter data (see, for instance, Refs. \cite{Shafieloo:2012ht,Seikel:2013fda,Cai:2015zoa,Cai:2015pia,Wang:2017jdm,Zhou:2019gda,Cai:2019bdh,Mukherjee:2020vkx,Gomez-Valent:2018hwc,Zhang:2018gjb,Aljaf:2020eqh,Li:2019nux,Liao:2019qoc} for a number of GP applications in cosmology).

\subsection{Gaussian Processes \label{sec:GP}}

As previously mentioned, GP extend the idea of a Gaussian distribution by defining a mean function $\mu(z)$ together with a two-point covariance function $\mathcal{C}(z,z')$ to form a GP as a continuous curve
\begin{equation}
	\xi(z)\sim\mathcal{GP}\left(\mu(z),\mathcal{C}(z,z')\right)\,,
\end{equation}
together with its associated error regions $\Delta\xi(z)$, thus resulting in $\xi(z) \pm \Delta \xi(z)$. Without losing generality, the mean can be set to zero for all points in the reconstruction, since each point is not very sensitive to this. For the redshifts $z^*$ of the reconstruction at which we do not have data points, we can define a kernel function for the covariance such that $\mathcal{C}\left(z^*,z^{*'}\right) = \mathcal{K}\left(z^*,z^{*'}\right)$, which will form the majority of points. The kernel will embody all the information about the strength of the correlations between these reconstructed values as well as the amplitude of the deviations from the mean \cite{Gomez-Valent:2018hwc}. The only generic property about this is that it must be a symmetric function. On the other hand, for observational data points $\tilde{z}$ we have available the associated errors and covariance matrix $\mathcal{D}\left(\tilde{z},\tilde{z}'\right)$ between the points, so that the covariance can be written as $\mathcal{C}\left(\tilde{z},\tilde{z}'\right) = \mathcal{K}\left(\tilde{z},\tilde{z}'\right) + \mathcal{D}\left(\tilde{z},\tilde{z}'\right)$ which will give information about the kernel. Lastly, observational points and reconstructed points will be correlated by the kernel alone through $\mathcal{C}\left(z^*,\tilde{z}'\right) = \mathcal{K}\left(z^*,\tilde{z}'\right)$ \cite{Busti:2014aoa}.

\begin{table}[t]
\setlength{\tabcolsep}{2.5pt}
\caption{\label{tab:c_kernel} Different GP reconstructions of $H_0$ with the Cauchy kernel function of Eq.(\ref{eq:cauchy}). The reconstructed values of $H_0$ are complemented by their distance (in units of $\sigma$) from literature priors.}
\centering
\begin{tabular}{@{}c c c c c c}
    \br
    {\small Data set(s)} & {\small $H_0$} & {\small $d(H_0,H_0^{\rm R})$} & {\small $d(H_0,H_0^{\rm TRGB})$} & {\small $d(H_0,H_0^{\rm HW})$} & {\small $d(H_0,H_0^{\rm P18})$} \\
    \mr
    {\footnotesize CC} & $69.396 \pm 5.1862$ & -0.8618 & -0.0732 & -0.7132 & 0.3831 \\
    {\footnotesize CC+SN} & $67.082 \pm 1.6819$ & -3.1566 & -1.0780 & -2.5619 & -0.1814 \\
    {\footnotesize CC+SN+BAO} & $66.179 \pm 1.4717$ & -3.8392 & -1.5173 & -3.1144 & -0.7858 \\
    \mr
    {\footnotesize CC+$H_0^{\rm R}$} & $73.802 \pm 1.3757$ & -0.1152 & 1.7187 & 0.2256 & 4.3738 \\
    {\footnotesize CC+SN+$H_0^{\rm R}$} & $72.056 \pm 1.0826$ & -1.1055 & 1.0404 & -0.6045 & 3.9046 \\
    {\footnotesize CC+SN+BAO+$H_0^{\rm R}$} & $71.166 \pm 1.0279$ & -1.6340 & 0.6377 & -1.0516 & 3.2943 \\
    \mr
    {\footnotesize CC+$H_0^{\rm TRGB}$} & $69.695 \pm 1.7603$ & -1.9168 & -0.0408 & -1.4524 & 1.2541 \\
    {\footnotesize CC+SN+$H_0^{\rm TRGB}$} & $68.508 \pm 1.2327$ & -2.9366 & -0.5749 & -2.2386 & 0.8330 \\
    {\footnotesize CC+SN+BAO+$H_0^{\rm TRGB}$} & $67.796 \pm 1.1512$ & -3.4101 & -0.9094 & -2.6275 & 0.3156 \\
    \mr
    {\footnotesize CC+$H_0^{\rm HW}$} & $73.003 \pm 1.6665$ & -0.4690 & 1.2755 & -0.1228 & 3.2205 \\
    {\footnotesize CC+SN+$H_0^{\rm HW}$} & $70.892 \pm 1.2087$ & -1.6830 & 0.4887 & -1.1323 & 2.6695 \\
    {\footnotesize CC+SN+BAO+$H_0^{\rm HW}$} & $69.895 \pm 1.1323$ & -2.2767 & 0.0434 & -1.6335 & 2.0160 \\
    \br
\end{tabular}
\end{table}

There exists a number of kernel function choices \cite{10.5555/1162254} where some behave slightly better in certain situations which remains an open discussion in the literature \cite{Seikel:2013fda}. In this work, we consider the effect of choosing a number of kernel functions, namely the general purpose square exponential
\begin{equation}\label{eq:square_exp}
	\mathcal{K}\left(z,\tilde{z}\right) = \sigma_f^2 \exp\left[-\frac{\left(z-\tilde{z}\right)^2}{2l_f^2}\right]\,,
\end{equation}
together with the Cauchy, Mat\'{e}rn and rational quadratic kernels which are respectively given by
\begin{eqnarray}
    \mathcal{K}\left(z,\tilde{z}\right) = &\sigma_f^2 \left[\frac{l_f}{\left(z-\tilde{z}\right)^2 + l_f^2}\right]\,,\label{eq:cauchy}\\
    \mathcal{K}\left(z,\tilde{z}\right) = &\sigma_f^2 \Bigg(1 + \frac{\sqrt{3}|z-\tilde{z}|}{l_f}\Bigg)\exp\left[-\frac{\sqrt{3}|z-\tilde{z}|}{l_f}\right]\,,\label{eq:Matern}\\
    \mathcal{K}\left(z,\tilde{z}\right) = &\sigma_f^2 \left[1 + \frac{(z-\tilde{z})^2}{2\alpha l_f^2}\right]^{-\alpha},\label{eq:rat_quad}
\end{eqnarray}
where $\sigma_f$, $l_f$ and $\alpha$ are the kernel hyperparameters which relate the strength and scope of the correlations between the reconstructed data points respectively \cite{2012JCAP...06..036S}. Here, the hyperparameters appear as constants and are called hyperparameters since their values point to the behaviour of the underlying function rather than a model that mimics this behaviour. The observational data being input into a GP then appears as a subset of Gaussian points that can be extended through the GP approach. This is achieved by maximizing the likelihood of the GP producing the observational data by estimating the underlying functional behavior for different values of the hyperparameters for particular kernel instances.

\begin{table}[t]
\setlength{\tabcolsep}{2.5pt}
\caption{\label{tab:m_kernel} Different GP reconstructions of $H_0$ with the Mat\'{e}rn kernel function of Eq.(\ref{eq:Matern}). The reconstructed values of $H_0$ are complemented by their distance (in units of $\sigma$) from literature priors.}
\centering
\begin{tabular}{@{}c c c c c c}
    \br
    {\small Data set(s)} & {\small $H_0$} & {\small $d(H_0,H_0^{\rm R})$} & {\small $d(H_0,H_0^{\rm TRGB})$} & {\small $d(H_0,H_0^{\rm HW})$} & {\small $d(H_0,H_0^{\rm P18})$} \\
    \mr
    {\footnotesize CC} & $68.434 \pm 5.0295$ & -1.0709 & -0.2545 & -0.9139 & 0.2045 \\
    {\footnotesize CC+SN} & $66.981 \pm 1.6798$ & -3.2049 & -1.1187 & -2.6052 & -0.2393 \\
    {\footnotesize CC+SN+BAO} & $66.139 \pm 1.4729$ & -3.8572 & -1.5337 & -3.1310 & -0.8110 \\
    \mr
    {\footnotesize CC+$H_0^{\rm R}$} & $73.777 \pm 1.3760$ & -0.1279 & 1.7078 & 0.2143 & 4.3560 \\
    {\footnotesize CC+SN+$H_0^{\rm R}$} & $72.016 \pm 1.0821$ & -1.1279 & 1.0223 & -0.6239 & 3.8728 \\
    {\footnotesize CC+SN+BAO+$H_0^{\rm R}$} & $71.148 \pm 1.0286$ & -1.6438 & 0.6292 & -1.0603 & 3.2767 \\
    \mr
    {\footnotesize CC+$H_0^{\rm TRGB}$} & $69.629 \pm 1.7597$ & -1.9462 & -0.0663 & -1.4791 & 1.2186 \\
    {\footnotesize CC+SN+$H_0^{\rm TRGB}$} & $68.457 \pm 1.2322$ & -2.964 & -0.5975 & -2.2626 & 0.7952 \\
    {\footnotesize CC+SN+BAO+$H_0^{\rm TRGB}$} & $67.772 \pm 1.1521$ & -3.4226 & -0.9203 & -2.6386 & 0.2959 \\
    \mr
    {\footnotesize CC+$H_0^{\rm HW}$} & $72.963 \pm 1.6668$ & -0.4875 & 1.2592 & -0.1396 & 3.1966 \\
    {\footnotesize CC+SN+$H_0^{\rm HW}$} & $70.840 \pm 1.2081$ & -1.7108 & 0.4657 & -1.1567 & 2.6312 \\
    {\footnotesize CC+SN+BAO+$H_0^{\rm HW}$} & $69.872 \pm 1.1328$ & -2.2888 & 0.0330 & -1.6442 & 1.9967 \\
    \br
\end{tabular}
\end{table}

Finally, the GP approach is model-independent in the context of a physical model and instead assumes a particular statistical kernel which dictates the correlation between the reconstructed points. In our analysis, we show that for Hubble data, this dependence is very weak and is almost totally model-independent for most intents and purposes, i.e. differing kernels produce the same results to a reasonable extent. In this sense, the GP approach is non-parametric in terms of physical models. In the following, we apply this approach to the case of Hubble data for several data sets.

\begin{table}[t]
\setlength{\tabcolsep}{2.5pt}
\caption{\label{tab:rq_kernel} Different GP reconstructions of $H_0$ with the rational quadratic kernel function of Eq.(\ref{eq:rat_quad}). The reconstructed values of $H_0$ are complemented by their distance (in units of $\sigma$) from literature priors.}
\centering
\begin{tabular}{@{}c c c c c c}
    \br
    {\small Data set(s)} & {\small $H_0$} & {\small $d(H_0,H_0^{\rm R})$} & {\small $d(H_0,H_0^{\rm TRGB})$} & {\small $d(H_0,H_0^{\rm HW})$} & {\small $d(H_0,H_0^{\rm P18})$} \\
    \mr
    {\footnotesize CC} & $70.672 \pm 5.4918$ & -0.5920 & 0.1502 & -0.4560 & 0.5933 \\
    {\footnotesize CC+SN} & $67.099 \pm 1.6865$ & -3.1438 & -1.0699 & -2.5515 & -0.1712 \\
    {\footnotesize CC+SN+BAO} & $66.195 \pm 1.4652$ & -3.8398 & -1.5129 & -3.1129 & -0.7782 \\
    \mr
    {\footnotesize CC+$H_0^{\rm R}$} & $73.851 \pm 1.3767$ & -0.0907 & 1.7391 & 0.2473 & 4.4042 \\
    {\footnotesize CC+SN+$H_0^{\rm R}$} & $72.081 \pm 1.0852$ & -1.0904 & 1.0514 & -0.5918 & 3.9179 \\
    {\footnotesize CC+SN+BAO+$H_0^{\rm R}$} & $71.620 \pm 1.0851$ & -1.3486 & 0.8388 & -0.8160 & 3.5320 \\
    \mr
    {\footnotesize CC+$H_0^{\rm TRGB}$} & $69.782 \pm 1.7636$ & -1.8761 & -0.0070 & -1.4159 & 1.2994 \\
    {\footnotesize CC+SN+$H_0^{\rm TRGB}$} & $68.523 \pm 1.2349$ & -2.9263 & -0.5679 & -2.2303 & 0.8430 \\
    {\footnotesize CC+SN+BAO+$H_0^{\rm TRGB}$} & $67.810 \pm 1.1469$ & -3.4074 & -0.9039 & -2.6237 & 0.3280 \\
    \mr
    {\footnotesize CC+$H_0^{\rm HW}$} & $73.077 \pm 1.6685$ & -0.4349 & 1.3042 & -0.0921 & 3.2593 \\
    {\footnotesize CC+SN+$H_0^{\rm HW}$} & $70.917 \pm 1.2118$ & -1.6674 & 0.4997 & -1.1194 & 2.6831 \\
    {\footnotesize CC+SN+BAO+$H_0^{\rm HW}$} & $69.911 \pm 1.1280$ & -2.2716 & 0.0504 & -1.6280 & 2.0348 \\
    \br
\end{tabular}
\end{table}

\subsection{Reconstruction of Hubble data}

We now apply the GP approach with the kernels specified in Eqs.(\ref{eq:cauchy})--(\ref{eq:rat_quad}) to a number of different $H(z)$ data sources, from which we reconstruct $H_0$. We do this using three principal sources of $H(z)$ data, namely cosmic chronometers (CC), supernovae of Type Ia (SN) and baryonic acoustic oscillation (BAO). CC are very efficient in obtaining $H(z)$ data at redshifts of $z \lesssim 2$ and do not rely on any cosmological models. They also avoid using Cepheids as distance scale indicators and are instead built on spectroscopic dating techniques. In this work, we only adopt those points in Ref.\cite{Moresco:2016mzx} that are independent of BAO observations in order to retain independence of cosmological models. 

For the SN data, we use a combination of the compressed Pantheon compilation \cite{Scolnic:2017caz} together with the CANDELS and CLASH Multi-cycle Treasury data \cite{Riess:2017lxs}. We use the Hubble rate parameter measurements of $E(z)=H(z)/H_0$ along with the corresponding correlation matrix, where only five of the reported six data points are adopted since the $z=1.5$ data point is not Gaussian--distributed (similar to Ref.\cite{Gomez-Valent:2018hwc}). In order to incorporate the SN data set in our GP analyses, we make use of an iterative numerical procedure \cite{Gomez-Valent:2018hwc} to determine an $H_0$ value. We first infer an $H_0$ value by applying GP to the CC data set only, and then we promote the SN $E(z)$ data points to the corresponding $H(z)=H_0E(z)$ values via a Monte Carlo routine. A number of successive GP reconstructions are applied on the combined CC + SN data set, until the resulting value of $H_0$ and its uncertainty converge to $\lesssim10^{-4}$.

We also include BAO data points from the Sloan Digital Sky Survey as reported in Refs.\cite{Alam:2016hwk,Bourboux:2017cbm,Zhao:2018gvb}, along with the corresponding correlation matrices. We follow a similar iterative technique for the inclusion of the BAO data set as the one adopted for the SN measurements. While BAO measurements are not entirely independent from $\Lambda$CDM, particularly due to the assumption of a fiducial radius of the sound horizon $r_d=147.78\,\mathrm{Mpc}$, within this context, they add perspective to the growing tension in the value of $H_0$. A final important point is that whenever a cosmological model is used, it is always assumed to be flat which at these redshifts would have a very small impact in any case. However, the latest Planck 2018 (P18) results report a spatial curvature density which is very small at $\Omega_k(z=0) = 0.001 \pm 0.002$ \cite{Aghanim:2018eyx}.

\begin{figure}[t]
\begin{center}
    \includegraphics[width=0.48\columnwidth]{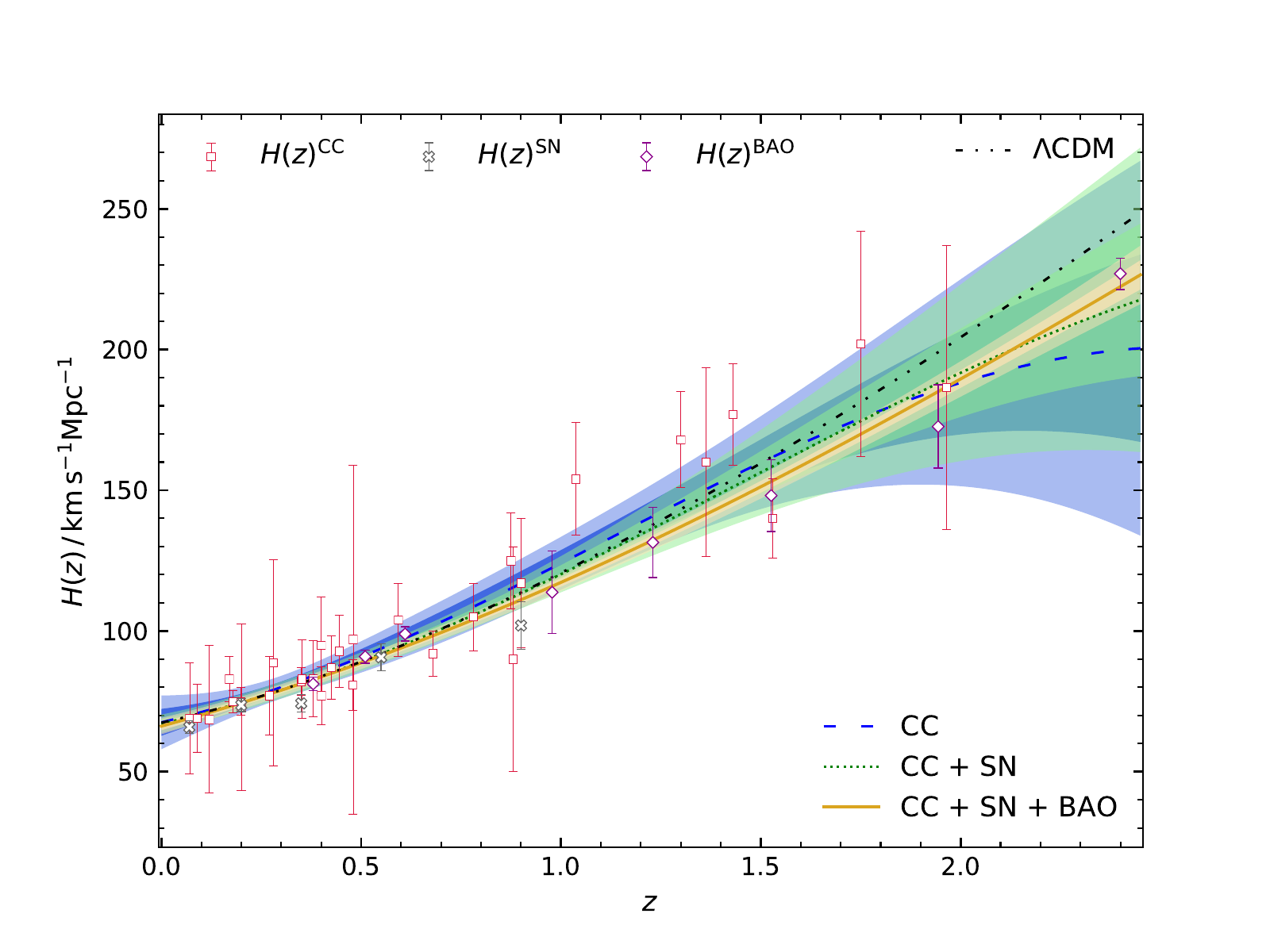}
    \includegraphics[width=0.48\columnwidth]{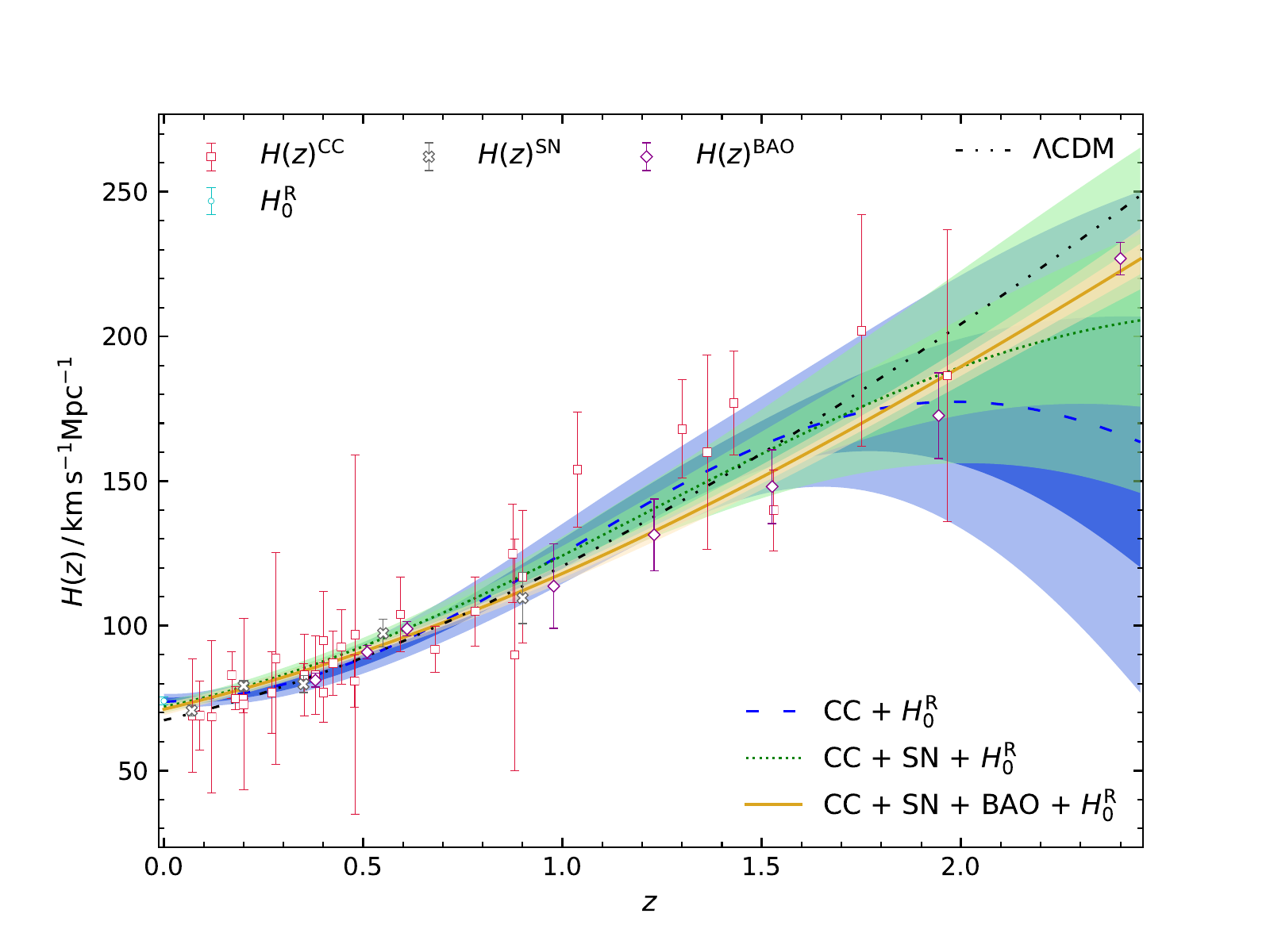}
    \includegraphics[width=0.48\columnwidth]{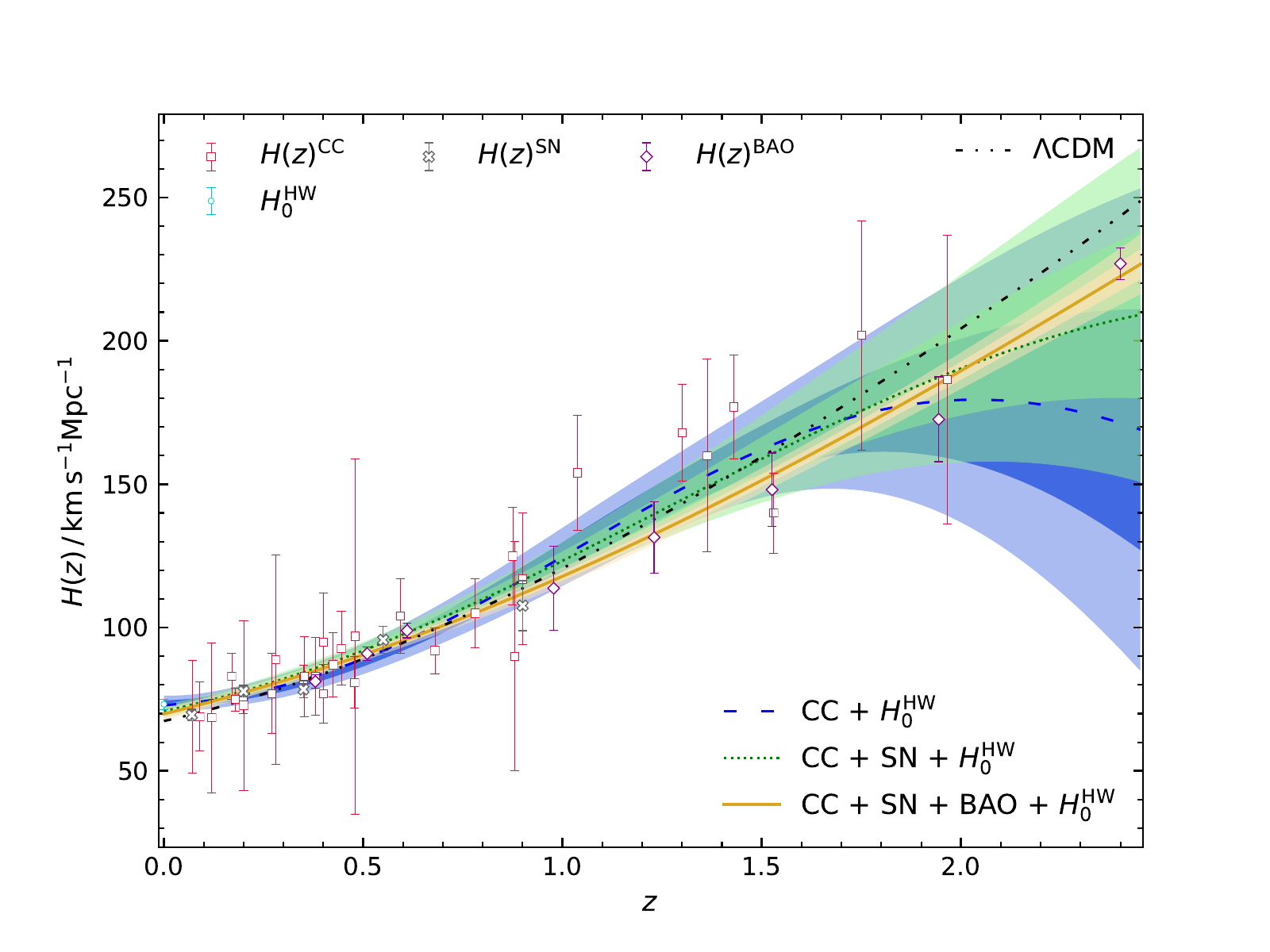}
    \includegraphics[width=0.475\columnwidth]{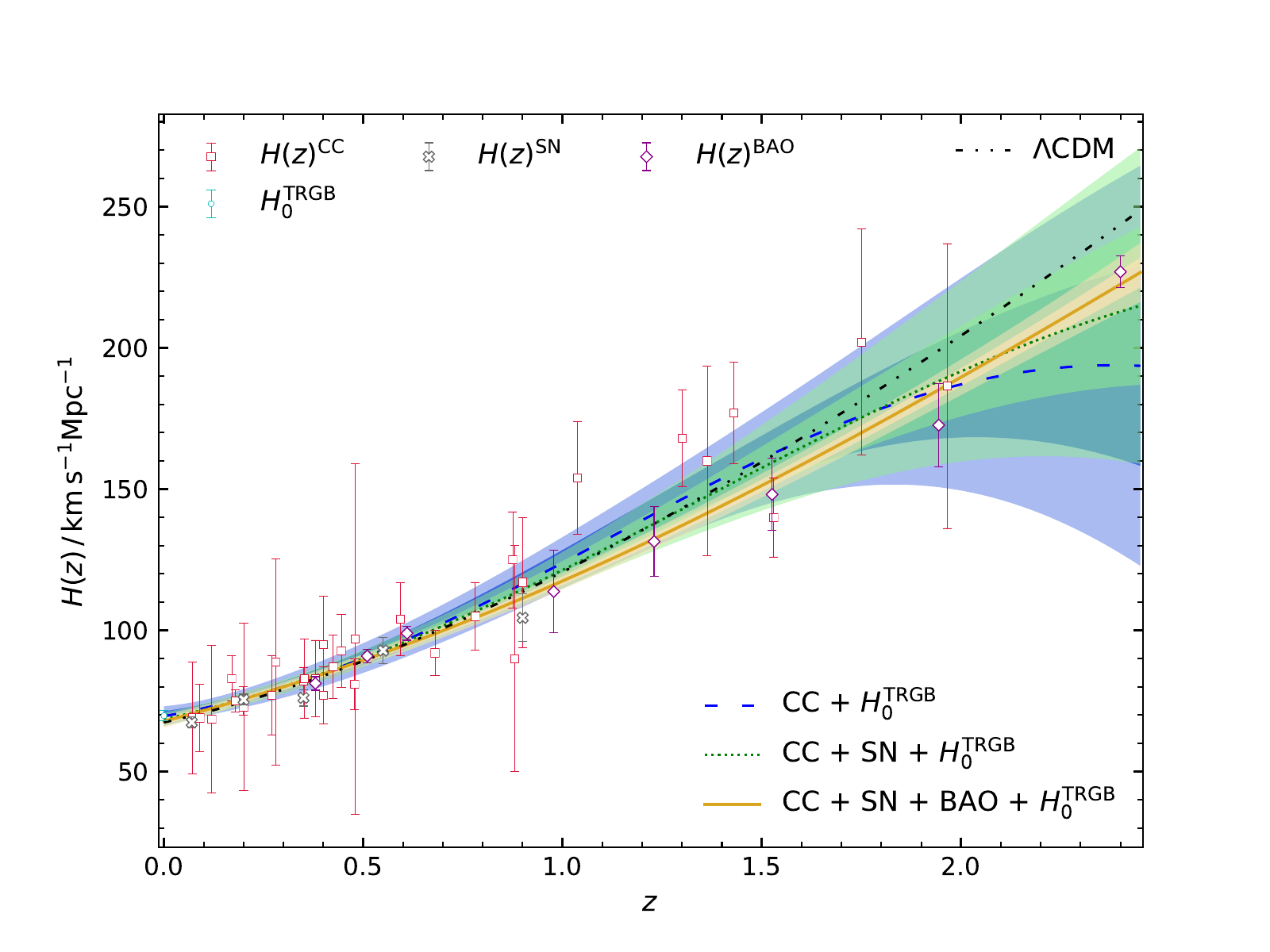}
    \caption{\label{fig:H_squaredexp}
    GP reconstructions of $H(z)$ with the squared exponential kernel function of Eq.(\ref{eq:square_exp}). The data sets along with the different $H_0^{}$ priors are indicated in each respective panel.
    }
\end{center}
\end{figure}

\begin{figure}[t]
\begin{center}
    \includegraphics[width=0.48\columnwidth]{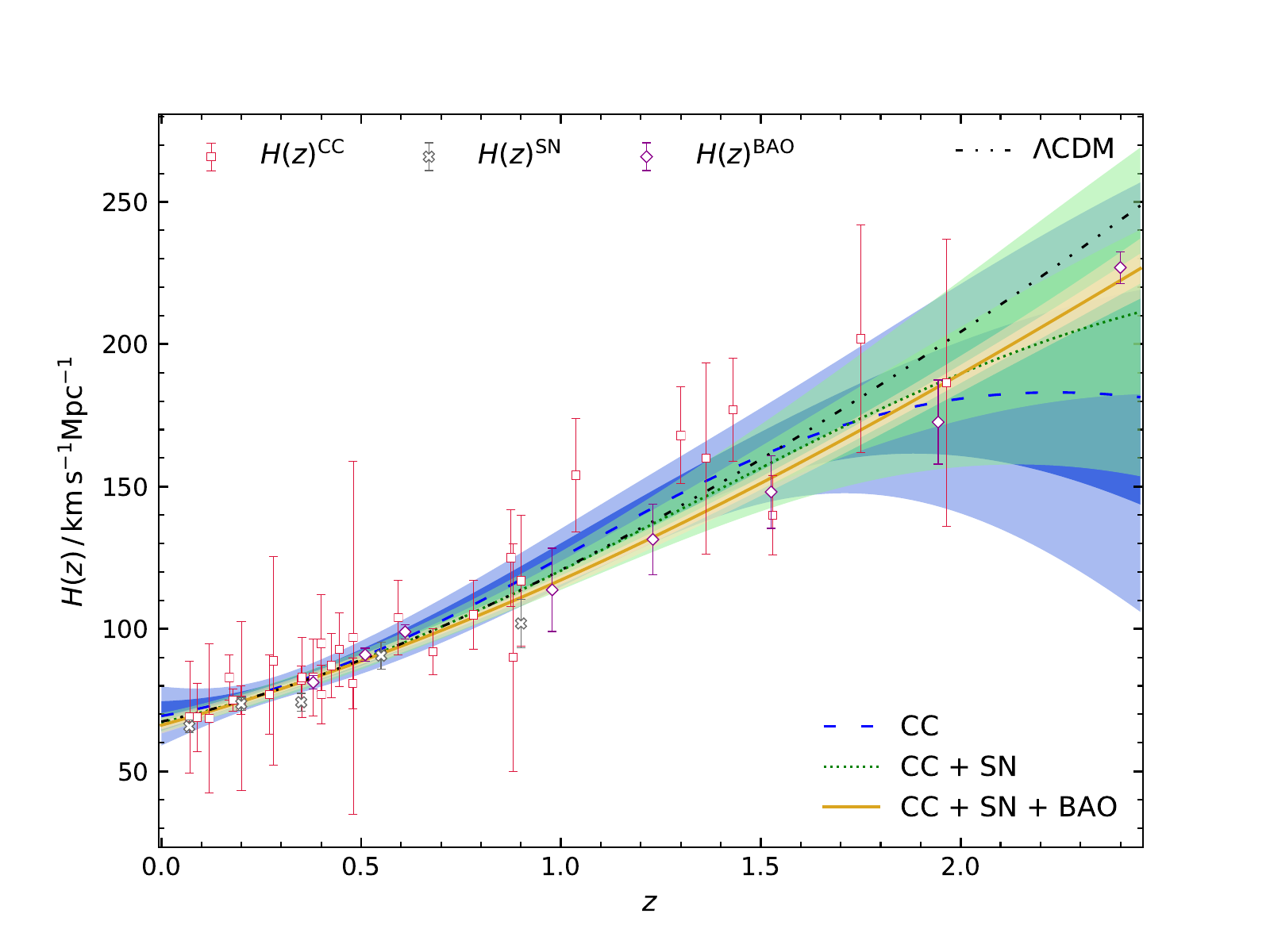}
    \includegraphics[width=0.48\columnwidth]{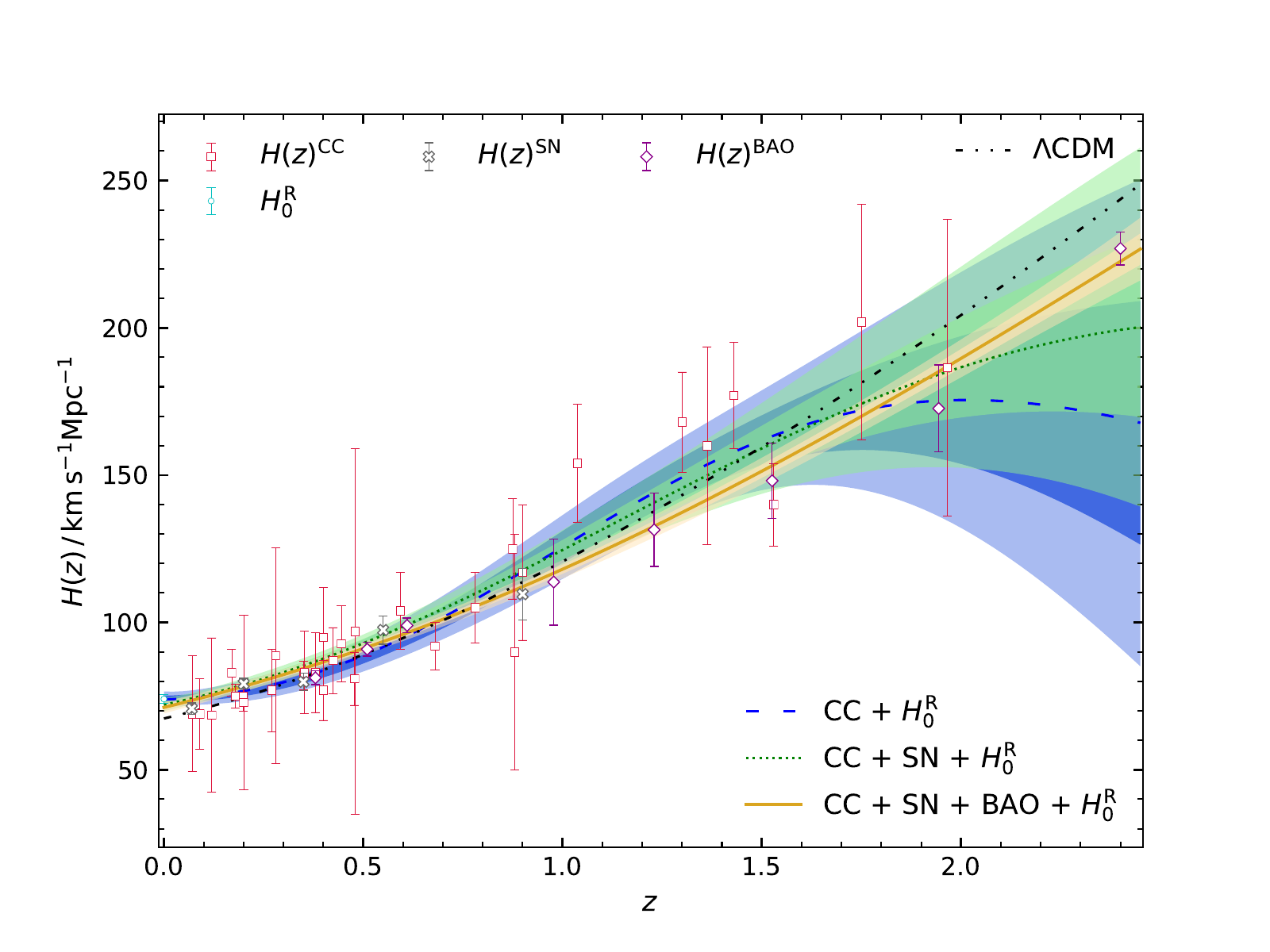}
    \includegraphics[width=0.48\columnwidth]{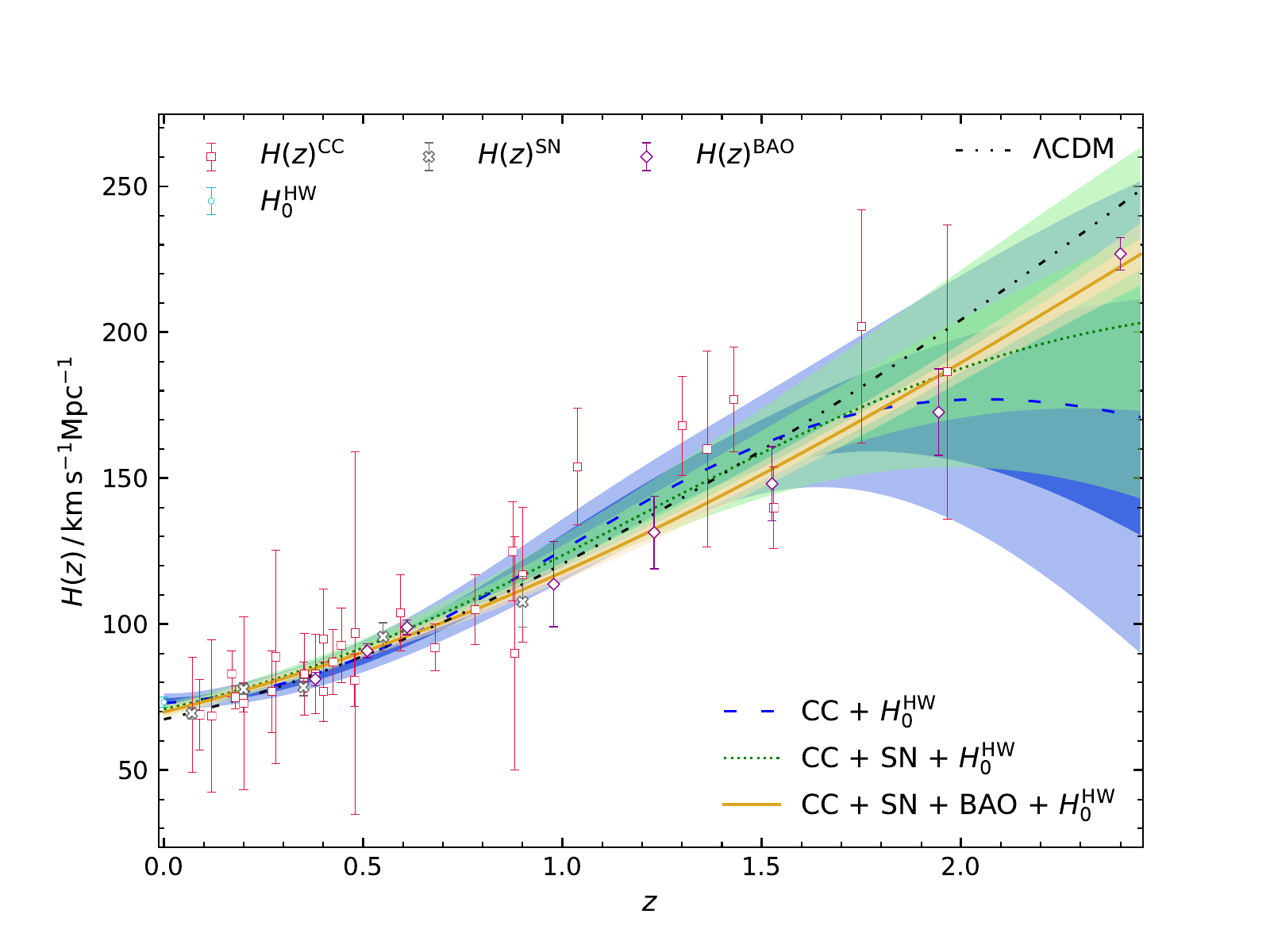}
    \includegraphics[width=0.48\columnwidth]{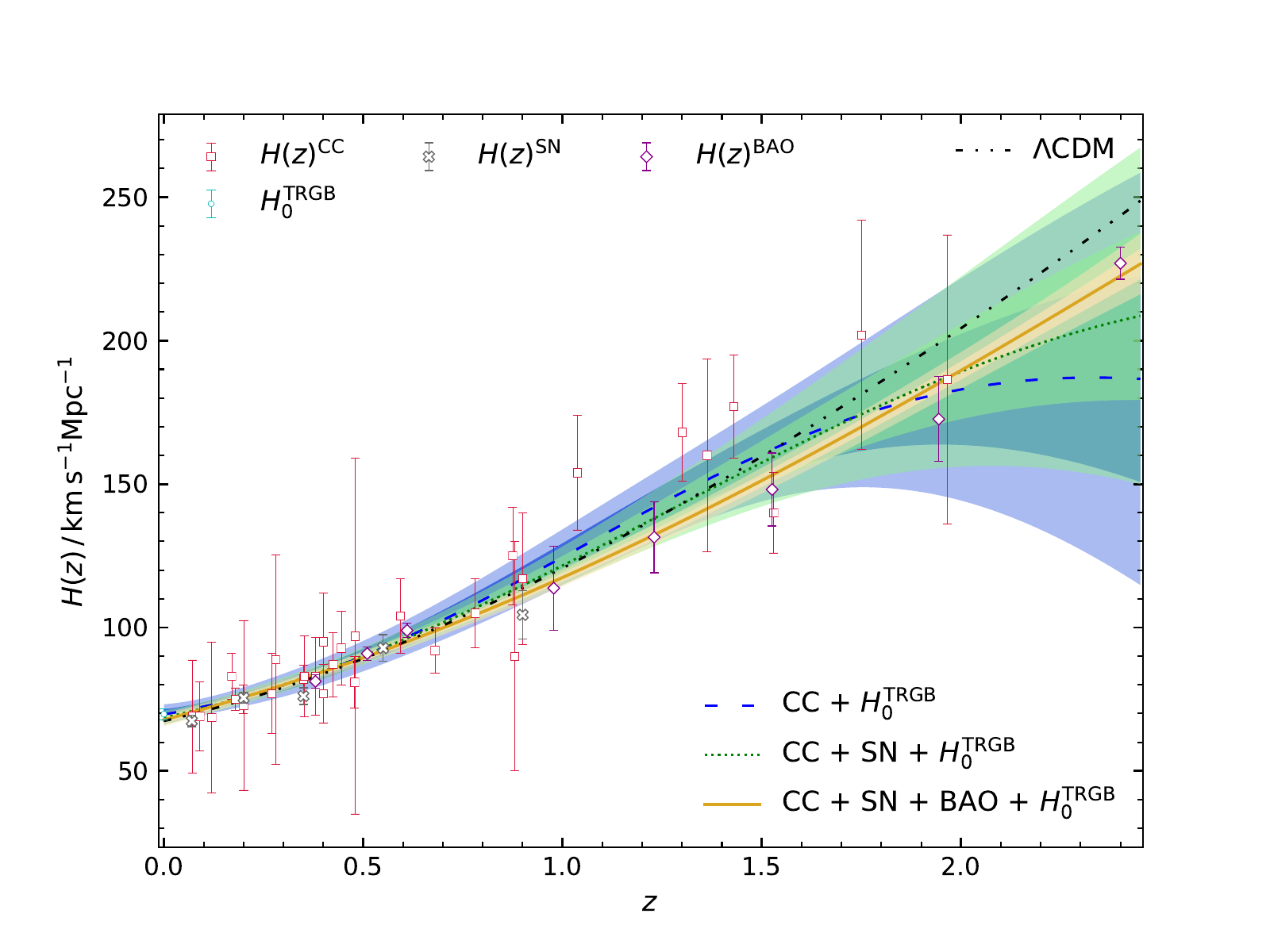}
    \caption{\label{fig:H_cauchy}
    GP reconstructions of $H(z)$ with the Cauchy kernel function of Eq.(\ref{eq:cauchy}). The data sets along with the different $H_0^{}$ priors are indicated in each respective panel.
    }
\end{center}
\end{figure}

\begin{figure}[t]
\begin{center}
    \includegraphics[width=0.48\columnwidth]{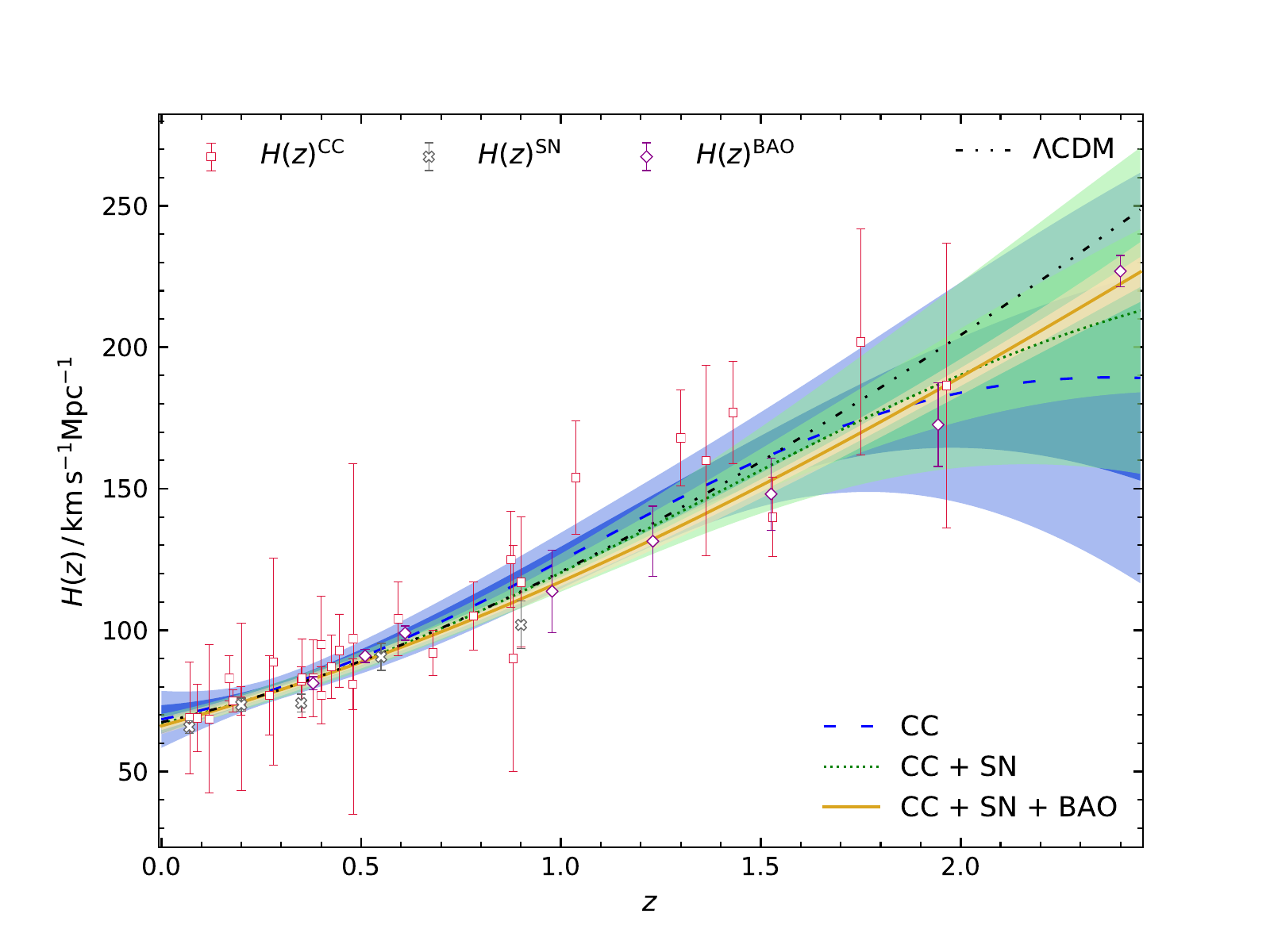}
    \includegraphics[width=0.48\columnwidth]{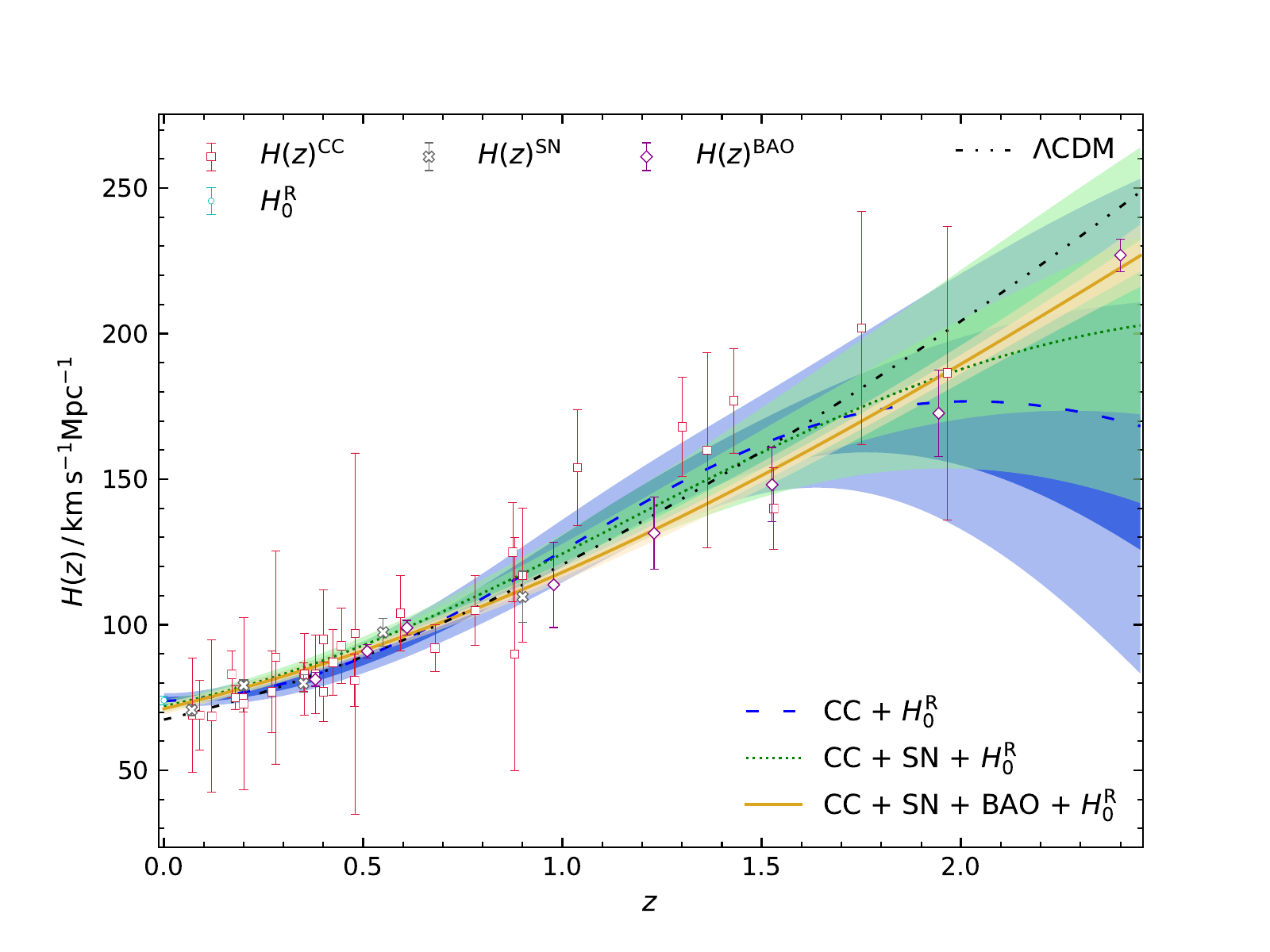}
    \includegraphics[width=0.48\columnwidth]{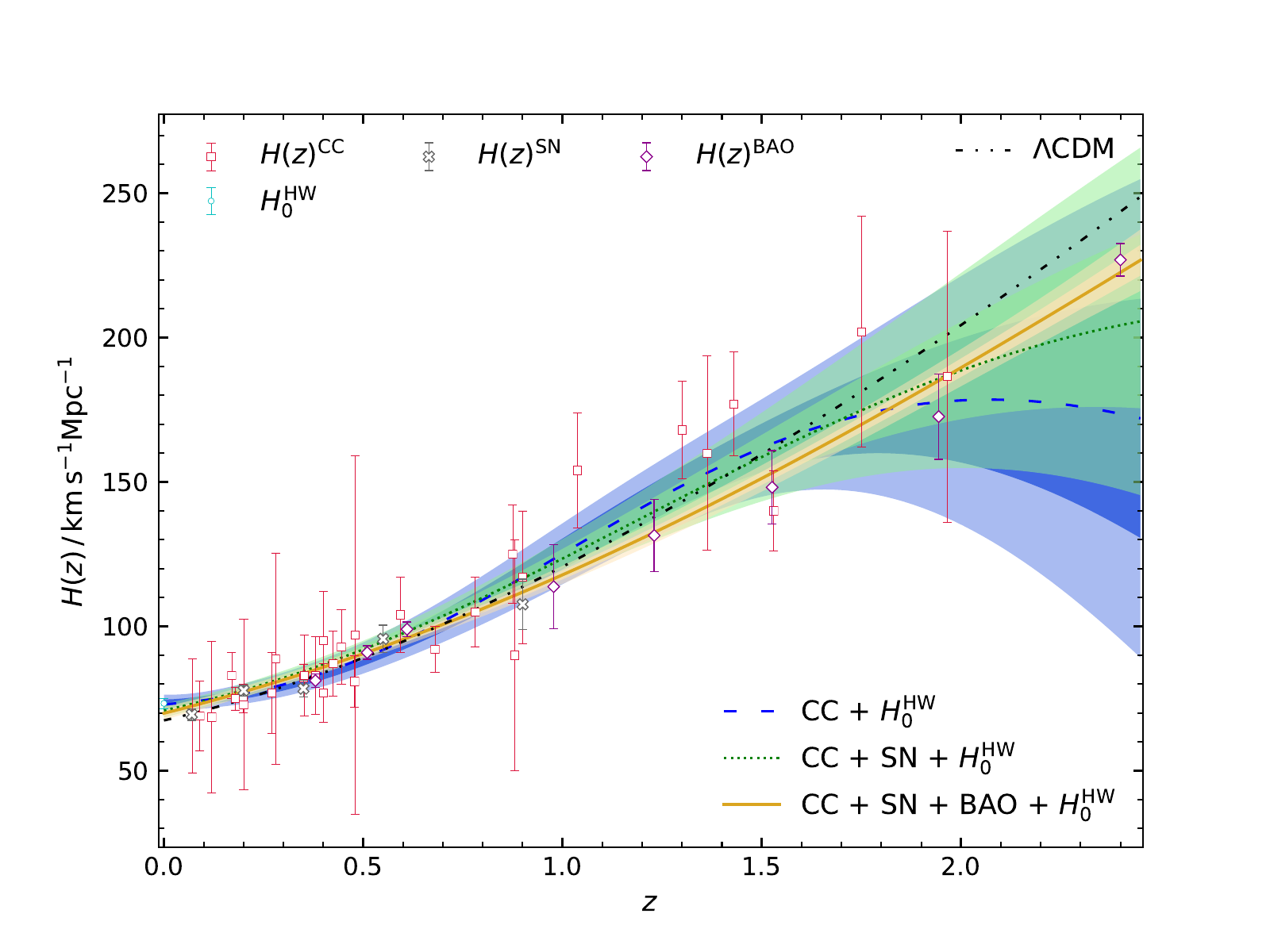}
    \includegraphics[width=0.48\columnwidth]{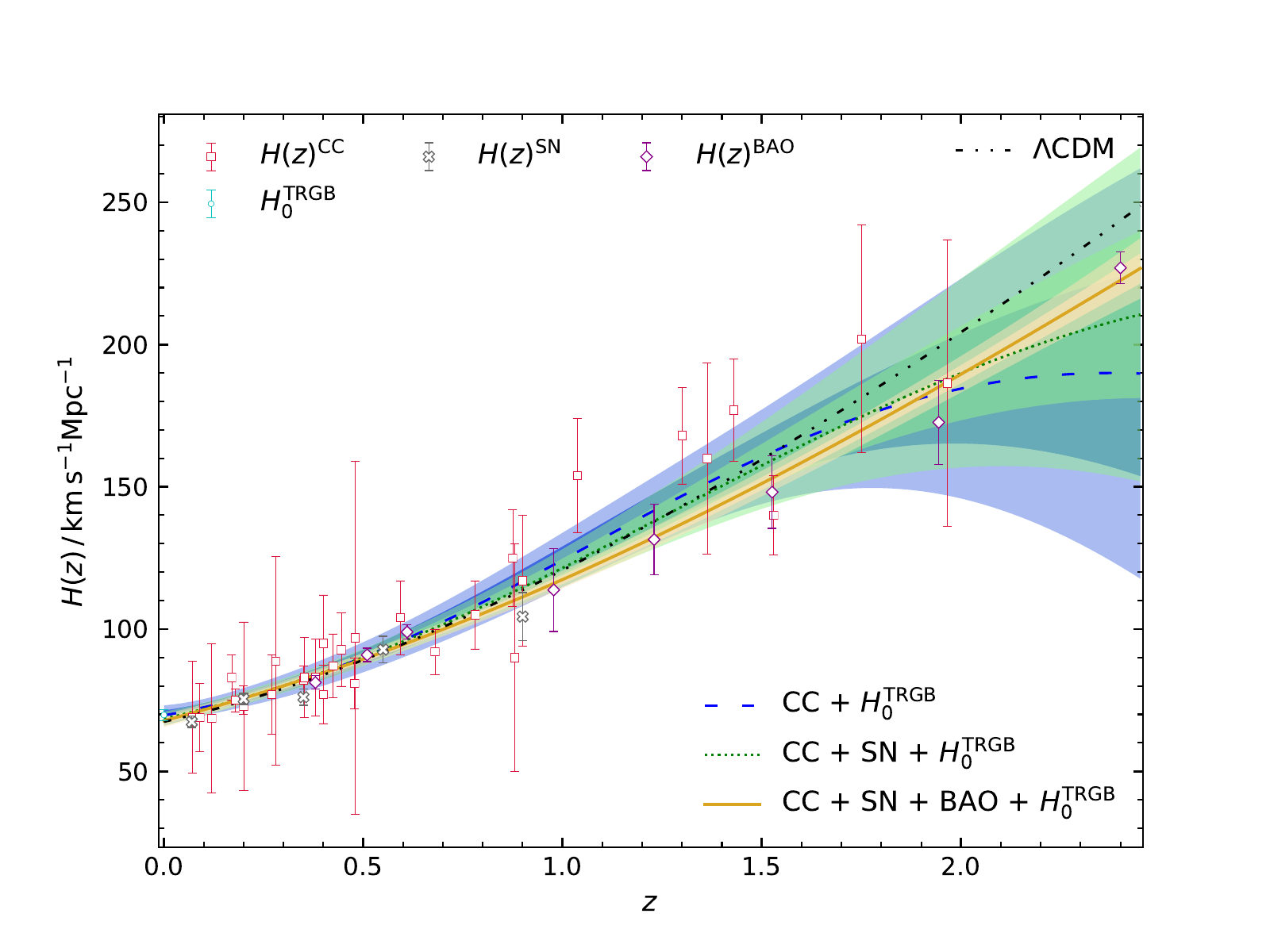}
    \caption{\label{fig:H_matern}
    GP reconstructions of $H(z)$ with the Mat\'{e}rn kernel function of Eq.(\ref{eq:Matern}). The data sets along with the different $H_0^{}$ priors are indicated in each respective panel.
    }
\end{center}
\end{figure}

\begin{figure}[t]
\begin{center}
    \includegraphics[width=0.48\columnwidth]{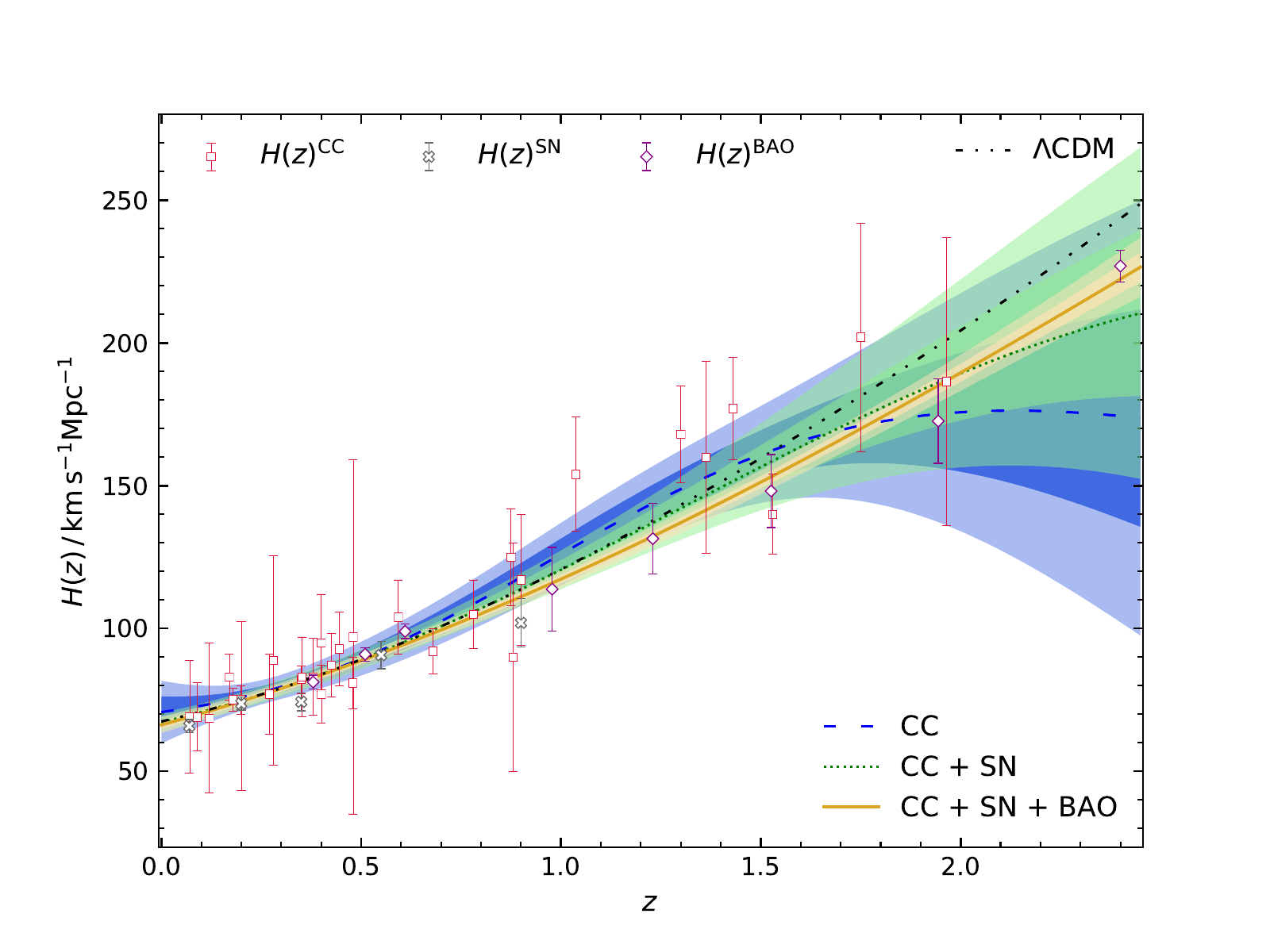}
    \includegraphics[width=0.48\columnwidth]{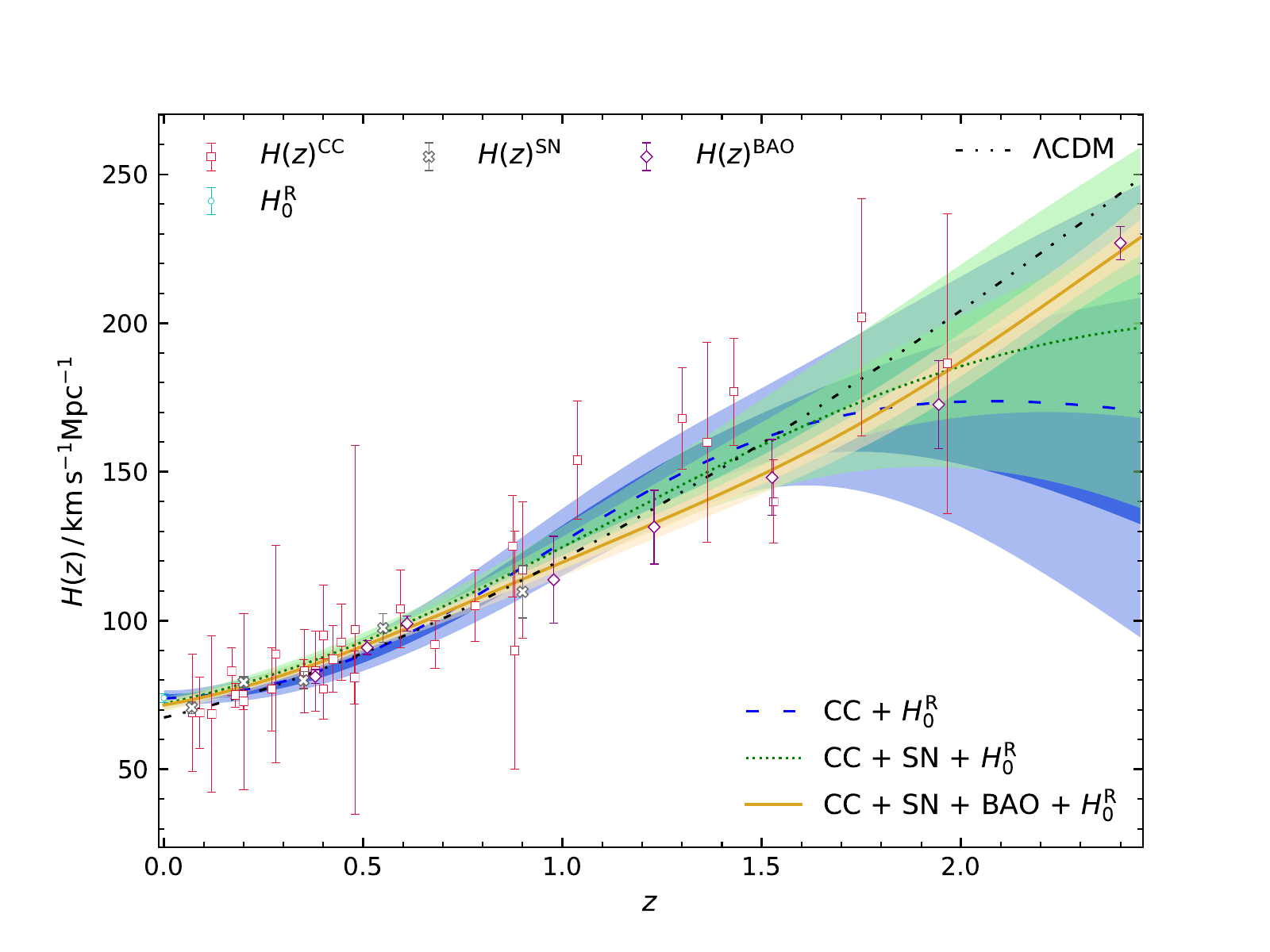}
    \includegraphics[width=0.48\columnwidth]{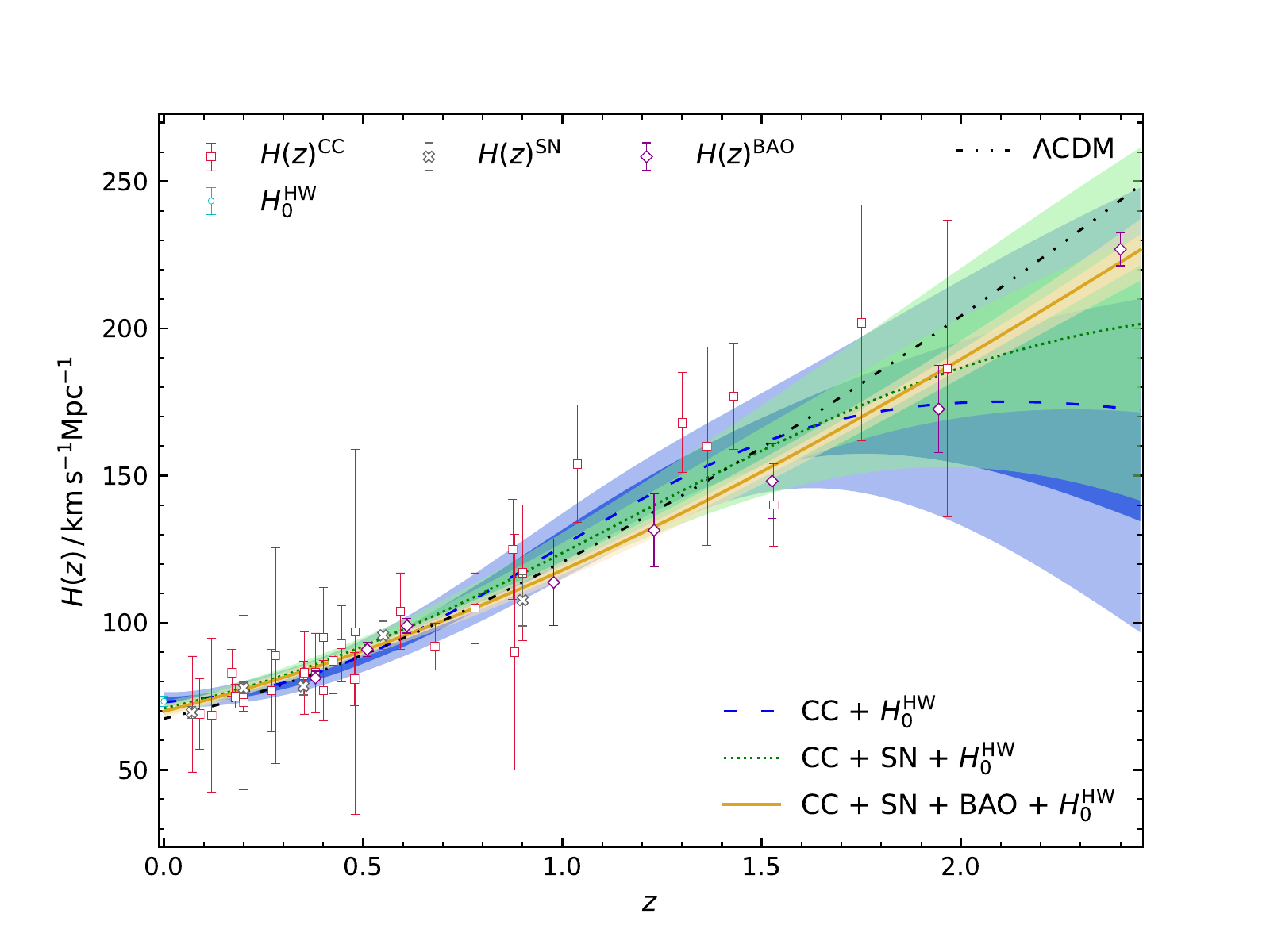}
    \includegraphics[width=0.48\columnwidth]{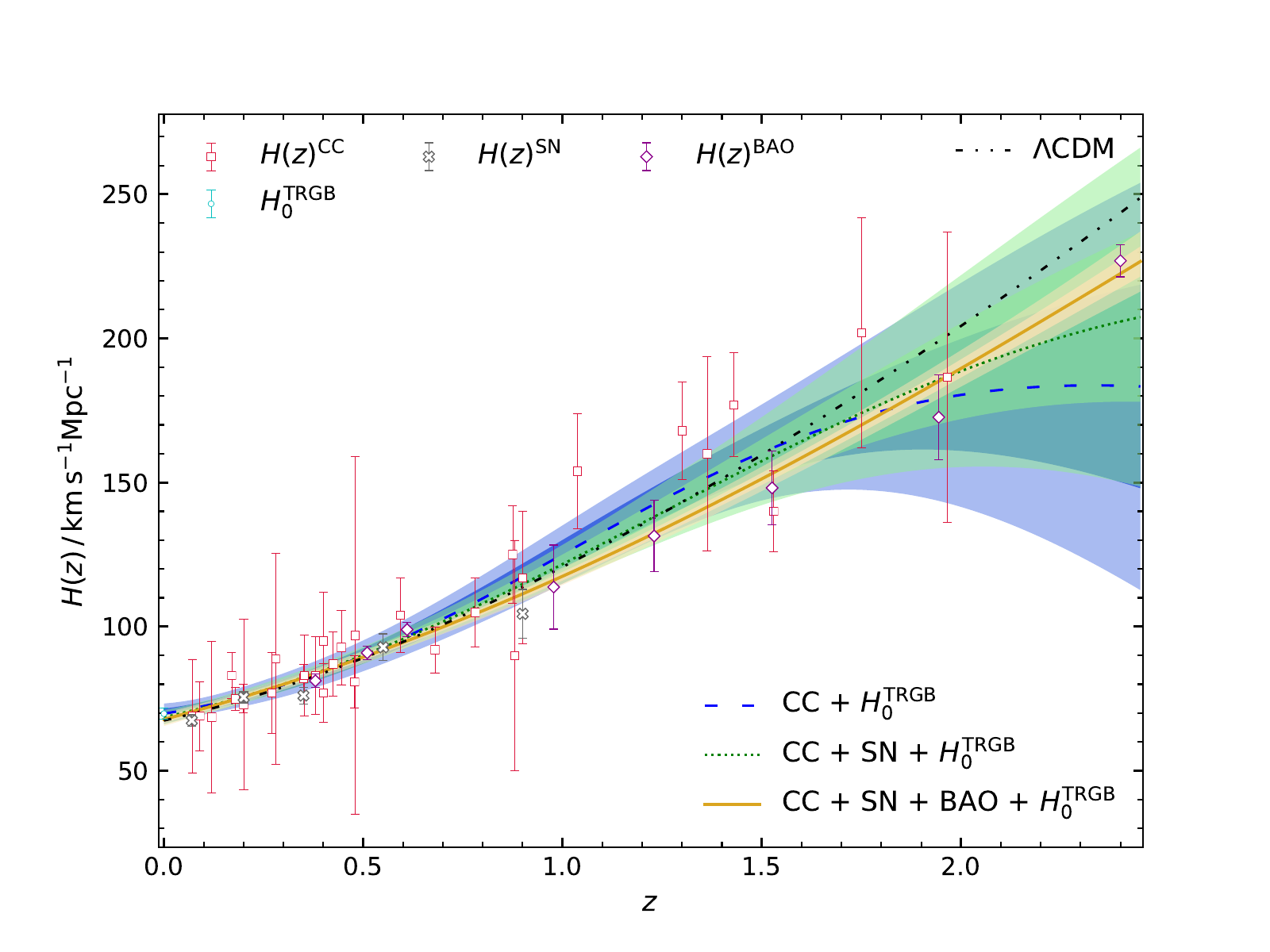}
    \caption{\label{fig:H_ratquad}
    GP reconstructions of $H(z)$ with the rational quadratic kernel function of Eq.(\ref{eq:rat_quad}). The data sets along with the different $H_0^{}$ priors are indicated in each respective panel.
    }
\end{center}
\end{figure}

\begin{figure}[t]
\begin{center}
    \includegraphics[width=0.48\columnwidth]{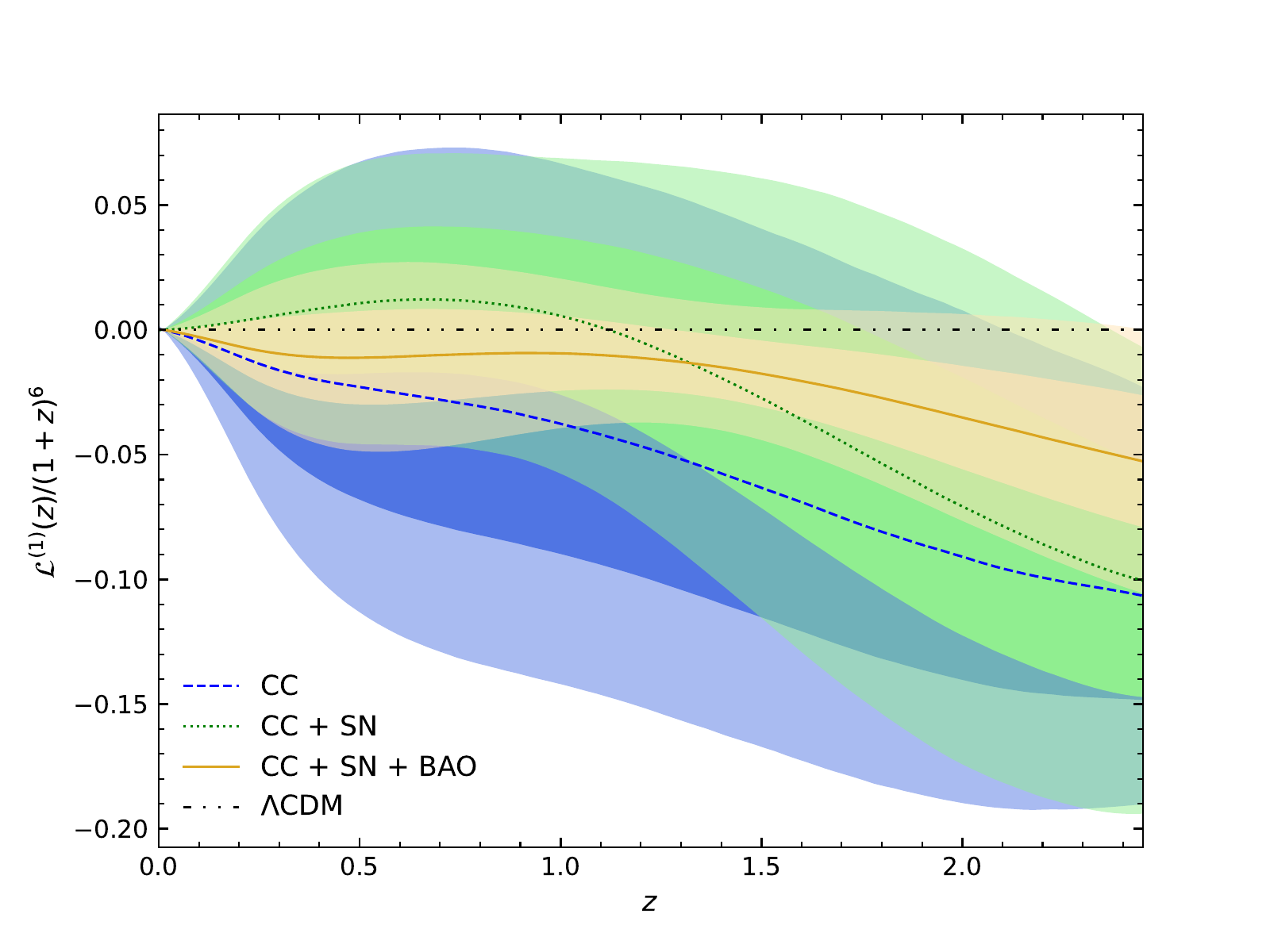}
    \includegraphics[width=0.48\columnwidth]{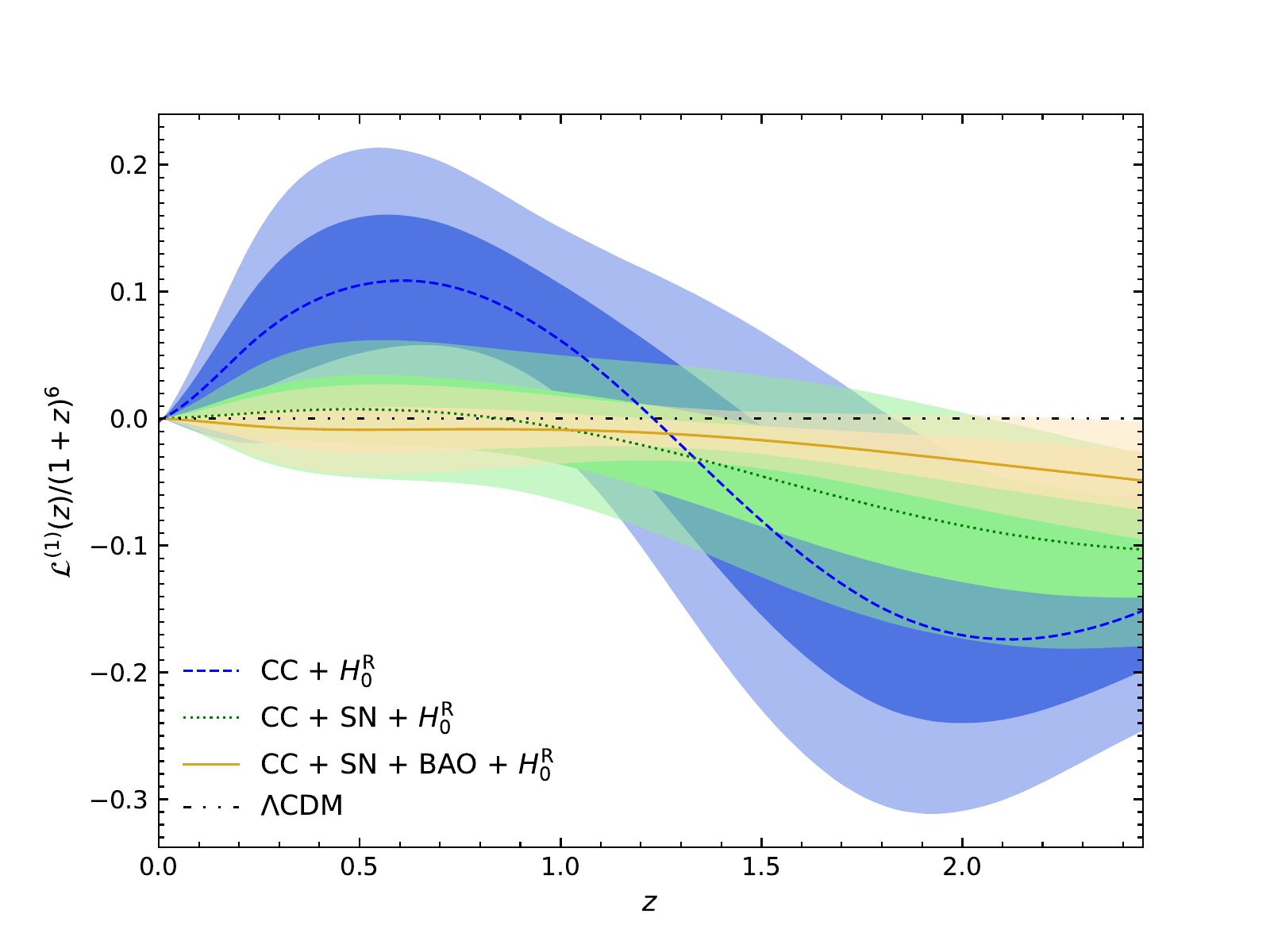}
    \includegraphics[width=0.48\columnwidth]{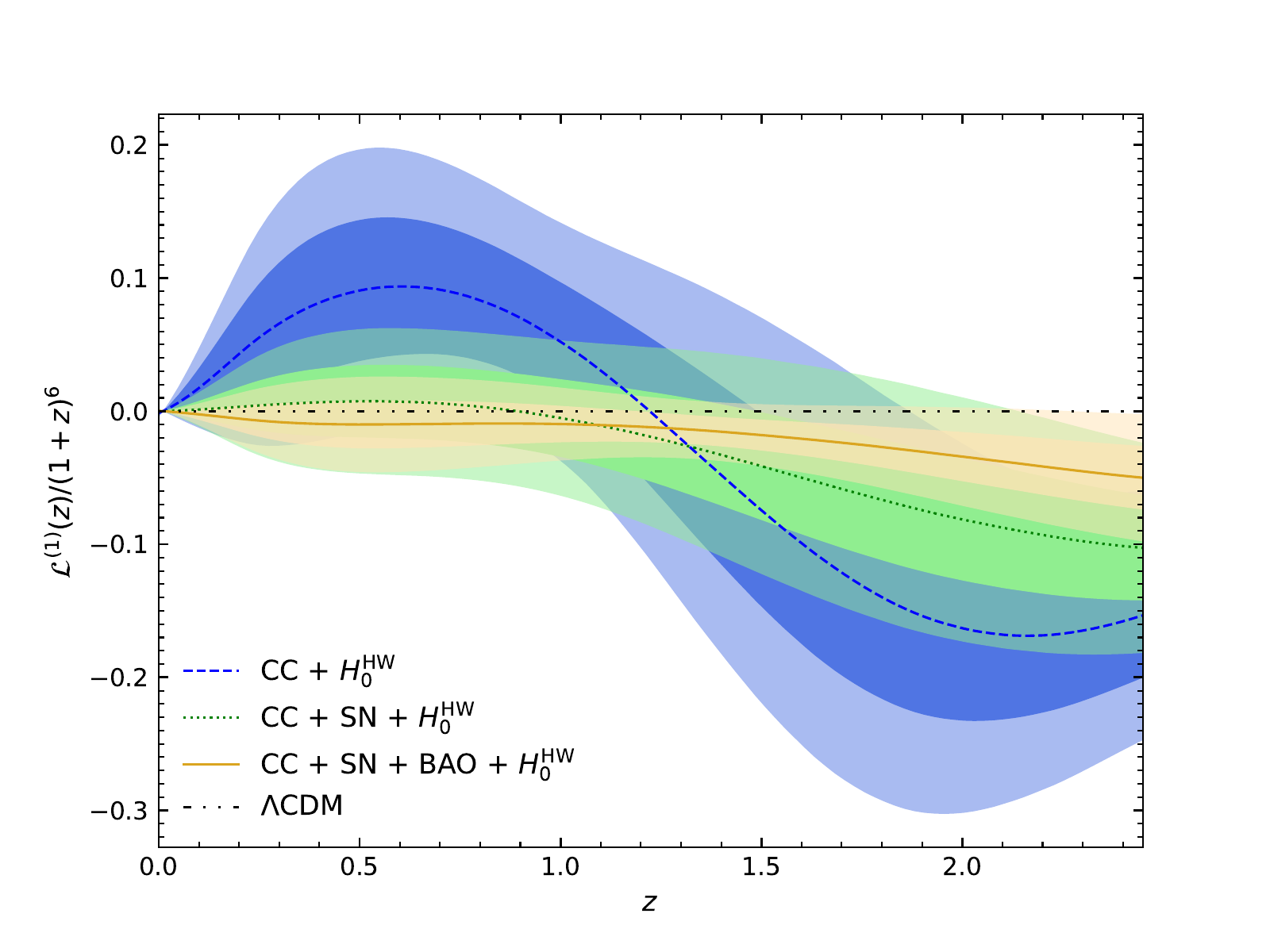}
    \includegraphics[width=0.485\columnwidth]{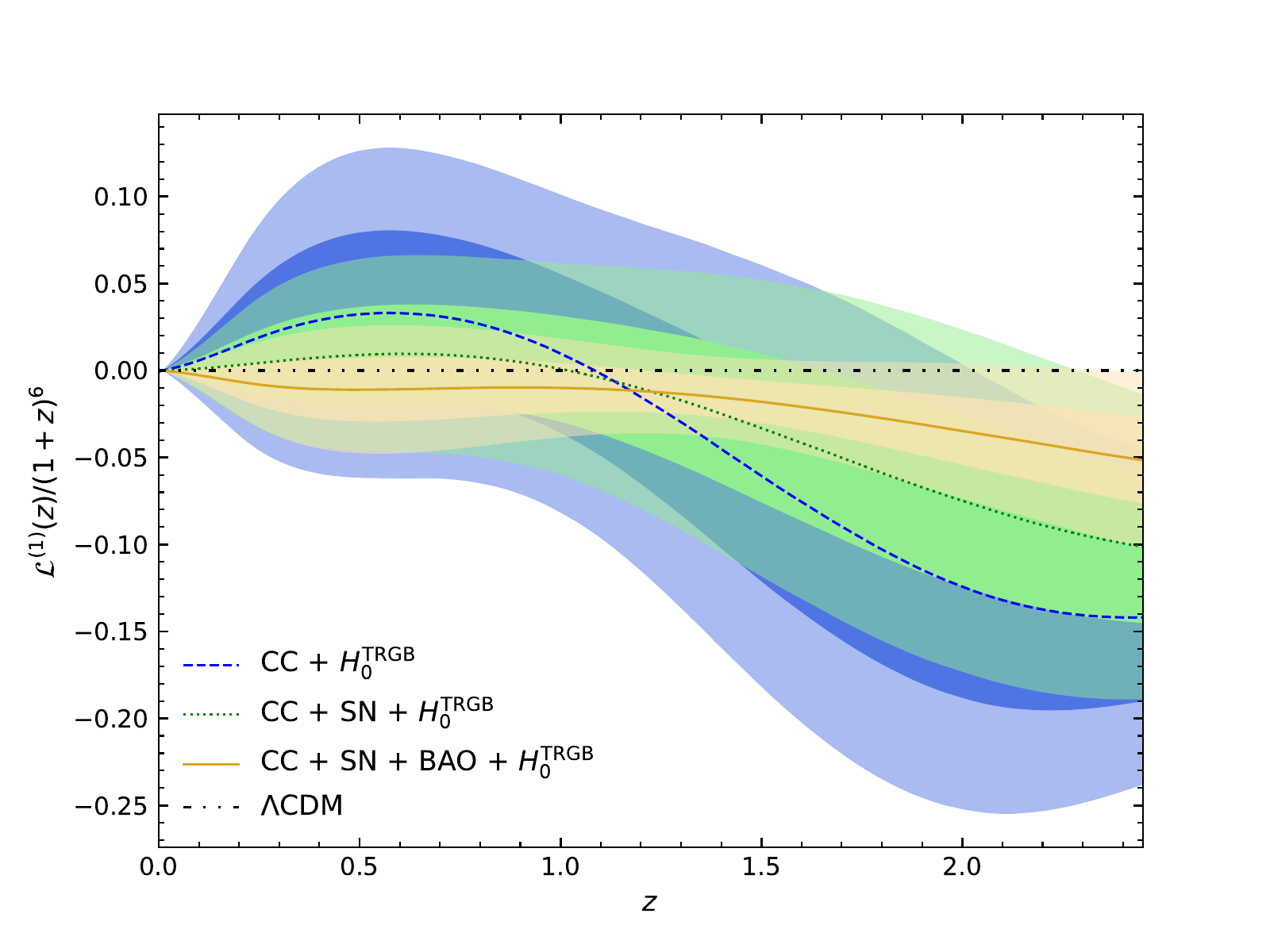}
    \caption{\label{fig:L1_squaredexp}
    GP reconstructions of $\mathcal{L}^{(1)}(z)/(1+z)^6$ with the squared exponential kernel function of Eq.(\ref{eq:square_exp}). The data sets along with the different $H_0^{}$ priors are indicated in each respective panel.
    }
\end{center}
\end{figure}

\begin{figure}[t]
\begin{center}
    \includegraphics[width=0.485\columnwidth]{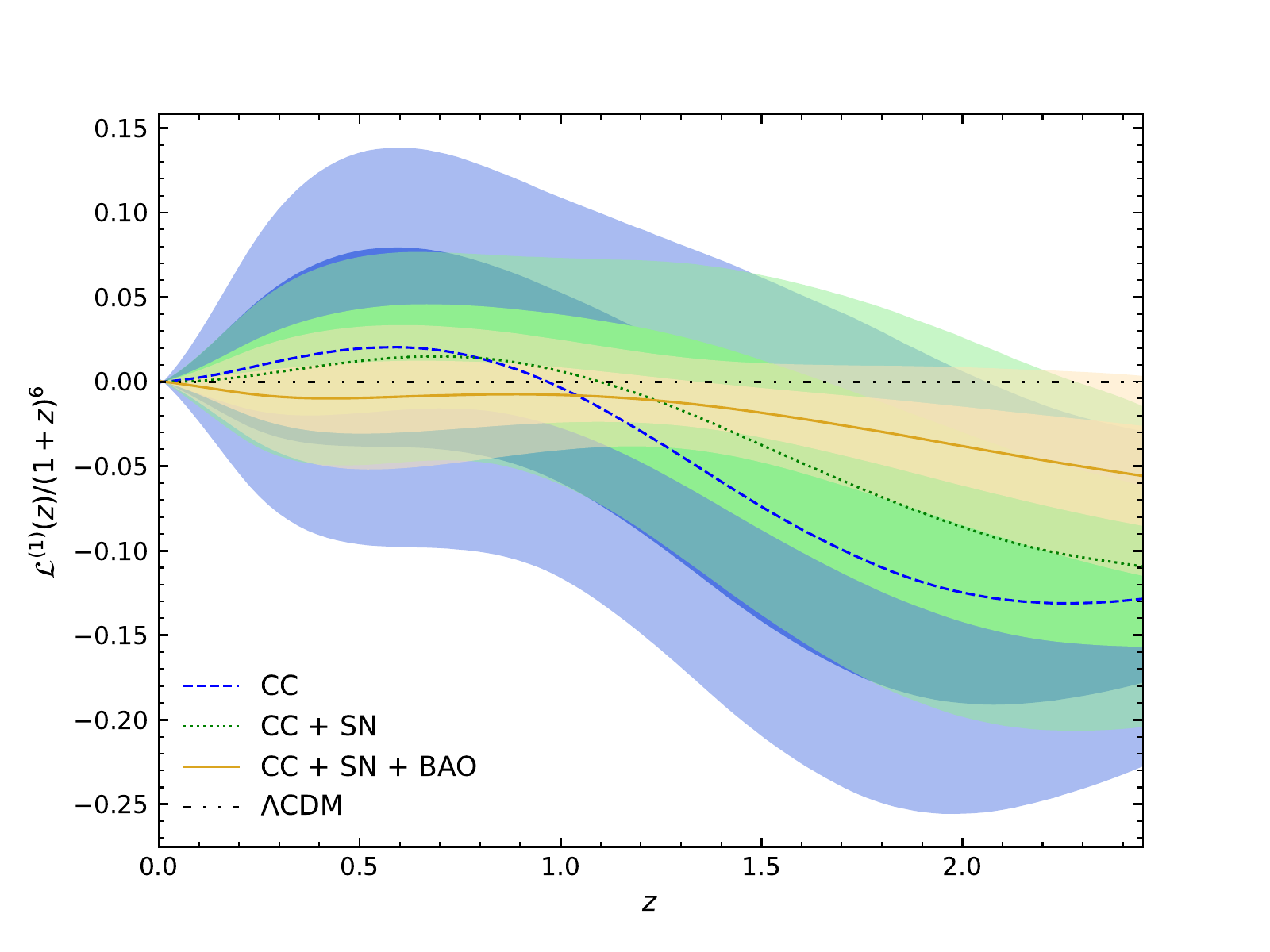}
    \includegraphics[width=0.48\columnwidth]{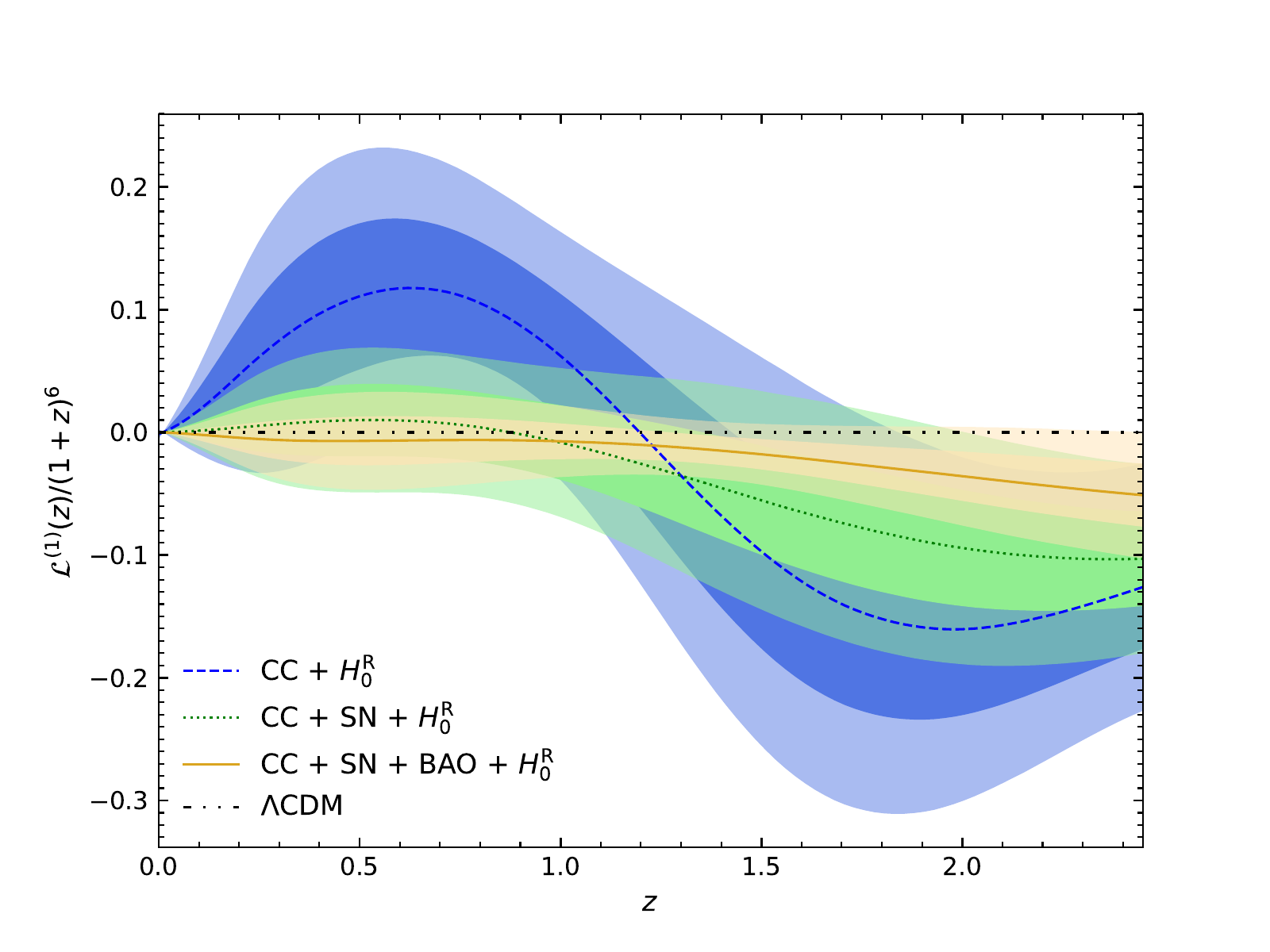}
    \includegraphics[width=0.48\columnwidth]{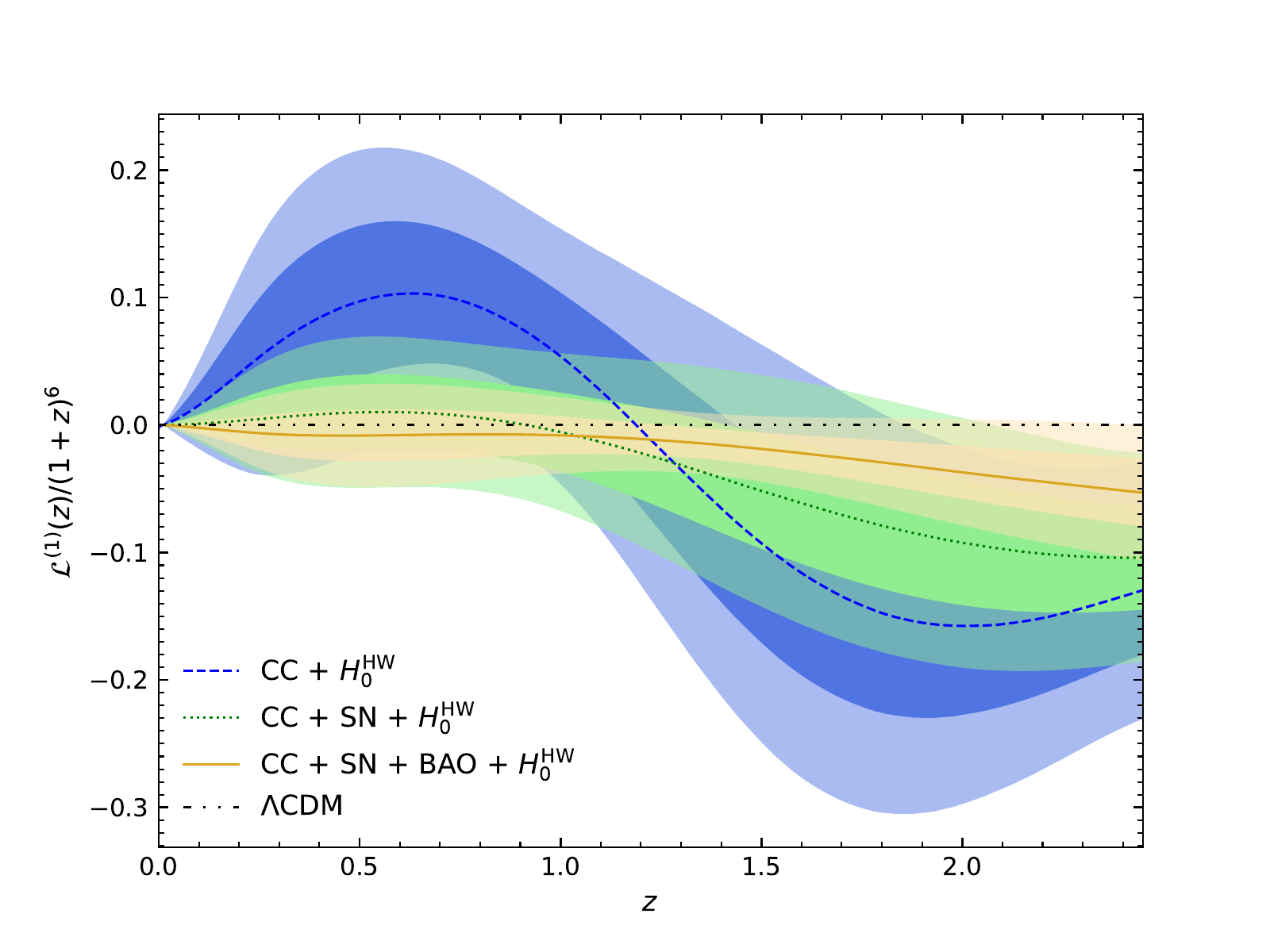}
    \includegraphics[width=0.482\columnwidth]{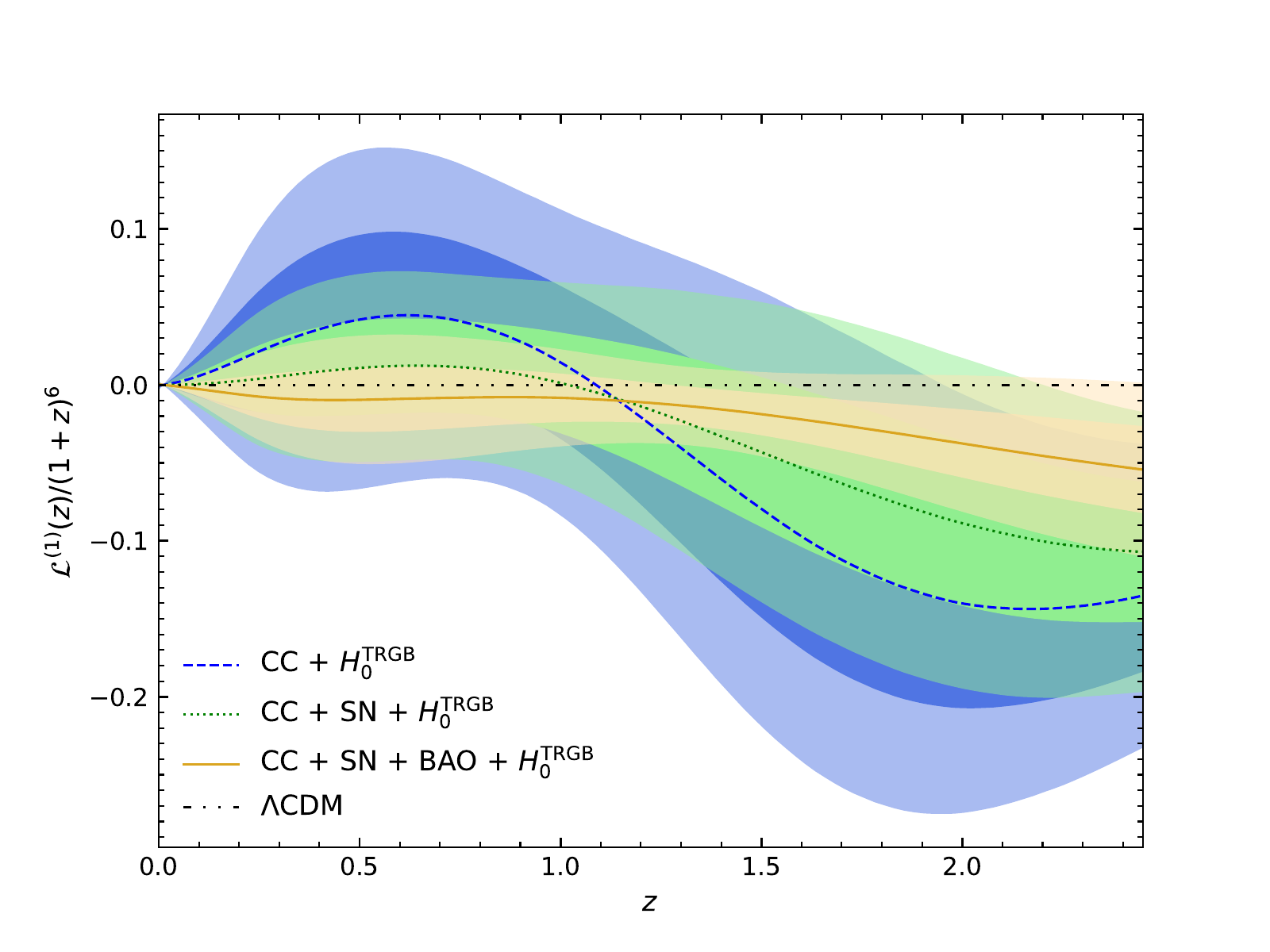}
    \caption{\label{fig:L1_cauchy}
    GP reconstructions of $\mathcal{L}^{(1)}(z)/(1+z)^6$ with the Cauchy kernel function of Eq.(\ref{eq:cauchy}). The data sets along with the different $H_0^{}$ priors are indicated in each respective panel.
    }
\end{center}
\end{figure}

\begin{figure}[t]
\begin{center}
    \includegraphics[width=0.485\columnwidth]{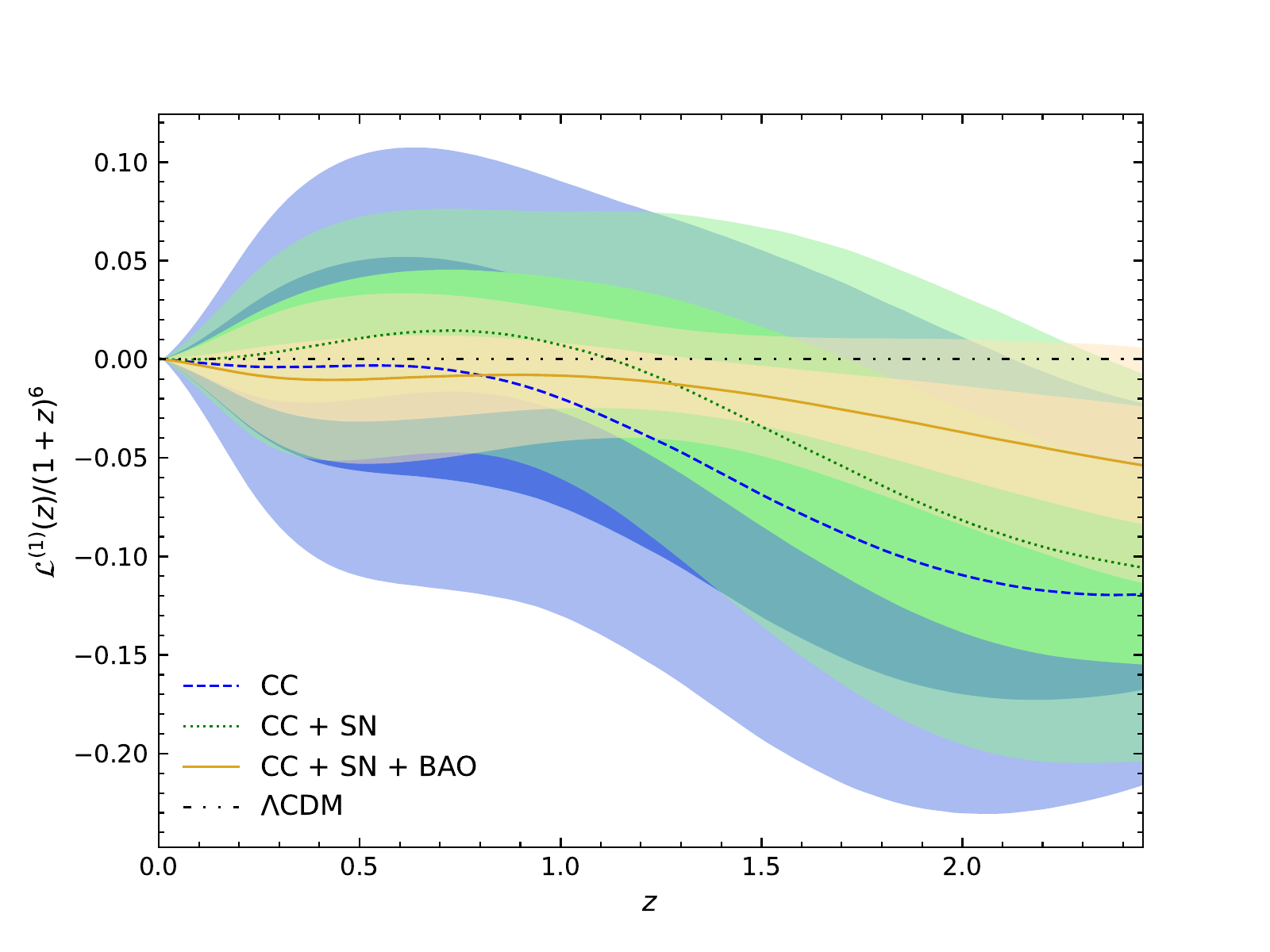}
    \includegraphics[width=0.48\columnwidth]{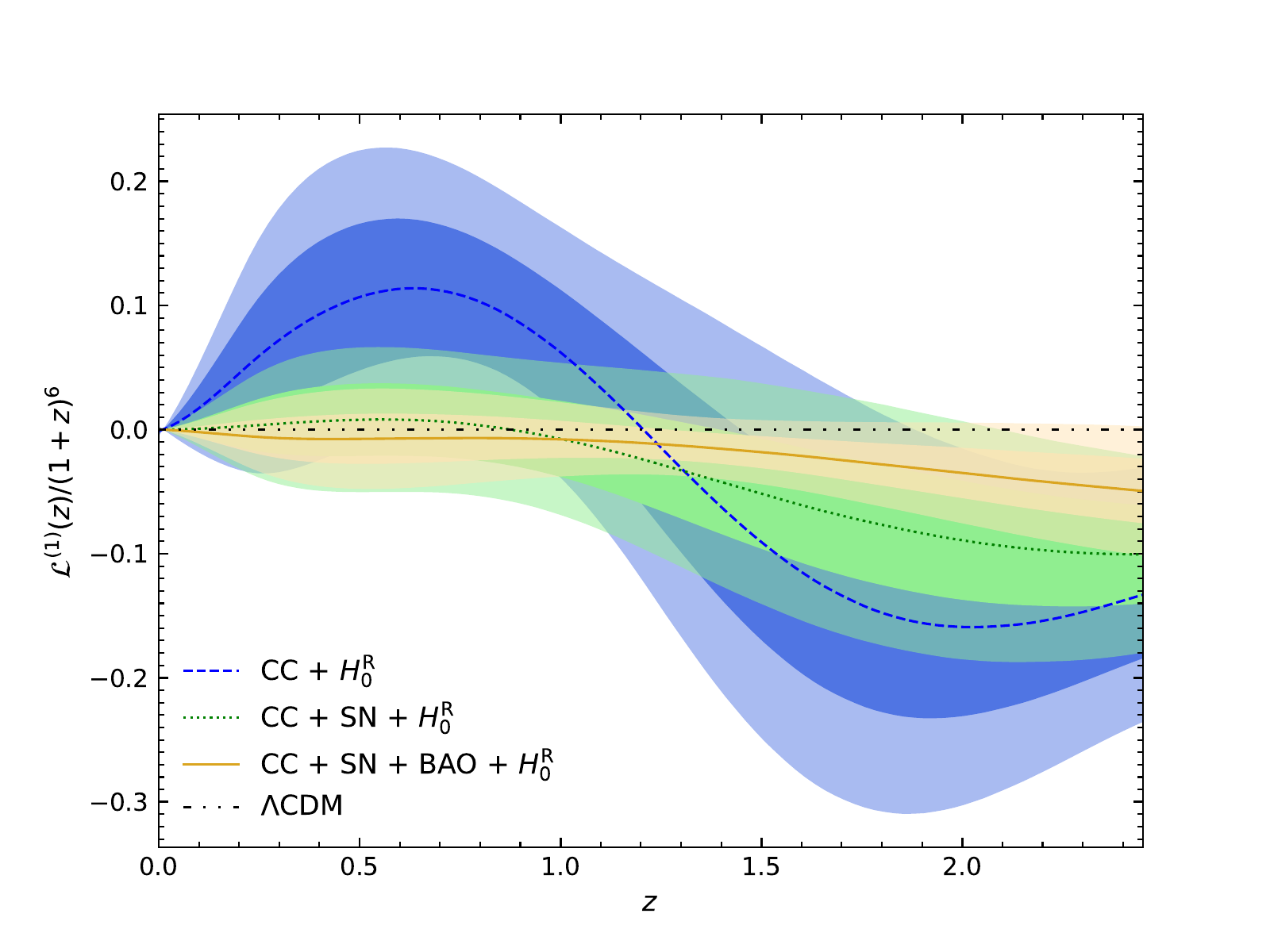}
    \includegraphics[width=0.48\columnwidth]{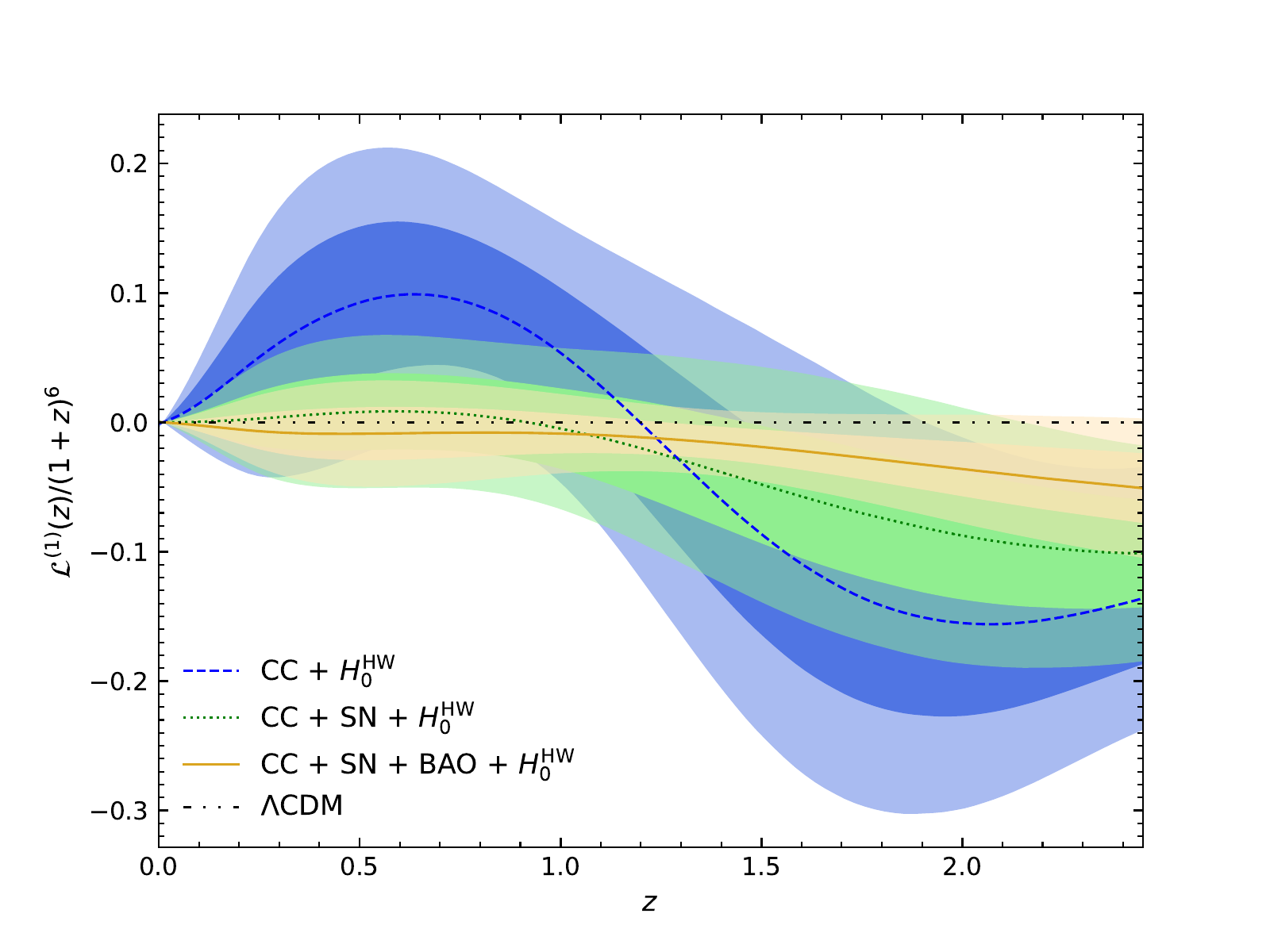}
    \includegraphics[width=0.48\columnwidth]{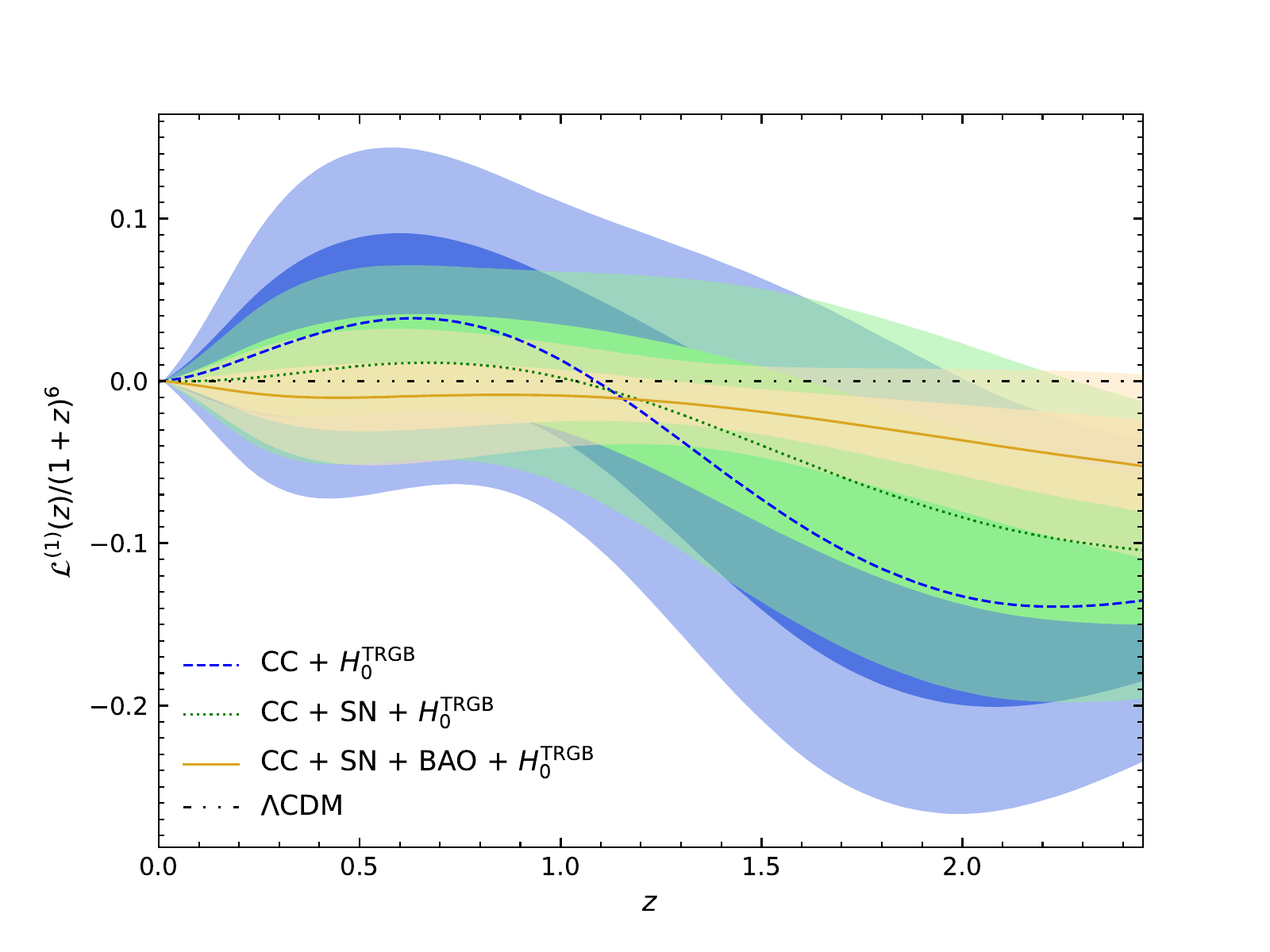}
    \caption{\label{fig:L1_matern}
    GP reconstructions of $\mathcal{L}^{(1)}(z)/(1+z)^6$ with the Mat\'{e}rn kernel function of Eq.(\ref{eq:Matern}). The data sets along with the different $H_0^{}$ priors are indicated in each respective panel.
    }
\end{center}
\end{figure}

\begin{figure}[t]
\begin{center}
    \includegraphics[width=0.48\columnwidth]{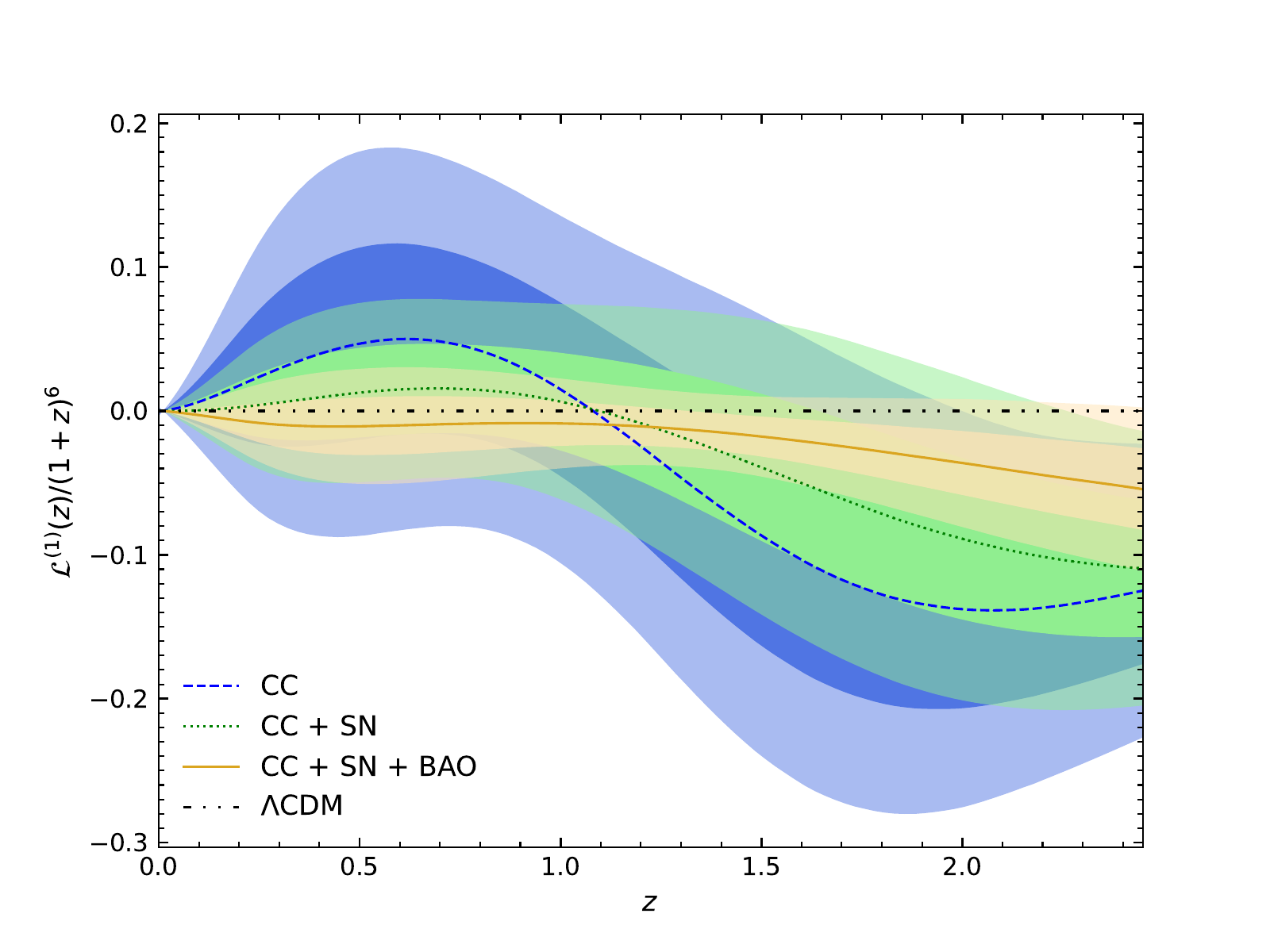}
    \includegraphics[width=0.48\columnwidth]{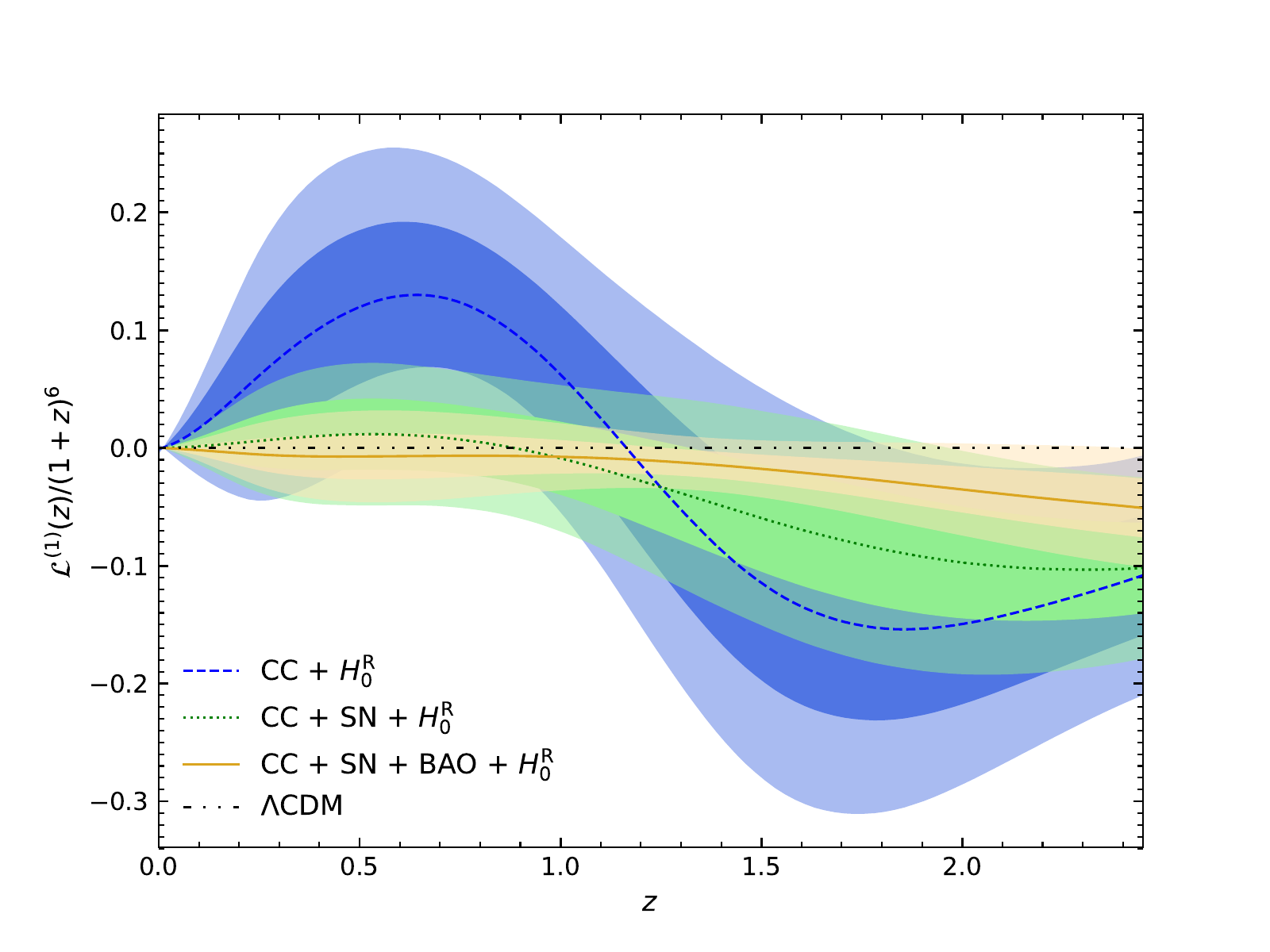}
    \includegraphics[width=0.48\columnwidth]{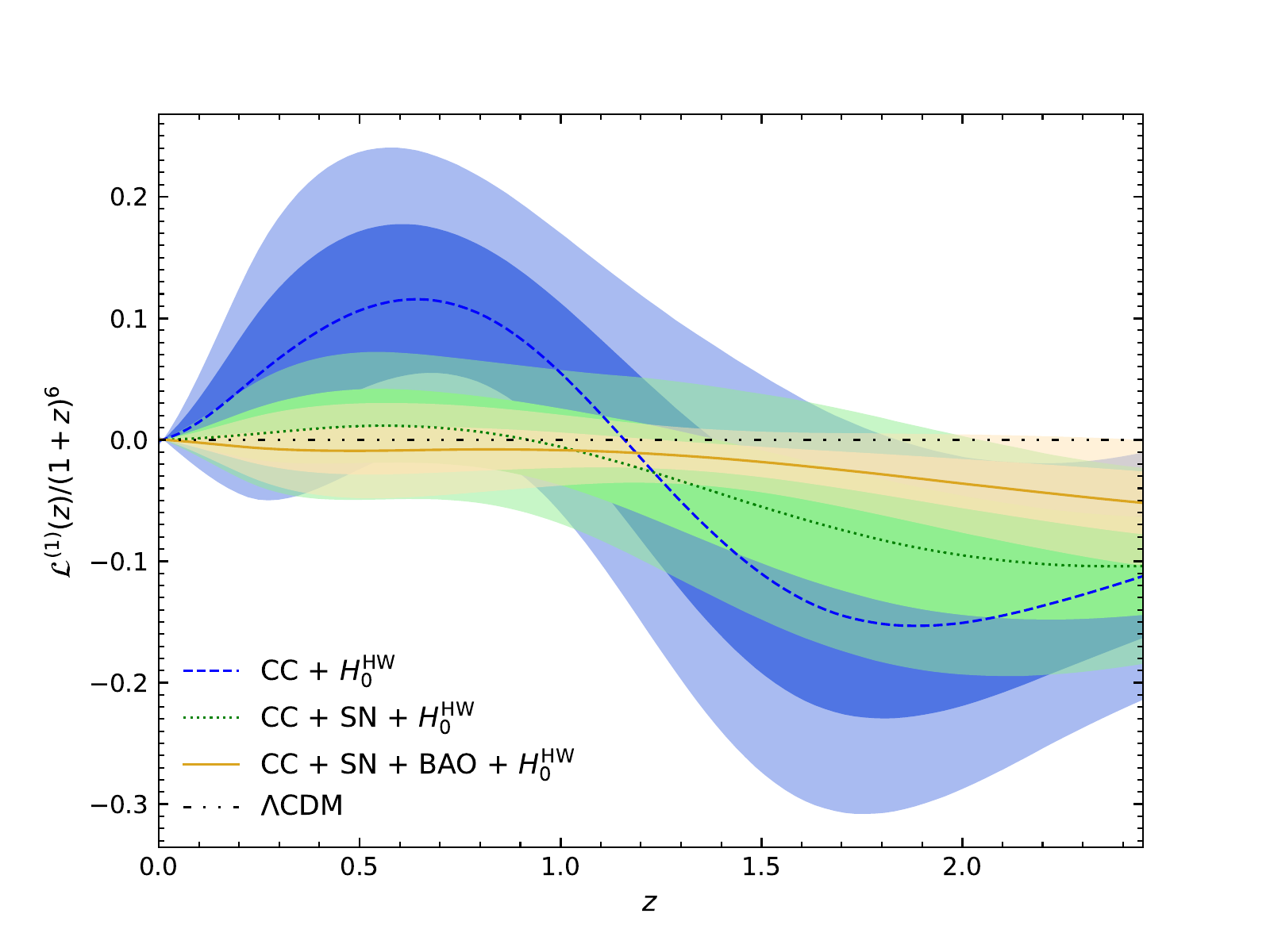}
    \includegraphics[width=0.48\columnwidth]{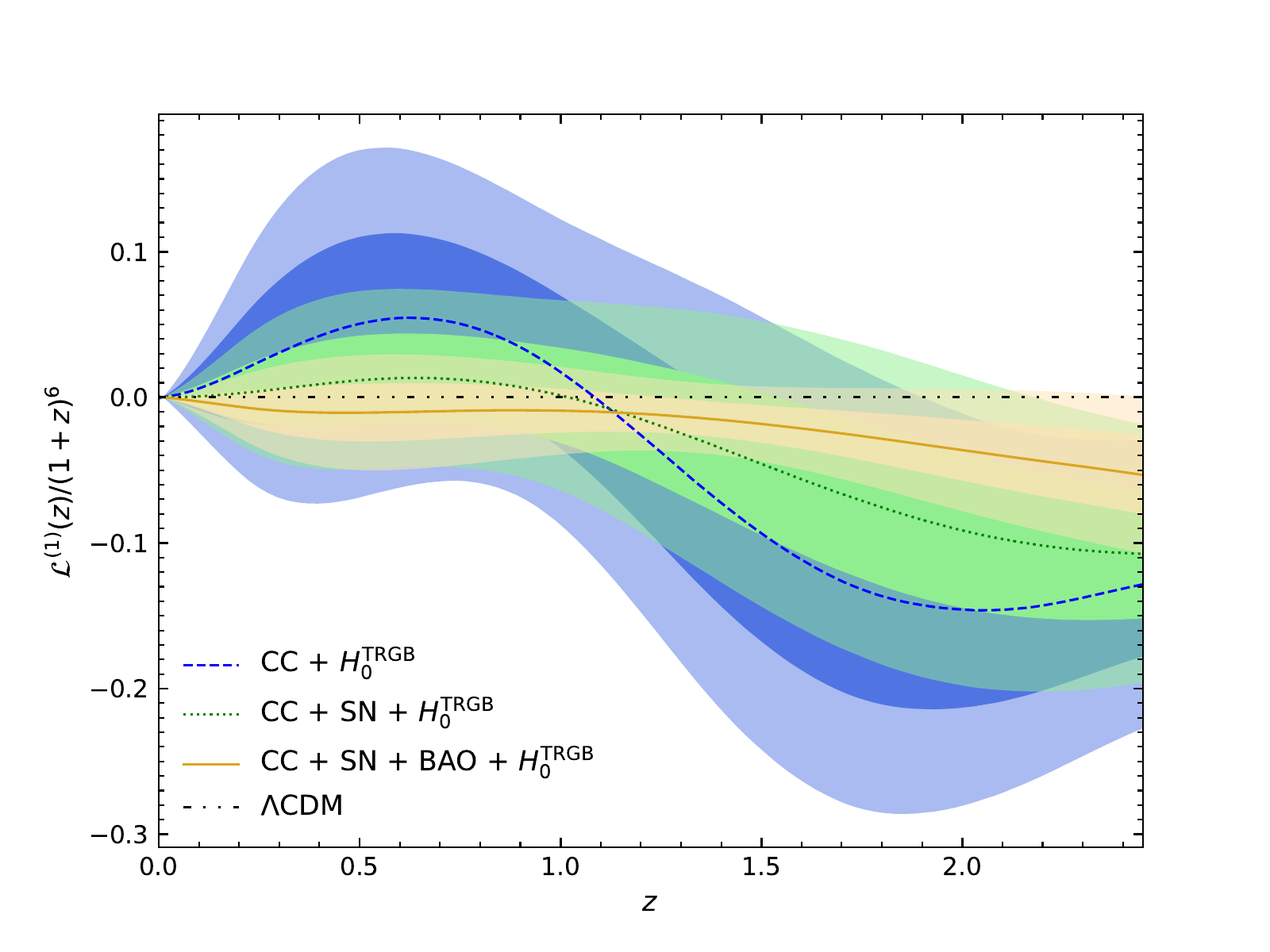}
    \caption{\label{fig:L1_ratquad}
    GP reconstructions of $\mathcal{L}^{(1)}(z)/(1+z)^6$ with the rational quadratic kernel function of Eq.(\ref{eq:rat_quad}). The data sets along with the different $H_0^{}$ priors are indicated in each respective panel.
    }
\end{center}
\end{figure}

Recently there has been an increase in the reported value of $H_0$ due to the growing tension in its value against the predicted value using the $\Lambda$CDM model with the latest cosmic microwave background (CMB) data from the Planck Mission \cite{Aghanim:2018eyx}. The Planck Collaboration report a low value of $67.4 \pm 0.5 \,{\rm km\, s}^{-1} {\rm Mpc}^{-1}$ while the Dark Energy Survey Collaboration give a similar value of $67.4^{+1.1}_{-1.2} \,{\rm km\, s}^{-1} {\rm Mpc}^{-1}$ \cite{PhysRevD.98.043526}. These predicted values of $H_0$ contrast with the cosmology independent estimates from late-time observations. The highest of these values is the Riess prior which is $H_0^{\rm R} = 74.22 \pm 1.82 \,{\rm km\, s}^{-1} {\rm Mpc}^{-1}$ \cite{Riess:2019cxk}. This value comes from long period observations of Cepheids in the Large Magellanic Cloud using the Hubble Space Telescope which has significantly reduced the uncertainty in this measurement. Another important recent announcement of $H_0$ is that from the H0LiCOW Collaboration \cite{Wong:2019kwg} which uses strong lensing from quasars and gives a value of $H_0^{\rm HW} = 73.3^{+1.7}_{-1.8} \,{\rm km\, s}^{-1} {\rm Mpc}^{-1}$. We also consider measurements using the tip of the red giant branch as a standard candle where $H_0^{\rm TRGB} = 69.8 \pm 1.9 \,{\rm km\, s}^{-1} {\rm Mpc}^{-1}$ \cite{Freedman:2019jwv}. While other measurements exist such as the novel approach of gravitational wave standard sirens \cite{Abbott:2017xzu}, the aforementioned measurements are the most representative model-independent values that exist to reasonable uncertainty in the literature.

In the present study, we perform a number of GP analyses using a combination of two choices, the first being the prior which is selected from having no prior, or one of the $H_0^{\rm R},\,H_0^{\rm HW},\,H_0^{\rm TRGB}$ values, while the second involves the choice of GP kernel. GP are model-independent but they do depend on the kernel hyperparameters. Given the level of precision in the discrepancy in $H_0$, we consider the kernels in Eqs.(\ref{eq:cauchy})--(\ref{eq:rat_quad}) in order to reduce any fine differences between these covariance functions on the estimation of $H_0$. To do this, we use a modified version of the public code GaPP (Gaussian Processes in Python)\footnote{\url{http://ascl.net/1303.027}} \cite{2012JCAP...06..036S} which implements the GP approach using these kernels.

We now apply the GP approach explained in subsection \ref{sec:GP} to the various sources of Hubble data together with the priors described here. The results pertaining to the value of $H_0$ are presented in Tables \ref{tab:se_kernel}--\ref{tab:rq_kernel} which contain the principal results for the square exponential, Cauchy, Mat\'{e}rn and rational quadratic kernels respectively. In each table, we present the GP inferred value of $H_0$ for the case of taking no prior, and the $H_0^{\rm R}$, $H_0^{\rm TRGB}$ and $H_0^{\rm HW}$ priors. In every case of this analysis, we determine the distance (in units of $\sigma$) between the GP determined value against the literature priors discussed above, so that this distance will be defined as
\begin{equation}
    d\left(H_{0,i},H_{0,j}\right) = \frac{H_{0,i} - H_{0,j}}{\sqrt{\sigma_i^2 + \sigma_j^2}}\,,
\end{equation}
where $H_{0,i}$ and $H_{0,j}$ are two respective values of the present value of the Hubble parameter together with their respective $1\sigma$ uncertainties $\sigma_i$ and $\sigma_j$.

The full results for each of the square exponential, Cauchy, Mat\'{e}rn and rational quadratic kernels are respectively presented in Figs. \ref{fig:H_squaredexp}--\ref{fig:H_ratquad}. These figures depict the wider GP reconstruction for the range within which the data appears. The GP approach is used for each set of priors for $H_0$ as shown in the sub-figures. Moreover, for every GP reconstruction, the $1\sigma$ and $2\sigma$ regions are shown. As a reference point, we present the $\Lambda$CDM behaviour in all instances. In addition to the reconstructions, each kernel GP is complemented by consistency tests for the $\Lambda$CDM model.

There also exist diagnostic tools to assess preferences in the reconstructions toward deviations from $\Lambda$CDM \cite{Yahya:2013xma,Zunckel:2008ti}. Considering the GR Friedmann equation for a cosmos filled with a dark fluid EoS $w(z)$
\begin{equation}
    \frac{H^2(z)}{H_0^2} = \Omega_m^0 \left(1+z\right)^3 + \Omega_{k}^0 \left(1+z\right)^2 + \Omega_{\Lambda}^0\exp\left[3\int_0^z \frac{1+w(z')}{1+z'} \mathrm{d}z'\right]\,,
\end{equation}
which can be rearranged to reconstruct the EoS of the dark fluid via
\begin{equation}
    w(z)=\frac{2(1+z)E(z)E'(z)-3E^2(z)}{3\left[E^2(z)-\Omega_m^0(1+z)^3\right]}\,,   
\end{equation}
where we have assumed spatial flatness. We report the GP reconstructions of $w(z)$ in \ref{app_par}, where it is clear that $w=-1$ is not excluded by the currently available data. However, we should point out that this reconstruction is dependent on the matter density parameter, which restricts us from constructing physical models. On the other hand, one could test the flat $\Lambda$CDM model by considering the following diagnostic redshift function
\begin{equation}\label{consis_test}
    \mathcal{O}_m^{(1)} (z) :=\frac{E^2(z)-1}{z(3+3z+z^2)}\,,
\end{equation}
which reduces to $\mathcal{O}_m^{(1)} (z)=\Omega_m^0$ in the $\Lambda$CDM scenario. We also report the GP reconstructions of $\mathcal{O}_m^{(1)}(z)$ in \ref{app_par}, where one could clearly notice that the considered data sets are in a very good agreement with the $\Lambda$CDM predictions, although the $H_0$ prior has a significant effect on the reconstruction and hence the viability of the concordance model of cosmology. A more effective diagnostic is the vanishing of the derivative of $\mathcal{O}_m^{(1)} (z)$, denoted by
\begin{equation}\label{diagnostic_for_H}
    \mathcal{L}^{(1)}(z)=3(1-E^2(z))(1+z)^2+2z(3+3z+z^2)E(z)E'(z)\,.
\end{equation}

Any deviation from $\mathcal{L}^{(1)}(z) = 0$ represents a deviation from $\Lambda$CDM, which makes $\mathcal{L}^{(1)}(z)$ a good diagnostic over which to assess the behaviour of the concordance model. For the square exponential, Cauchy, Mat\'{e}rn and rational quadratic kernels used in this study, the $\mathcal{L}^{(1)}(z)$ diagnostic is presented for each reconstruction in Figs. \ref{fig:L1_squaredexp}--\ref{fig:L1_ratquad}.
In the scenario where no prior is assumed, to a fairly consistent degree, the square exponential kernel produces lower values of $H_0$ while the CC data always produce higher reconstructed $H_0$ values which occur due to the data being placed at lower values of redshift. This is brought out in tables \ref{tab:se_kernel}--\ref{tab:rq_kernel}. However, in all cases, the results remain within the $1\sigma$ confidence levels across all the kernels for their respective reconstructions. Throughout, the $H_0^{\rm R}$ prior produces the highest values of $H_0$ with the rational quadratic giving the highest $H_0$ within that subset which is achieved when the CC data are taken alone. The impact of BAO data in all instances is to reduce the reconstructed value of $H_0$ which holds through for each of the priors and kernels. This property follows from the fact that BAO data appears at higher redshifts and acts to favour smaller uncertainties at those redshifts. The $H_0^{\rm HW}$ prior produces similar but lower values for $H_0$ when compared with the $H_0^{\rm R}$ prior since their values are fairly close. The lowest inferred values of $H_0$ with a prior are found with the $H_0^{\rm TRGB}$ prior, since this is the lowest of the three. In fact, it is the $H_0^{\rm TRGB}$ prior that gives the lowest tension with the recent results by the Planck Collaboration. This feature is brought out in the last column of tables \ref{tab:se_kernel}--\ref{tab:rq_kernel} which illustrates the distance between the reconstructed $H_0$ and $H_0^{\rm P18}$, which produces the lowest discrepancy for the TRGB setting.

In Fig. \ref{fig:H_squaredexp}, the square exponential kernel GP reconstructions are shown for the redshift range of the full data set. In all cases, the BAO data reduce the $1\sigma$ and $2\sigma$ uncertainties at higher redshifts since the other data sets do not feature points in that regime. In fact, in the cases of CC and CC+SN, the $\Lambda$CDM theoretical prediction only deviates into the $2\sigma$ uncertainty region for these high values of redshift, and only outside of both when the BAO data are included. The BAO data are dependent on the concordance model of cosmology, and so one would expect it to produce issues of this kind since the other two data sets are independent of cosmological models. The remainder of the reconstruction regions remain in close range of the $\Lambda$CDM prediction. This is further exposed by the diagnostic consistency test shown in Fig. \ref{fig:L1_squaredexp} where a number of regions mark slight deviations from $\Lambda$CDM. However, it is the BAO data set that exposes this deviation at high redshifts. Another interesting feature of these diagnostic tests is that as with the $H_0$ reconstructions in table \ref{tab:se_kernel}, the $H_0^{\rm R}$ prior brings about the largest deviation of the reconstructions.

Concerning the $1\sigma$ confidence regions, a similar picture unfolds for the GP Hubble reconstructions and diagnostic tests for the Cauchy, Mat\'{e}rn and rational quadratic kernels which are respectively shown in Figs. \ref{fig:H_cauchy},\ref{fig:L1_cauchy}, Figs.  \ref{fig:H_matern},\ref{fig:L1_matern} and Figs. \ref{fig:H_ratquad},\ref{fig:L1_ratquad}. The resonance between these GP reconstructions over the kernel choices is a crucial property to the model independence of the analysis which shows that while dependent on a covariance model, the resulting Hubble parameter evolution is independent of a cosmological model.

In the above analyses, the GP approach was used to reconstruct the Hubble parameter history of the evolution of the Universe with a focus on the inferred value of $H_0$. However, another value of growing importance is that of the redshift value at which the Universe transitioned from a decelerating cosmos to its present state of acceleration, $z_{\rm t}$ \cite{Peebles:2002gy}. This second probe of dark energy provides further details on its properties and possible information on its eventual interpretation which may include new physics beyond $\Lambda$CDM. Given that the transition occurs at a different point in the evolution of the Universe, this may also reveal any potential time dependence of dark energy, as well as the evolution of the ratio of matter to dark energy along the cosmic timeline \cite{Capozziello:2015rda}. In the $\Lambda$CDM model, for a flat FLRW cosmology, the transition redshift turns out as \cite{peebles1993principles}
\begin{equation}
    z_{\rm t} = \left[\frac{2\left(1-\Omega_{m}^0\right)}{\Omega_m^0}\right]^{1/3} - 1\,,
\end{equation}
which for P18 values gives $z_t\simeq0.63$.

Given that we use the GP approach to reconstruct the Hubble parameter beyond the indicative $\Lambda$CDM value of the transition redshift, we can also make a determination of this value for each of the reconstructions that form part of the study up to this point. These results are reported in Table \ref{tab:kernels_z_t} where, respectively, we give the reconstructed transition redshifts for the square exponential, Cauchy, Mat\'{e}rn and rational quadratic kernels for the various data sets and prior combinations. These values are inferred from the GP reconstructed deceleration parameter (which are illustrated in \ref{app_par})
\begin{equation}
    q(z)=(1+z)\frac{H'(z)}{H(z)}-1\,,
\end{equation}
from which the transition time is straightforwardly inferred.

Similar to the inferred values of $H_0$ reported in Tables \ref{tab:se_kernel}--\ref{tab:rq_kernel}, the $H_0^{\rm R}$ prior produces the most extreme reconstructed parameter values of the transition redshift. In this case, the Riess prior produces the lowest values of $z_{\rm t}$ which results from the fact that a higher prior would imply a lower redshift turning point for the acceleration of the Universe. While this occurs consistently for all three respective data sets, the lowest transition redshift occurs for the CC data set since this produces the highest reconstructed $H_0$ value. To a lesser extent, the $H_0^{\rm HW}$ prior produces the next lowest inferred $z_{\rm t}$ reconstructions followed by the $H_0^{\rm TRGB}$ prior which is expected since $H_0^{\rm TRGB}<H_0^{\rm HW}$. Another important similarity with the $H_0$ reconstructions is that since the CC, CC+SN and CC+SN+BAO produce the leading values of $H_0$ in that order, then they will produce, by and large, the leading $z_{\rm t}$ in ascending order since the transition will occur closest to the present time. Finally, on the issue of uncertainties, since the $\Lambda$CDM-dependent BAO data sets are characterised by values of the Hubble parameter at the highest redshifts, they produce the lowest uncertainties, particularly with respect to the CC inferred errors.

\begin{table}[t]
    \setlength\extrarowheight{.2em}
    \setlength{\tabcolsep}{2.5pt}
    \caption{\label{tab:kernels_z_t} Values of $z_{\rm t}$ along with their corresponding $1\sigma$ uncertainties inferred via the GP reconstructions of $q(z)$ with different data sets and kernel functions as specified in Eqs.(\ref{eq:square_exp})--(\ref{eq:rat_quad}).}
    \centering
    \begin{tabular}{@{}c c c c c}
        \br
        \multirow{2}{*}{Data set(s)} & \multicolumn{4}{c}{$z_\mathrm{t}$} \\
        \cline{2-5}
        & Square exponential & Cauchy & Mat\'{e}rn & Rational quadratic \\
        \mr
        {\small CC} & $0.616^{+ \mathrm{nan} }_{- 0.137 }$ & $0.572^{+ 0.267 }_{- 0.118 }$ & $0.591^{+ \mathrm{nan} }_{- 0.129 }$ & $0.553^{+ 0.211 }_{- 0.110}$ \\
        {\small CC+SN} & $0.607^{+ 0.132 }_{- 0.084 }$ & $0.600^{+ 0.132 }_{- 0.087 }$ & $0.605^{+ 0.131 }_{- 0.089 }$ & $0.598^{+ 0.133 }_{- 0.088 }$ \\
        {\small CC+SN+BAO} & $0.667^{+ 0.095 }_{- 0.075 }$ & $0.658^{+ 0.105 }_{- 0.081 }$ & $0.659^{+ 0.108 }_{- 0.082 }$ & $0.664^{+ 0.101 }_{- 0.079 }$ \\
        \mr
        {\small CC+$H_0^{\rm R}$} & $0.551^{+ 0.112 }_{- 0.079 }$ & $0.540^{+ 0.114 }_{- 0.081 }$ & $0.546^{+ 0.115 }_{- 0.082 }$ & $0.528^{+ 0.114 }_{- 0.082 }$ \\
        {\small CC+SN+$H_0^{\rm R}$} & $0.702^{+ 0.246 }_{- 0.112 }$ & $0.687^{+ 0.268 }_{- 0.114 }$ & $0.694^{+ 0.263 }_{- 0.116 }$ & $0.679^{+ \mathrm{nan} }_{- 0.114 }$ \\
        {\small CC+SN+BAO+$H_0^{\rm R}$} & $0.783^{+ 0.107 }_{- 0.088 }$ & $0.770^{+ 0.119 }_{- 0.095 }$ & $0.772^{+ 0.123 }_{- 0.097 }$ & $0.783^{+ 0.118 }_{- 0.095 }$ \\
        \mr
        {\small CC+$H_0^{\rm TRGB}$} & $0.573^{+ 0.165 }_{- 0.096 }$ & $0.552^{+ 0.161 }_{- 0.099 }$ & $0.564^{+ 0.167 }_{- 0.102 }$ & $0.537^{+ 0.159 }_{- 0.098 }$ \\
        {\small CC+SN+$H_0^{\rm TRGB}$} & $0.633^{+ 0.149 }_{- 0.091 }$ & $0.624^{+ 0.148 }_{- 0.094 }$ & $0.629^{+ 0.148 }_{- 0.094 }$ & $0.622^{+ 0.148 }_{- 0.094 }$ \\
        {\small CC+SN+BAO+$H_0^{\rm TRGB}$} & $0.703^{+ 0.099 }_{- 0.079 }$ & $0.693^{+ 0.109 }_{- 0.085 }$ & $0.695^{+ 0.112 }_{- 0.087 }$ & $0.699^{+ 0.106 }_{- 0.084 }$ \\
        \mr
        {\small CC+$H_0^{\rm HW}$} & $0.556^{+ 0.120 }_{- 0.083 }$ & $0.544^{+ 0.122 }_{- 0.085 }$ & $0.551^{+ 0.124 }_{- 0.087 }$ & $0.531^{+ 0.121 }_{- 0.085 }$ \\
        {\small CC+SN+$H_0^{\rm HW}$} & $0.679^{+ 0.196 }_{- 0.104 }$ & $0.666^{+ 0.197 }_{- 0.107 }$ & $0.671^{+ 0.200 }_{- 0.108 }$ & $0.661^{+ 0.201 }_{- 0.107 }$ \\
        {\small CC+SN+BAO+$H_0^{\rm HW}$} & $0.752^{+ 0.106 }_{- 0.085 }$ & $0.740^{+ 0.116 }_{- 0.091 }$ & $0.743^{+ 0.119 }_{- 0.094 }$ & $0.745^{+ 0.113 }_{- 0.091 }$ \\
        \br
    \end{tabular}
\end{table}

\section{Reconstruction of \texorpdfstring{$f(T)$}{fT} Gravity}
\label{sec:GPfT}

\begin{figure}[t!]
\begin{center}
    \includegraphics[width=0.48\columnwidth]{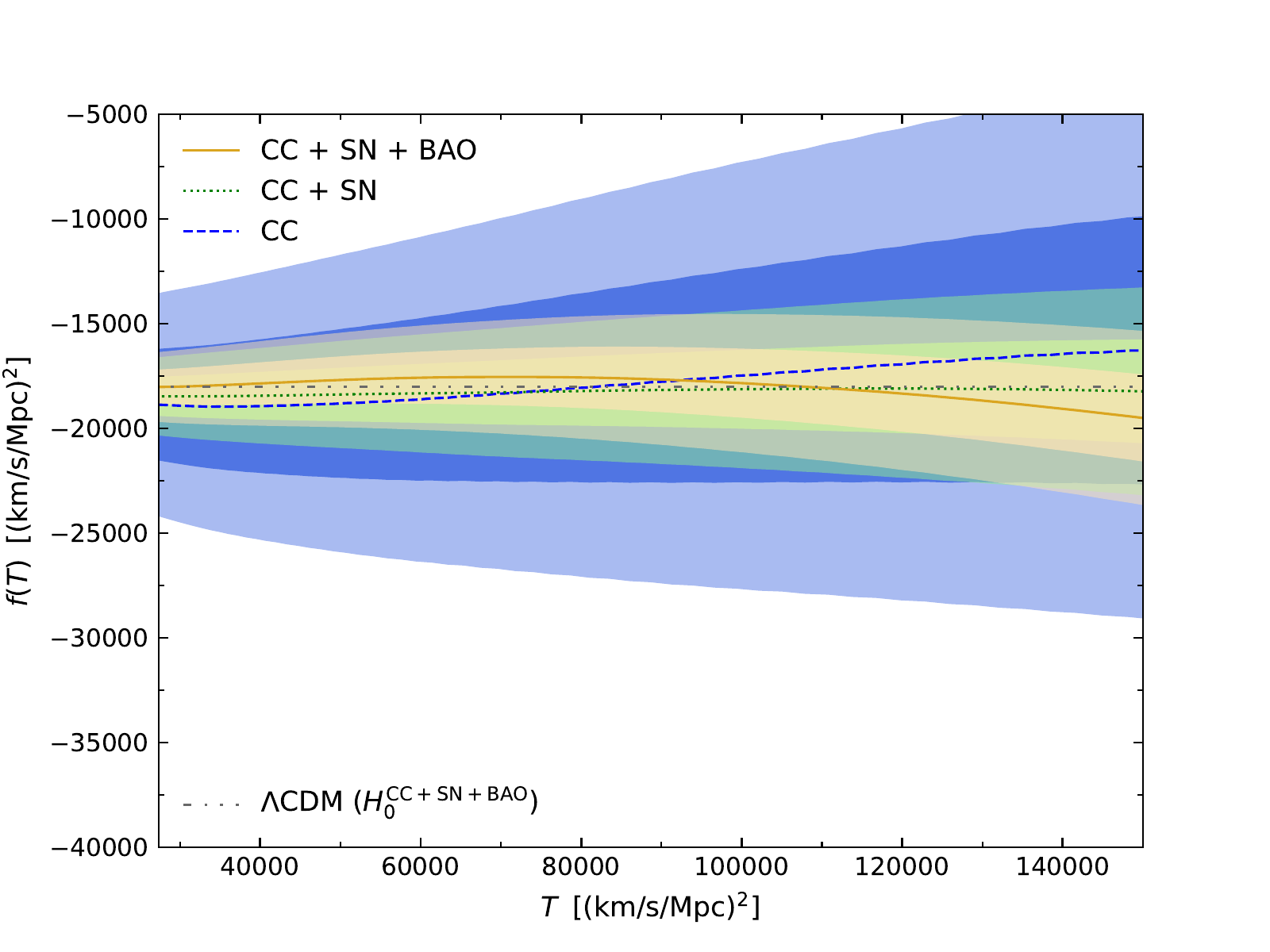}
    \includegraphics[width=0.48\columnwidth]{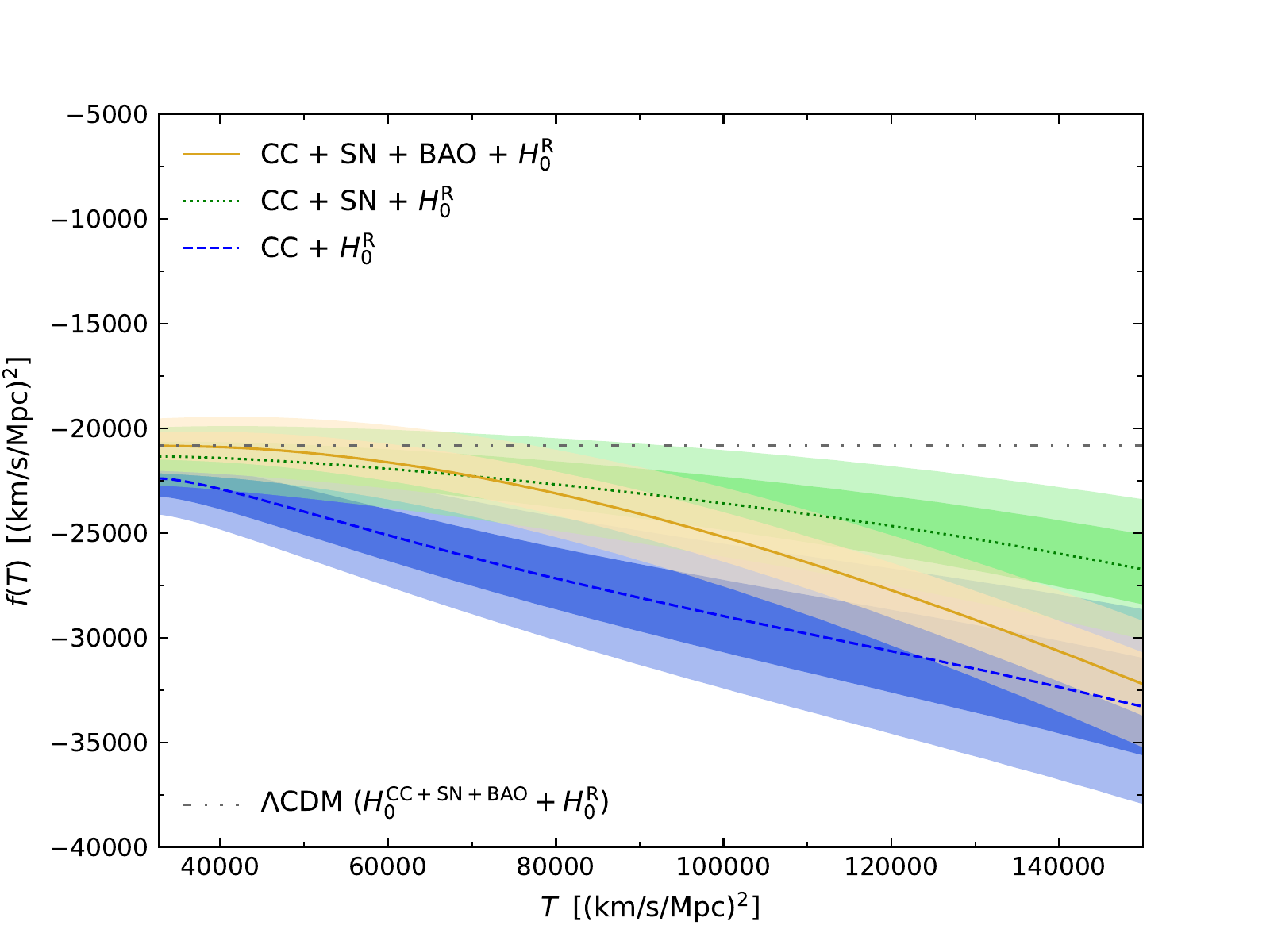}
    \includegraphics[width=0.48\columnwidth]{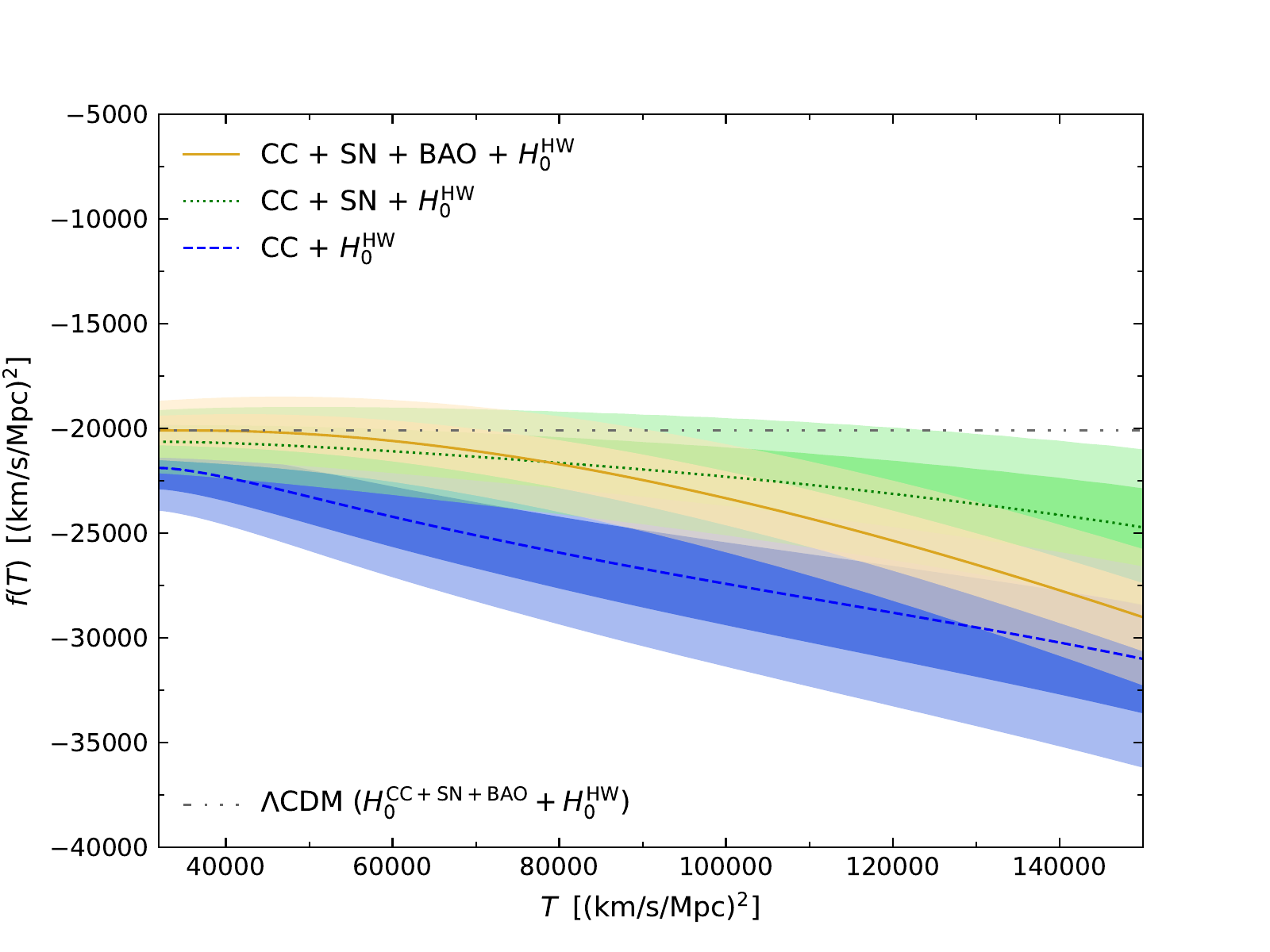}
    \includegraphics[width=0.475\columnwidth]{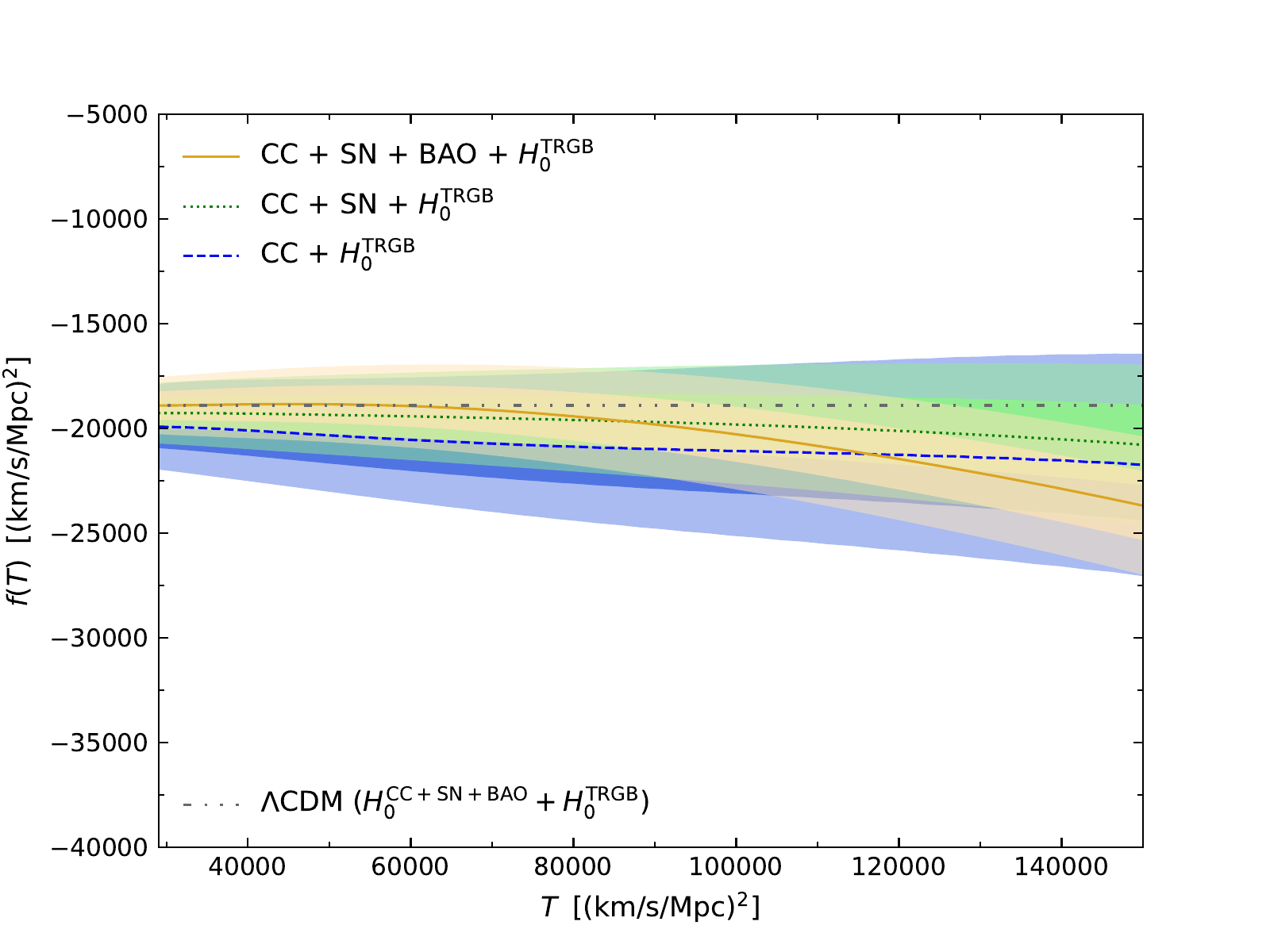}
    \caption{\label{fig:fT_squaredexp}
    GP reconstructions of $f(T)$ with the squared exponential kernel function of Eq.(\ref{eq:square_exp}). The data sets along with the different $H_0^{}$ priors are indicated in each respective panel.
    }
\end{center}
\end{figure}

\begin{figure}[t!]
\begin{center}
    \includegraphics[width=0.48\columnwidth]{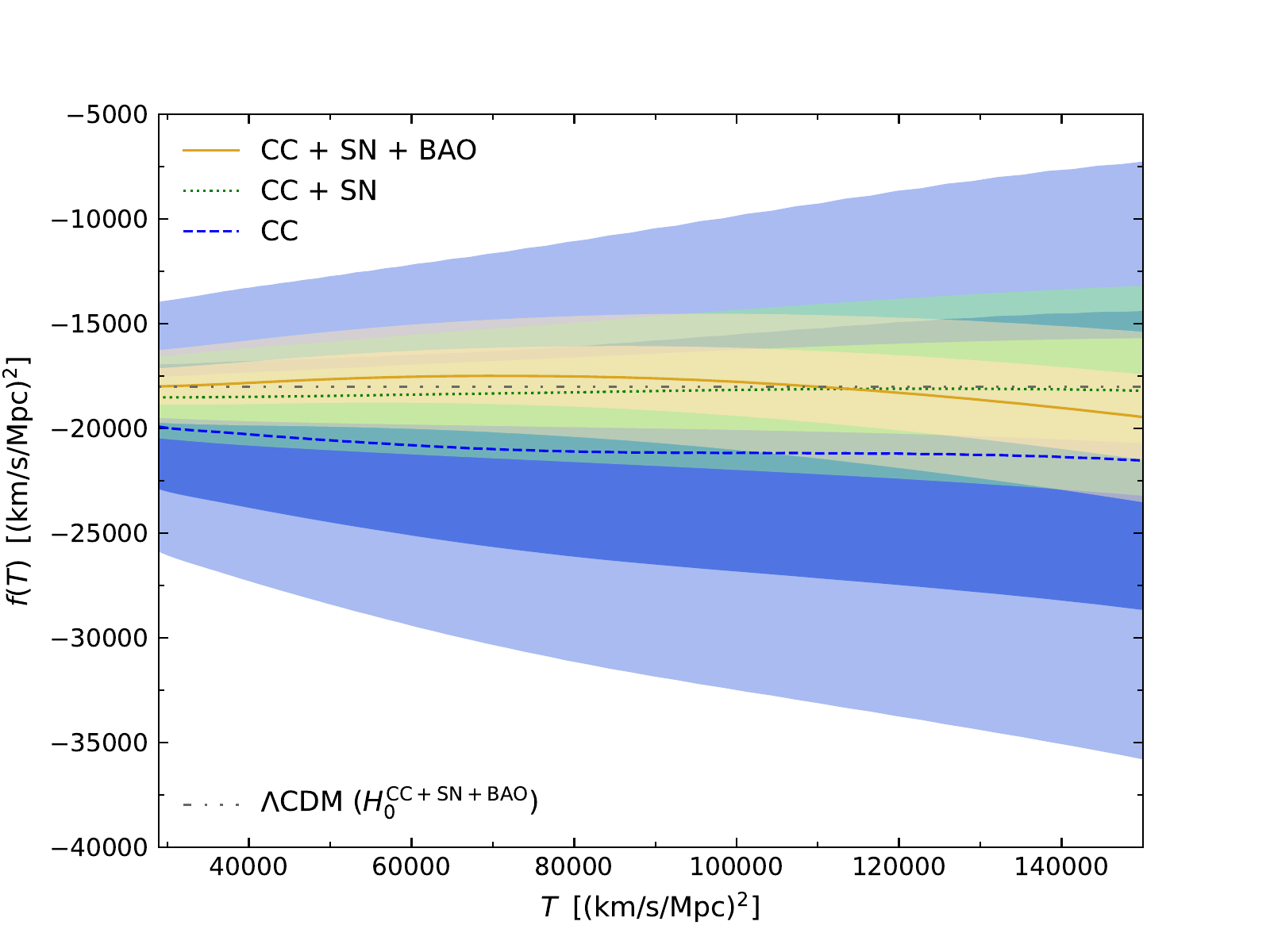}
    \includegraphics[width=0.48\columnwidth]{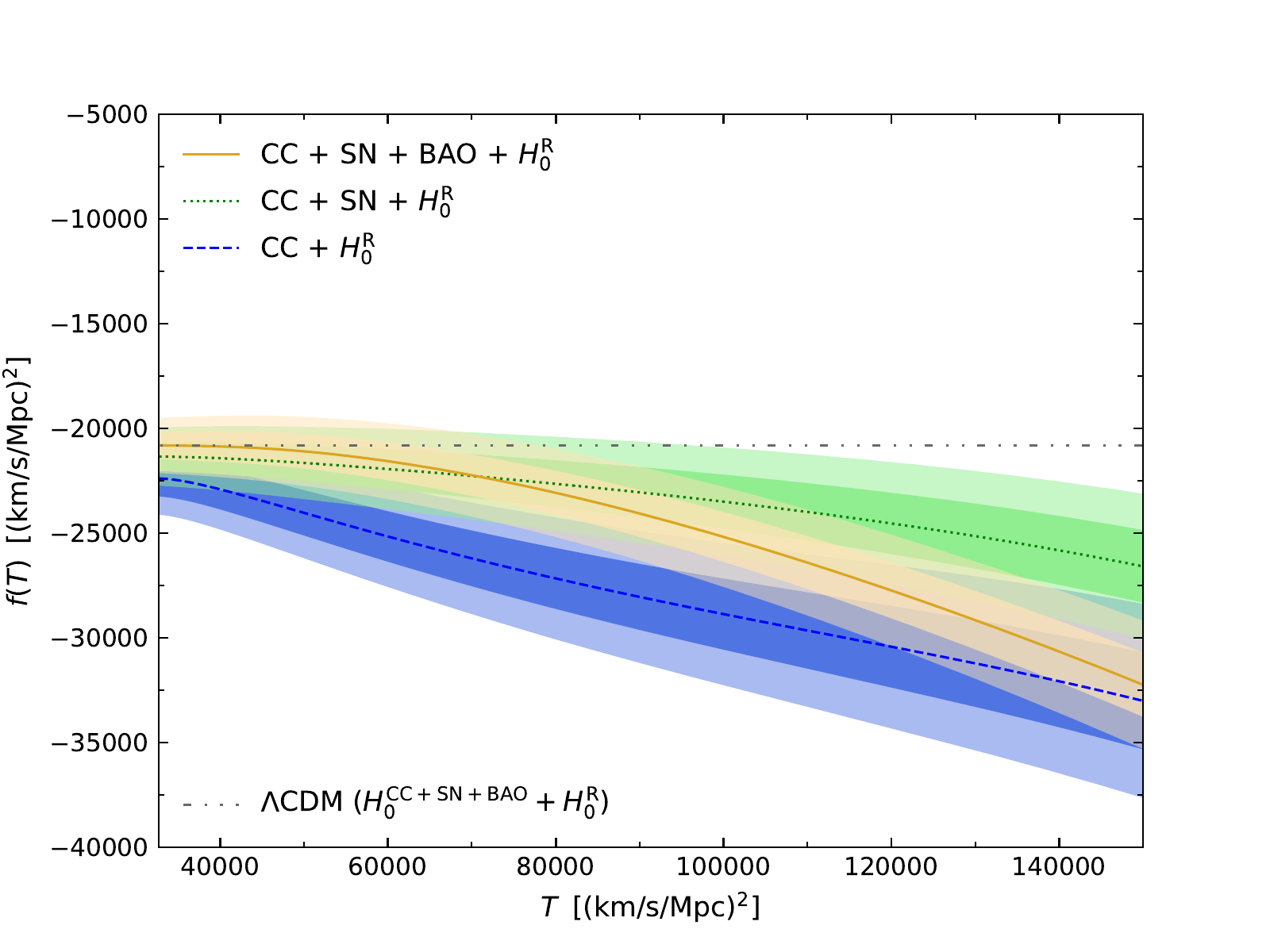}
    \includegraphics[width=0.48\columnwidth]{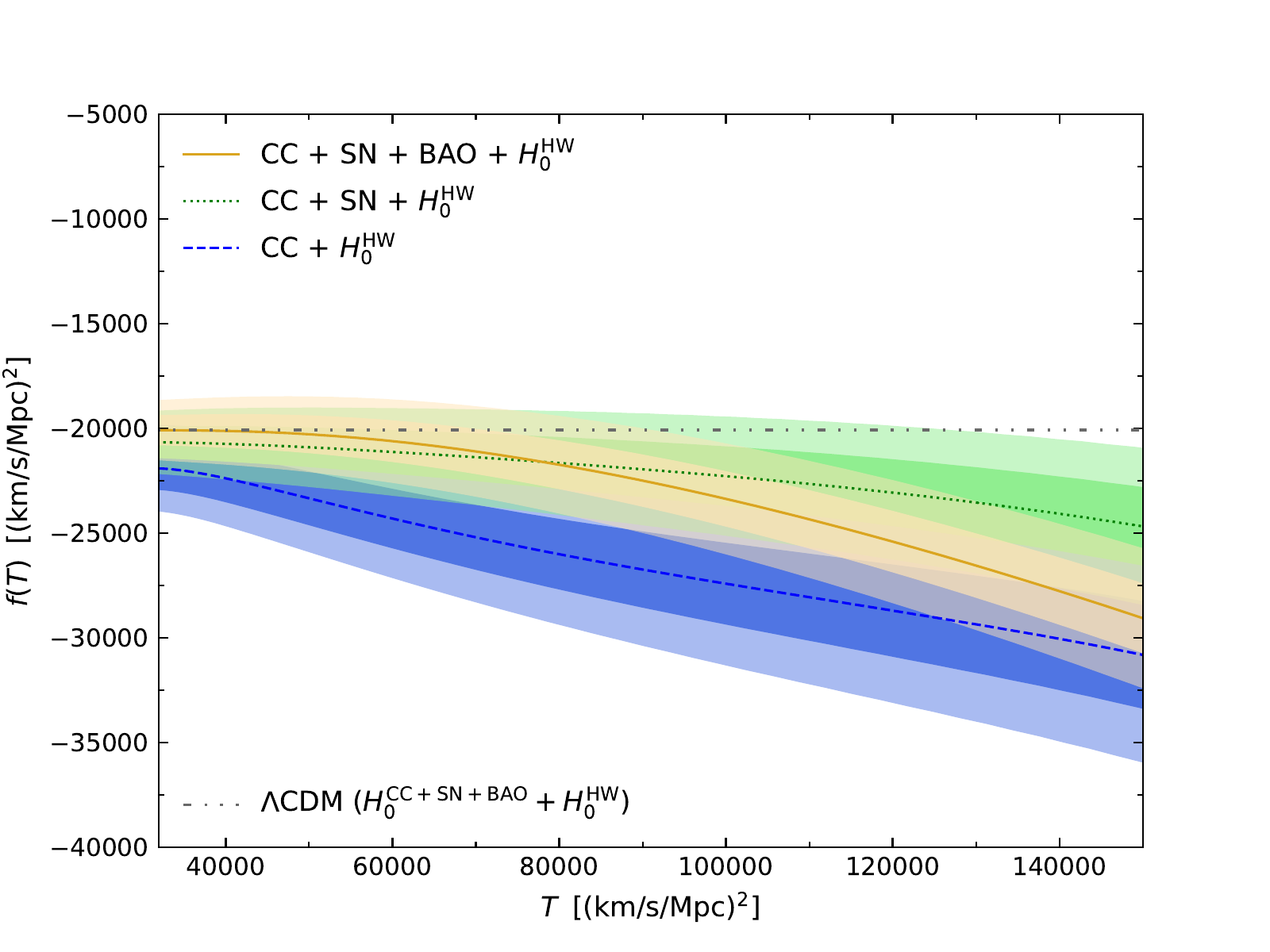}
    \includegraphics[width=0.475\columnwidth]{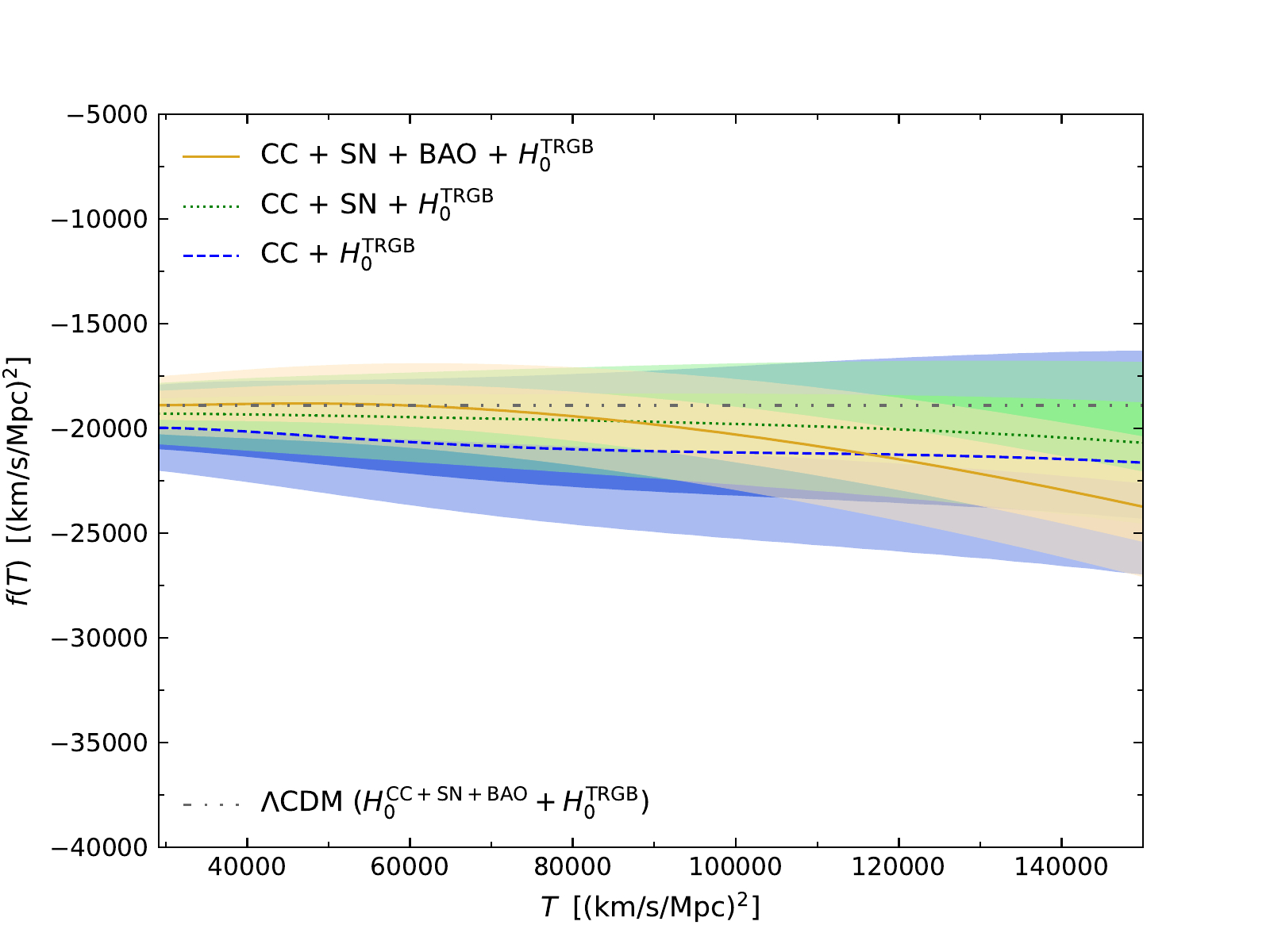}
    \caption{\label{fig:fT_cauchy}
    GP reconstructions of $f(T)$ with the Cauchy kernel function of Eq.(\ref{eq:cauchy}). The data sets along with the different $H_0^{}$ priors are indicated in each respective panel.
    }
\end{center}
\end{figure}

\begin{figure}[t!]
\begin{center}
    \includegraphics[width=0.48\columnwidth]{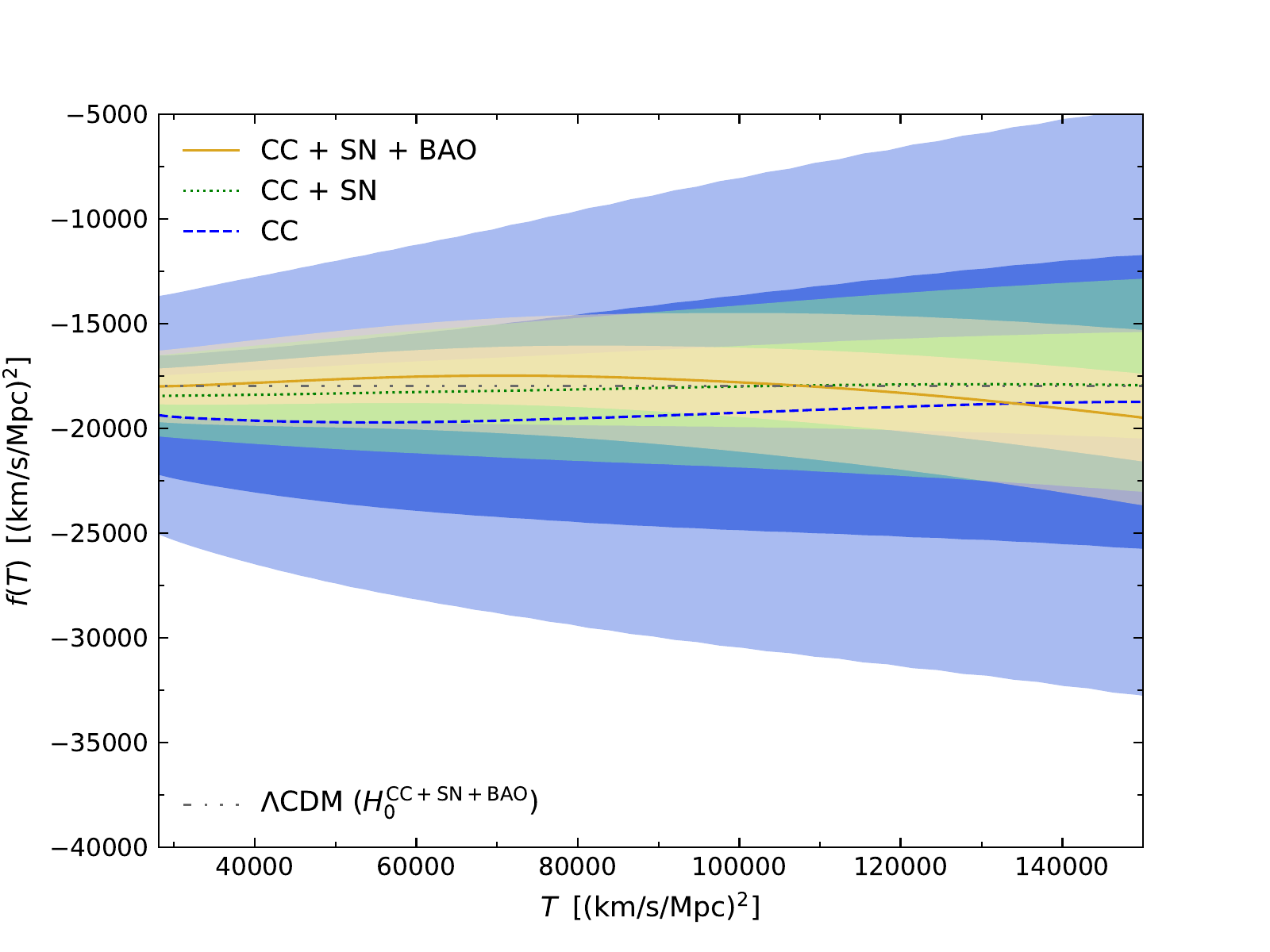}
    \includegraphics[width=0.48\columnwidth]{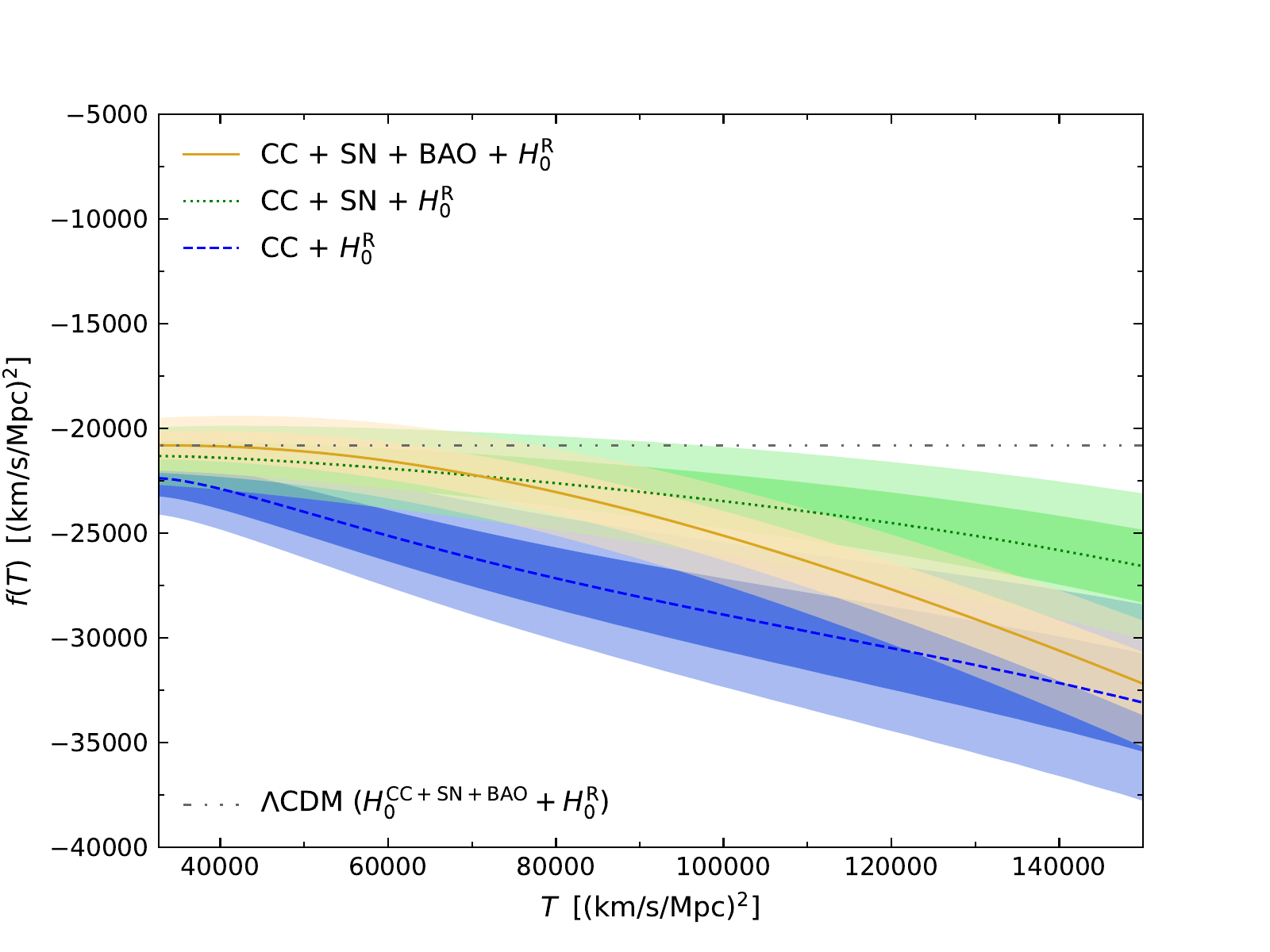}
    \includegraphics[width=0.48\columnwidth]{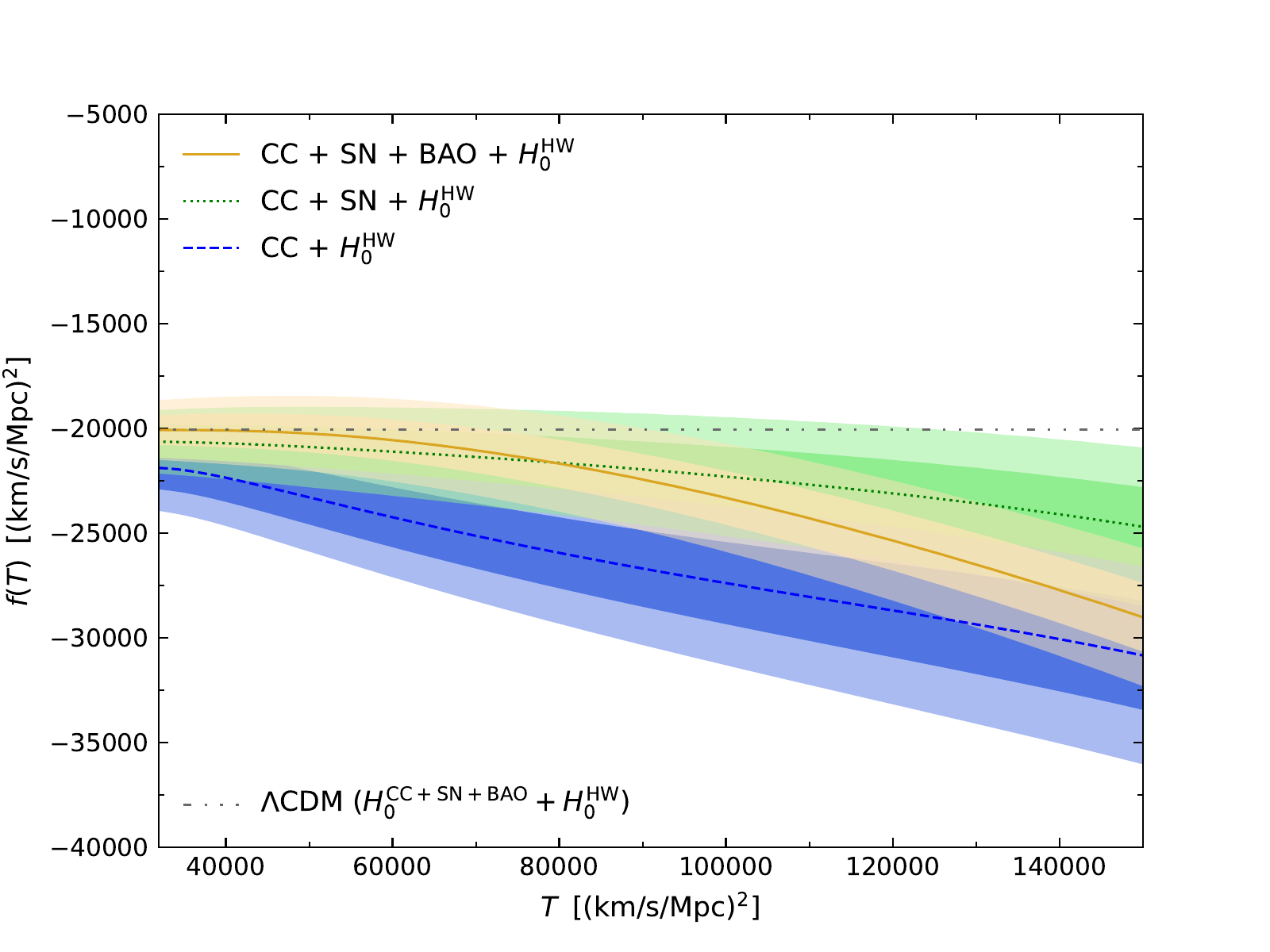}
    \includegraphics[width=0.475\columnwidth]{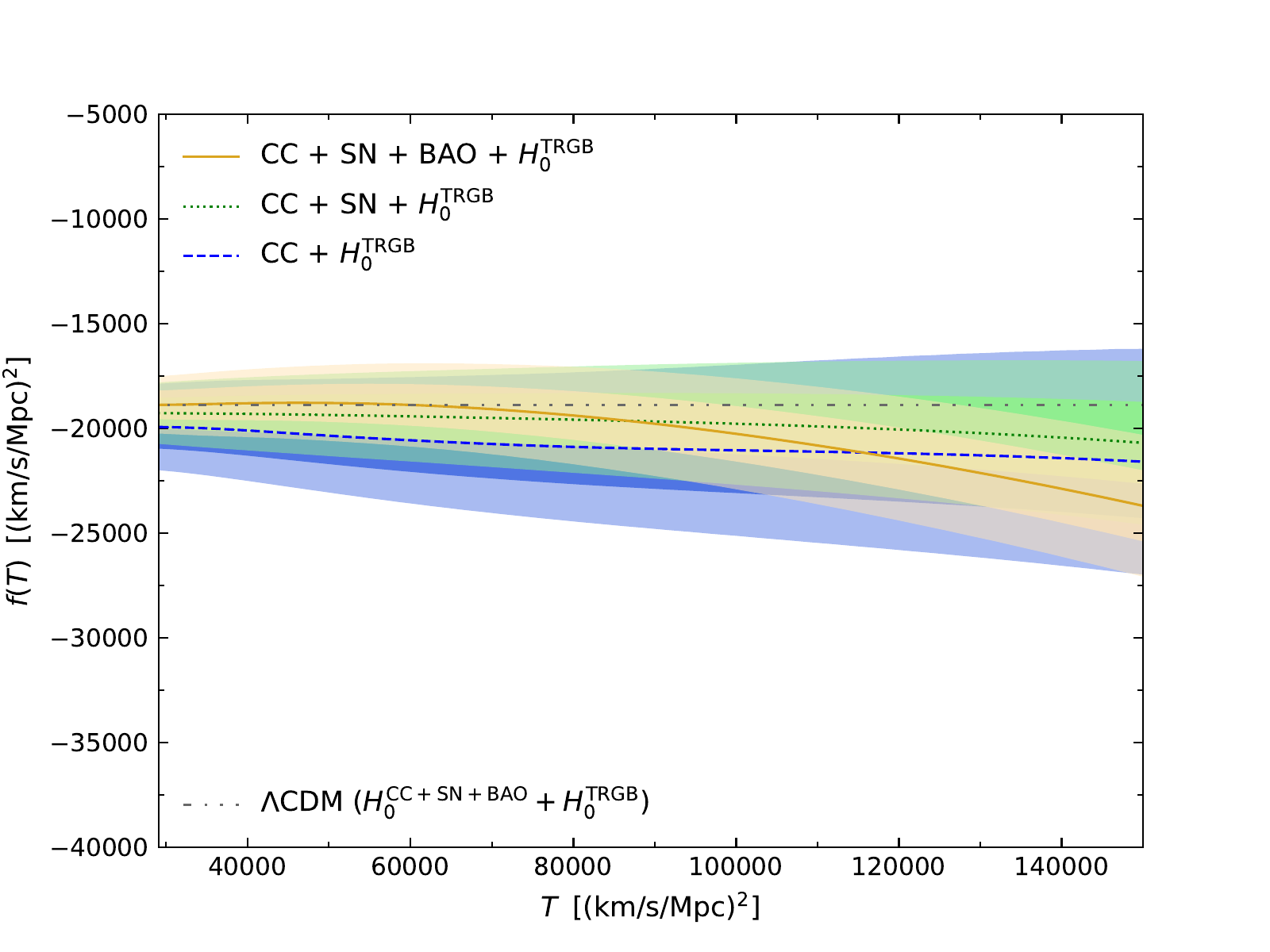}
    \caption{\label{fig:fT_matern}
    GP reconstructions of $f(T)$ with the Mat\'{e}rn kernel function of Eq.(\ref{eq:Matern}). The data sets along with the different $H_0^{}$ priors are indicated in each respective panel.
    }
\end{center}
\end{figure}

\begin{figure}[t!]
\begin{center}
    \includegraphics[width=0.48\columnwidth]{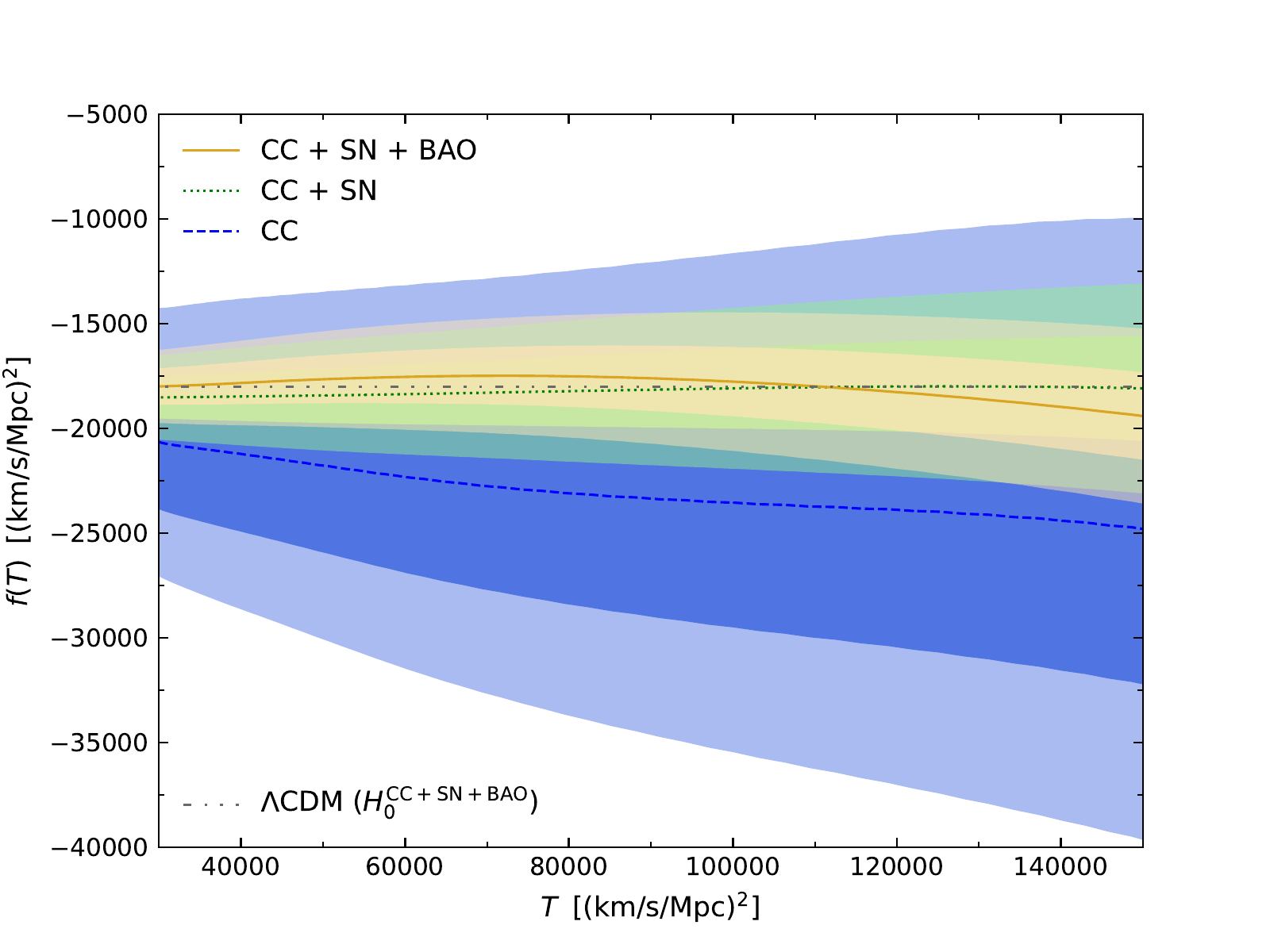}
    \includegraphics[width=0.48\columnwidth]{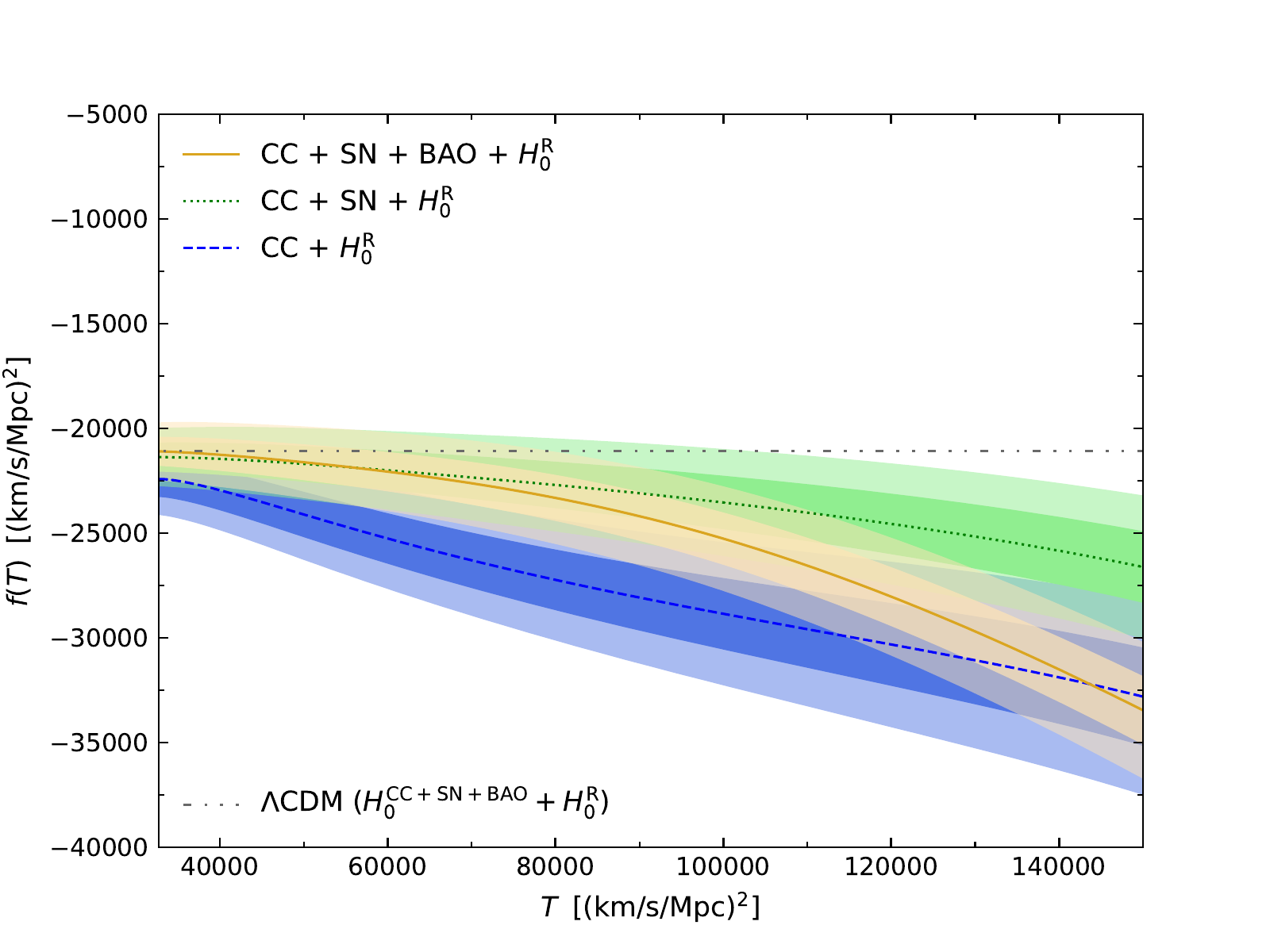}
    \includegraphics[width=0.48\columnwidth]{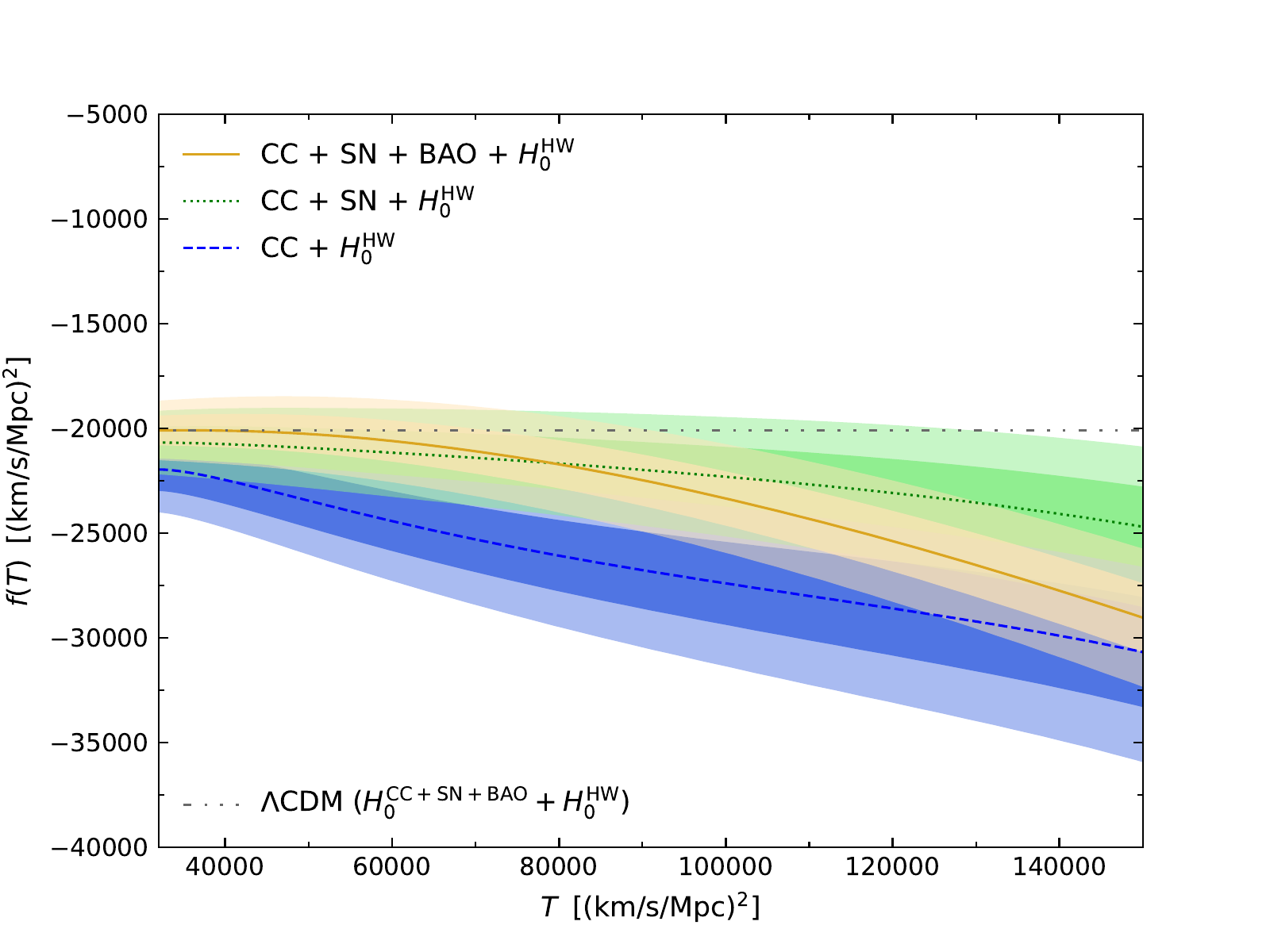}
    \includegraphics[width=0.475\columnwidth]{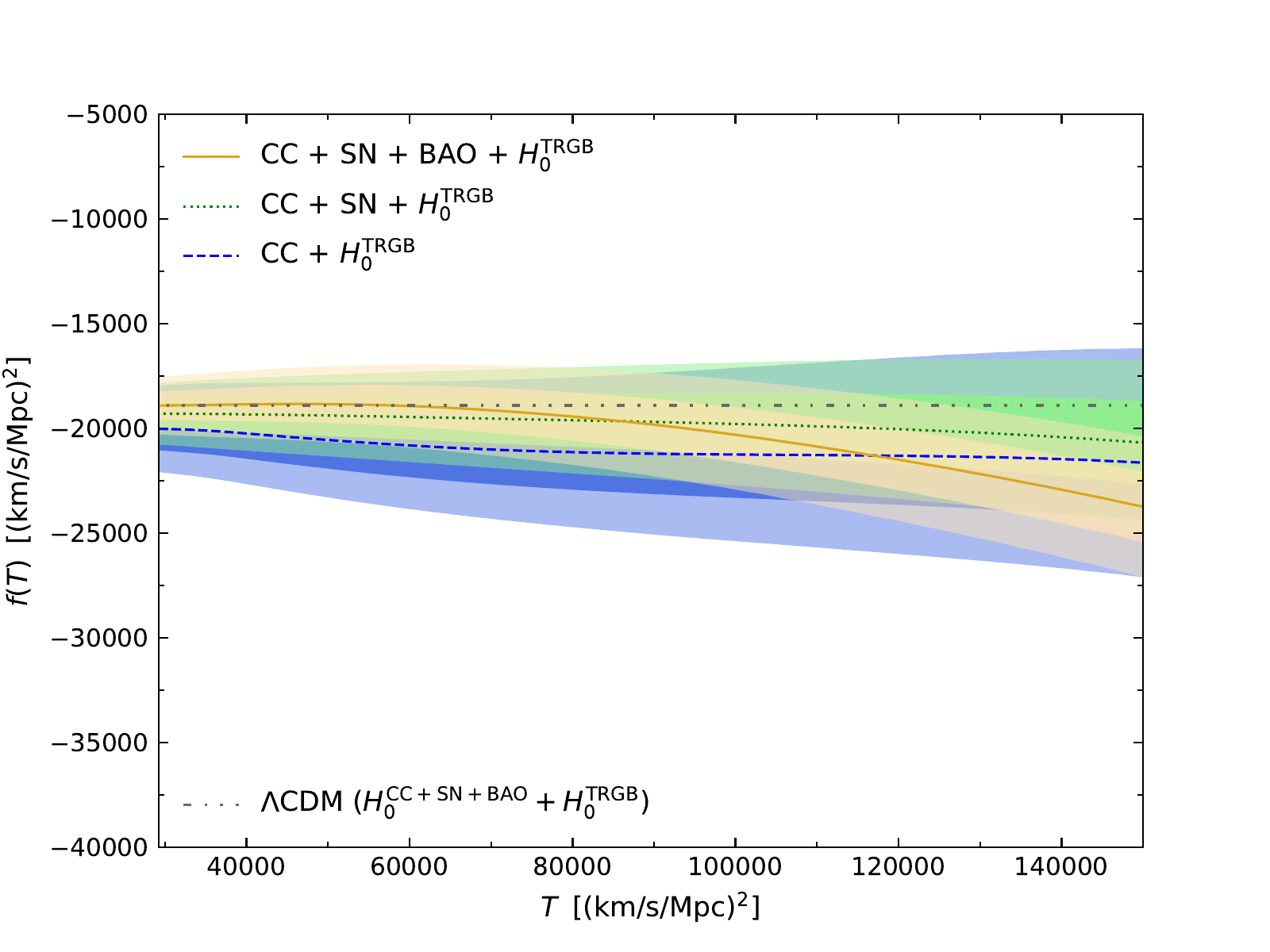}
    \caption{\label{fig:fT_rat_quad}
    GP reconstructions of $f(T)$ with the rational quadratic kernel function of Eq.(\ref{eq:rat_quad}). The data sets along with the different $H_0^{}$ priors are indicated in each respective panel.
    }
\end{center}
\end{figure}

The GP approach has thus far been used to reconstruct less noisy behaviours for observational data. Here, that reconstructed data will be used to construct data-driven models of $f(T)$ gravity, which is described in section \ref{sec:TG}. This means that the cosmological model-independent data that was reconstructed by the GP approach will now be used in conjunction with a general $f(T)$ dominated Universe without assuming a specific form of the arbitrary Lagrangian in Eq.(\ref{f_T_Lagrangian}). Similar to Ref.\cite{Cai:2019bdh}, we do this via an arbitrary $f(T)$ Lagrangian, but here we differentiate between the different data sets and moreover show how the different kernels reconstruct the Lagrangian.

The linchpin of this analysis rests on the relation between the $f(T)$ gravity scalar $T$ and the Hubble parameter which is represented by Eq.(\ref{Tor_sca_flrw}) which relates the GP reconstructions in section \ref{sec:GPH0} with the cosmological dynamics of $f(T)$ gravity. The cosmological dynamics of $f(T)$ gravity can be exposed by the Friedmann equation in Eq.(\ref{Friedmann_eq}).

In order to express the Friedmann equation in terms of redshift dependence alone, we first convert the Lagrangian derivative term so that
\begin{equation}
    f_T = \frac{\mathrm{d}f/\mathrm{d}z}{\mathrm{d}T/\mathrm{d}z} = \frac{f'(z)}{T'(z)}\,,
\end{equation}
where $f'(z) = \mathrm{d}f/\mathrm{d}z$ and $T'(z)= 12HH'$ are redshift derivatives. The immediate follow up becomes the issue of handling $f'(z)$ in the analysis which is tackled in this analysis through the central differencing method as
\begin{equation}
    f'(z_i) \simeq \frac{f(z_{i+1}) - f(z_{i-1})}{z_{i+1} - z_{i-1}}\,,
\end{equation}
since this will produce smaller uncertainties $\mathcal{O}(\Delta z^2)$ rather than $\mathcal{O}(\Delta z)$ which occur for the forward and backward differencing methods, where $\Delta z = z_{i+1} - z_{i-1}$. This method produces a numerical propagation equation for $f(z)$ given by
\begin{equation}\label{prop_eq_f_T}
    f(z_{i+1}) = f(z_{i-1}) + 2\left(z_{i+1} - z_{i-1}\right) \frac{H'(z_i)}{H(z_i)} \left(3H^2(z_i) + \frac{f(z_i)}{2} - 3 H_0^2 \Omega_{m}^0 \left(1+z_i\right)^3\right)\,,
\end{equation}
where the propagation equation parameters $H_0$ and $\Omega_{m}^{0}$ are selected from the corresponding GP reconstruction within the P18 and $H_0$ priors, respectively. While superior in terms of having lower associated uncertainties, the central differencing requires two initial conditions to be employed which we form as follows:
\begin{enumerate}
    \item Friedmann equation boundary condition: Evaluating the Friedmann equation in Eq.(\ref{Friedmann_eq}) at $z=0$ gives
    \begin{equation}\label{f_T_boundary_condition}
        f(z=0) \simeq 16\pi G \rho_{m}^0 - 6 H_0^2 = 6H_0^2\left(\Omega_{m}^0 - 1\right)\,,
    \end{equation}
    where we have imposed that $\Lambda$CDM dominates at present times, i.e. $f_T(z=0) \simeq 0$. This again relies on the same parameters as the propagation equation itself;
    \item The second boundary condition can be obtained by using the forward differencing method through
    \begin{equation}\label{f_T_2nd_boundary_condition}
        f'(z_i) \simeq \frac{f(z_{i+1}) - f(z_{i})}{z_{i+1} - z_{i}}\,,
    \end{equation}
    that leads to the equation
    \begin{equation}\label{f_T_boundary_condition_2}
        f(z_{i+1}) = f(z_{i}) + 6 \left(z_{i+1} - z_{i}\right) \frac{H'(z_{i})}{H(z_{i})} \left[H^2(z_{i}) + \frac{f(z_{i})}{6} - H_0^2 \Omega_{m}^0 \left(1+z_{i}\right)^3\right]\,,
    \end{equation}
    which straightforwardly leads to the necessary second boundary condition. 
\end{enumerate}
The propagation of the $f(T)$ function is complemented by its associated Monte Carlo error propagation which produces the 1$\sigma$ and 2$\sigma$ uncertainties.

Together the propagation equation in Eq.(\ref{prop_eq_f_T}) along with the boundary conditions in (i) and (ii) can express the redshift dependent Lagrangian $f(z)$ in terms of $z$ in a model-independent way. Similarly, the corresponding torsion scalar can be associated with each of the redshift values in question through the Hubble parameter relation in Eq.(\ref{Tor_sca_flrw}). In this way, the Lagrangian function $f(T)$ can be plotted as a function of the torsion scalar $T$.

For each of the kernel and prior choices provided in Figs. \ref{fig:H_squaredexp}--\ref{fig:H_ratquad}, the $f(T)$ reconstructions against the torsion scalar are illustrated in Figs. \ref{fig:fT_squaredexp}--\ref{fig:fT_rat_quad} where the $1\sigma$ and $2\sigma$ regions are shown in every case. We mention that the $\Lambda$CDM paradigm appears as a constant in these plots with a value of $f(T)\rightarrow6H_0^2(\Omega_m^0-1)$, which is denoted by horizontal lines in Figs. \ref{fig:fT_squaredexp}--\ref{fig:fT_rat_quad}.

As we observe from Figs. \ref{fig:fT_squaredexp}--\ref{fig:fT_rat_quad}, for all kernels and considered datasets, the $\Lambda$CDM scenario lies inside the reconstructed region. Nevertheless, the GP reconstruction procedure shows a slight tendency of $f'(T)$ to negative values, i.e. to $f(T)$ forms that are slightly decreasing functions of $T$. This is the main result of the present work and the aforementioned feature needs to be taken into account in the $f(T)$ model building.

In the case where no prior is used for the GP reconstruction, the $f(T)$ evolution remains within the 2$\sigma$ confidence region of $\Lambda$CDM for the breadth of the evolution interval, and mostly within the 1$\sigma$ region for the data set combinations. The furthest propagated line to the $\Lambda$CDM is the Hubble parameter reconstructed from the combined data that includes the BAO data set. On the other hand, once the priors are included the situation changes drastically, with the $H_0^{\rm R}$ prior favouring a slight deviation from $\Lambda$CDM over a portion of the cosmic evolution for all datasets. The same situation, but to a lesser extent, occurs for the $H_0^{\rm HW}$ prior, with the $H_0^{\rm TRGB}$ prior being the only one to not affect this propagation in a significant way. These $f(T)$ propagations show that the datasets alone favour an $f(T)$ that only slightly deviates from $\Lambda$CDM within the redshift region being probed, while the propagations that do contain literature priors prefer a stronger deviation from $\Lambda$CDM, however the latter is inside the allowed regions.

The same general situation is found for the Cauchy, Mat\'{e}rn and rational quadratic kernels in Figs. \ref{fig:fT_cauchy}--\ref{fig:fT_rat_quad}. This reinforces the general conclusions that all dataset combinations alone suggest a slight deviation from $\Lambda$CDM within this redshift region.

\section{Conclusions}
\label{sec:conclusions}

Gaussian processes offer an approach to reconstruct the underlying functional behaviour of a variable of a stochastic process. In the cosmological setting under investigation GP represent an effective tool to treat cosmological parameters, such as the Hubble parameter whose present value has been the focus of intense debate in recent years. Furthermore, we utilise the method first suggested in Ref.\cite{Cai:2019bdh}, where GP are used to reconstruct the arbitrary Lagrangian in theories beyond $\Lambda$CDM without imposing a prior ansatz form of this function. This approach could offer a novel technique by which theories beyond $\Lambda$CDM are constructed.

In order to accomplish this procedure, we applied the GP approach to the Friedmann equations of $f(T)$ gravity, and also reconstructed the Hubble function in a model-independent way. We considered four kernels in order to examine whether any dependence on the statistical model exists, which was indeed verified at 1$\sigma$ confidence level. As observed in Tables \ref{tab:se_kernel}--\ref{tab:rq_kernel}, we consider the case of CC, CC+SN, and CC+SN+BAO datasets separately for these kernels, with separate $H_0^{\rm R}$, $H_0^{\rm TRGB}$ and $H_0^{\rm HW}$ priors. In particular, we found the most conservative values of $H_0$ with the CC+SN+BAO dataset combination. However, it is worth nothing that the BAO data does have some dependence on the $\Lambda$CDM model. Also, in these GP reconstructions, the highest values of $H_0$ came from CC data, with CC+SN having the effect of lowering the tendency of the Hubble parameter behaviour at intermediate redshift values. Moreover, the highest values of the $H_0$ parameter were inferred from the $H_0^{\rm R}$ prior, since it is the highest prior value which is even larger than the instance of no priors. The Hubble parameter behaviours are plotted in Figs. \ref{fig:H_squaredexp}--\ref{fig:H_ratquad} for the full range of redshifts under investigation together with the 1$\sigma$ and 2$\sigma$ confidence levels.

In Figs. \ref{fig:L1_cauchy}--\ref{fig:L1_ratquad} we depicted the corresponding diagnostic test for the $\Lambda$CDM paradigm, which is developed in Eq.(\ref{diagnostic_for_H}).
The reconstructions remained within the 1$\sigma$ confidence levels for low redshifts but this situation changed as we explored high redshifts across the various dataset choices and kernel combinations. Another interesting result was the value of the transition redshift, which denotes the point at which the Universe transitioned from a decelerating to an accelerating phase, which was shown in table \ref{tab:kernels_z_t}. This is determined using the corresponding GP reconstructed deceleration parameters in \ref{app_par}, where we also presented the dark energy EoS.

Given the Hubble parameter GP reconstructions we proceeded to determine how this feeds into the model-independent reconstruction of the $f(T)$ function. We laid out the numerical propagation equation for the $f(T)$ arbitrary function along with the boundary conditions necessary for this setup. Hence, we were able to present the model-independent reconstructed region for the $f(T)$ form as a function of $T$. We showed that for all kernels and considered datasets, the $\Lambda$CDM scenario lies inside the reconstructed region, however there is a slight tendency towards $f(T)$ forms that are slightly decreasing functions of $T$, and this is more intense at higher redshifts. Hence, this feature needs to be taken into account in the current and future $f(T)$ model building.

As final remark, it is worth noticing that combining the present GP procedure with cosmography, based on suitable polynomial functions, could be a robust approach to obtain a self-consistent cosmic history for large ranges of redshift range. For instance, in Ref.\cite{Capozziello:2019cav} examples of cosmological models derived from modified gravity are reconstructed by cosmography. In particular, for $f(T)$ gravity models, slight deviations from $\Lambda$CDM are obtained at large redshifts. In a forthcoming paper, GP approach and cosmography will be both considered to constrain reliable cosmological models.

\ack\label{sec:acknowledgements}
This article is based upon work from CANTATA COST (European Cooperation in Science and Technology) action CA15117, EU Framework Programme Horizon 2020. Additionally it is supported in part by the USTC Fellowship for international professors. The authors would like to acknowledge networking support by the COST Action CA18108 and funding support from Cosmology@MALTA which is supported by the University of Malta.

\appendix
\section{Gaussian processes reconstruction of the deceleration parameter, of \texorpdfstring{$\mathcal{O}_m^{(1)}(z)$}{}, and of the dark energy equation of state}\label{app_par}
\setcounter{section}{1}

\begin{figure}[t!]
\begin{center}
    \includegraphics[width=0.46\columnwidth]{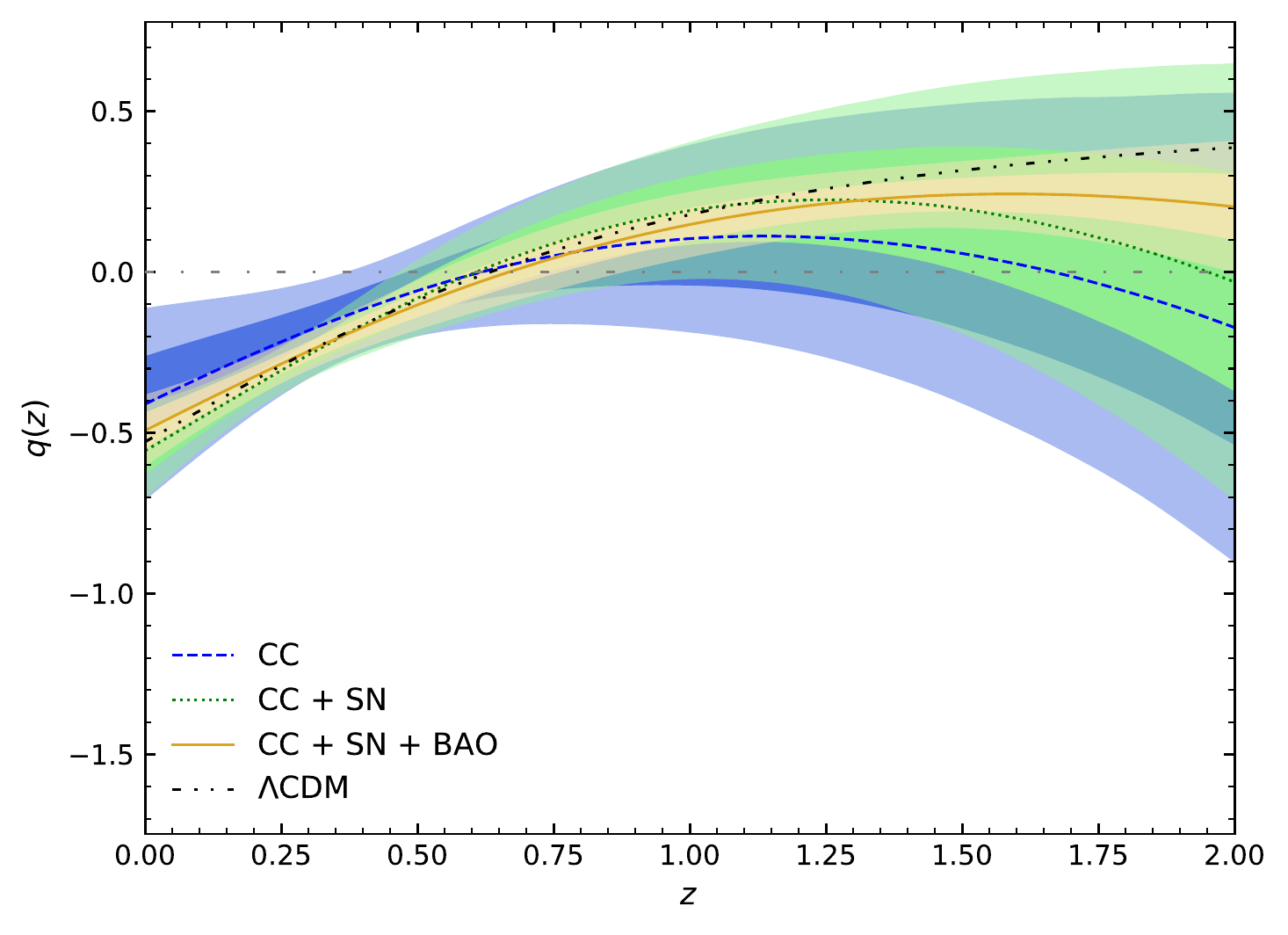}
    \includegraphics[width=0.46\columnwidth]{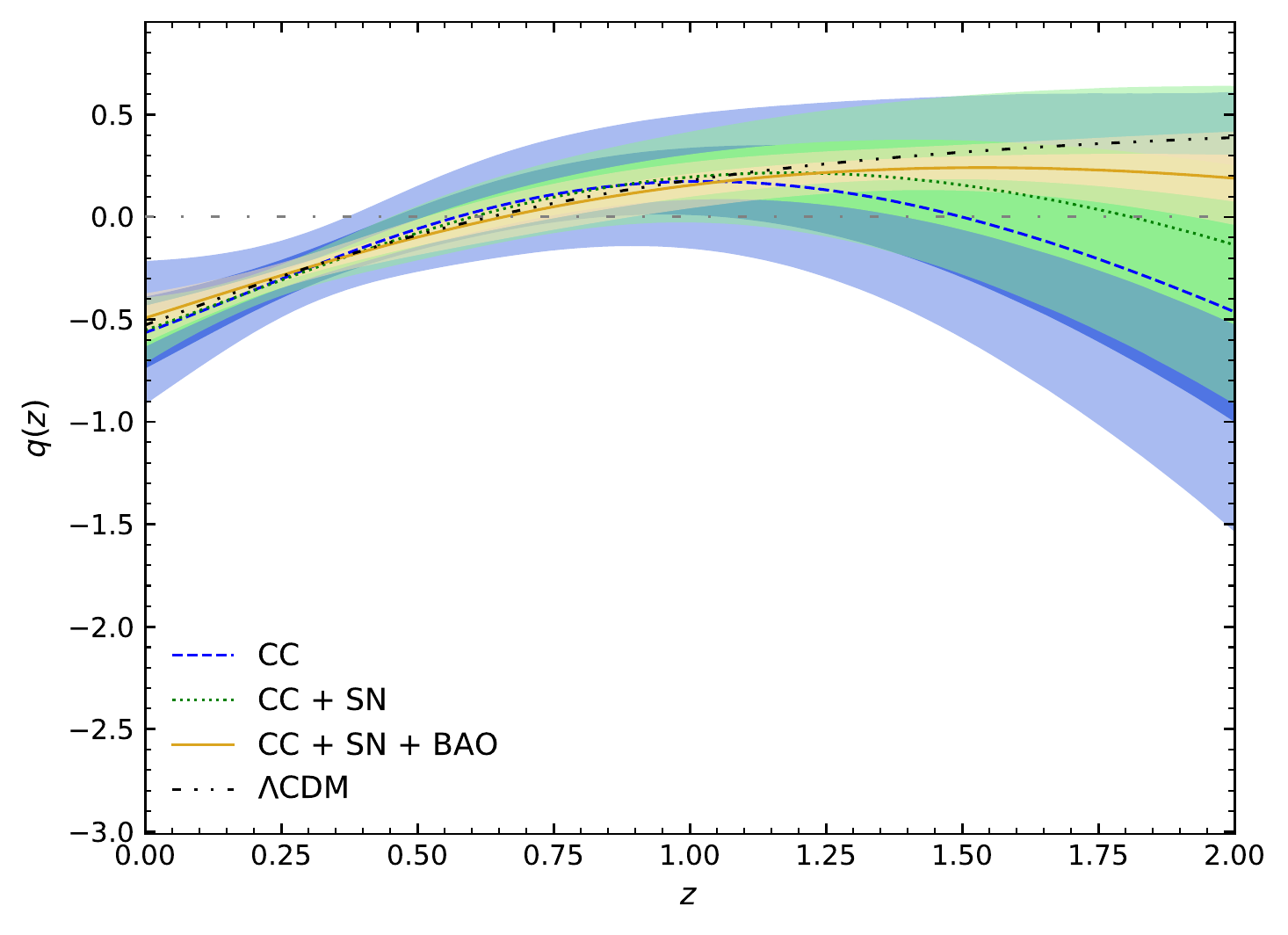}
    \includegraphics[width=0.46\columnwidth]{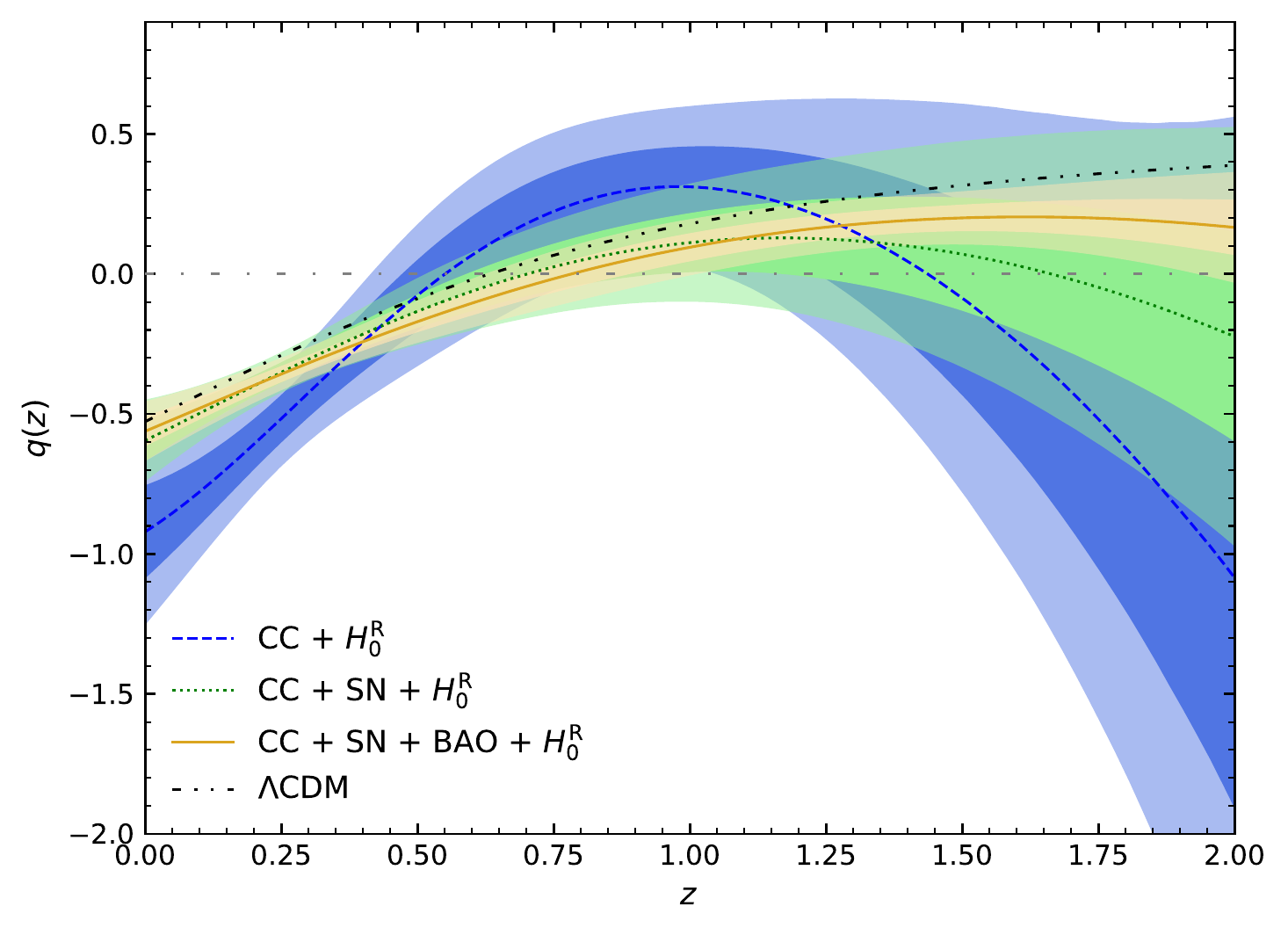}
    \includegraphics[width=0.46\columnwidth]{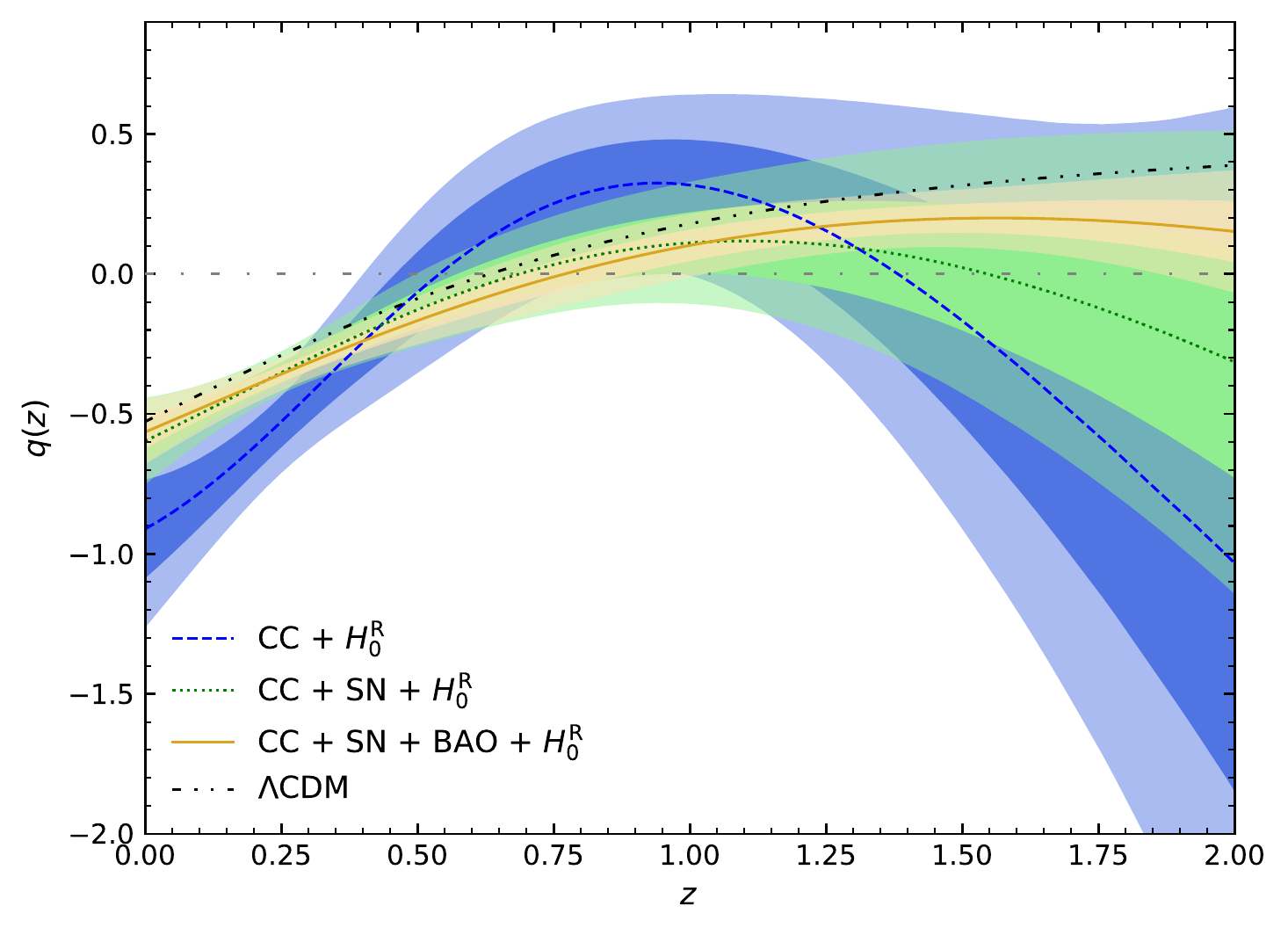}
    \includegraphics[width=0.46\columnwidth]{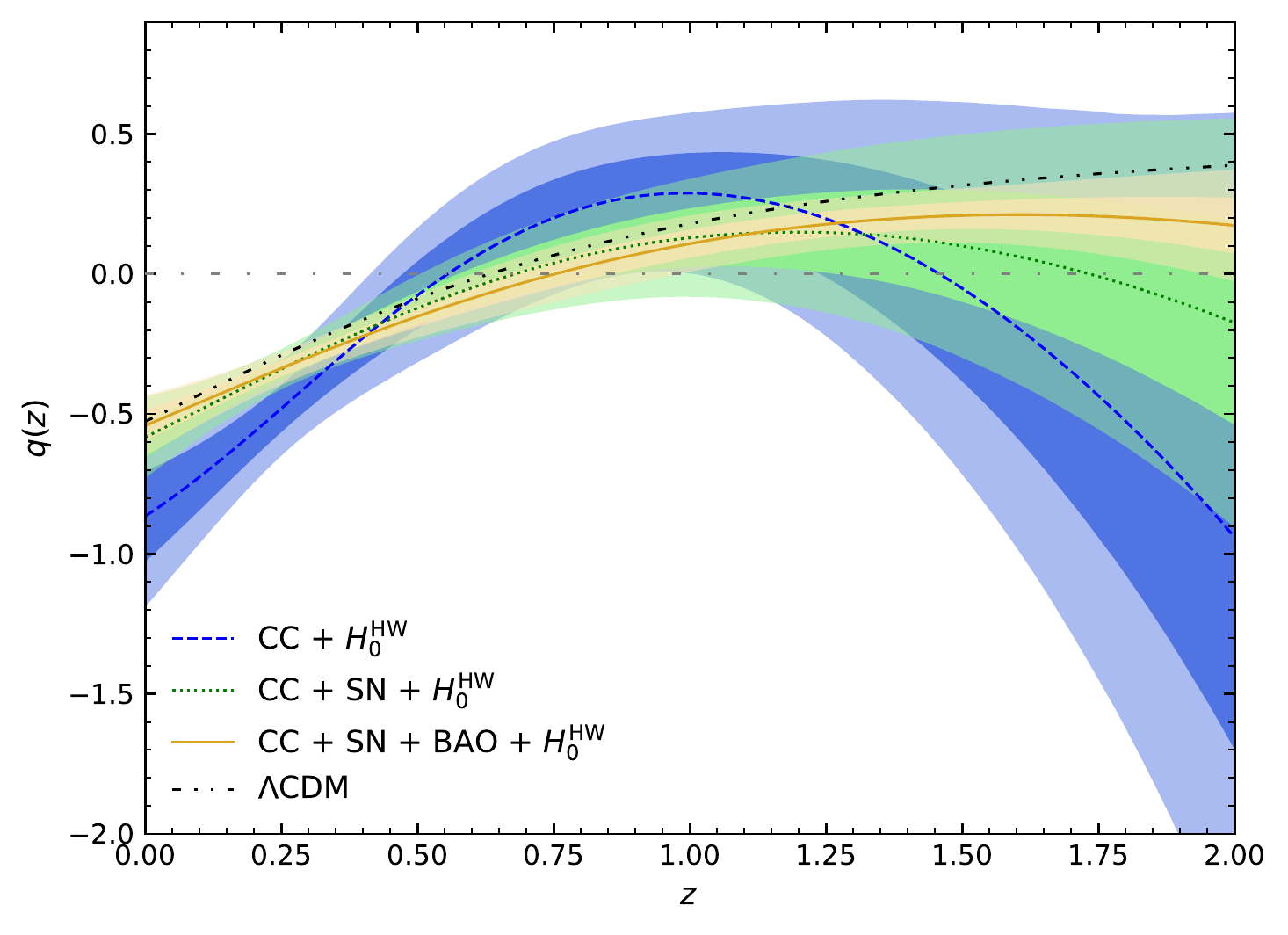}
    \includegraphics[width=0.46\columnwidth]{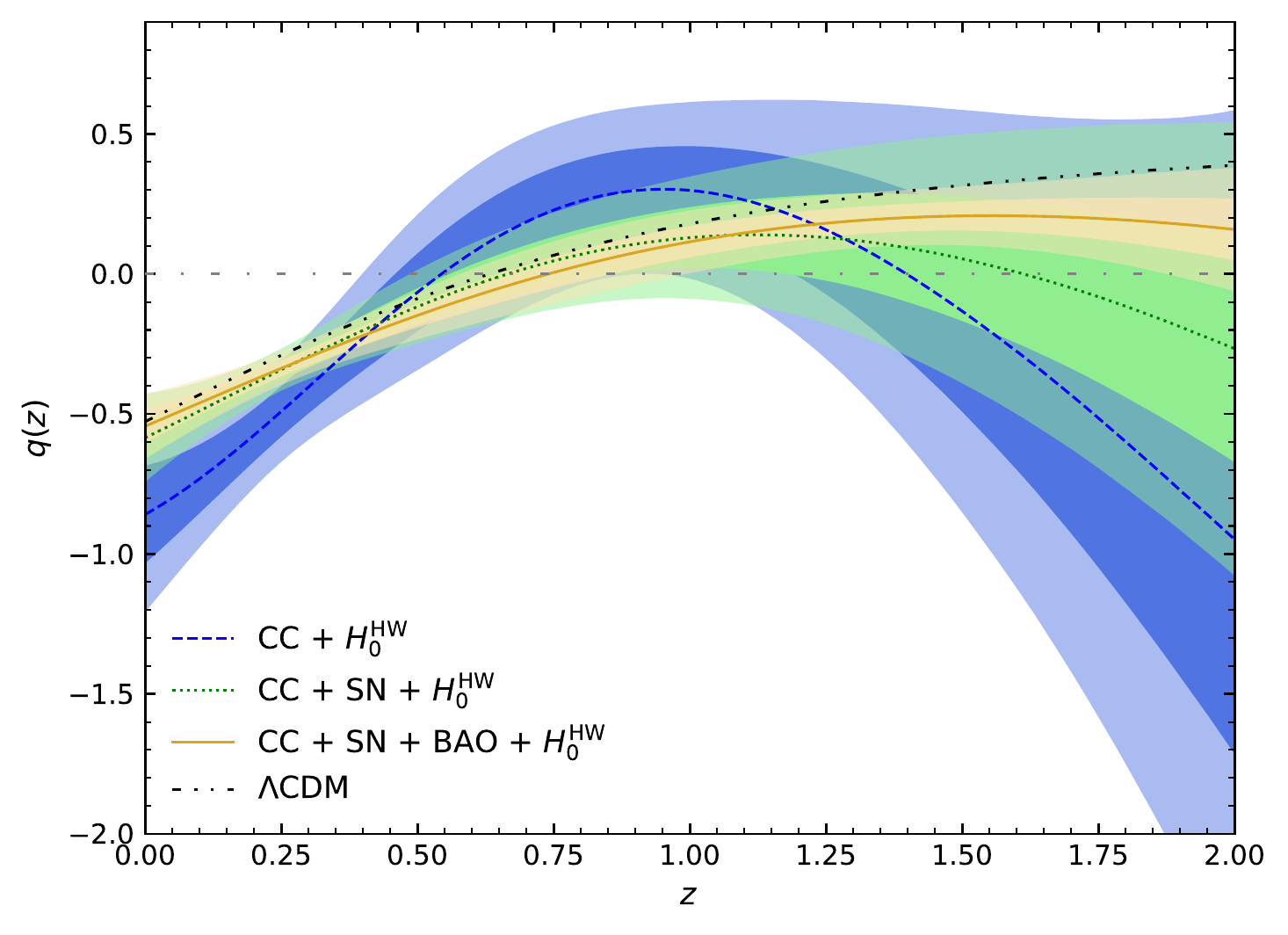}
    \includegraphics[width=0.46\columnwidth]{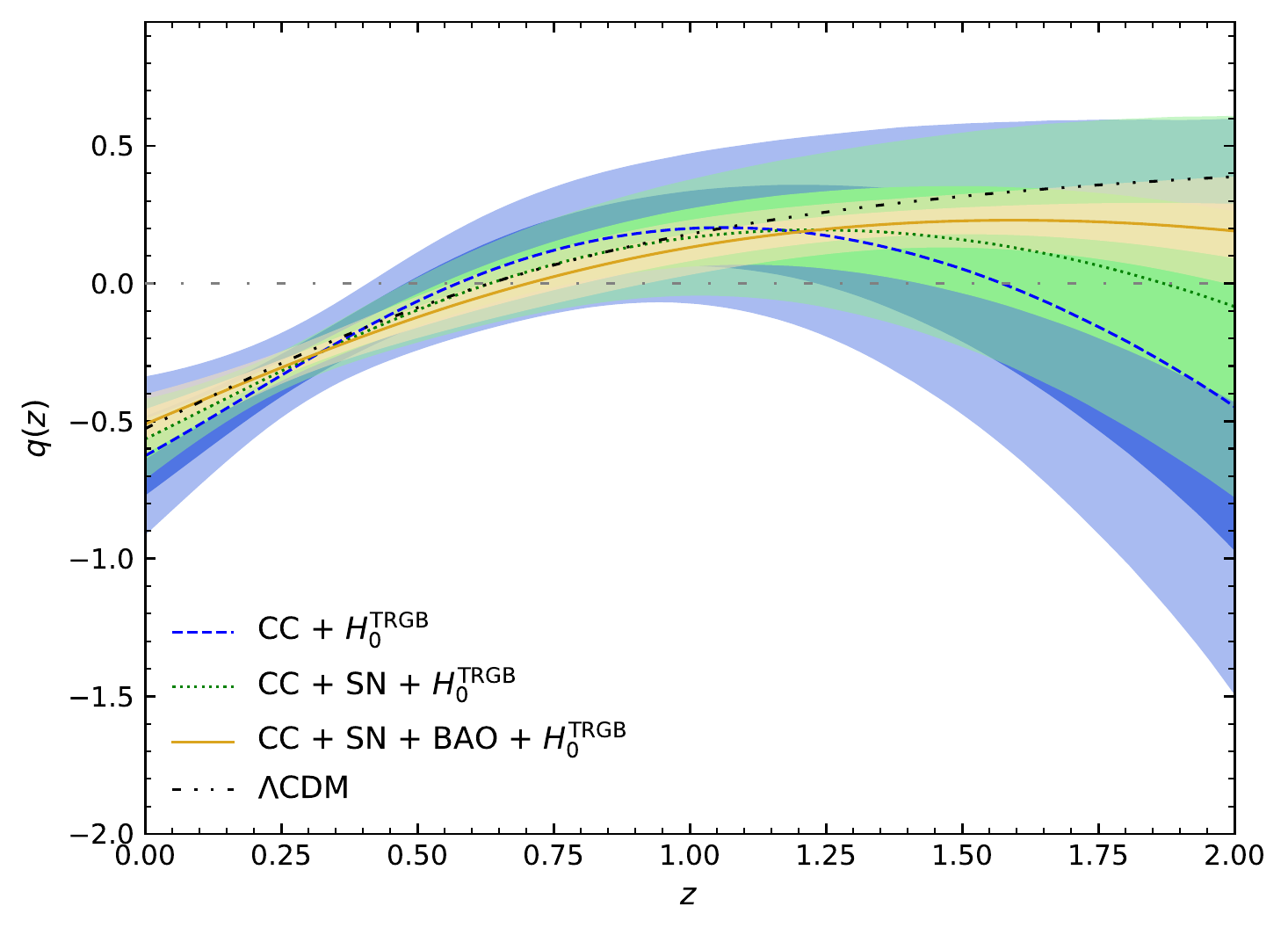}
    \includegraphics[width=0.46\columnwidth]{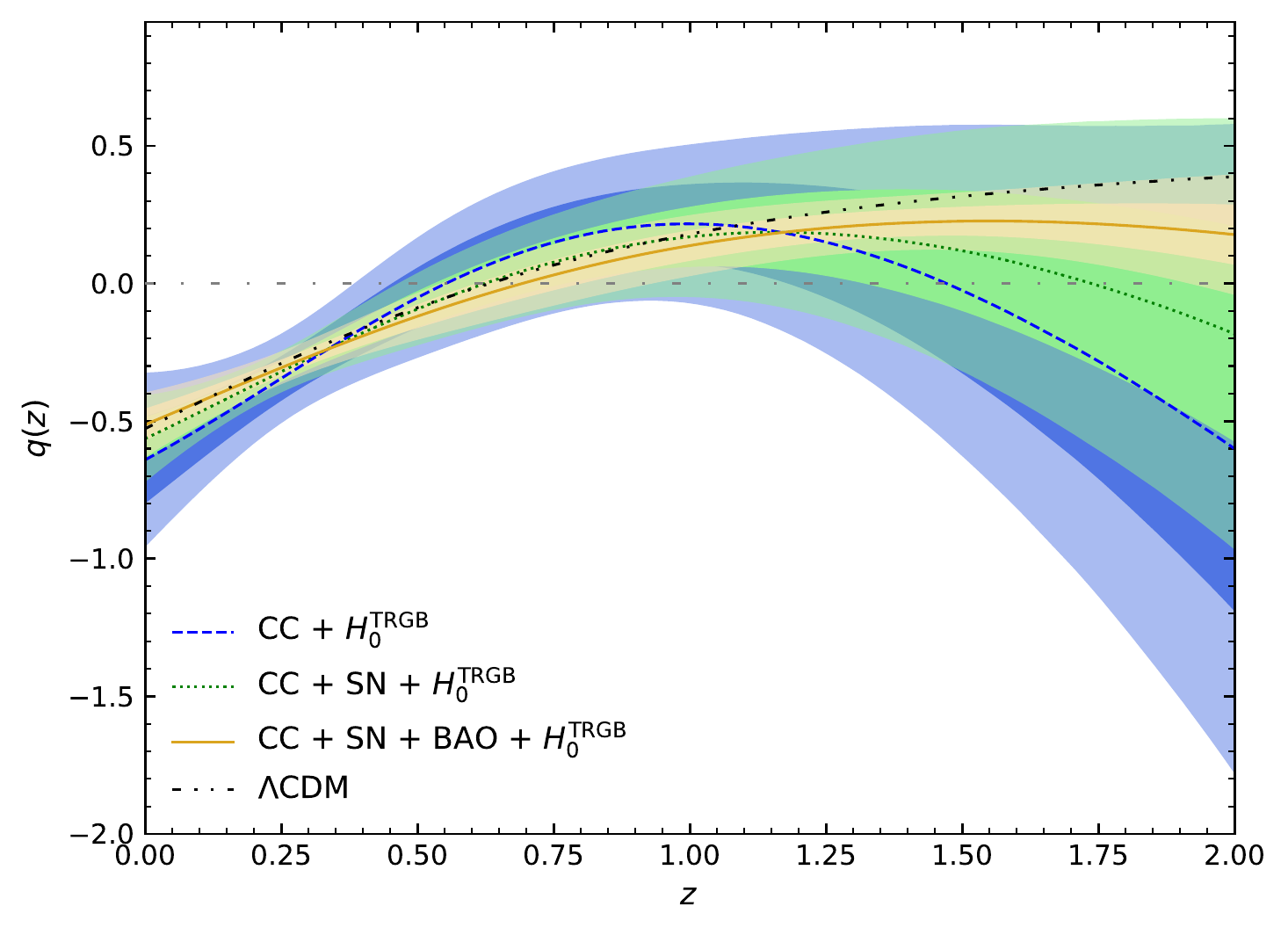}
    \caption{\label{fig:qz_squaredexp_cauchy}
    GP reconstructions of $q(z)$ with the squared exponential (left) and Cauchy (right) kernel functions, along with the $\Lambda$CDM prediction.
    }
\end{center}
\end{figure}

\begin{figure}[t!]
\begin{center}
    \includegraphics[width=0.46\columnwidth]{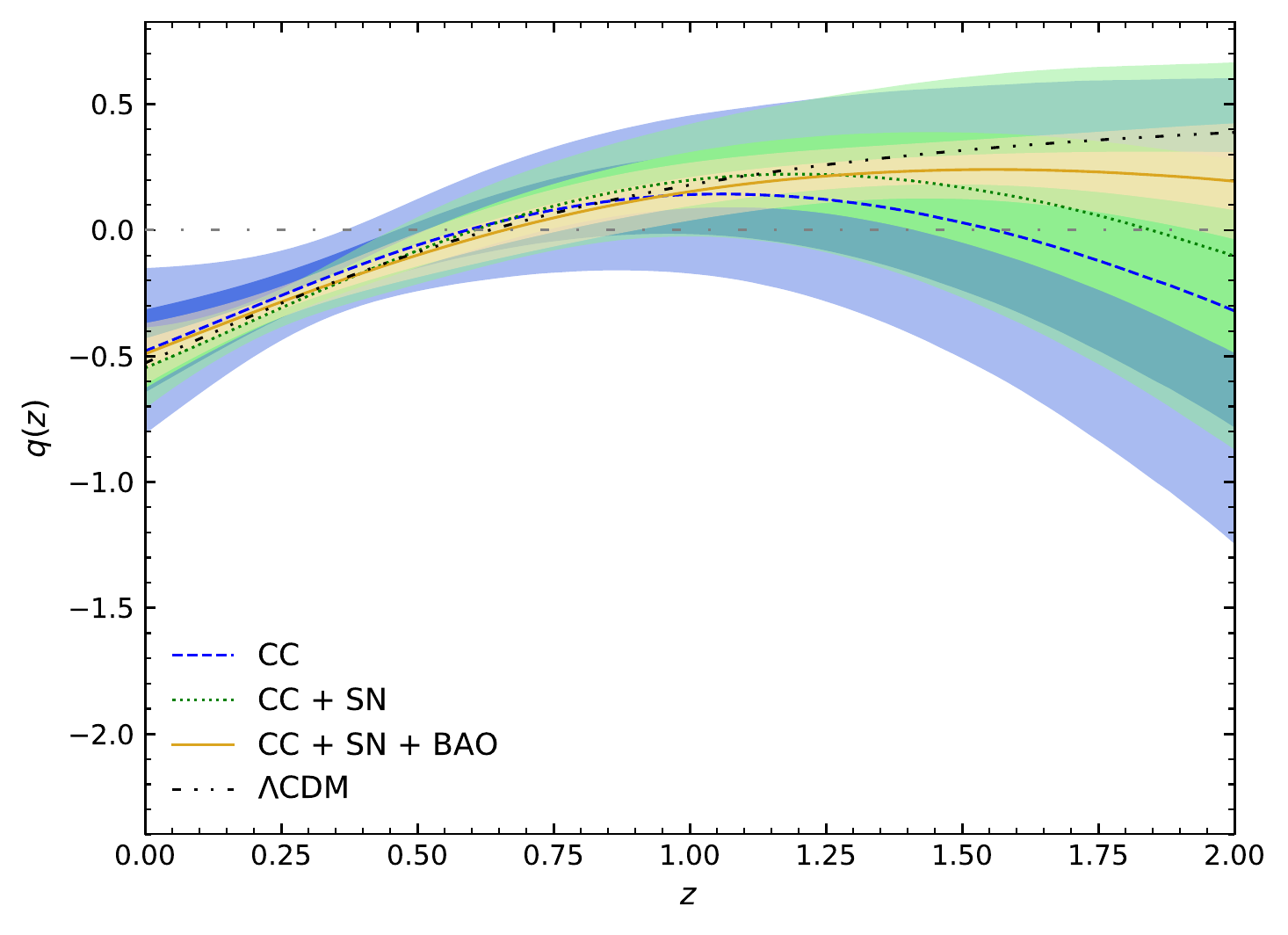}
    \includegraphics[width=0.46\columnwidth]{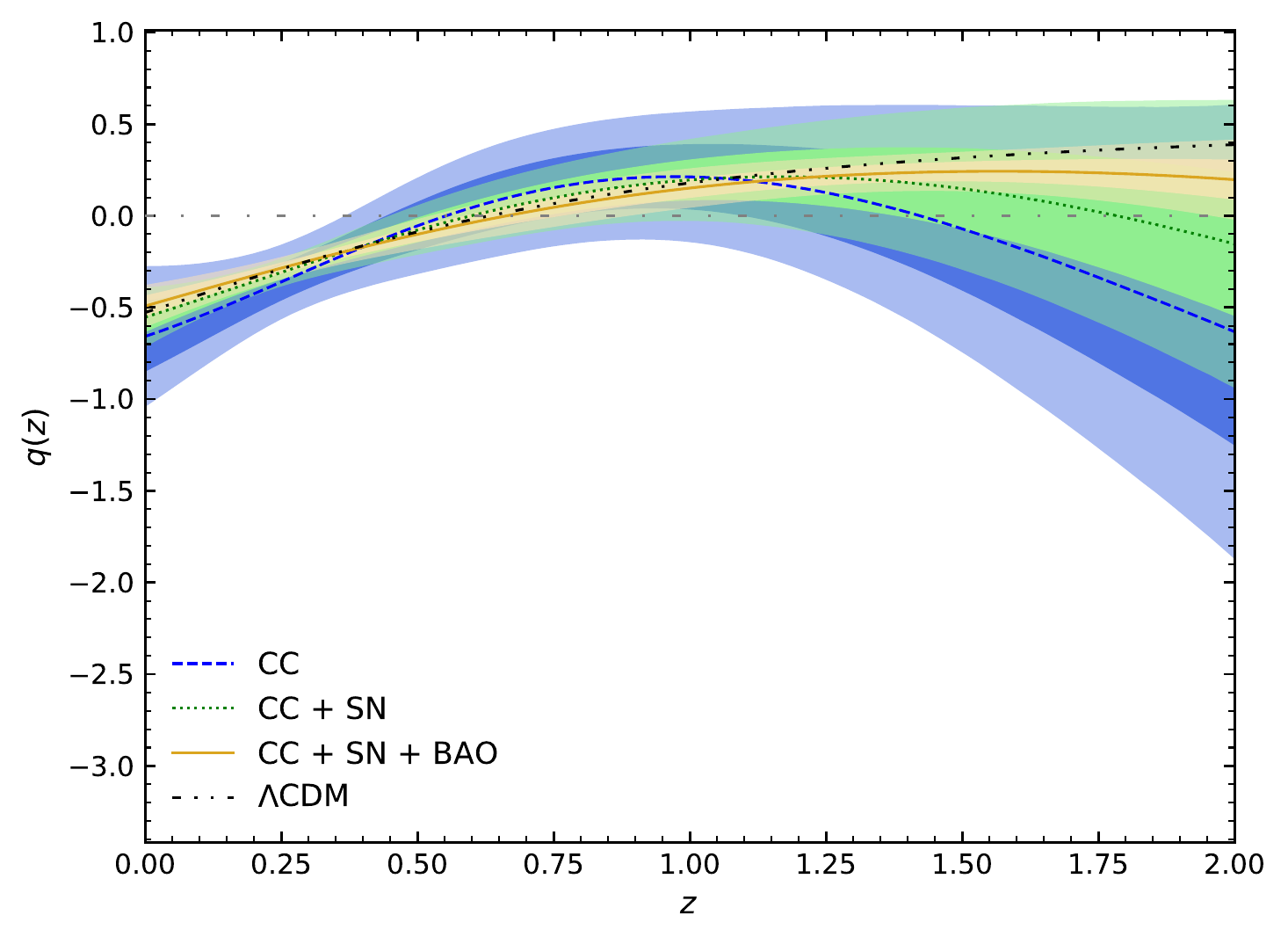}
    \includegraphics[width=0.46\columnwidth]{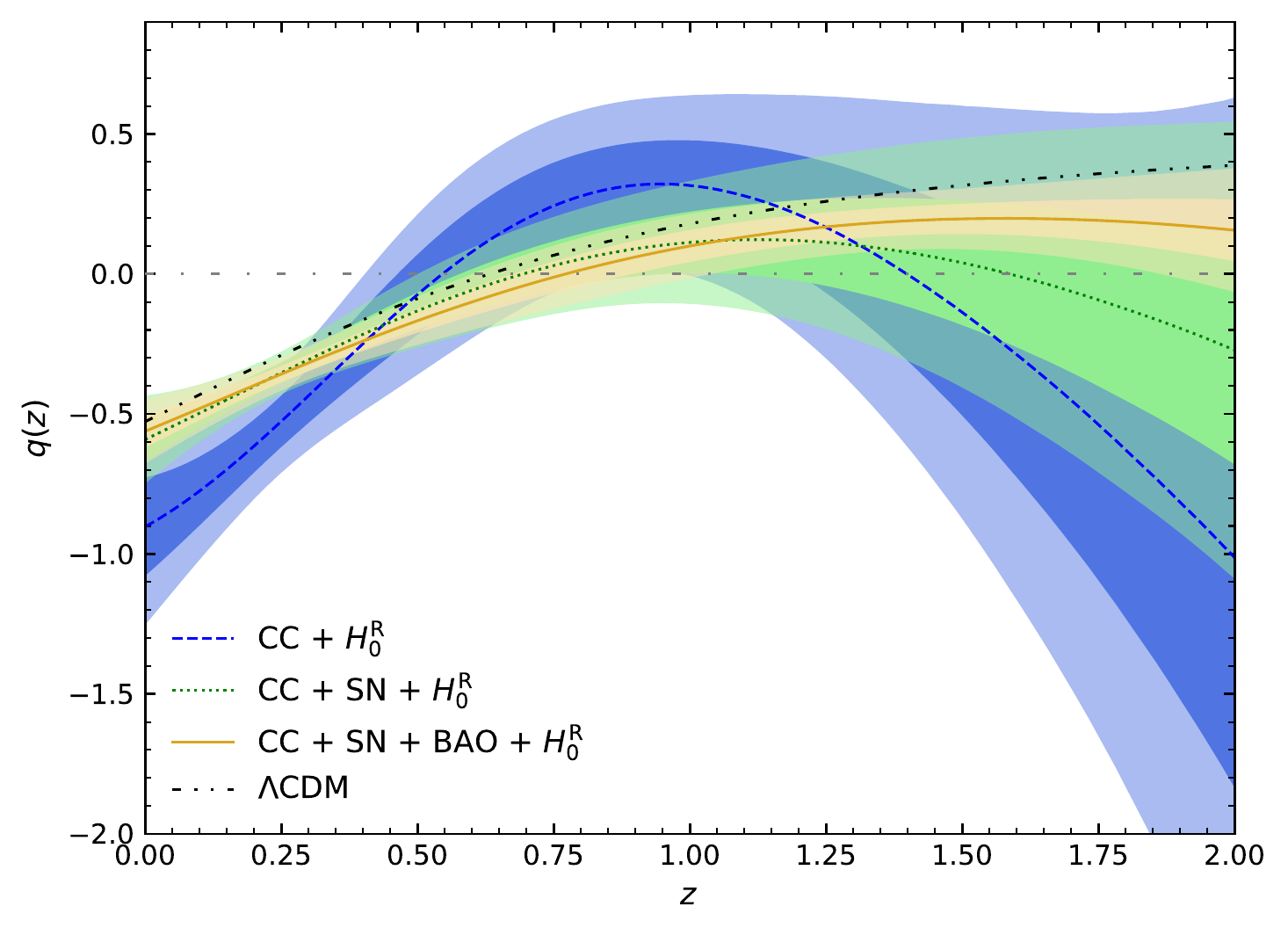}
    \includegraphics[width=0.46\columnwidth]{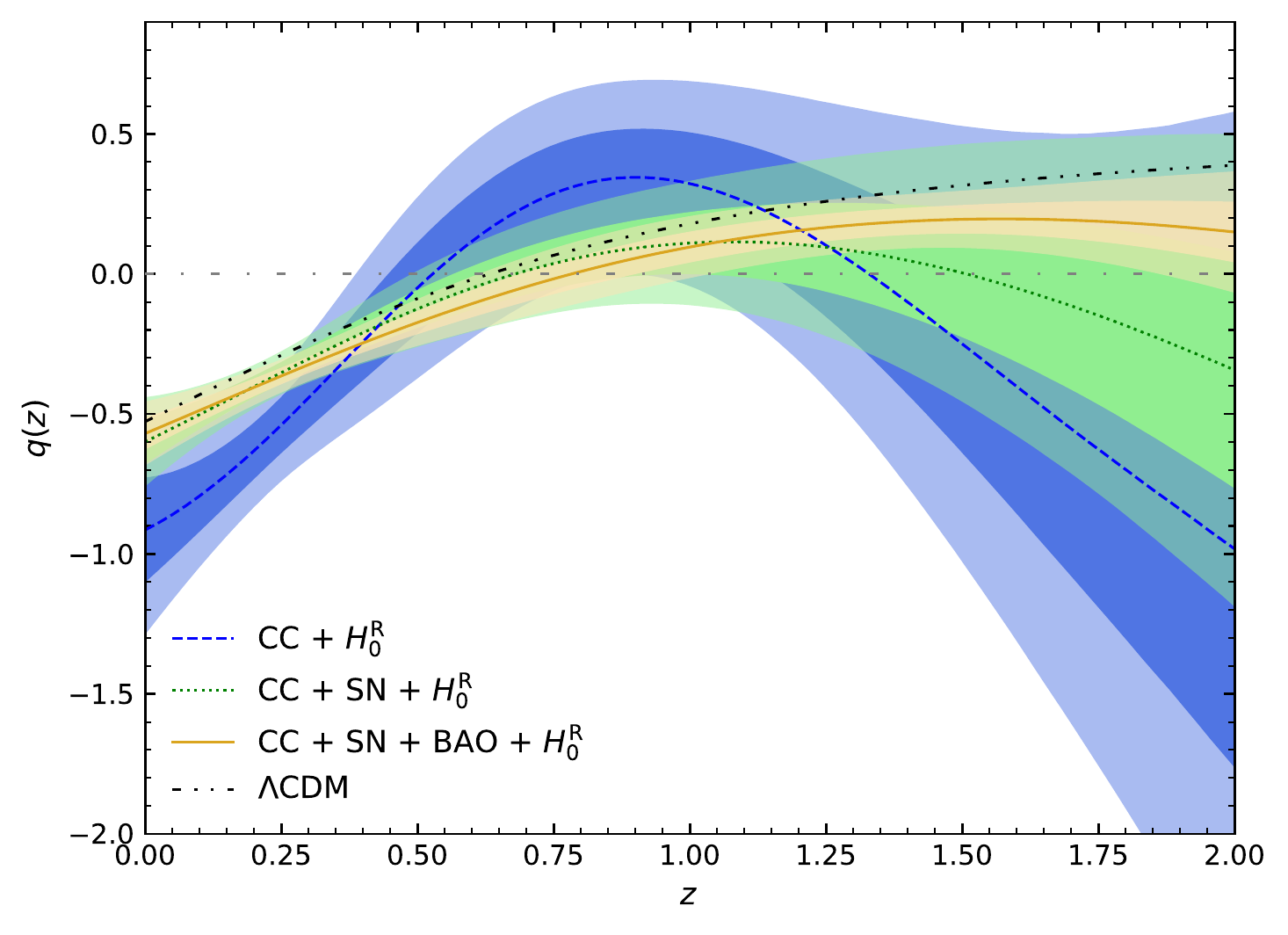}
    \includegraphics[width=0.46\columnwidth]{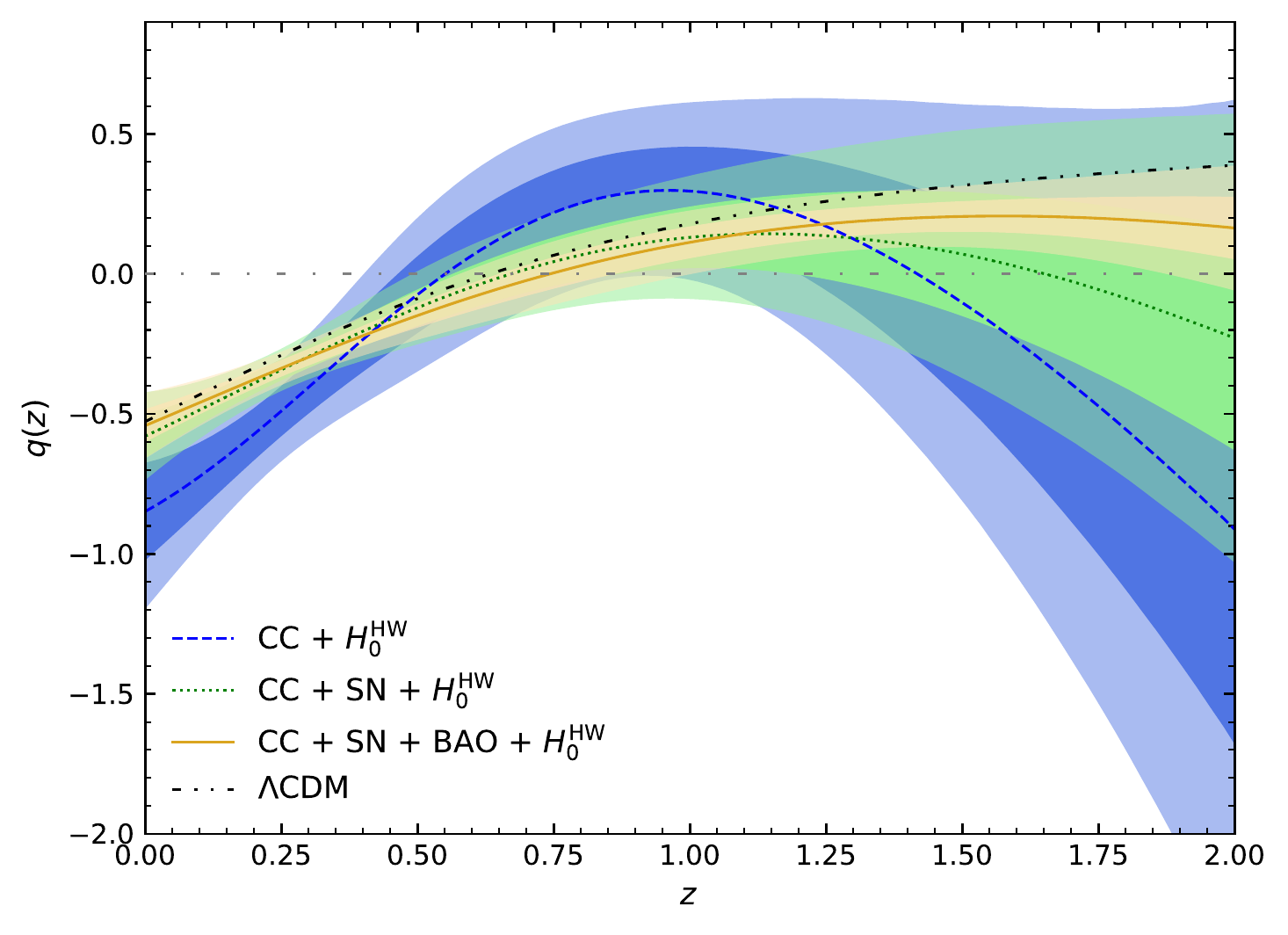}
    \includegraphics[width=0.46\columnwidth]{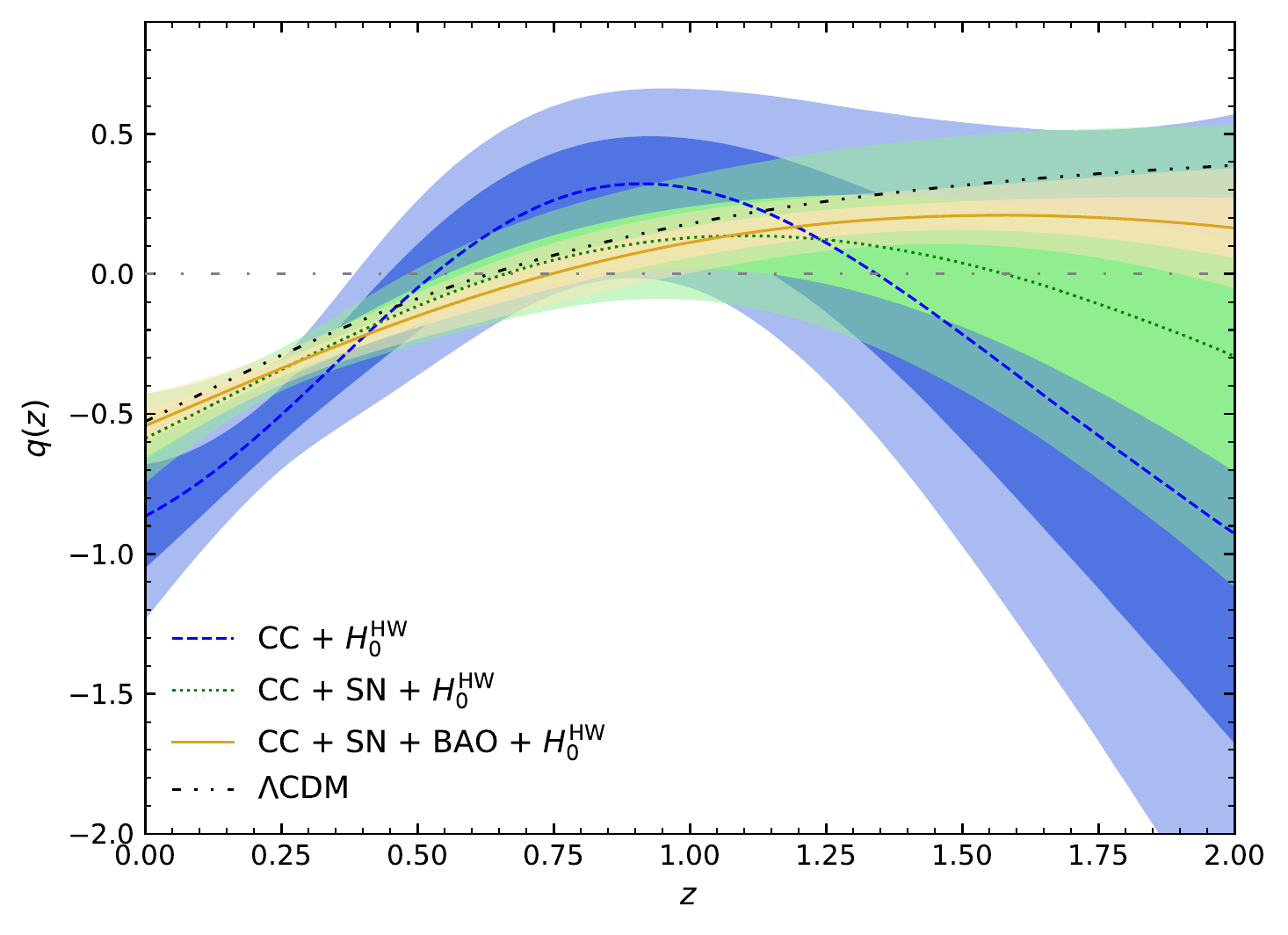}
    \includegraphics[width=0.46\columnwidth]{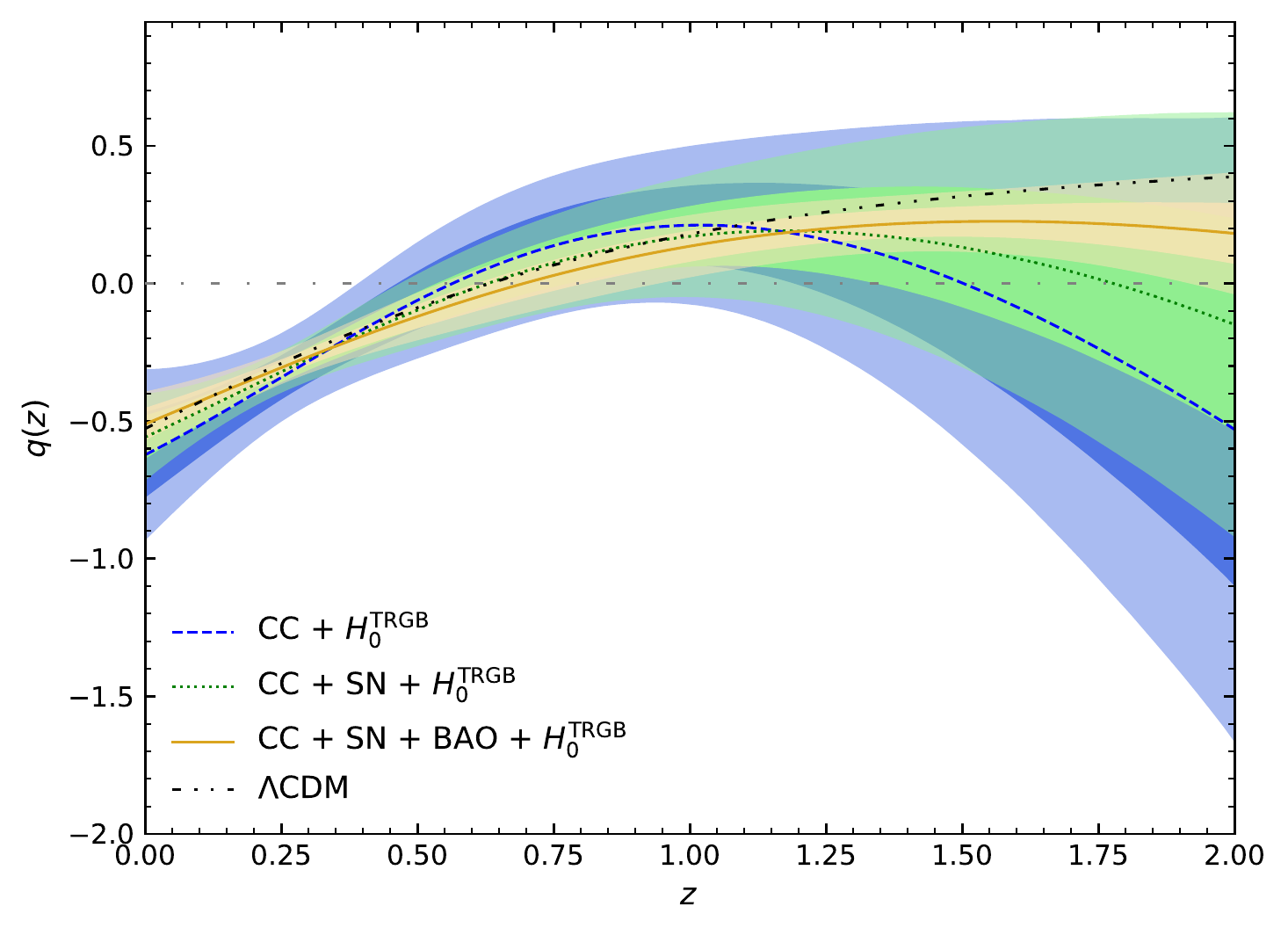}
    \includegraphics[width=0.46\columnwidth]{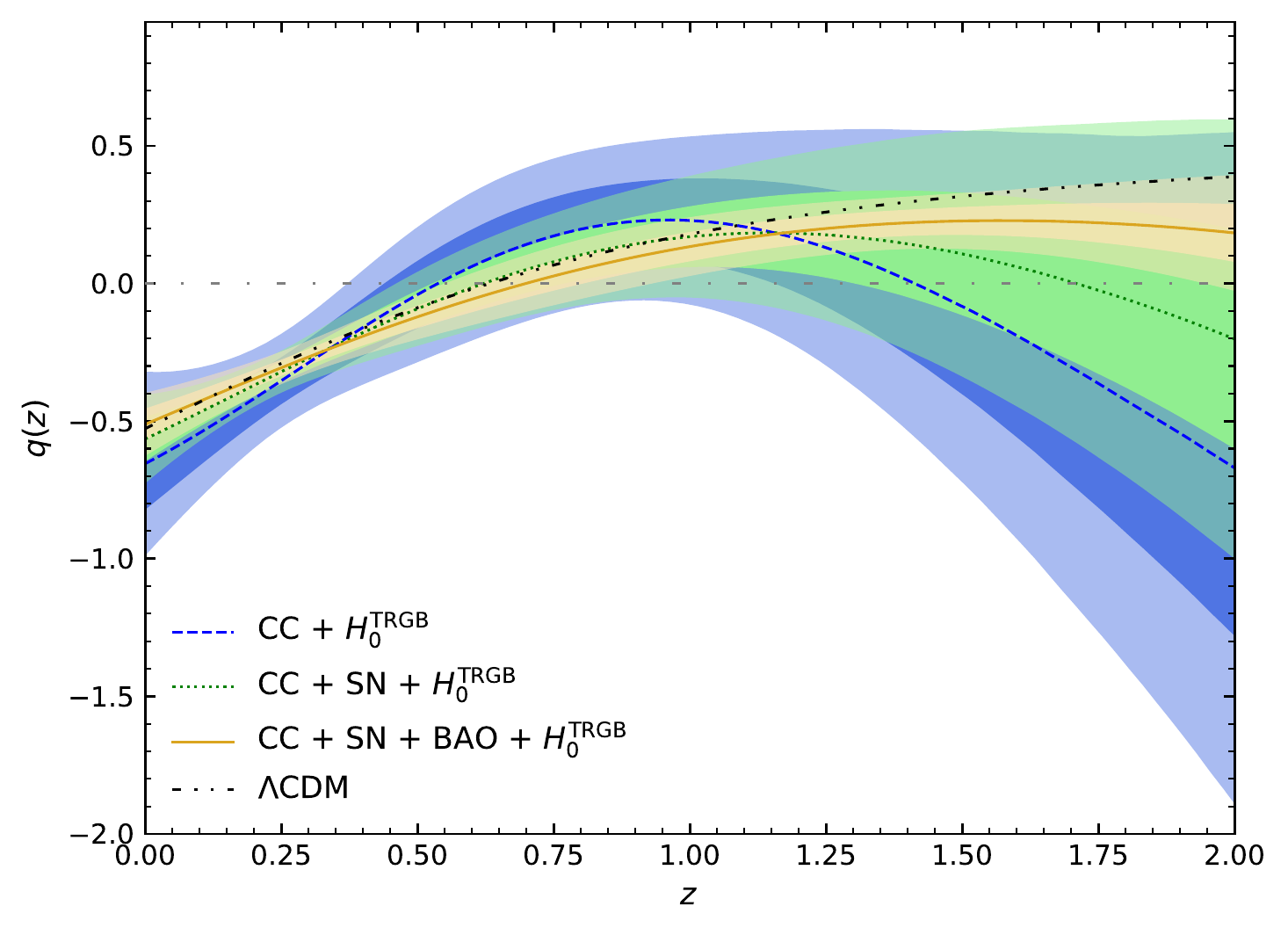}
    \caption{\label{fig:qz_mat_rat_quad}
    GP reconstructions of $q(z)$ with the Mat\'{e}rn (left) and rational quadratic (right) kernel functions, along with the $\Lambda$CDM prediction.
    }
\end{center}
\end{figure}

\begin{figure}[t!]
\begin{center}
    \includegraphics[width=0.445\columnwidth]{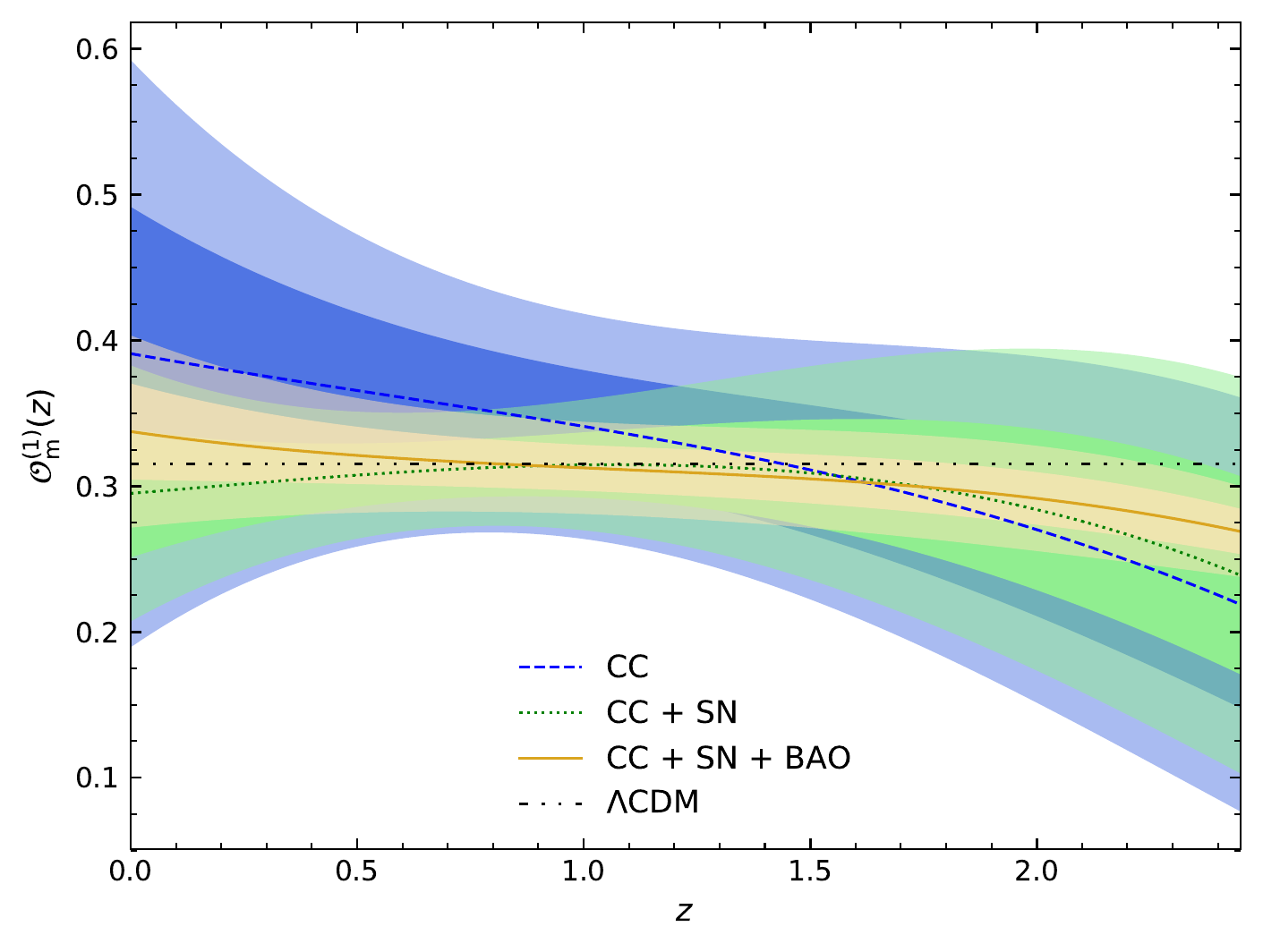}
    \includegraphics[width=0.445\columnwidth]{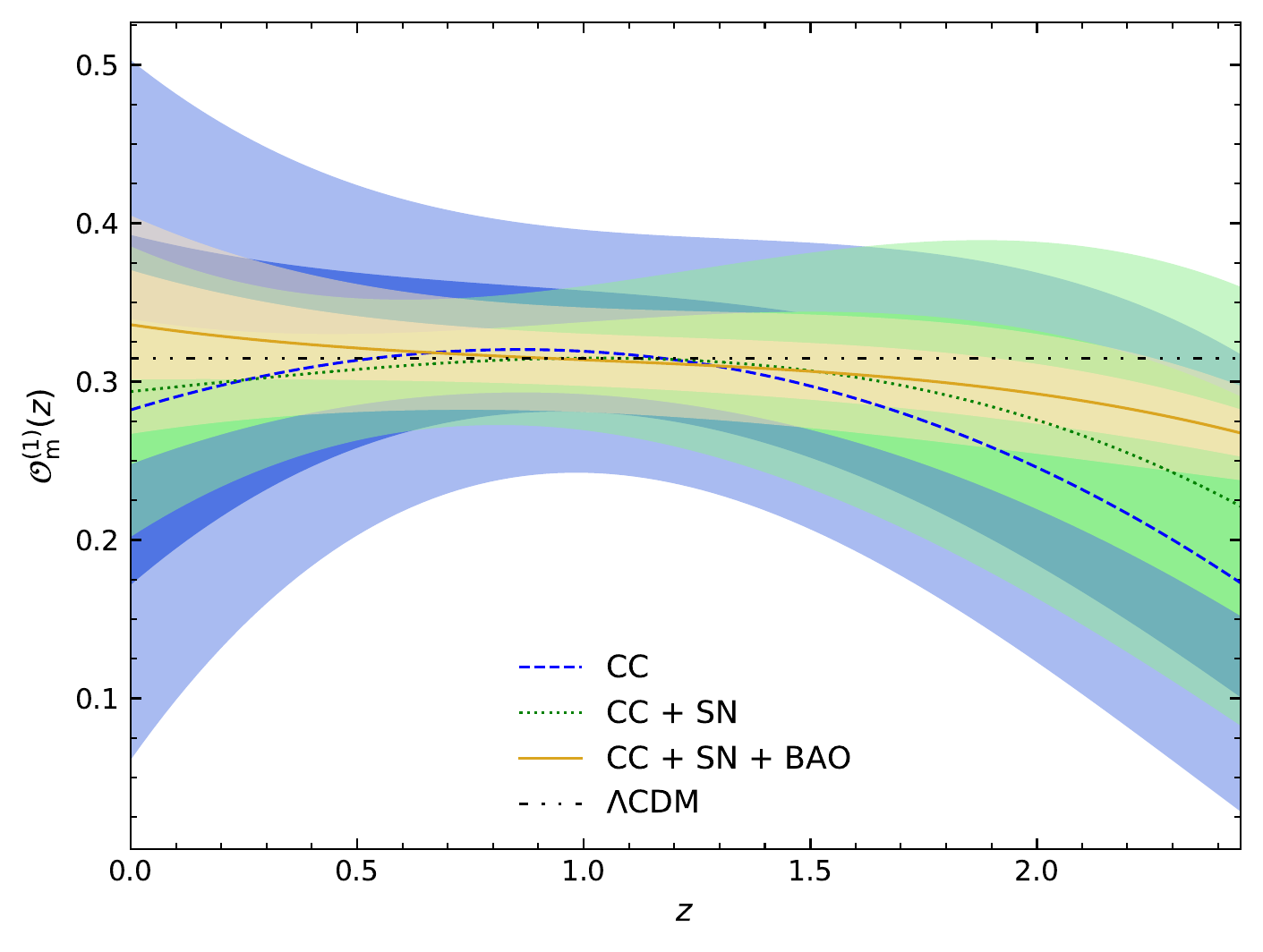}
    \includegraphics[width=0.445\columnwidth]{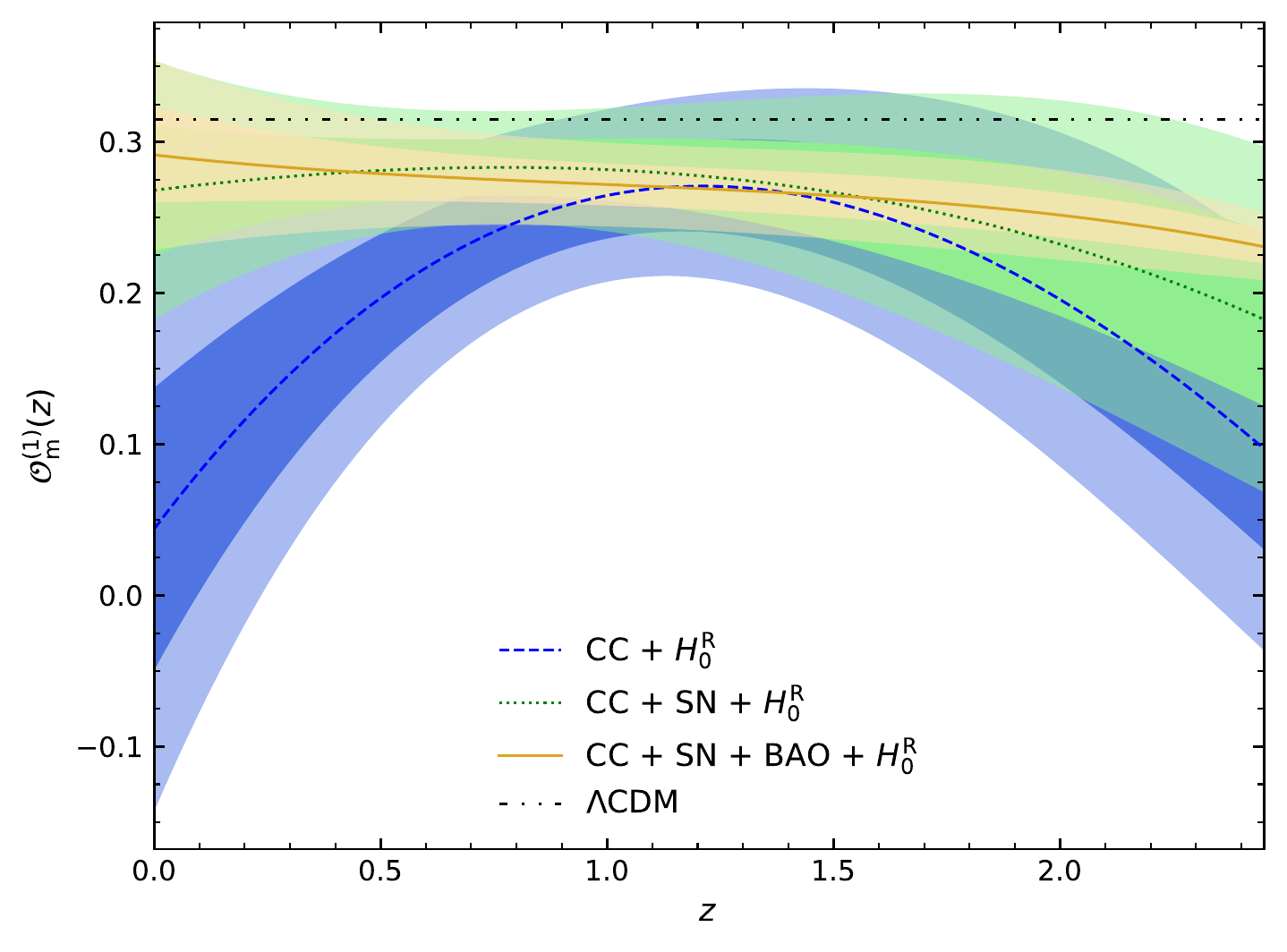}
    \includegraphics[width=0.445\columnwidth]{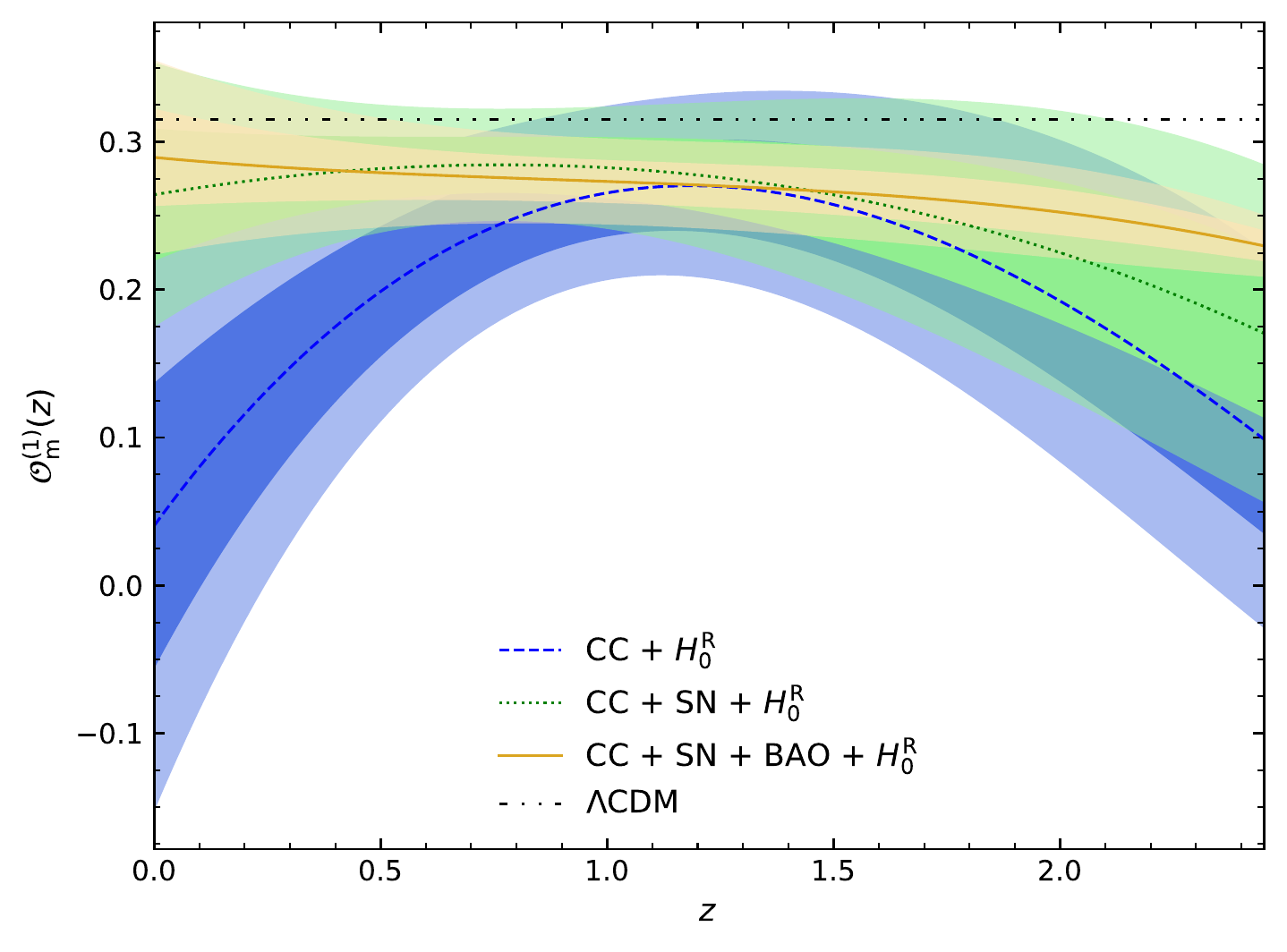}
    \includegraphics[width=0.445\columnwidth]{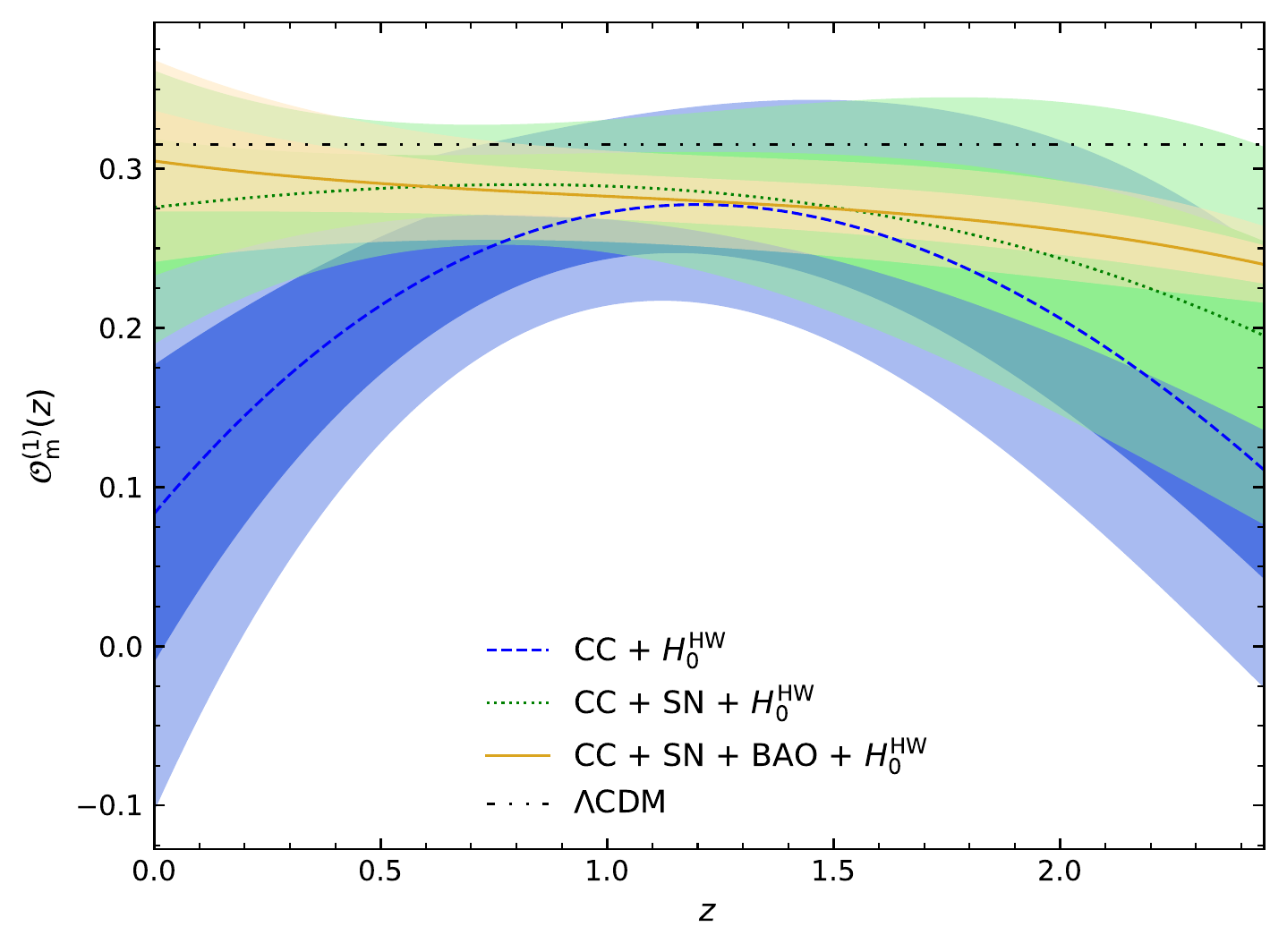}
    \includegraphics[width=0.445\columnwidth]{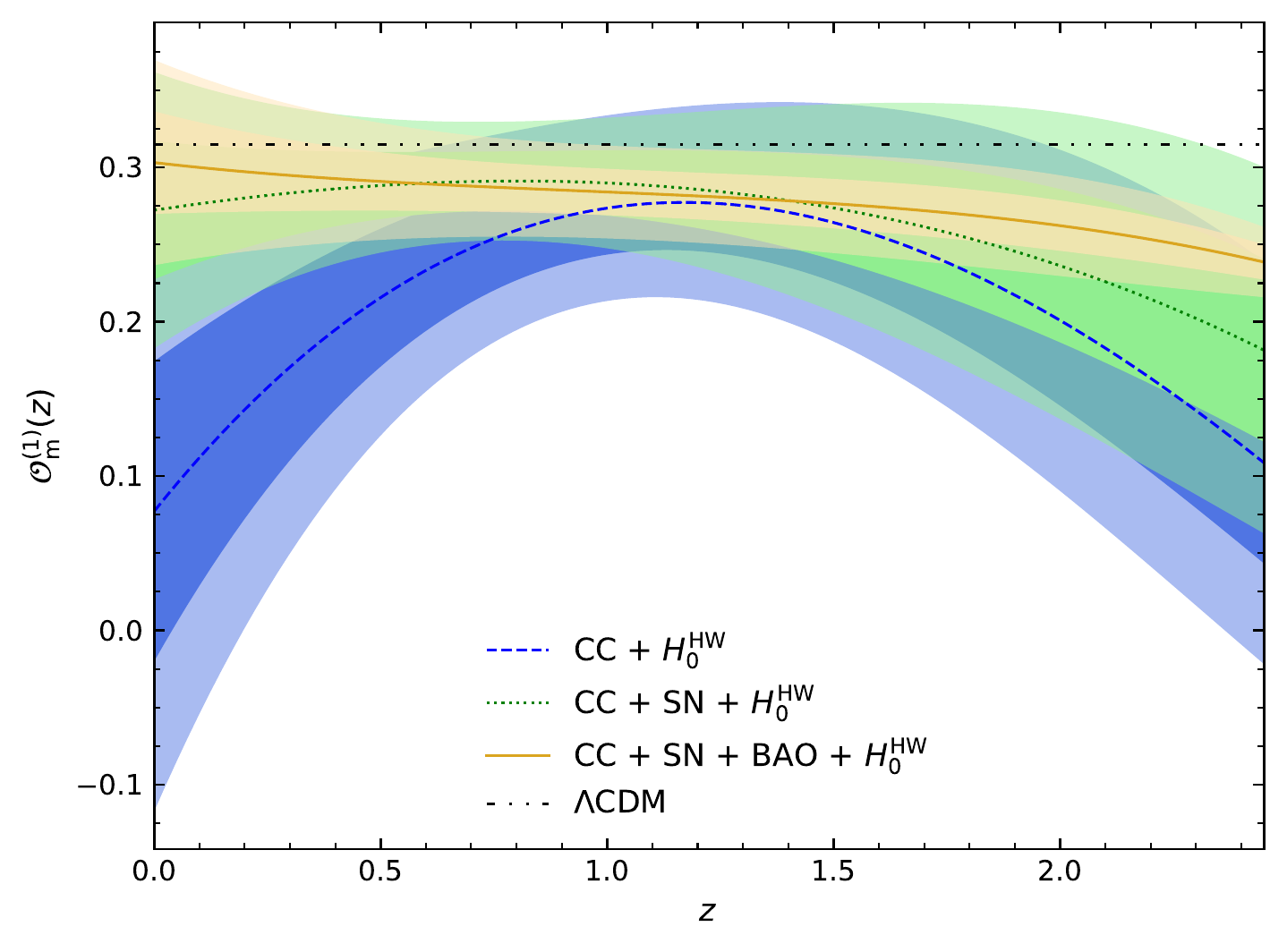}
    \includegraphics[width=0.445\columnwidth]{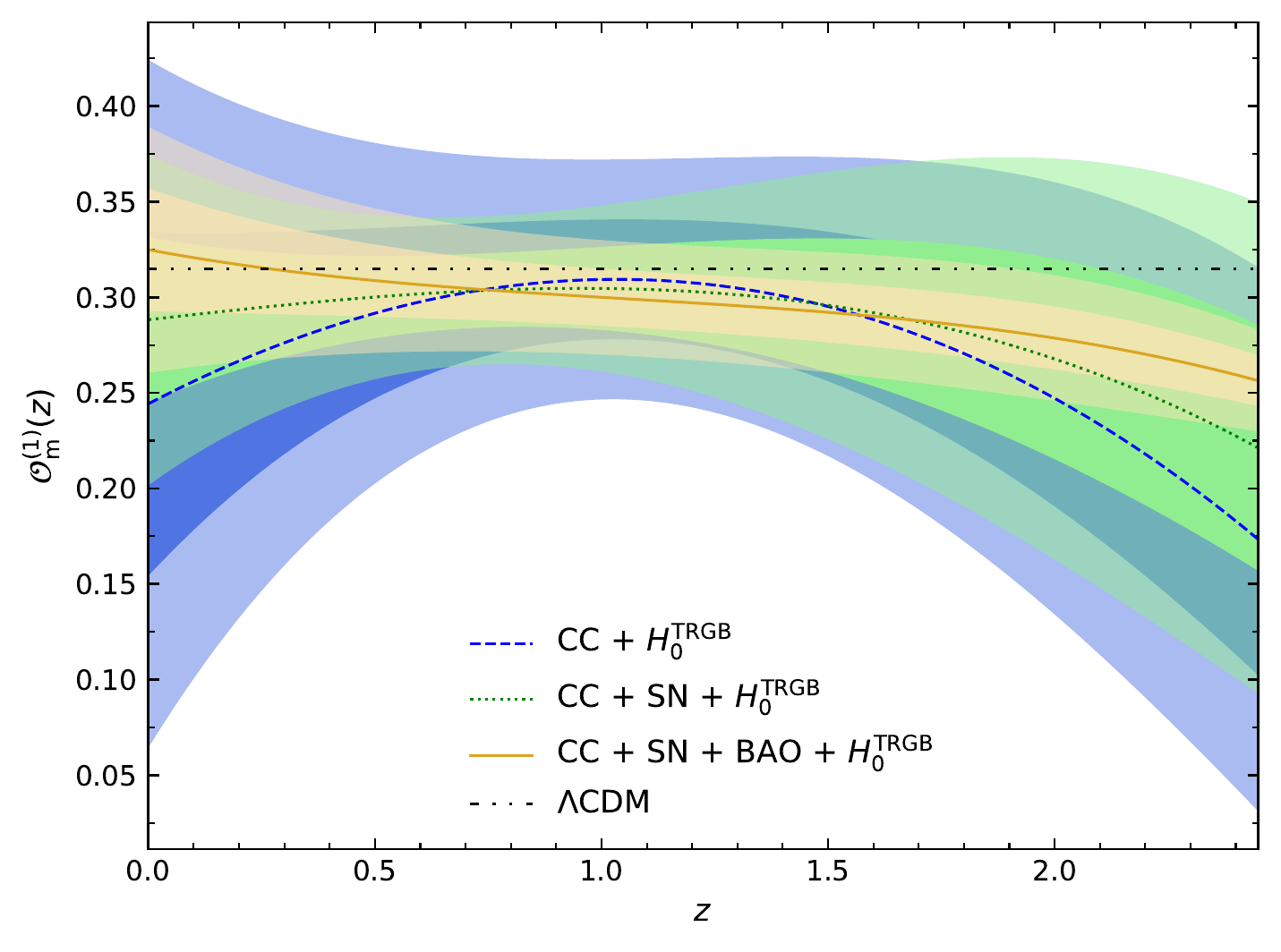}
    \includegraphics[width=0.445\columnwidth]{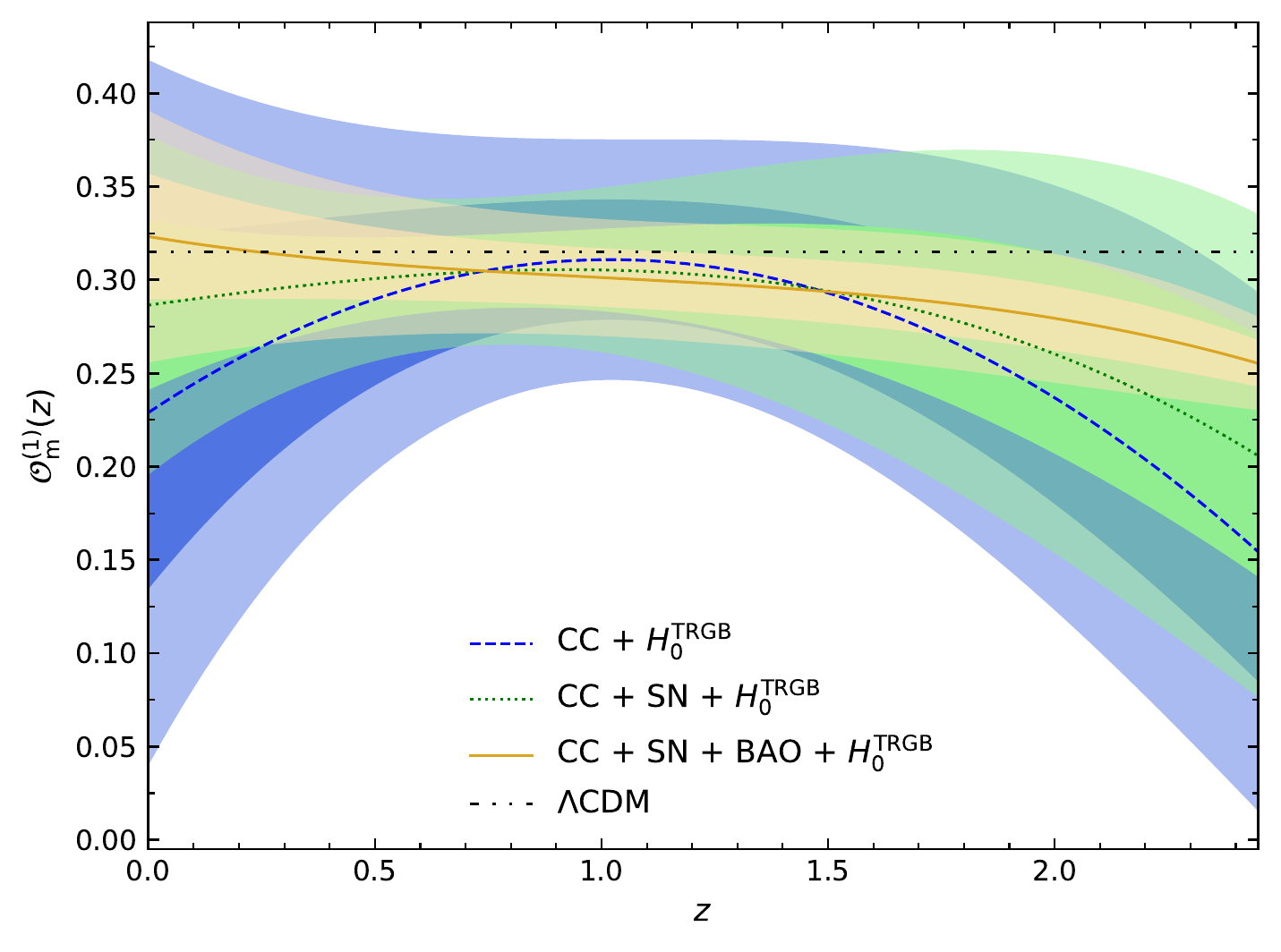}
    \caption{\label{fig:Om1_squaredexp_cauchy}
    GP reconstructions of $\mathcal{O}_m^{(1)}(z)$ with the squared exponential (left) and Cauchy (right) kernel functions, along with the $\Lambda$CDM prediction.
    }
\end{center}
\end{figure}

\begin{figure}[t!]
\begin{center}
    \includegraphics[width=0.445\columnwidth]{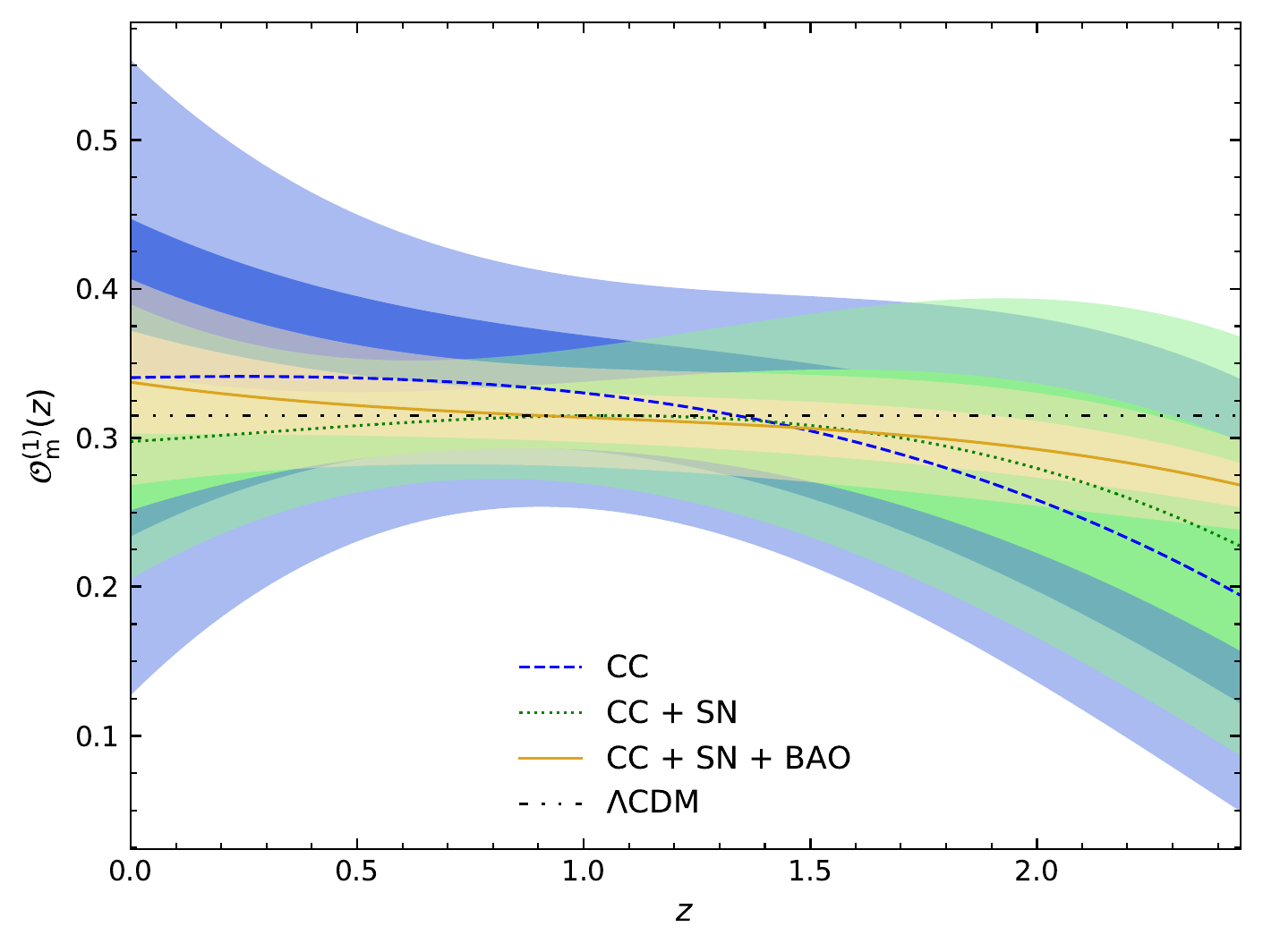}
    \includegraphics[width=0.445\columnwidth]{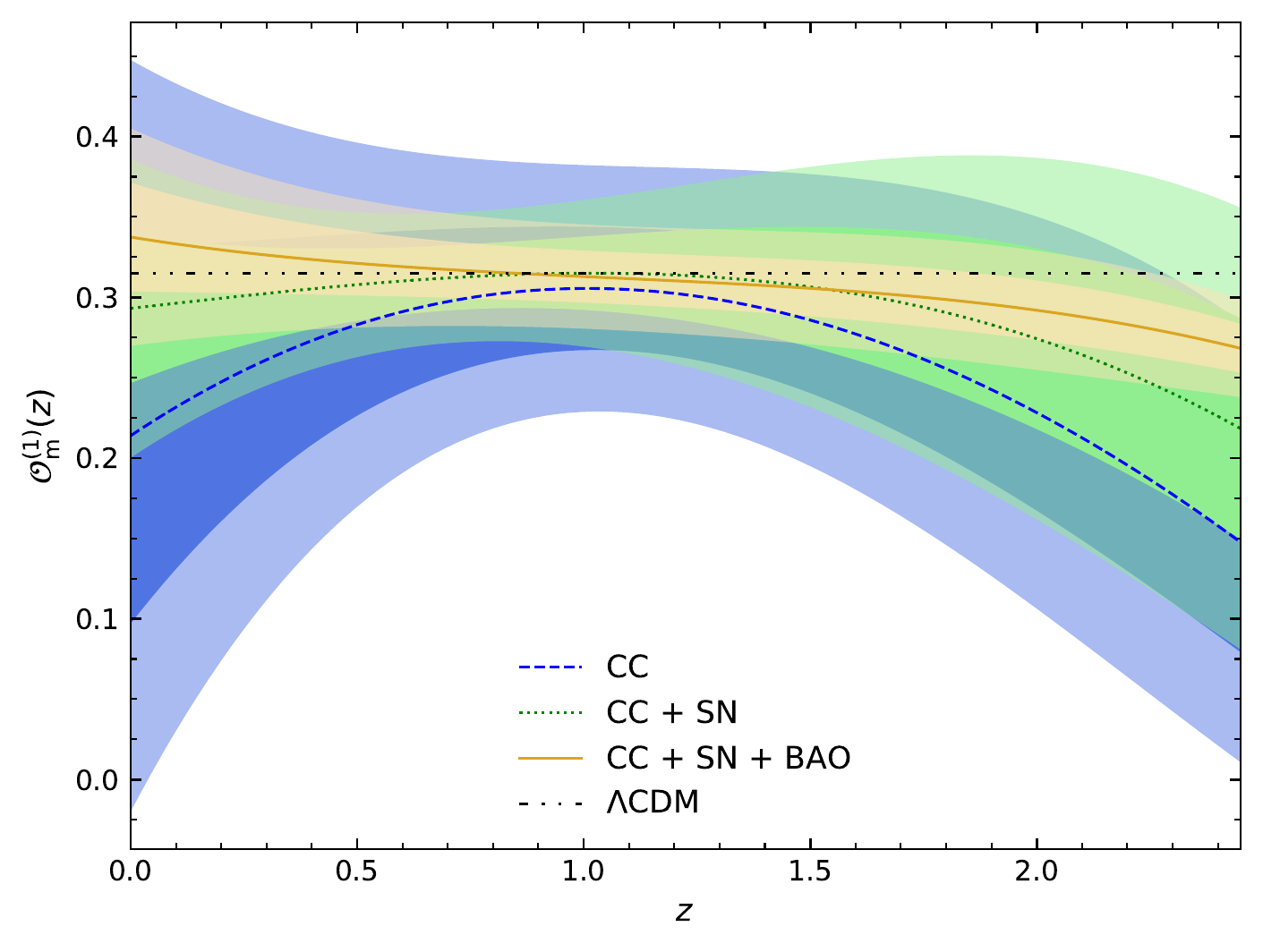}
    \includegraphics[width=0.445\columnwidth]{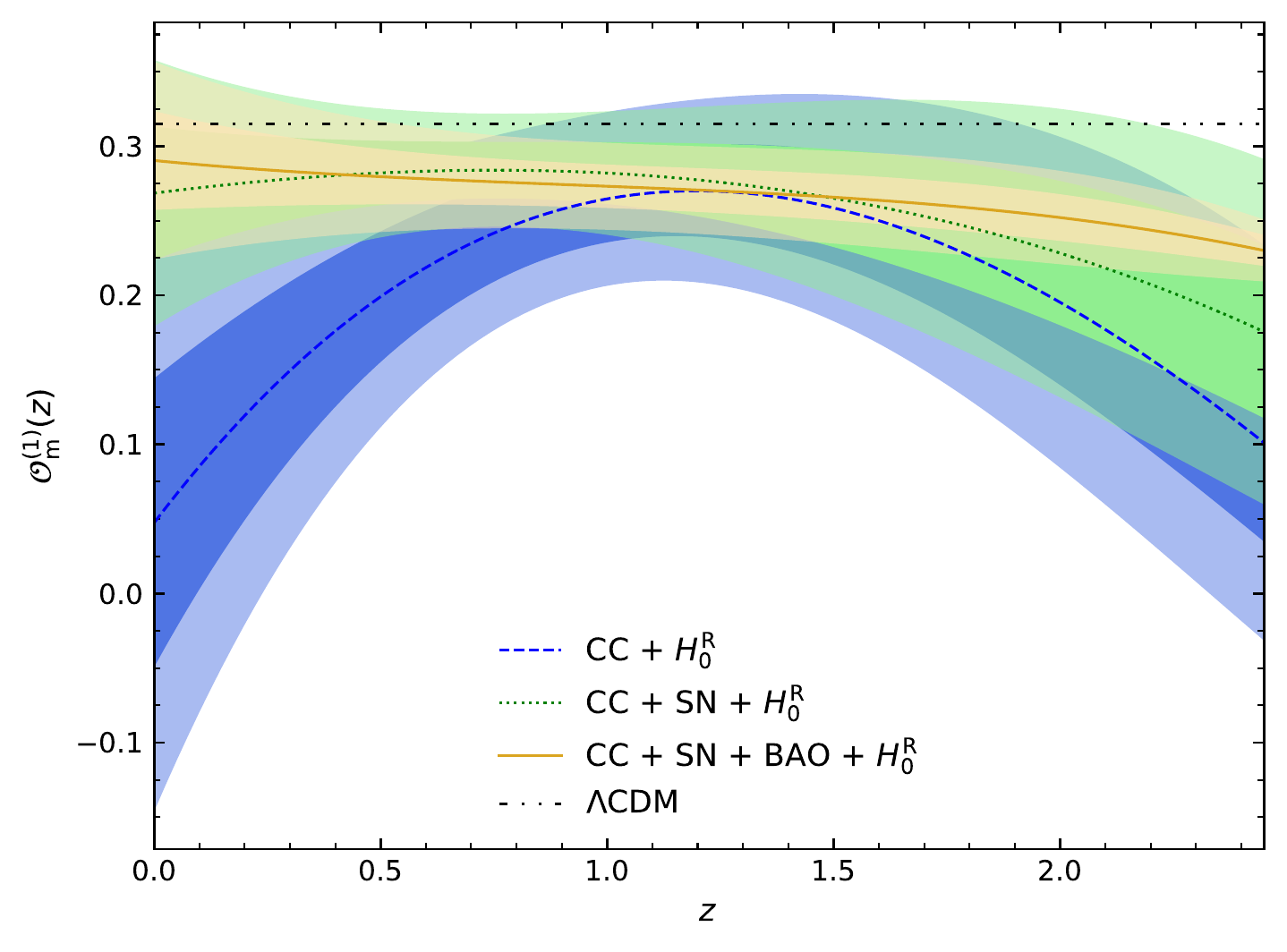}
    \includegraphics[width=0.445\columnwidth]{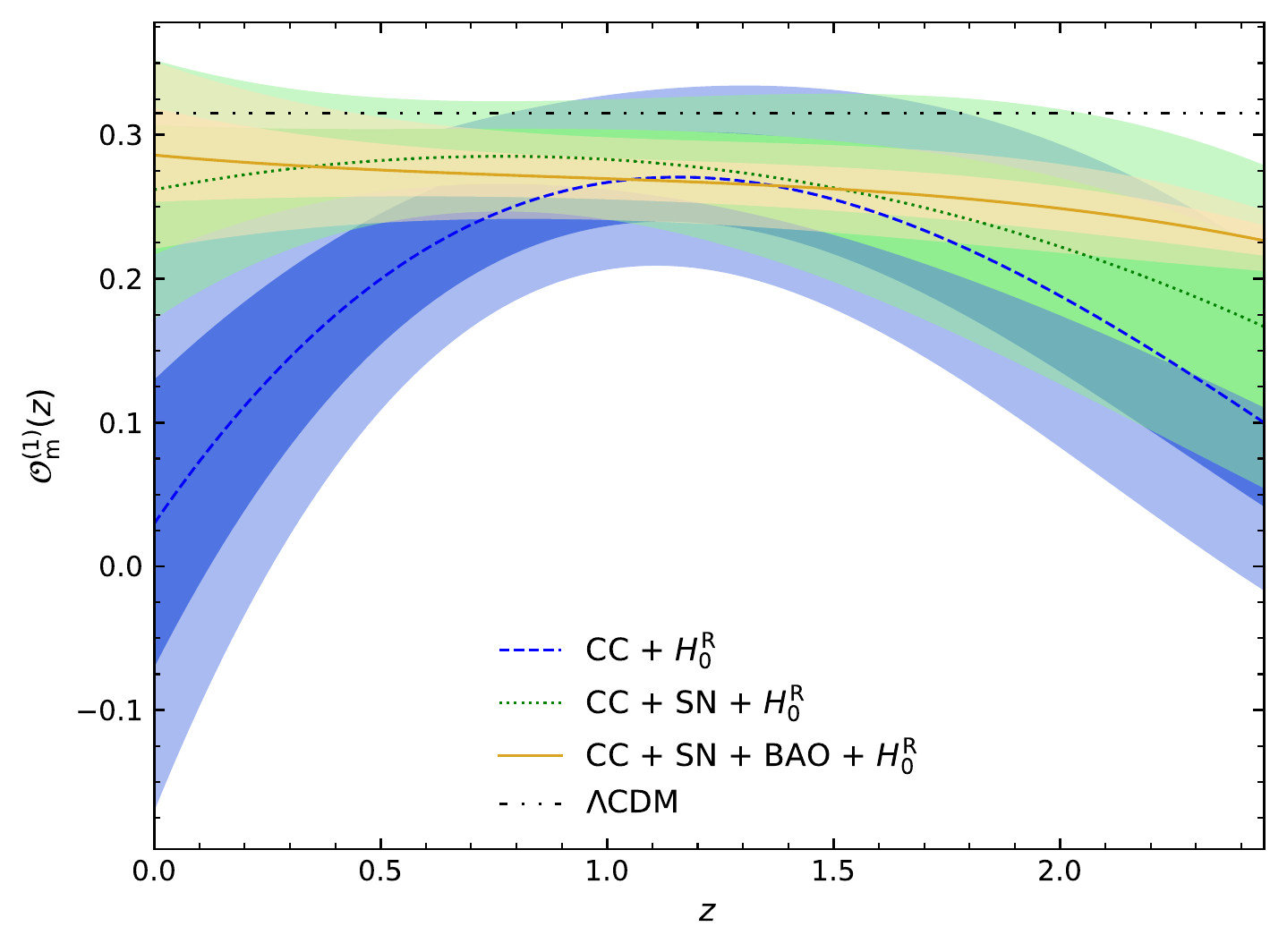}
    \includegraphics[width=0.445\columnwidth]{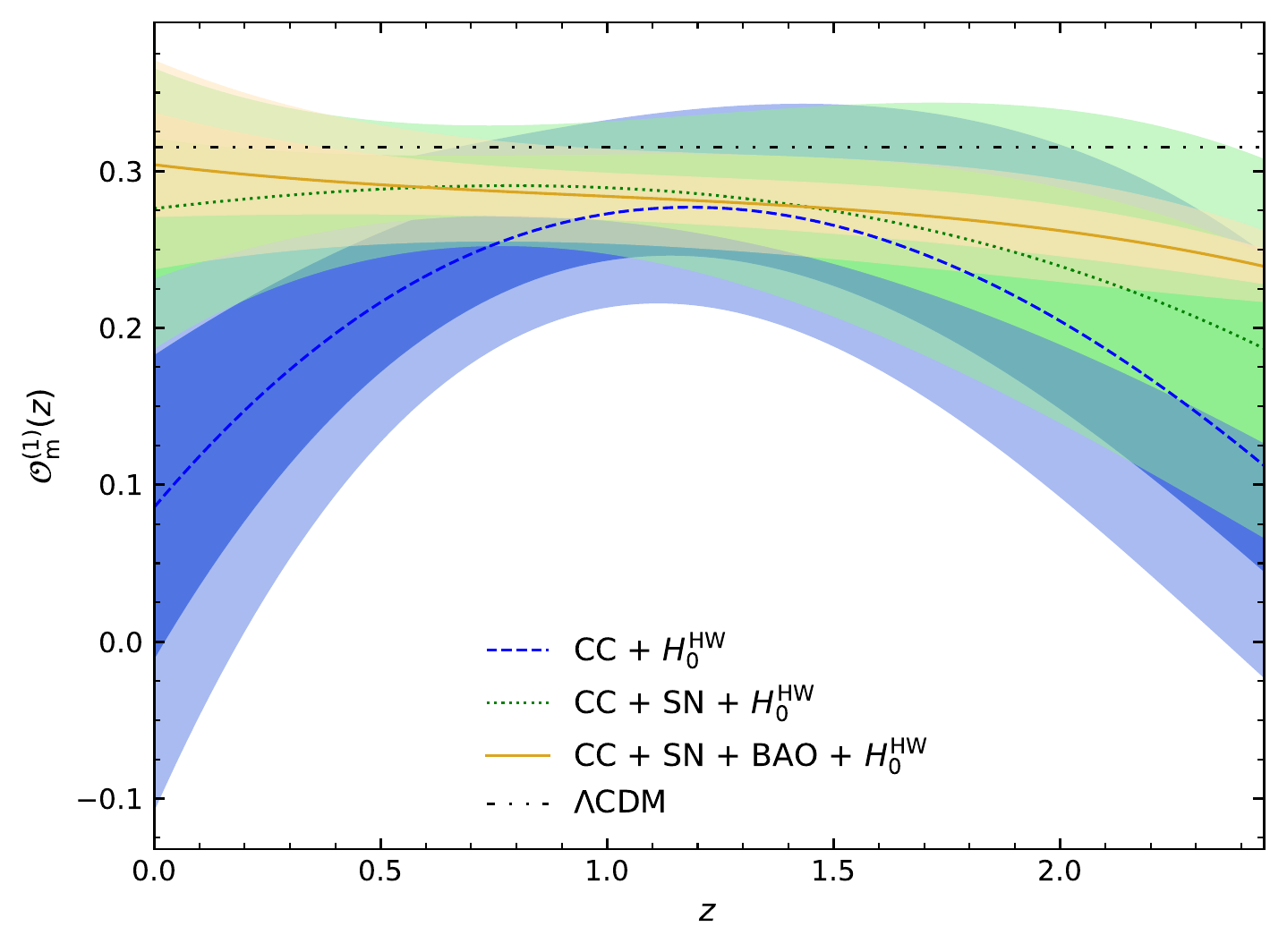}
    \includegraphics[width=0.445\columnwidth]{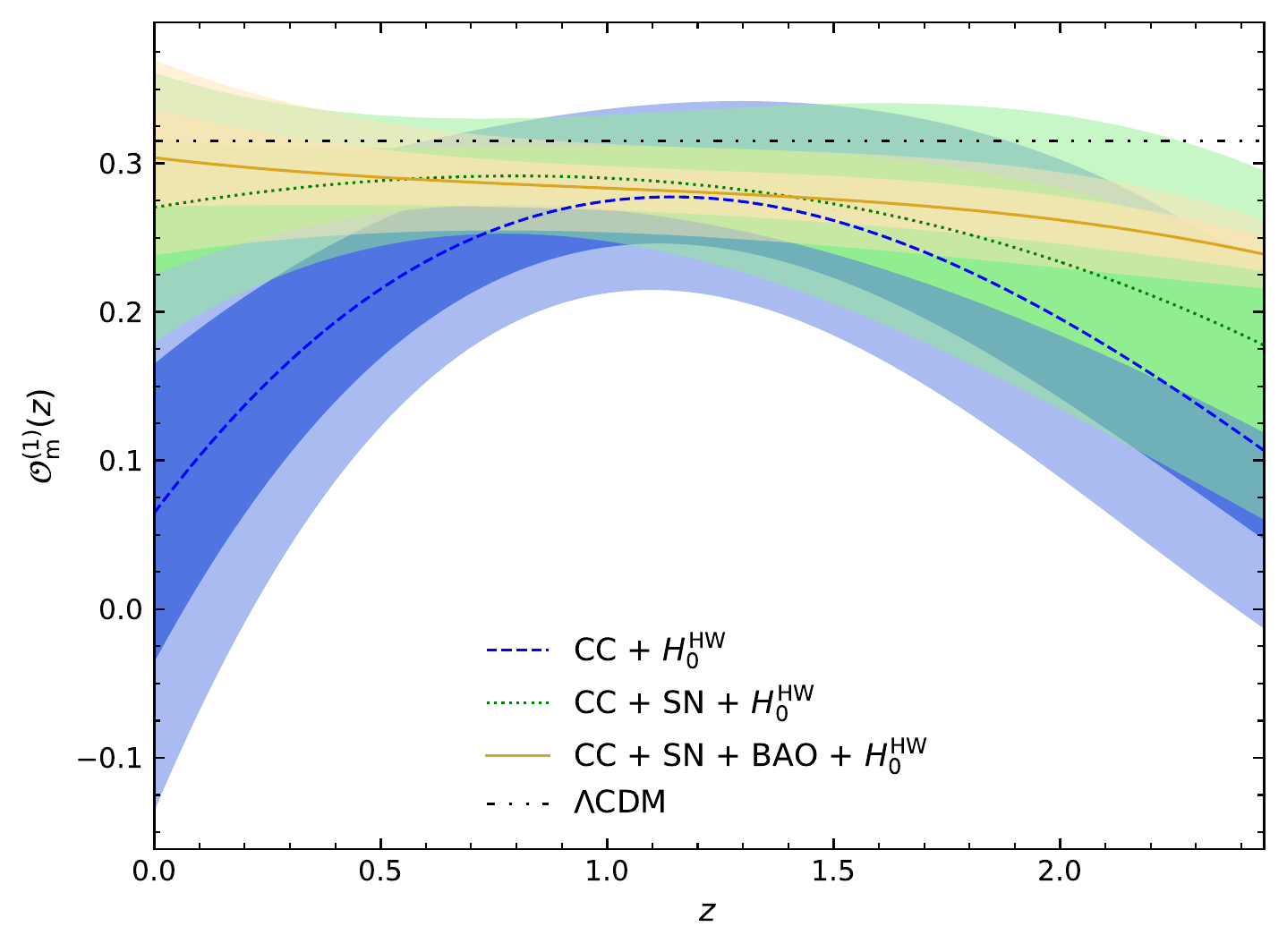}
    \includegraphics[width=0.445\columnwidth]{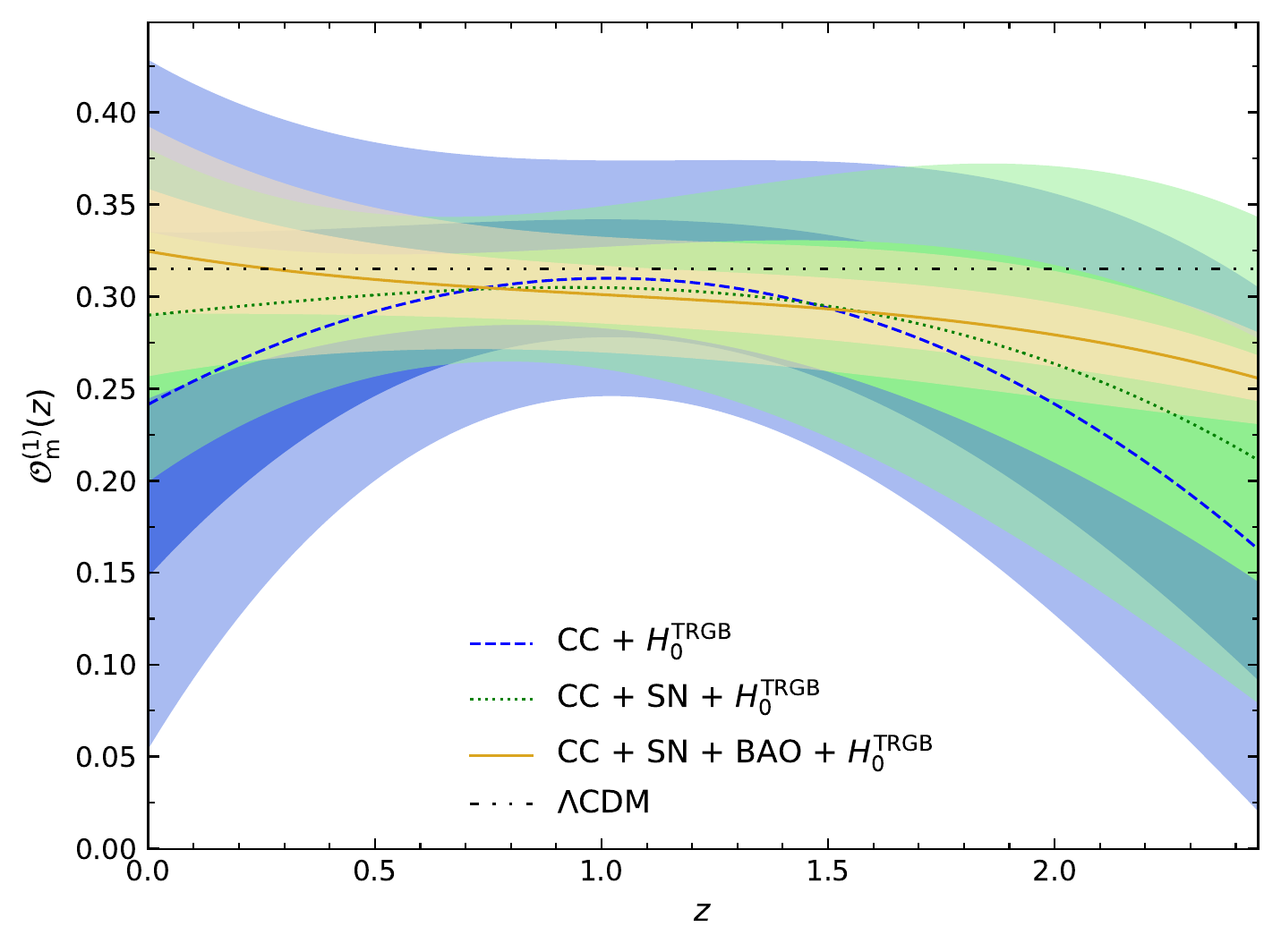}
    \includegraphics[width=0.445\columnwidth]{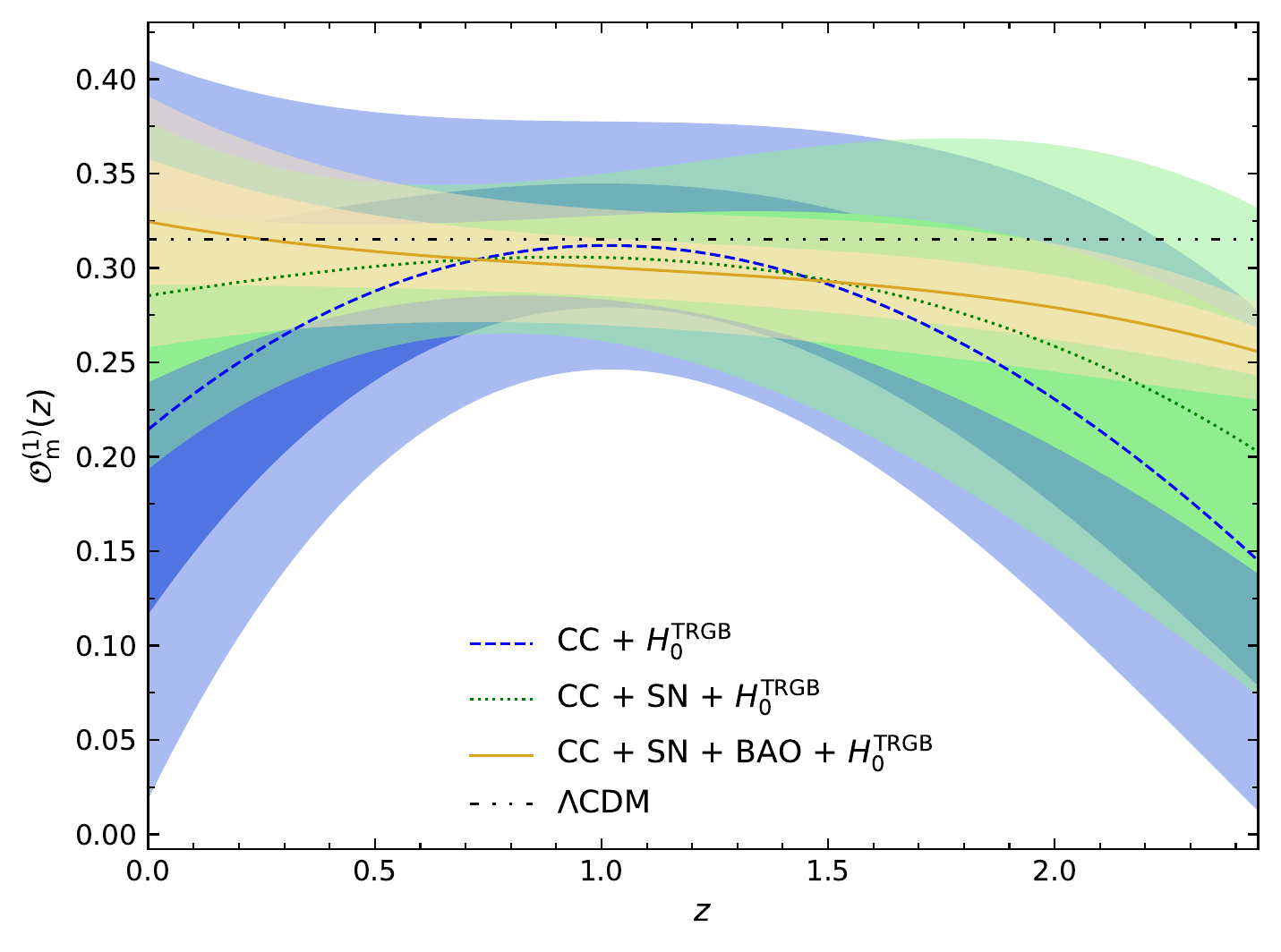}
    \caption{\label{fig:Om1_mat_rat_quad}
    GP reconstructions of $\mathcal{O}_m^{(1)}(z)$ with the Mat\'{e}rn (left) and rational quadratic (right) kernel functions, along with the $\Lambda$CDM prediction.
    }
\end{center}
\end{figure}

\begin{figure}[t!]
\begin{center}
    \includegraphics[width=0.445\columnwidth]{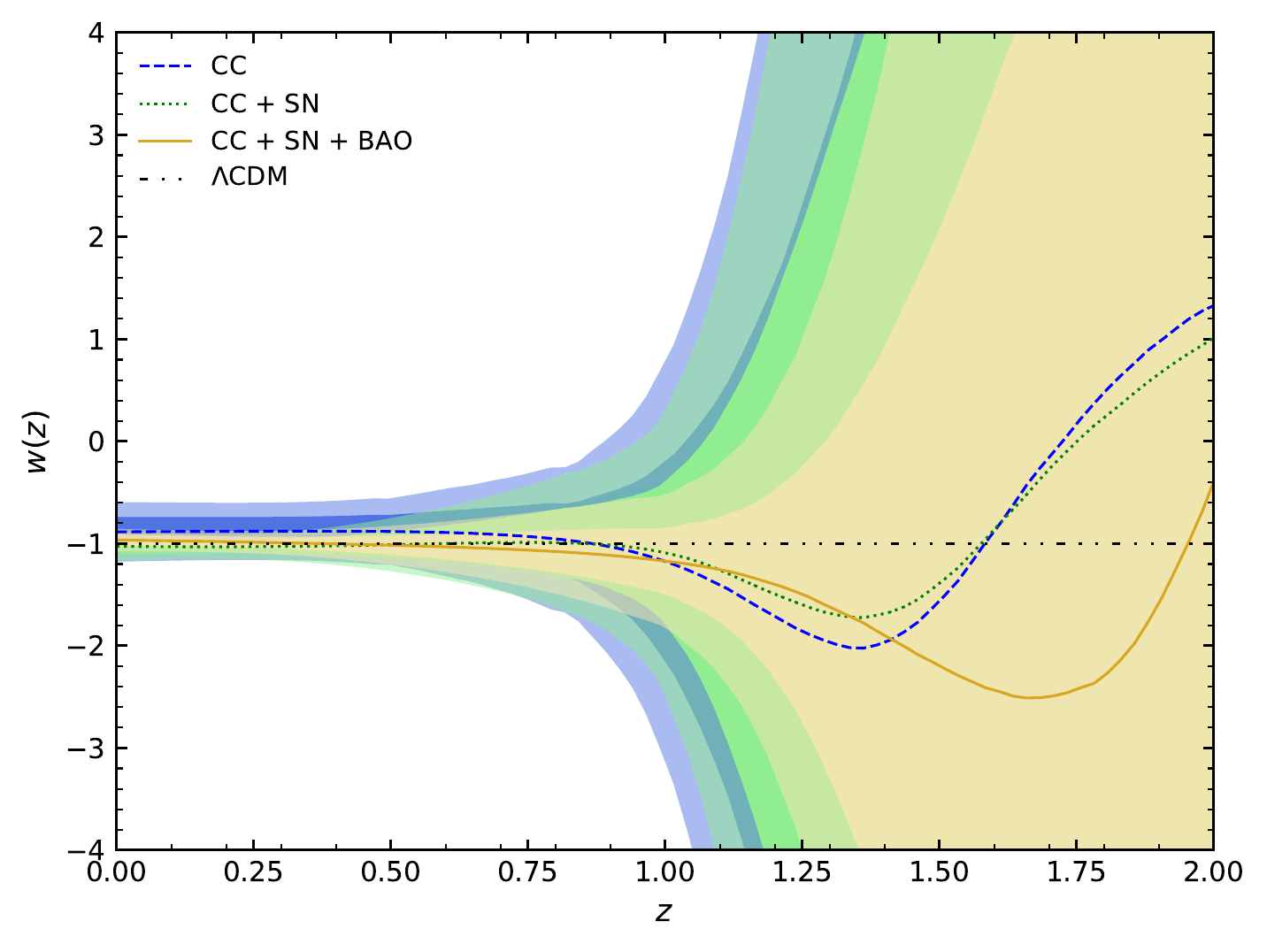}
    \includegraphics[width=0.445\columnwidth]{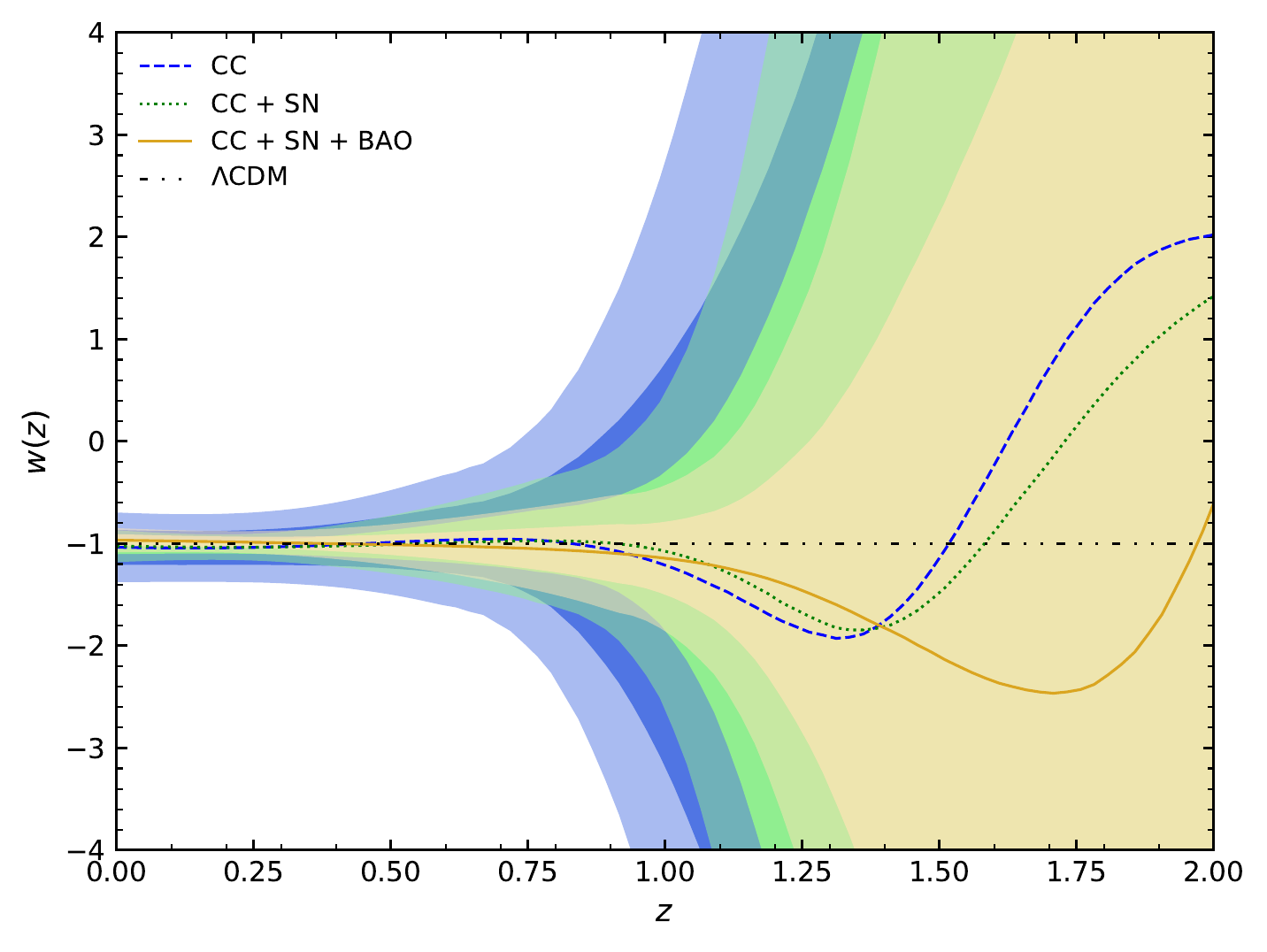}
    \includegraphics[width=0.445\columnwidth]{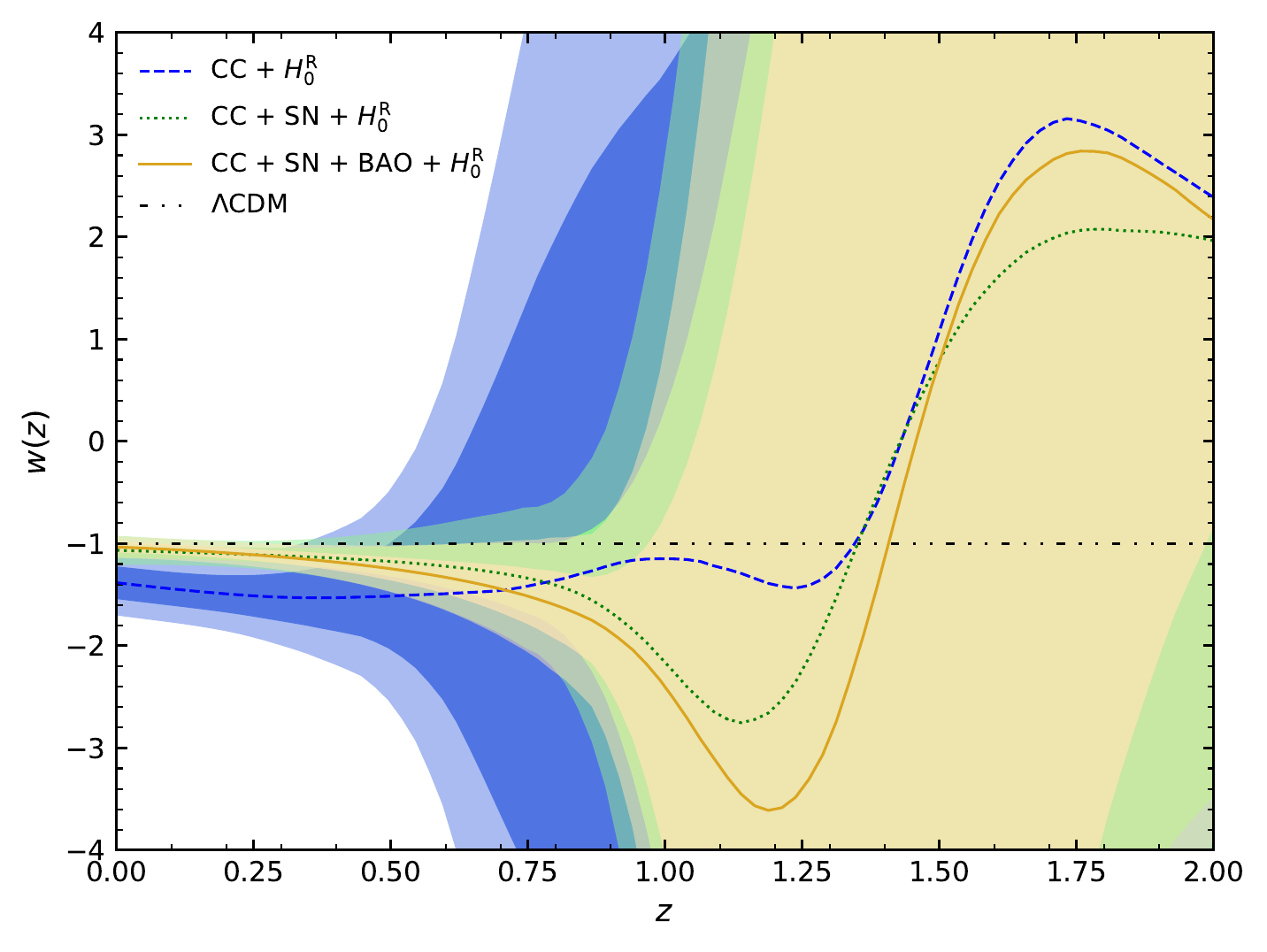}
    \includegraphics[width=0.445\columnwidth]{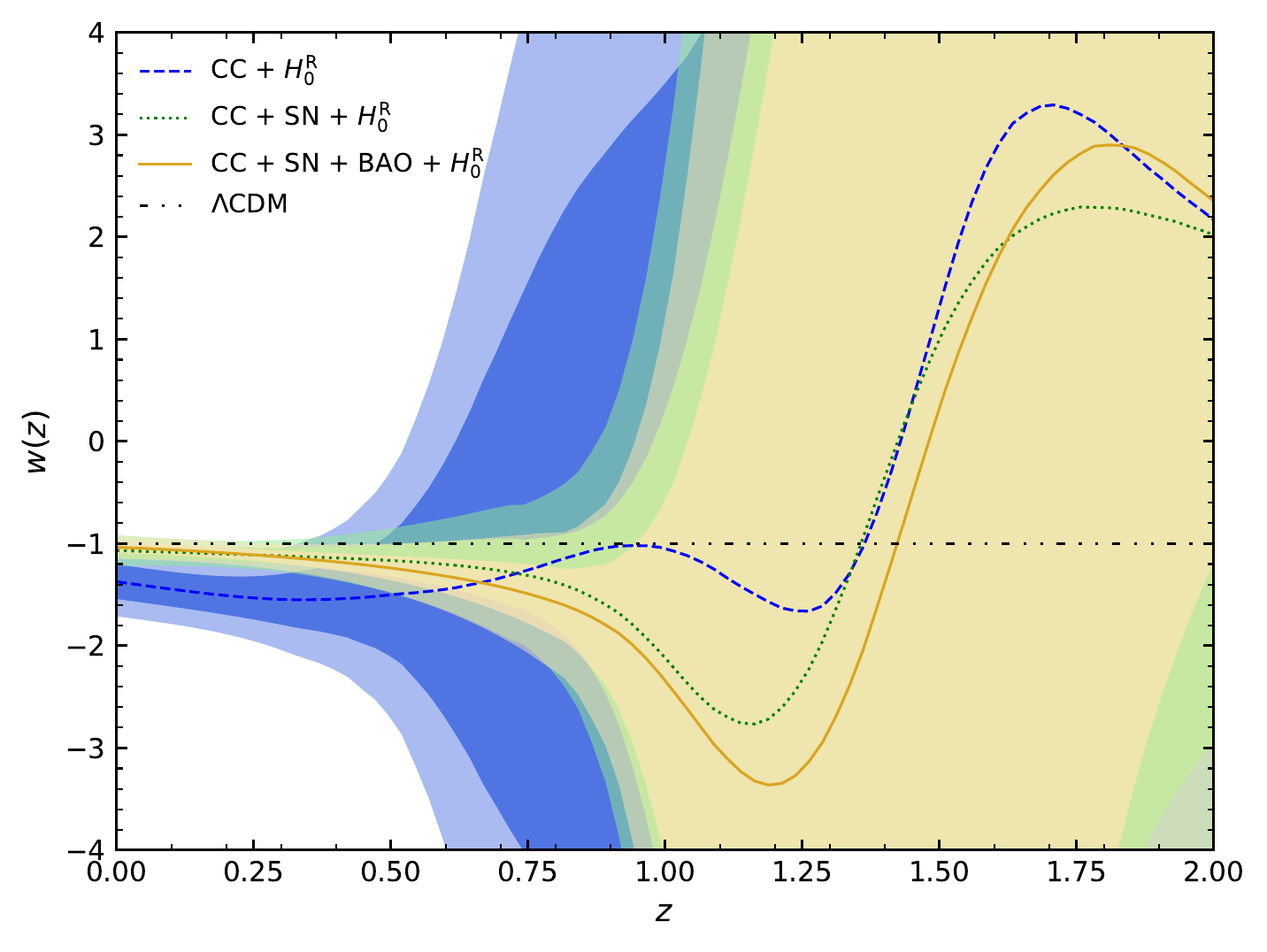}
    \includegraphics[width=0.445\columnwidth]{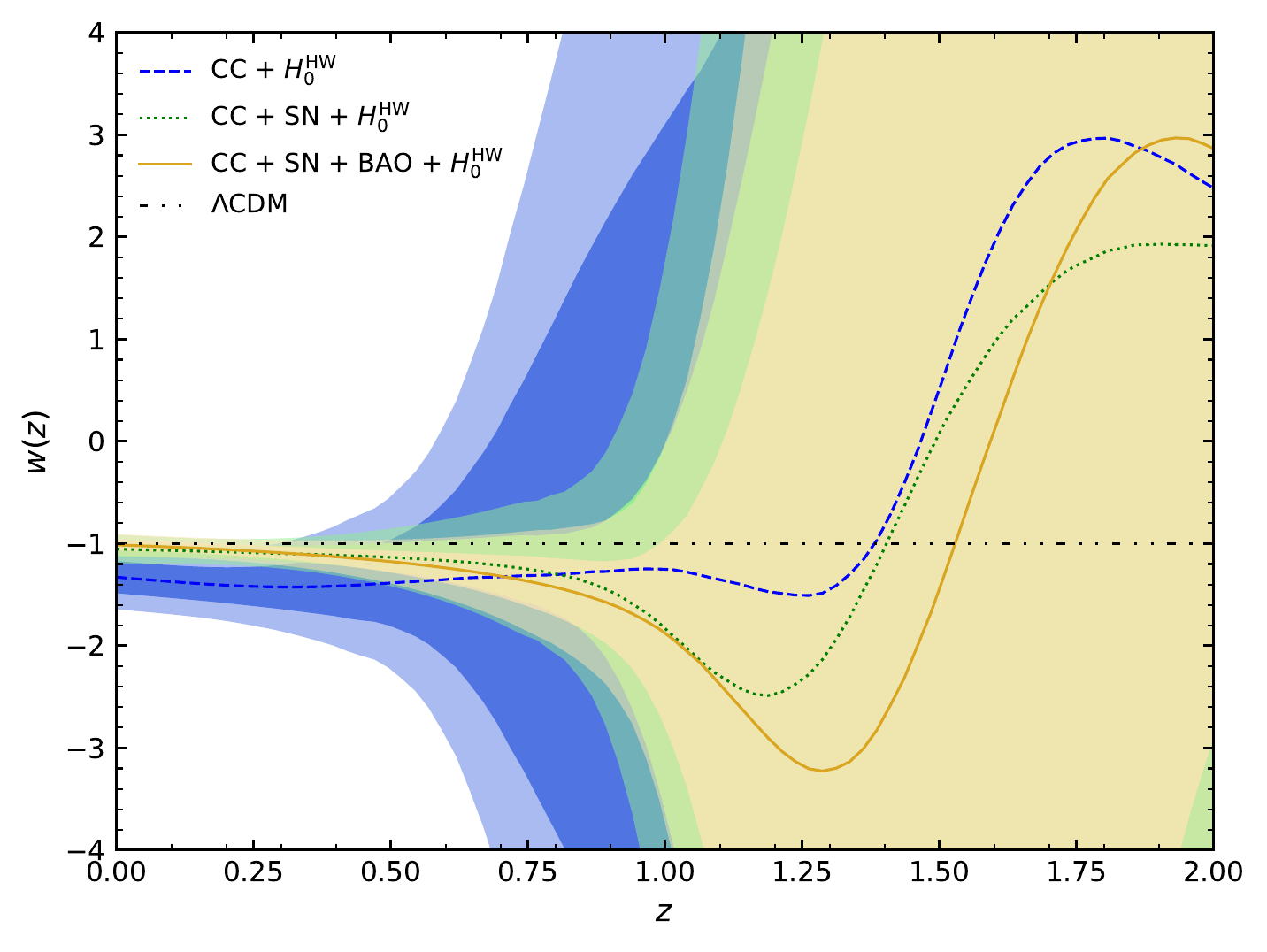}
    \includegraphics[width=0.445\columnwidth]{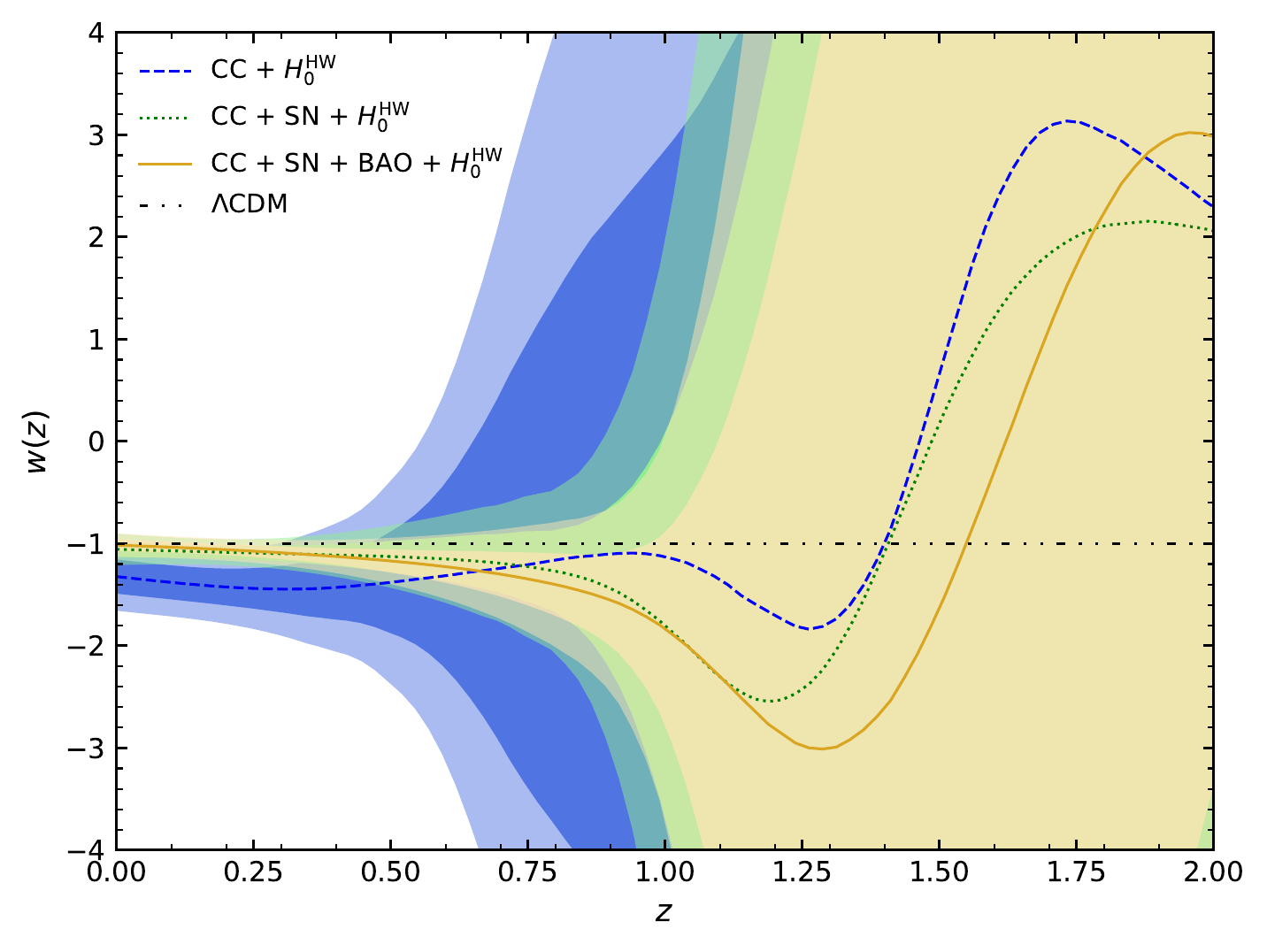}
    \includegraphics[width=0.445\columnwidth]{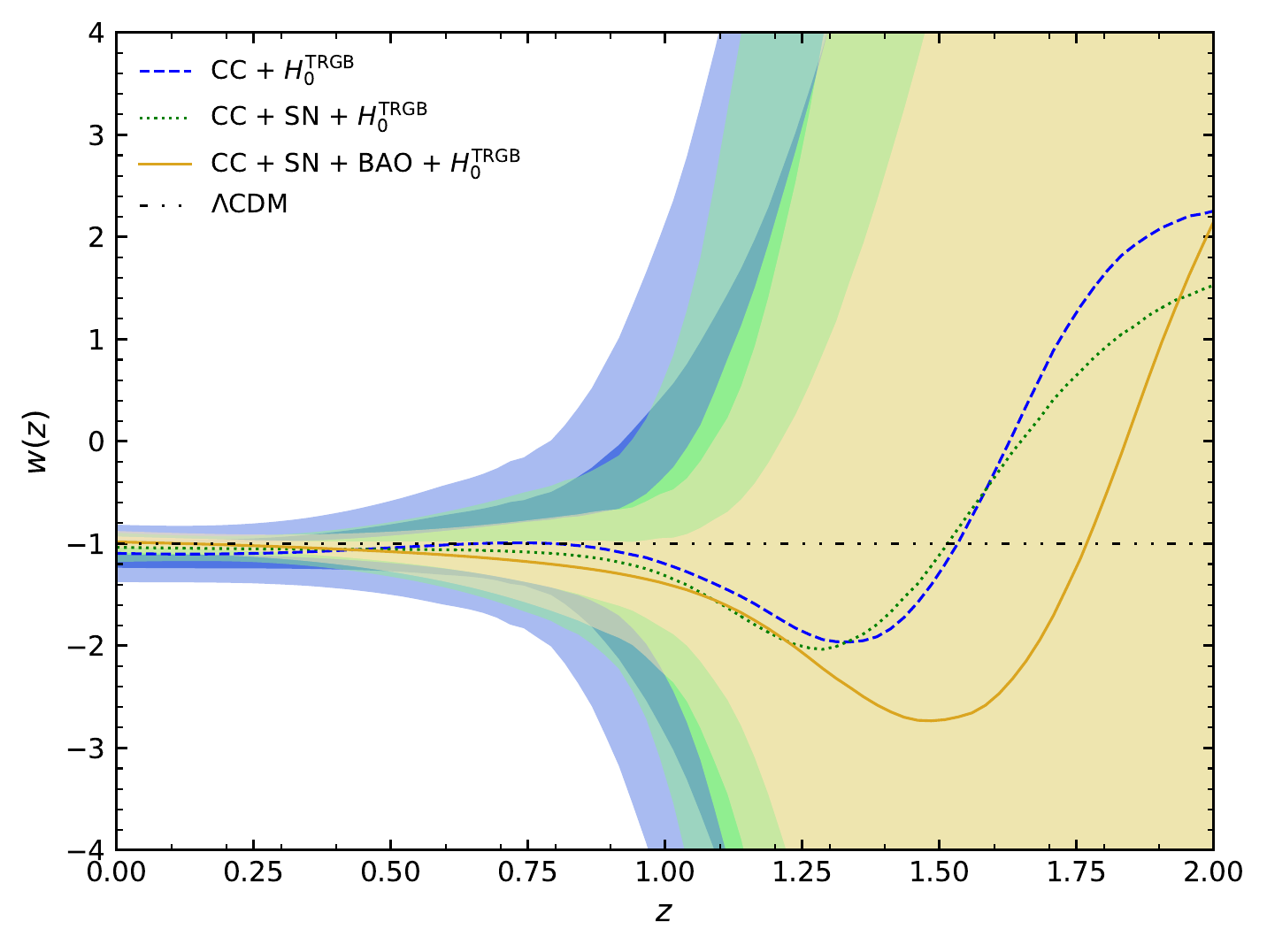}
    \includegraphics[width=0.445\columnwidth]{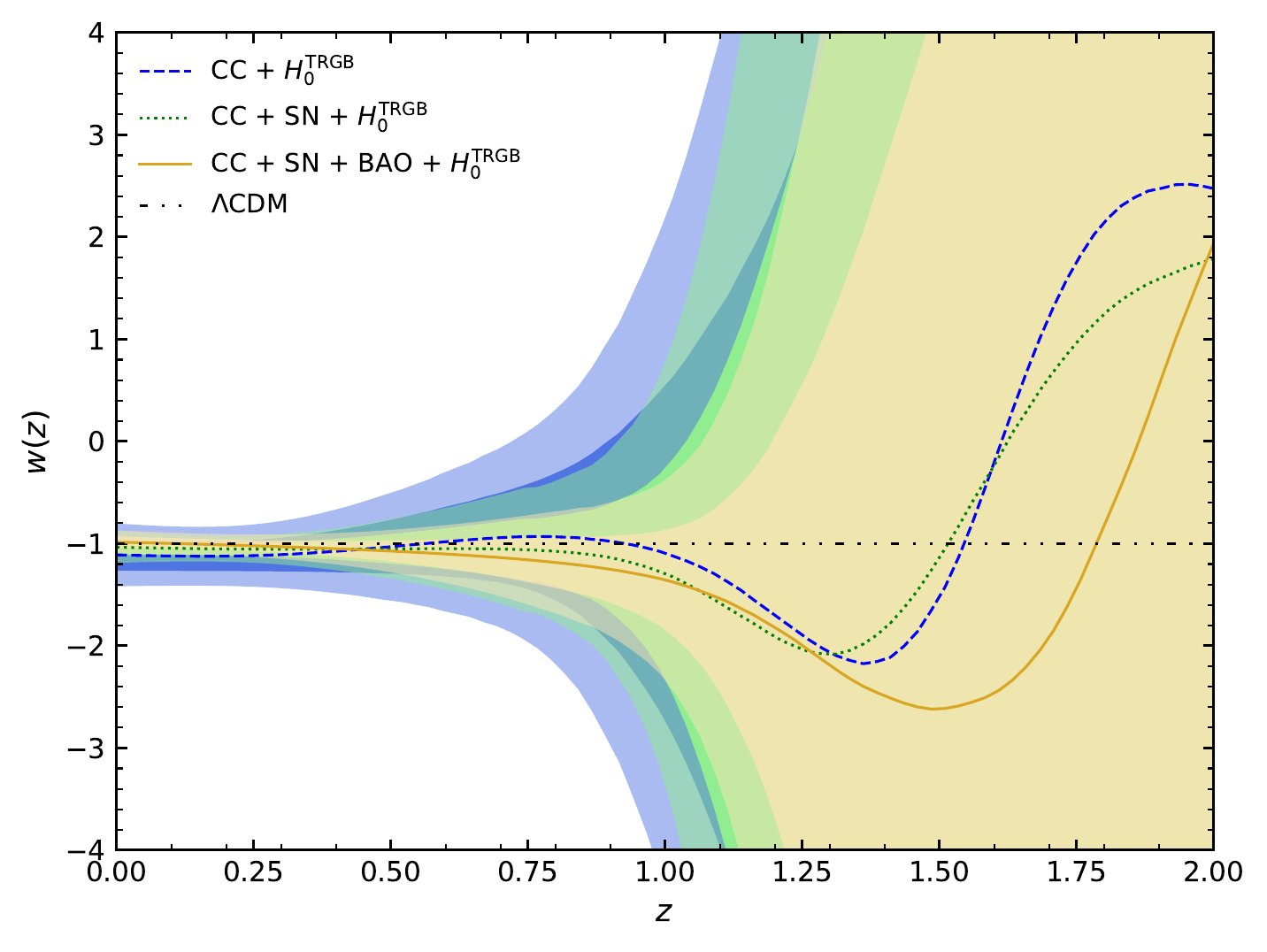}
    \caption{\label{fig:wz_squaredexp_cauchy}
    GP reconstructions of $w(z)$ with the squared exponential (left) and Cauchy (right) kernel functions, along with the $\Lambda$CDM prediction.
    }
\end{center}
\end{figure}

\begin{figure}[t!]
\begin{center}
    \includegraphics[width=0.445\columnwidth]{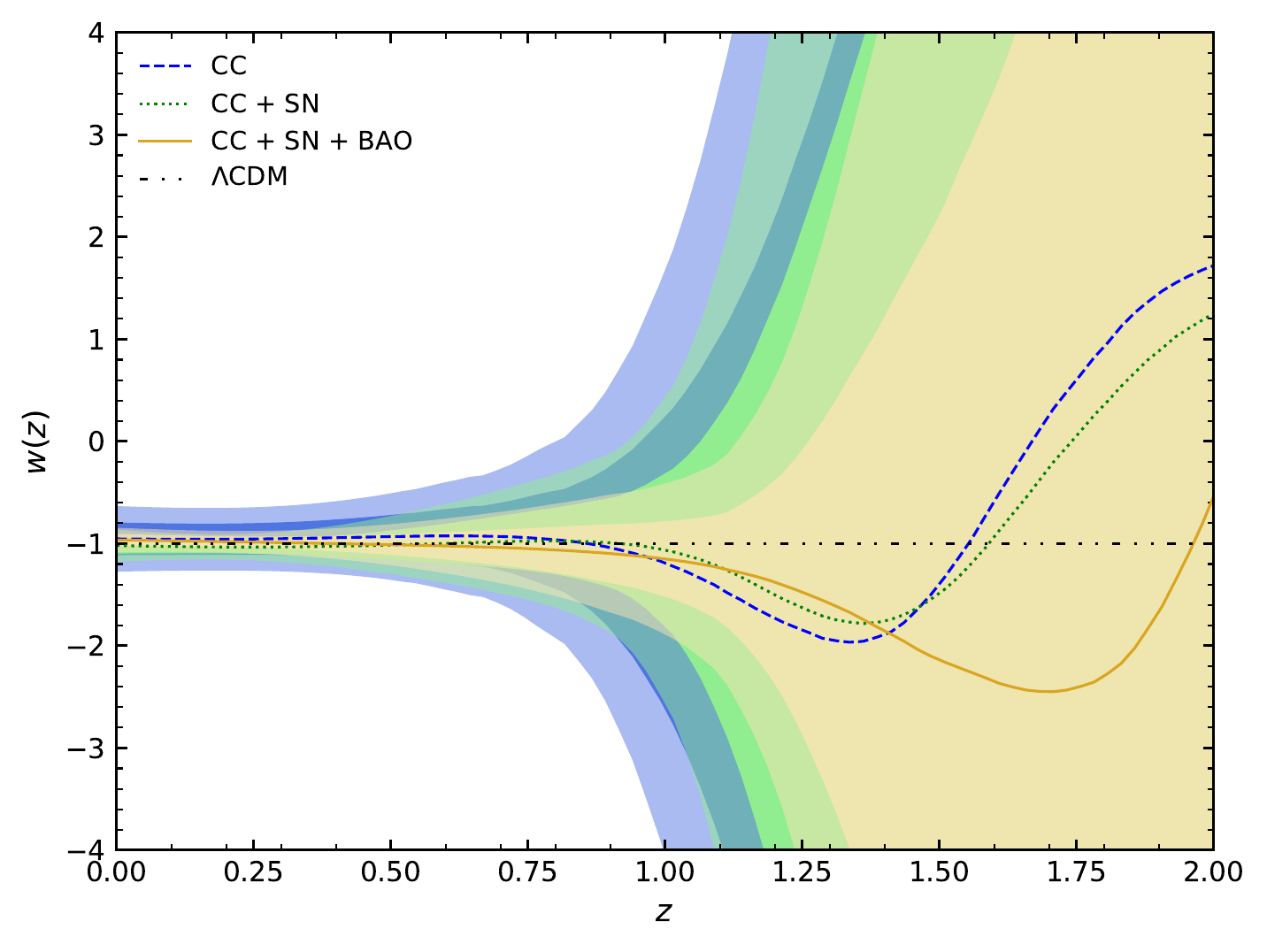}
    \includegraphics[width=0.445\columnwidth]{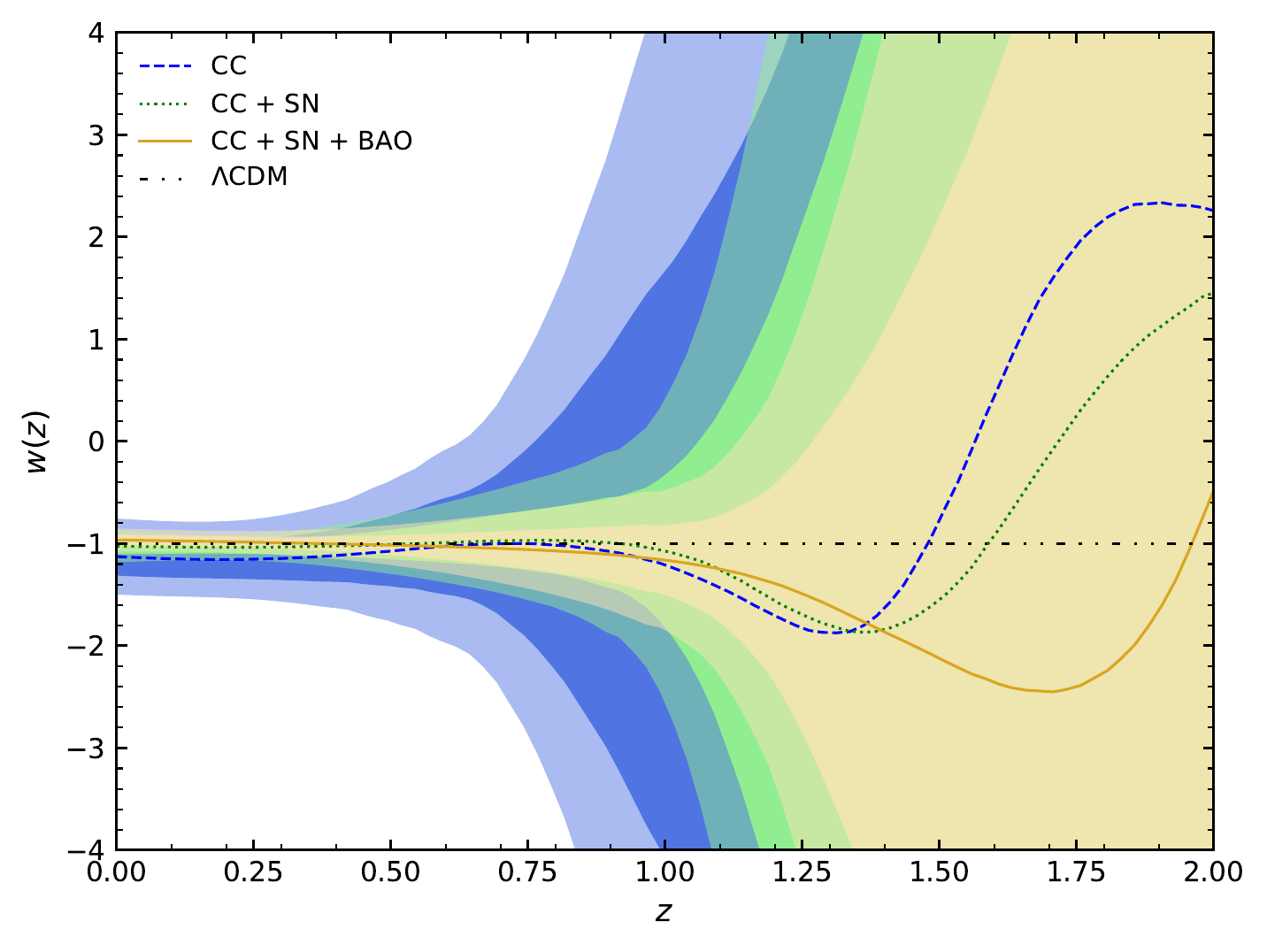}
    \includegraphics[width=0.445\columnwidth]{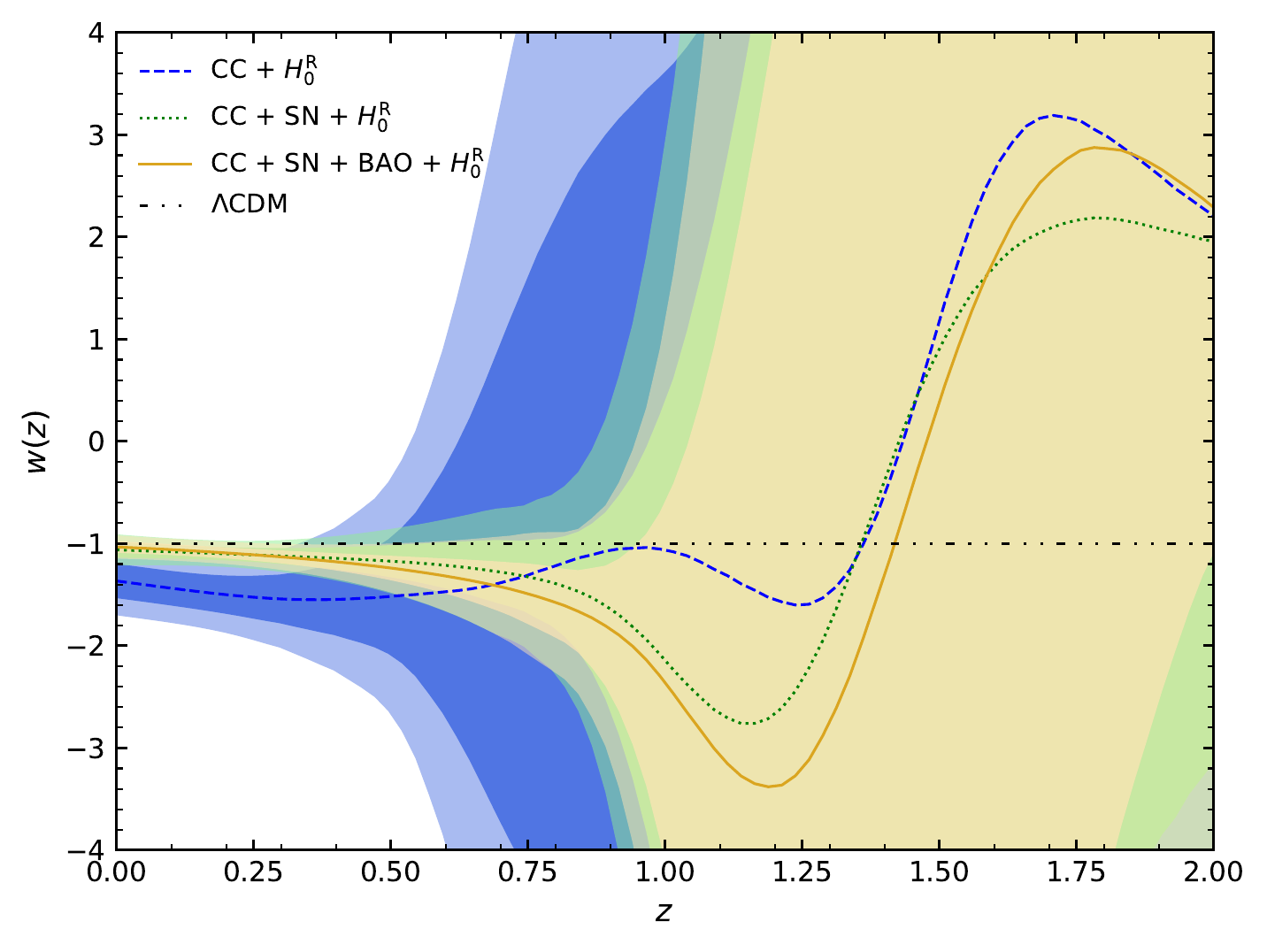}
    \includegraphics[width=0.445\columnwidth]{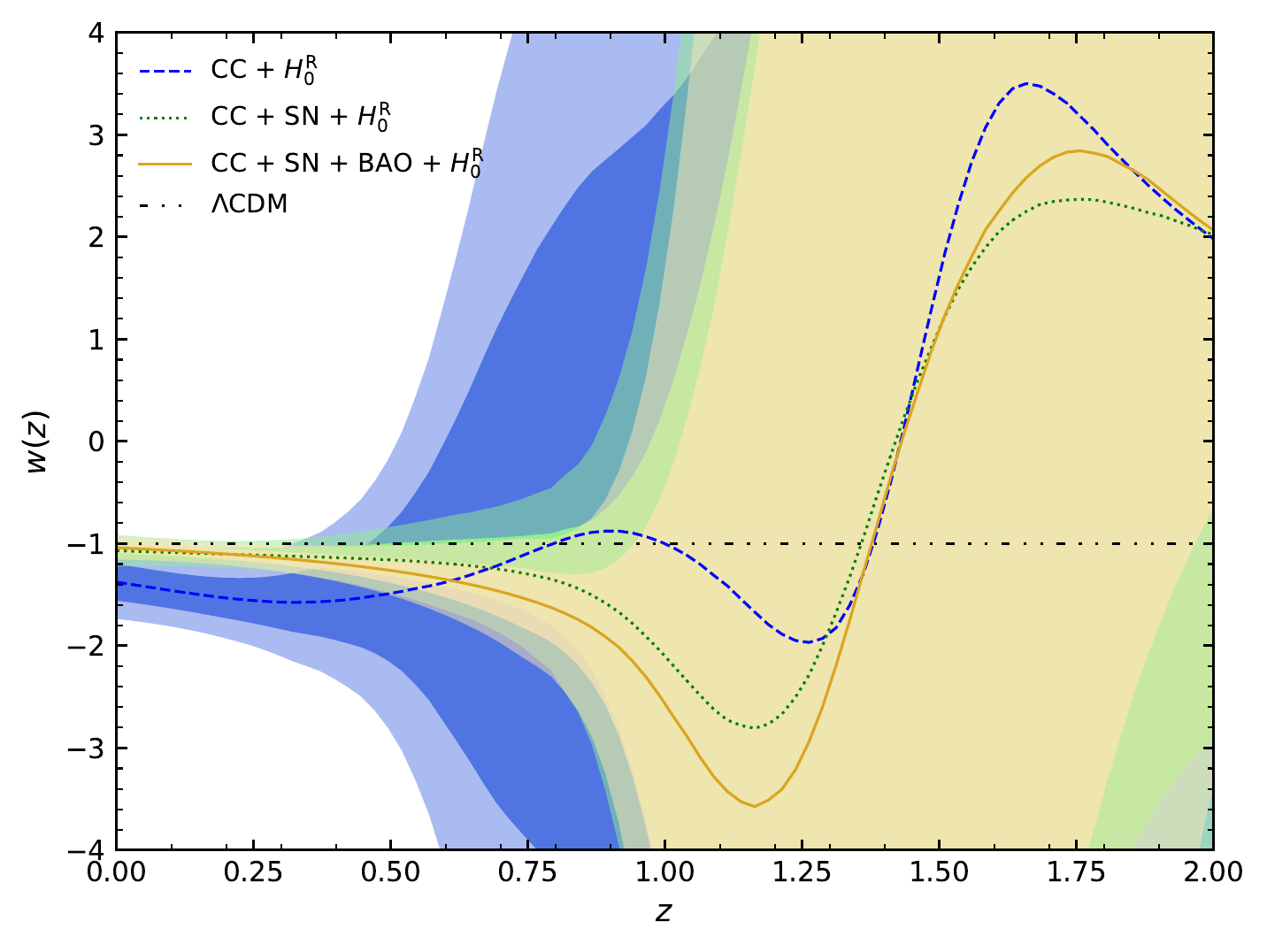}
    \includegraphics[width=0.445\columnwidth]{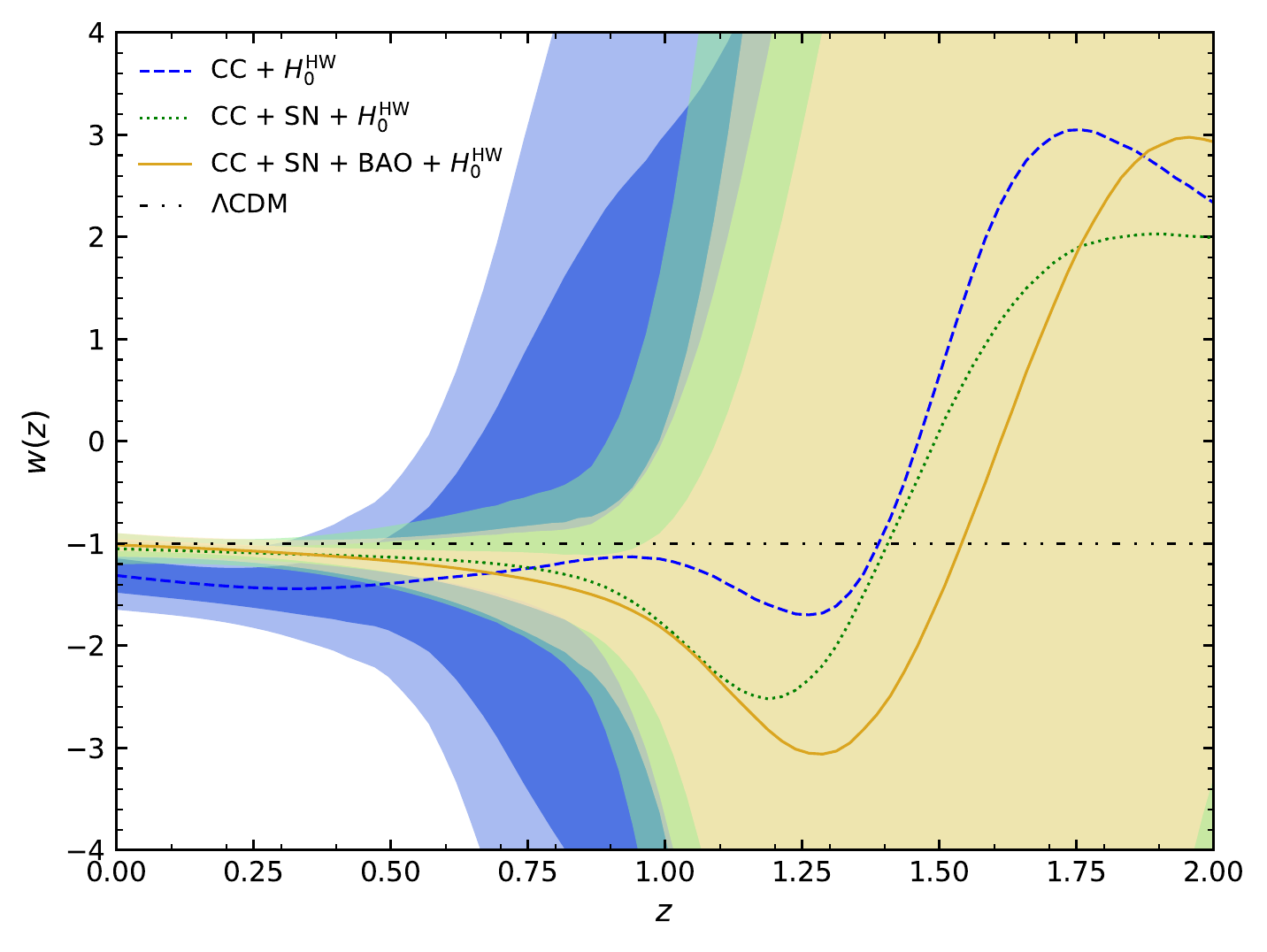}
    \includegraphics[width=0.445\columnwidth]{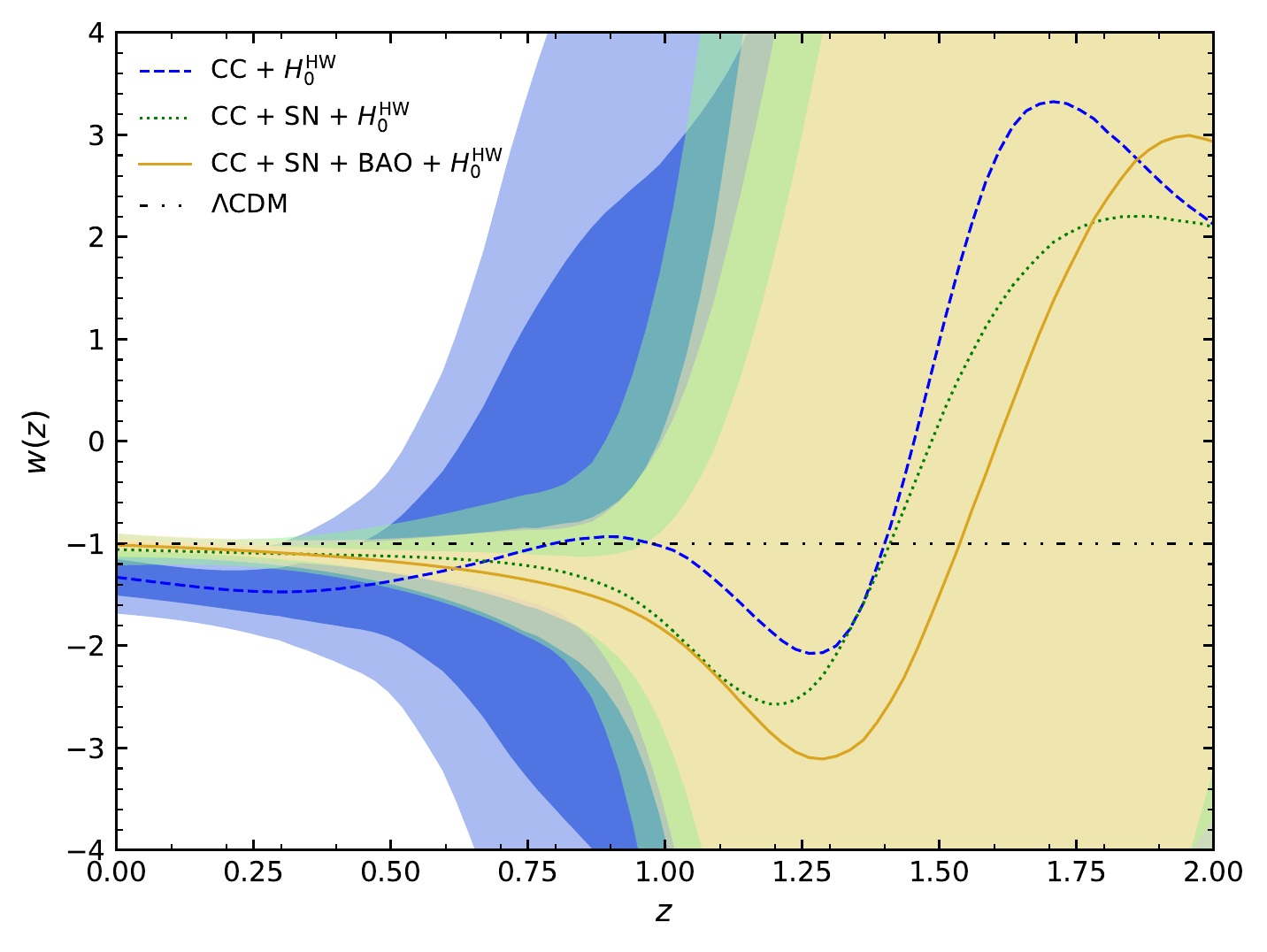}
    \includegraphics[width=0.445\columnwidth]{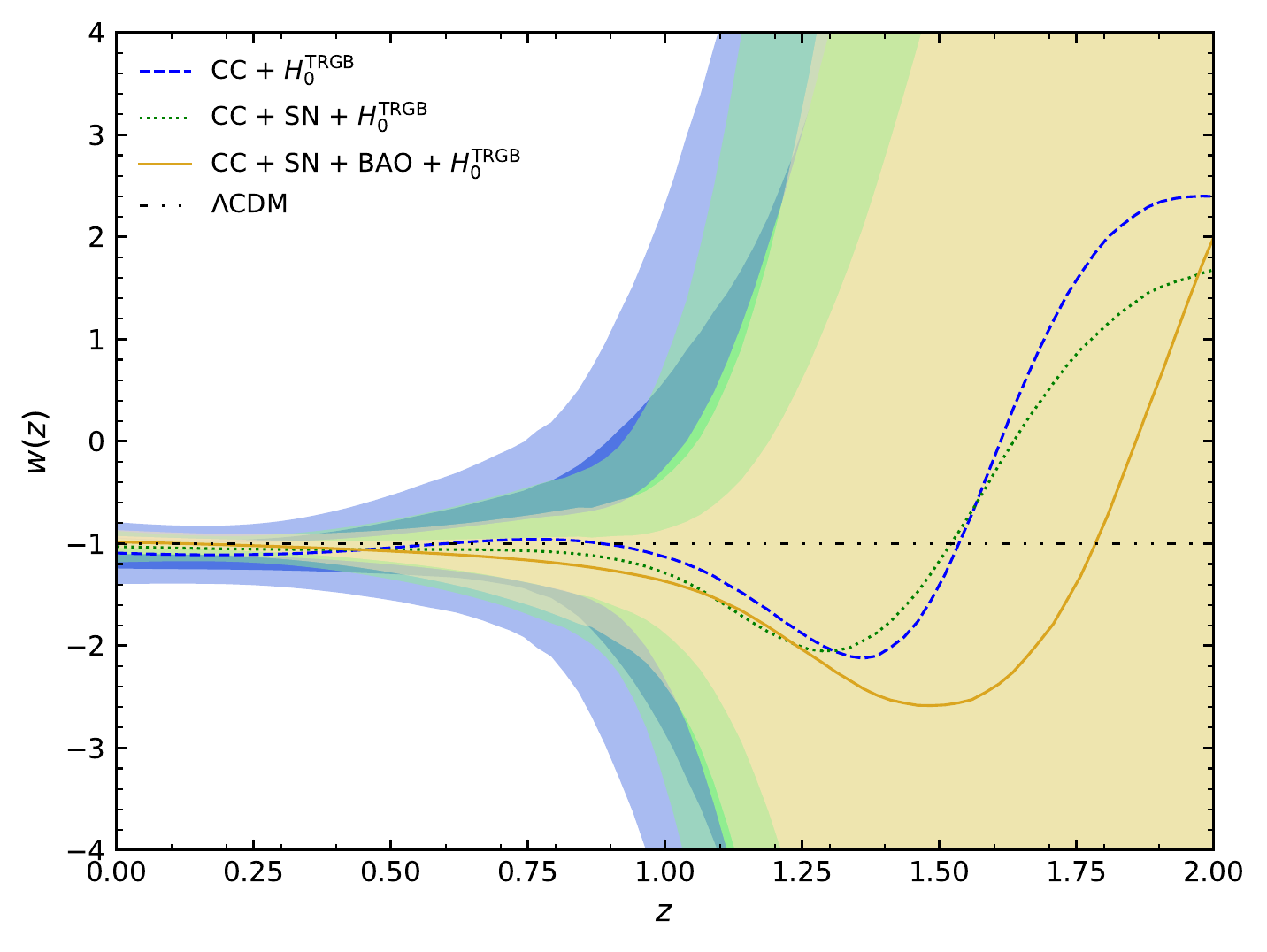}
    \includegraphics[width=0.445\columnwidth]{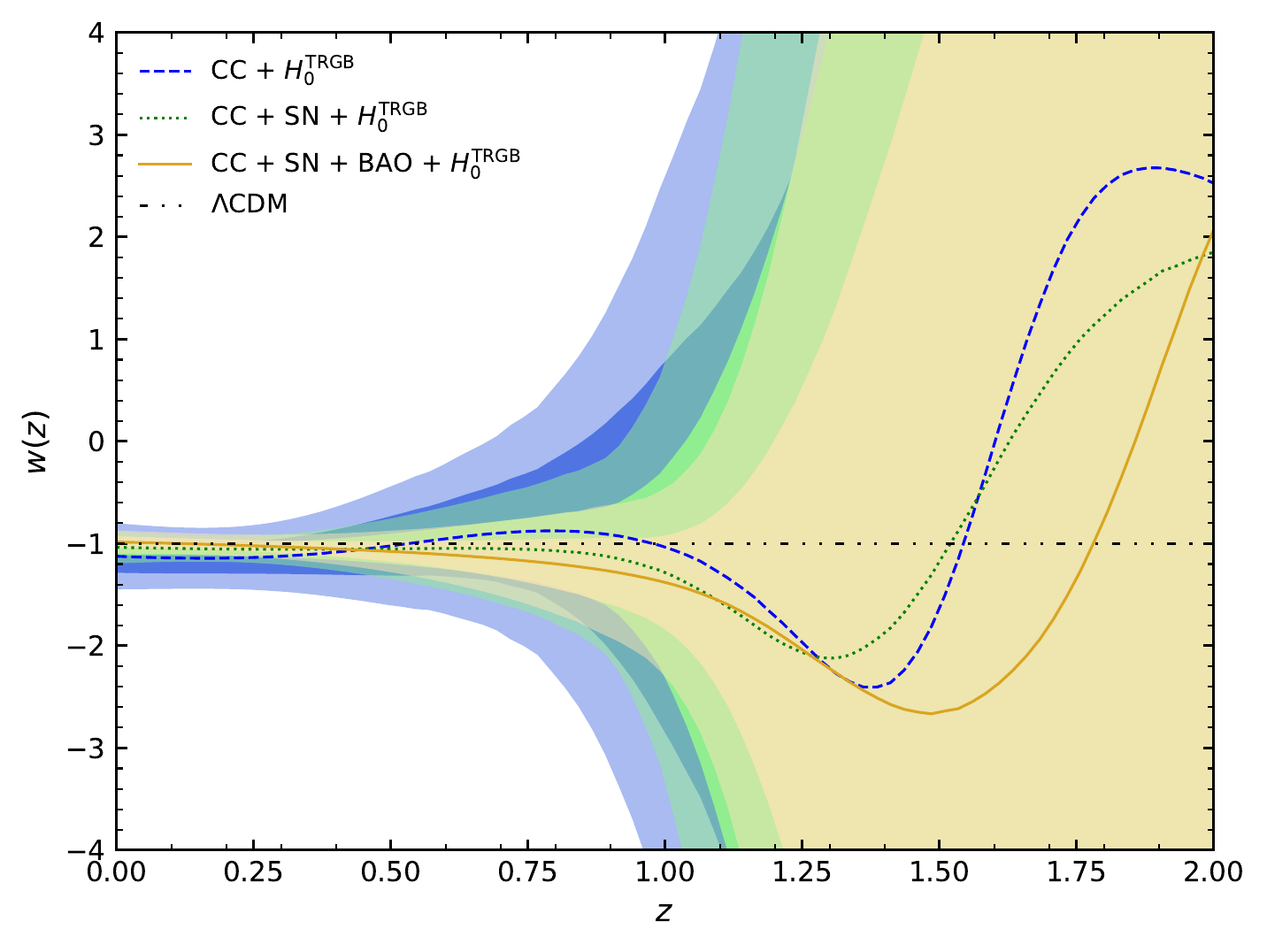}
    \caption{\label{fig:wz_mat_rat_quad}
    GP reconstructions of $w(z)$ with the Mat\'{e}rn (left) and rational quadratic (right) kernel functions, along with the $\Lambda$CDM prediction.
    }
\end{center}
\end{figure}

In this appendix we present the GP reconstructions of the deceleration parameter and the dark energy EoS for the kernel and prior combinations explored in section \ref{sec:GPH0}. In Figs. \ref{fig:qz_squaredexp_cauchy}--\ref{fig:qz_mat_rat_quad}, the $q(z)$ parameter is shown with the 1$\sigma$ and 2$\sigma$ confidence levels. As in the case of the Hubble parameter, the uncertainties increase drastically for the CC data set but reduce for the CC+SN data set and reduce even further for the full CC+SN+BAO data set combination across the redshift axis. Similarly, the largest difference to the no prior scenario is found for the case of the $H_0^{\rm R}$ prior due to it being the largest literature value of $H_0$.

The diagnostic used in section \ref{sec:GPH0} to produce Figs. \ref{fig:L1_squaredexp}--\ref{fig:L1_ratquad} are ultimately based on the consistency test routed in Eq.(\ref{consis_test}). In Figs. \ref{fig:Om1_squaredexp_cauchy}--\ref{fig:Om1_mat_rat_quad} the core consistency tests for $\mathcal{O}_m^{(1)}(z)$ are plotted. Naturally, we find the same general behaviour as in the $\mathcal{L}^{(1)}(z)$ case with strong agreement for low redshifts and divergences occurring in many cases at higher redshifts where the $\Lambda$CDM value falls outside of the 2$\sigma$ uncertainty in many cases. The exception to this is for the case of the $H_0^{\rm R}$ and $H_0^{\rm HW}$ priors for the CC data set which also diverges for low redshifts.

Finally, in Figs. \ref{fig:wz_squaredexp_cauchy}--\ref{fig:wz_mat_rat_quad} we present the dark energy GP EoS $\omega(z)$ reconstructions. In this instance, we observe consistent agreement with observations for low redshifts. However, as in other works on $\omega(z)$ reconstructions \cite{Seikel:2013fda,2012JCAP...06..036S}, the uncertainties are divergent for higher redshifts resulting in very low confidence levels.

\section*{References}
\bibliographystyle{JHEP}

\providecommand{\href}[2]{#2}\begingroup\raggedright\endgroup


\end{document}